\documentclass[12pt]{article}
\usepackage{bm}       
\usepackage{graphicx}
\usepackage{hyperref}      
\usepackage{verbatim}
\usepackage{changepage}
\usepackage{appendix}
\usepackage[margin=2.5cm]{geometry}
\usepackage[width=.85\textwidth]{caption}
\usepackage{morefloats}
\usepackage{amssymb,amsmath}

\begin{document}

\def\bibname{References}
\bibliographystyle{CMS}

\raggedbottom

\pagenumbering{roman}

\parindent=0pt
\parskip=8pt
\setlength{\evensidemargin}{0pt}
\setlength{\oddsidemargin}{0pt}
\setlength{\marginparsep}{0.0in}
\setlength{\marginparwidth}{0.0in}
\marginparpush=0pt


\pagenumbering{arabic}

\renewcommand{\arraystretch}{1.25}
\addtolength{\arraycolsep}{-3pt}

\newcommand{\ifb}{fb$^{-1}$}
\newcommand{\tauptaum}{\ensuremath{\tau^+\tau^-}}
\newcommand{\tautau}{\ensuremath{\tau\tau}}

\begin{center}\begin{boldmath}

\Huge
\textbf{Heavy Higgs Scalars at Future Hadron Colliders (A Snowmass Whitepaper)}
\normalsize



\begin{center}


{Eric Brownson$^{\bf 1}$}
{Nathaniel Craig$^{\bf 2}$}
{Ulrich Heintz$^{\bf 3}$}
{Gena Kukartsev$^{\bf 3}$}
{Meenakshi Narain$^{\bf 3}$}
{Neeti Parashar$^{\bf 4}$}
{John Stupak III$^{\bf 4}$}

{
$^{\bf 1}$  University of Puerto Rico, Mayaguez, PR\\
$^{\bf 2}$  Rutgers University, Piscataway, NJ\\
$^{\bf 3}$  Brown University, Providence, RI\\
$^{\bf 4}$  Purdue University Calumet, Hammond, IN\\
}

\end{center}


\end{boldmath}\end{center}

 
\abstract{
We investigate the prospects for discovery or exclusion of additional Higgs scalars at the 14 TeV and 33 TeV LHC in the context of theories with two Higgs doublets. We focus on the modes with the largest production rates at hadron colliders, namely gluon fusion production of a heavy CP-even scalar $H$ or a heavy CP-odd pseudoscalar $A$. We consider the sensitivity of the decay channels $H \to ZZ \to 4\ell$, and $A \to Zh$ with $Z\to\ell^+\ell^-$ and $h \to b \bar b$ or $h\to\tauptaum$.
}


\section{Introduction}
\label{sec:Intro}

A boson with properties consistent with those predicted for the Standard Model~(SM) Higgs boson has been discovered at the LHC~\cite{CMSHiggsDisco,ATLASHiggsDisco}.  The SM contains the simplest mechanism of spontaneous Electroweak Symmetry Breaking~(EWSB) which can give mass to the SM gauge bosons and fermions.  However, it is entirely possible that some non-minimal mechanism is responsible for the generation of mass.  A large class of non-minimal extensions of the SM Higgs mechanism are described at low energies by a Two Higgs Doublet Model~(2HDM)~\cite{2HDM,2HDMReview}, in which there is an additional Higgs doublet relative to the SM Higgs sector.  Such a model provides an effective theory description for many natural EWSB extensions, such as the Higgs sector of the Minimal Supersymmetric Standard Model~(MSSM)~\cite{MSSM,SUSY}, Twin Higgs models~\cite{twinHiggs0,twinHiggs}, and certain classes of Composite Higgs models~\cite{compositeHiggs0,compositeHiggs,compositeHiggs2}.

In the 2HDM, after EWSB there remain 5 physical Higgs bosons: two CP-even scalars, $h$ and $H$; one CP-odd pseudoscalar, $A$; and a charged pair, $H^{\pm}$.  The general parameter space of the 2HDM is vast, but there are several well-motivated asumptions which allow simplification.  If, as suggested by the experimenatally observed level of CP violation in nature, the 2HDM is forced to conserve CP at tree level, then after EWSB there are nine free parameters in the scalar potential.  A useful basis takes as these nine parameters the four physical Higgs masses, $m_h, m_H, m_A$, and $m_{H^{\pm}}$; two angles, $\alpha$ and $\beta$; and three couplings, $\lambda_5, \lambda_6$, and $\lambda_7$.  To simplify the parameter space, the (tree-level) MSSM values of the three scalar couplings are used, $\lambda_5=\lambda_6=\lambda_7=0$. In addition, the absence of new contributions to tree-level flavor violation may be guaranteed by enforcing discrete symmetries that constrain a given species of right-handed quarks or leptons to couple to only one Higgs doublet in the ultraviolet; this restriction leads to four types of 2HDM that classify the possible couplings.   With this choice, the couplings of each physical Higgs to SM states are determined entirely by the angles $\alpha$ and $\beta$, and the type of 2HDM.  

\begin{figure}[htbpp]
\begin{center}
\includegraphics[width=0.4\columnwidth,height=0.4\textheight,keepaspectratio=true]{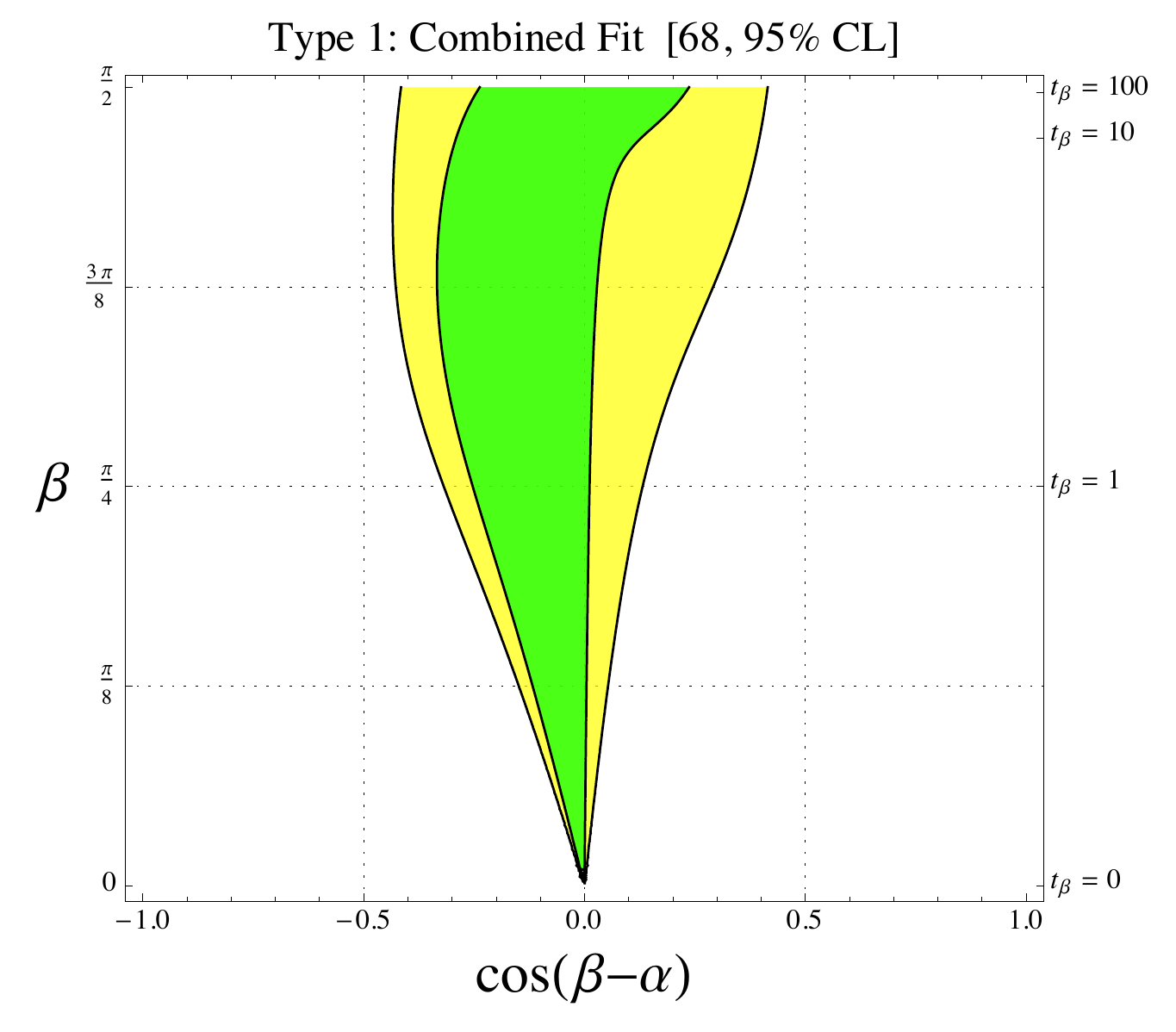}
\includegraphics[width=0.4\columnwidth,height=0.4\textheight,keepaspectratio=true]{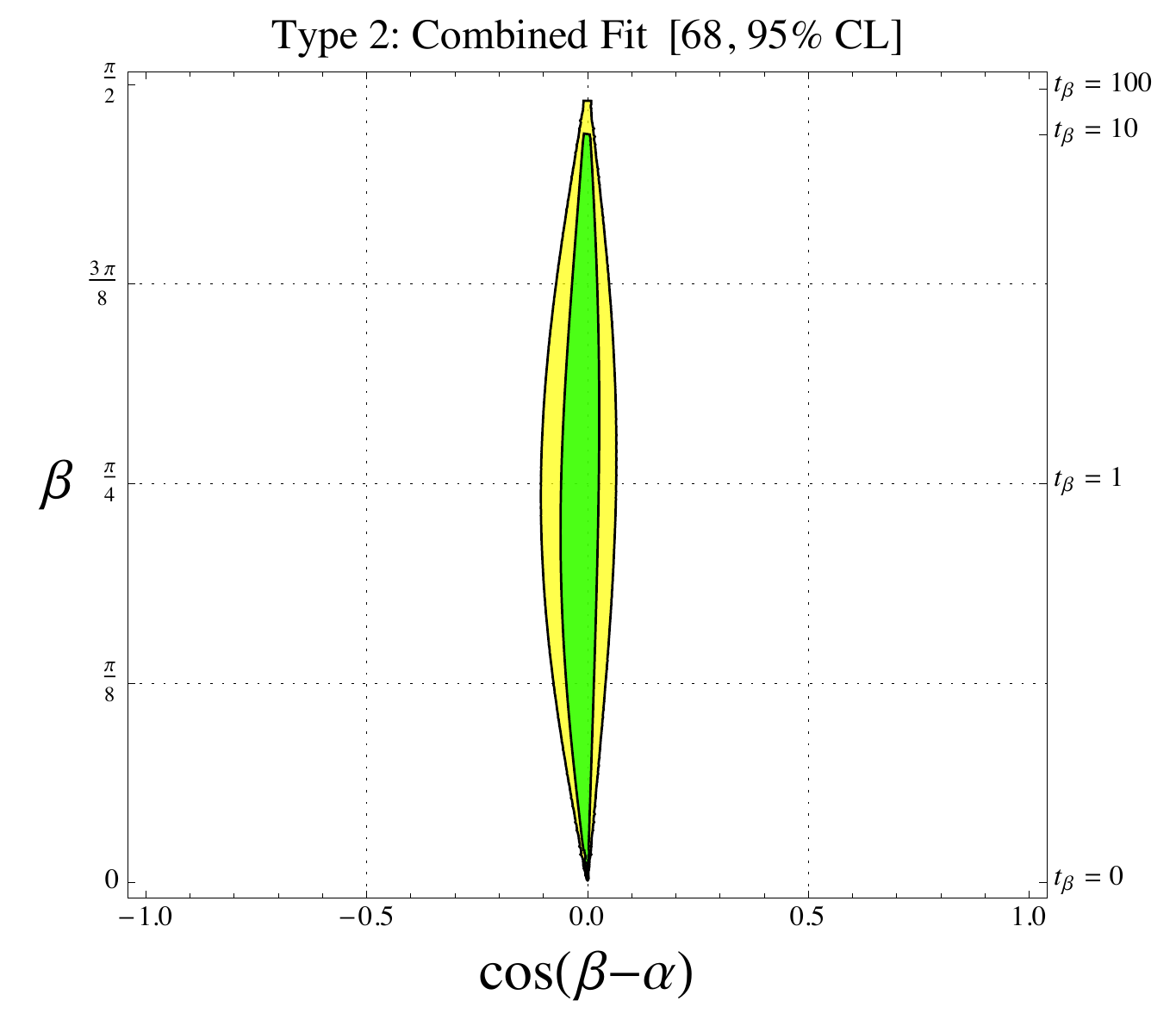}
\caption{68\% and 95\% CL allowed regions of parameter space for type I~(left) and type II~(right) 2HDMs~\cite{constraints}.}
\label{fig:constraints}
\end{center}
\end{figure}

Current measurements~\cite{CMSHiggsCouplings,ATLASHiggsCouplings} of the various couplings of the observed Higgs-like state can be used to constrain the allowed parameter space of the 2HDM.  Based on such an approach~\cite{constraints,constraints2,chen1,chen2}, the regions which are allowed at 68\% and 95\% Confidence Level~(CL) are shown in Figure~\ref{fig:constraints} for type I and II 2HDMs (the allowed regions for type III and IV 2HDMs are parametrically similar to type I and type II, respectively). Here the allowed regions of parameter space are shown in terms of the angle $\beta$ and the combination $\cos(\beta - \alpha)$. In the so-called alignment limit $\cos(\beta - \alpha) \to 0$, the couplings of the light CP-even scalar $h$ become identical to the SM Higgs. Especially in the type II 2HDM, the value of $| \mathrm{cos}\left(\beta - \alpha\right)|$ -- and hence deviations from SM-like Higgs couplings -- is required to be small by the current coupling measurements.


\section{Background and Signal Simulation}
\label{sec:Simulation}

Background Monte Carlo~(MC) samples are generated centrally for many common backgrounds~\cite{snowmassBackgrounds,snowmassOSG},  for use in a variety of Snowmass studies.  Events are generated with MadGraph~\cite{Madgraph}.  Pythia~\cite{Pythia} is used for fragmentation and hadronization.  Finally, Delphes~\cite{Delphes} is used to perform fast detector simulation for a generic ``LHC-like'' detector~\cite{snowmassDet}.  The naming convention used for backgrounds collectively denotes bosons ($W^{\pm}$, $Z$, $\gamma$, and in some samples the SM Higgs boson) by ``B".  Charged and neutral leptons are collectively denoted by ``L".  Light quarks and anti-quarks (including bottom) are denoted by ``j".  The top quark and anti-quark are denoted by ``t".

\begin{figure}[htbp]
\begin{center}
\includegraphics[width=0.4\columnwidth,height=0.4\textheight,keepaspectratio=true]{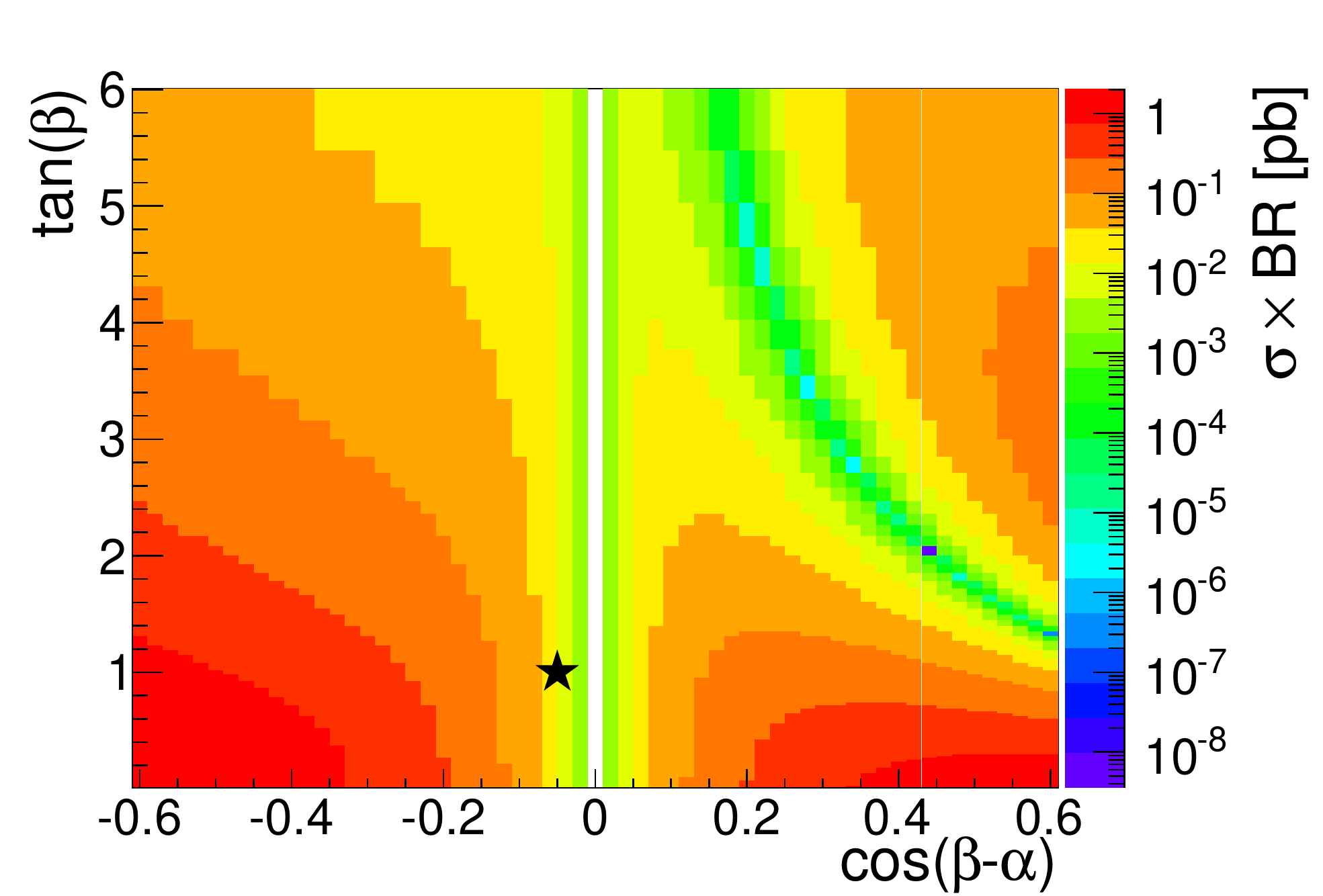}
\includegraphics[width=0.4\columnwidth,height=0.4\textheight,keepaspectratio=true]{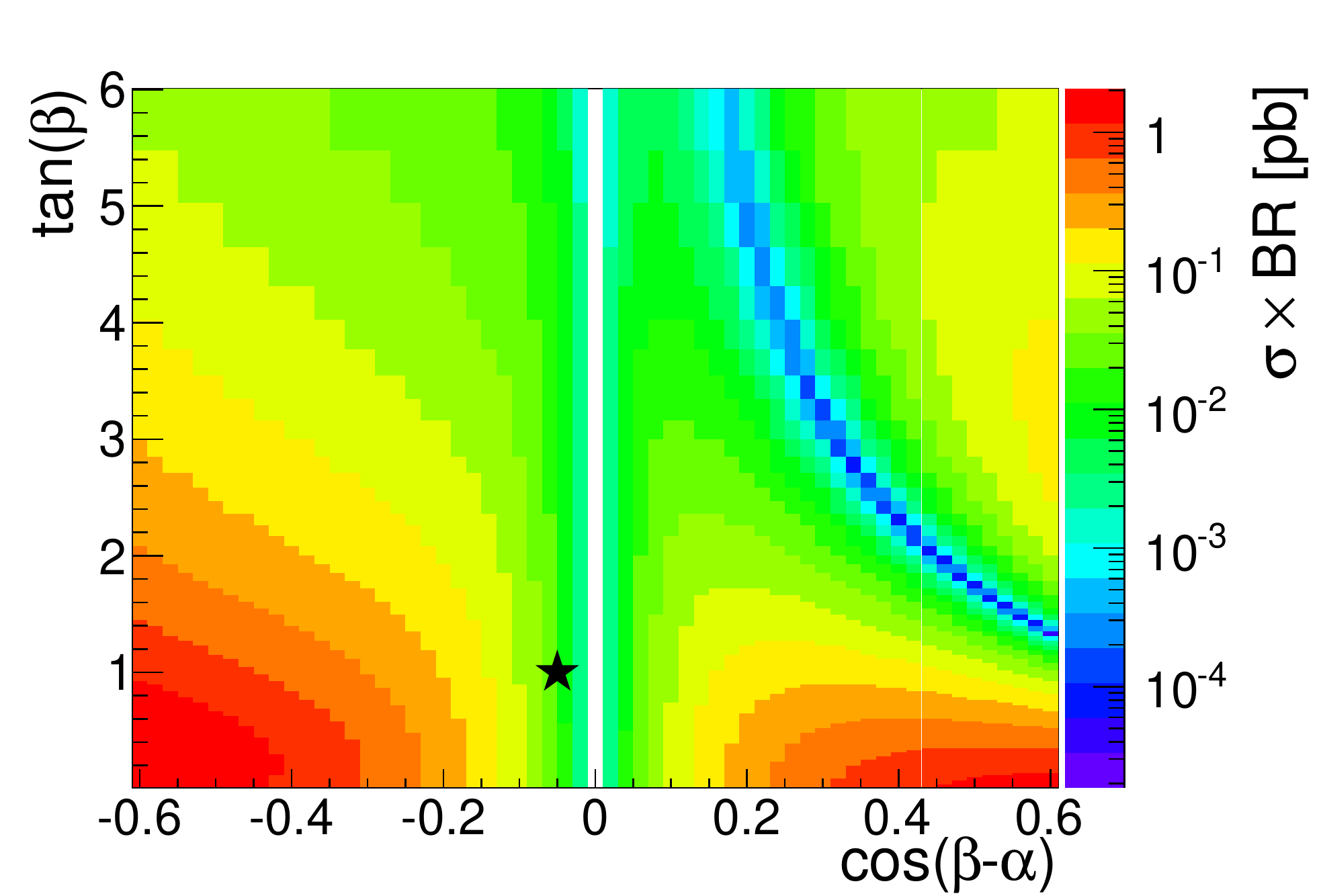}
\includegraphics[width=0.4\columnwidth,height=0.4\textheight,keepaspectratio=true]{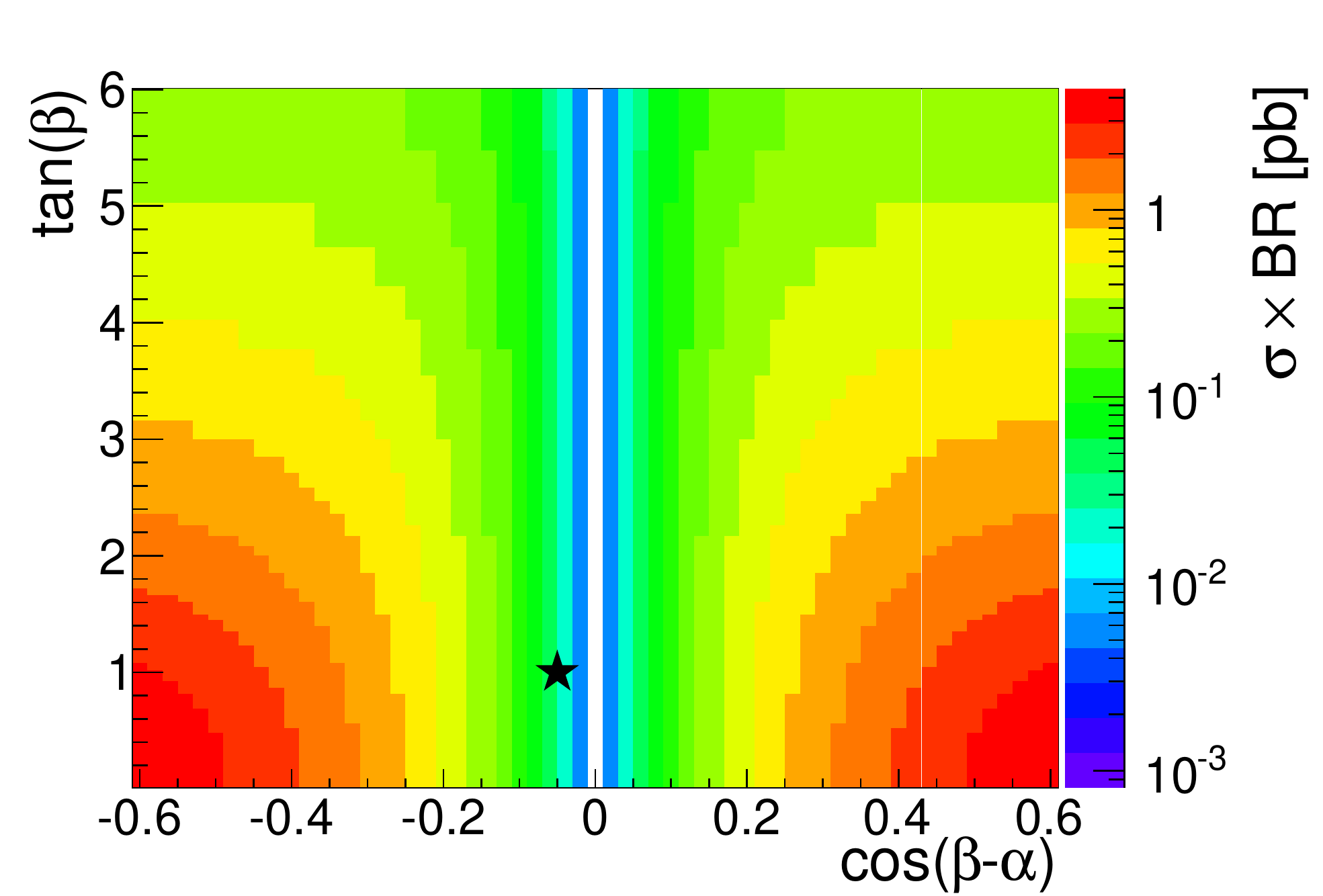}
\includegraphics[width=0.4\columnwidth,height=0.4\textheight,keepaspectratio=true]{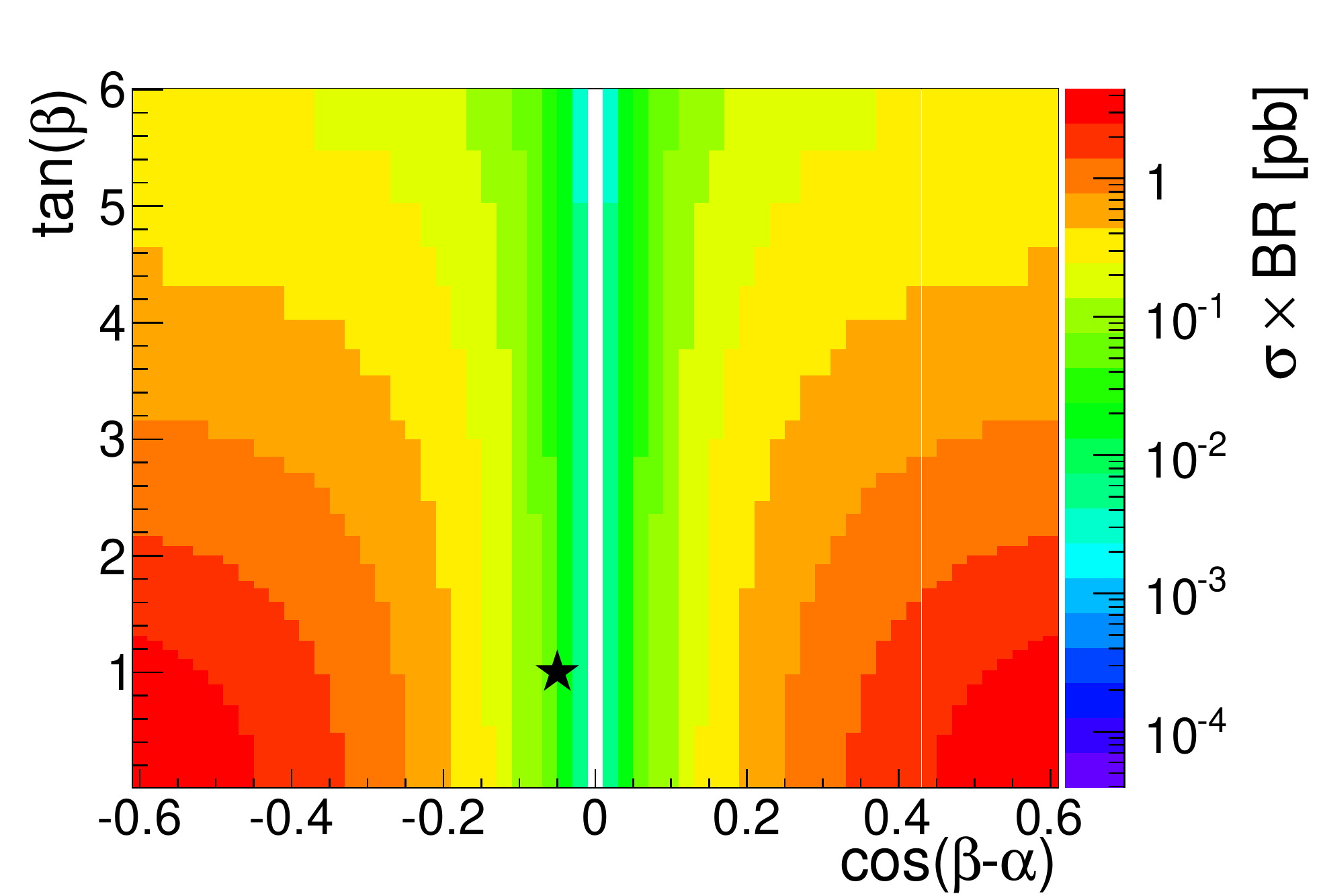}
\caption{Cross section times branching ratio for $pp\rightarrow H\rightarrow ZZ$~(top) and $pp\rightarrow A\rightarrow Zh$~(bottom) at $\sqrt{s}=14$~TeV in the tan$(\beta)$ versus cos($\beta - \alpha$) plane, for type I 2HDM (left) and type II 2HDM (right).  All plots assume $m(H/A)=500$ GeV.}
\label{fig:sigmaBR14tanBeta}
\end{center}
\end{figure}

\begin{figure}[htbp]
\begin{center}
\includegraphics[width=0.4\columnwidth,height=0.4\textheight,keepaspectratio=true]{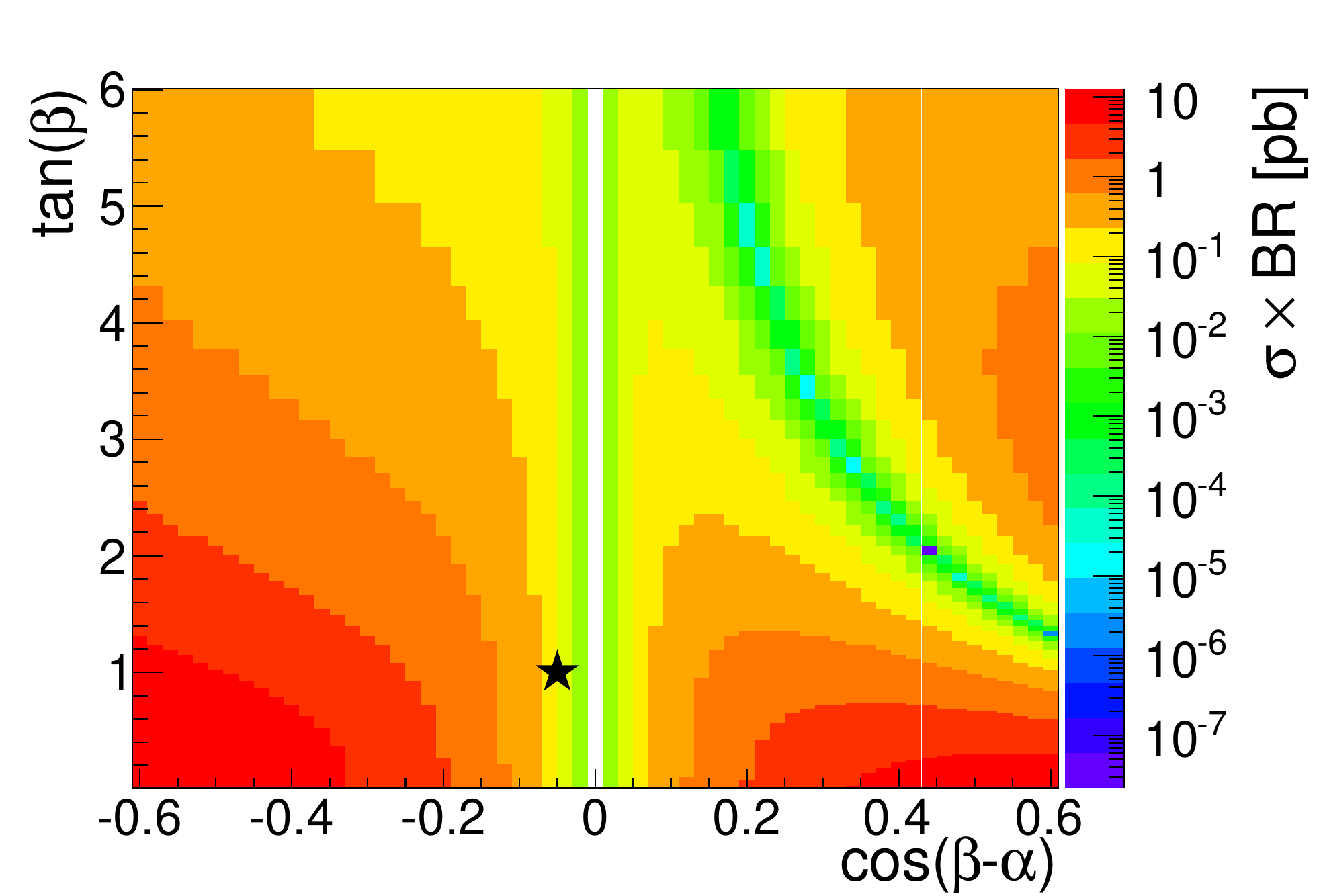}
\includegraphics[width=0.4\columnwidth,height=0.4\textheight,keepaspectratio=true]{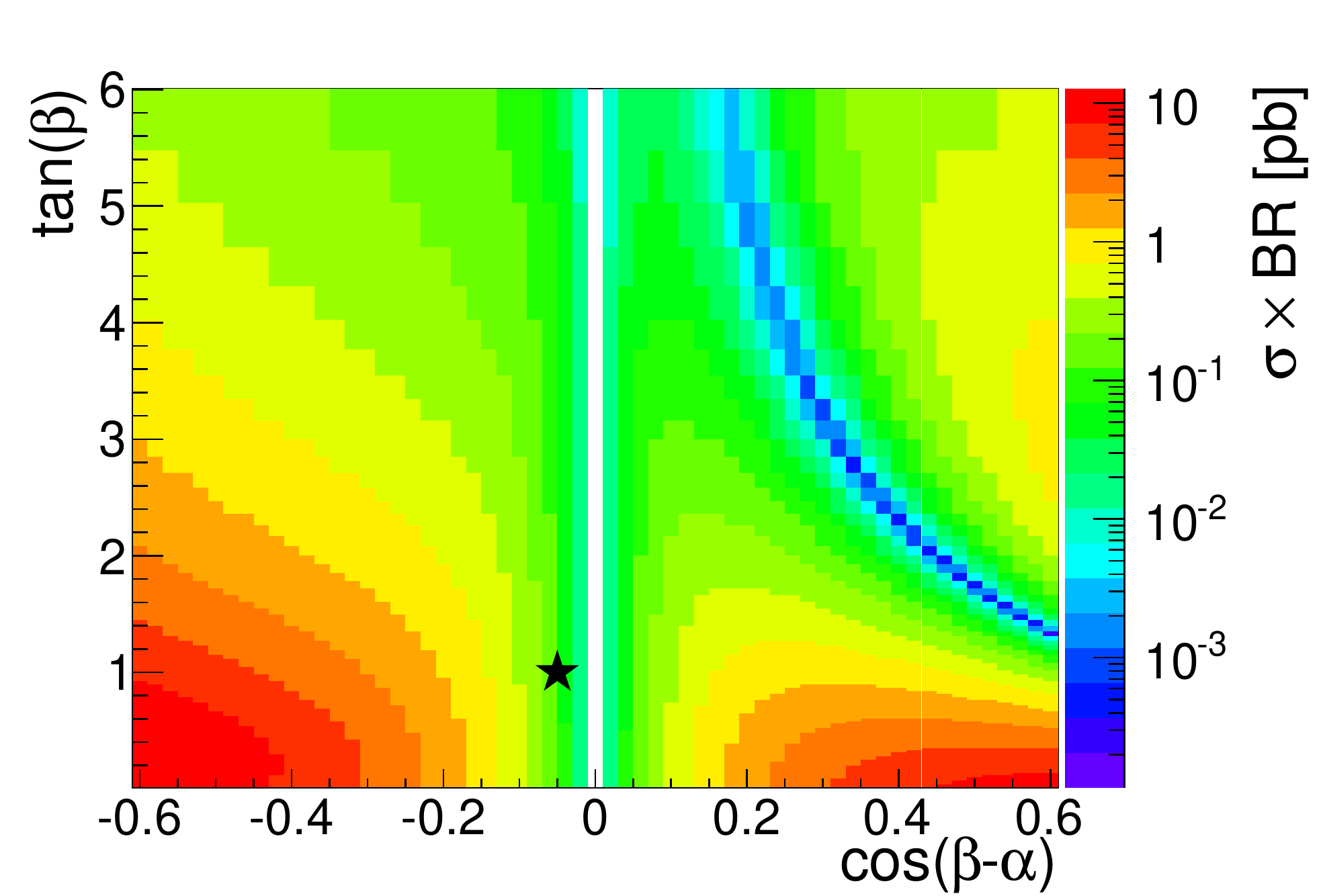}
\includegraphics[width=0.4\columnwidth,height=0.4\textheight,keepaspectratio=true]{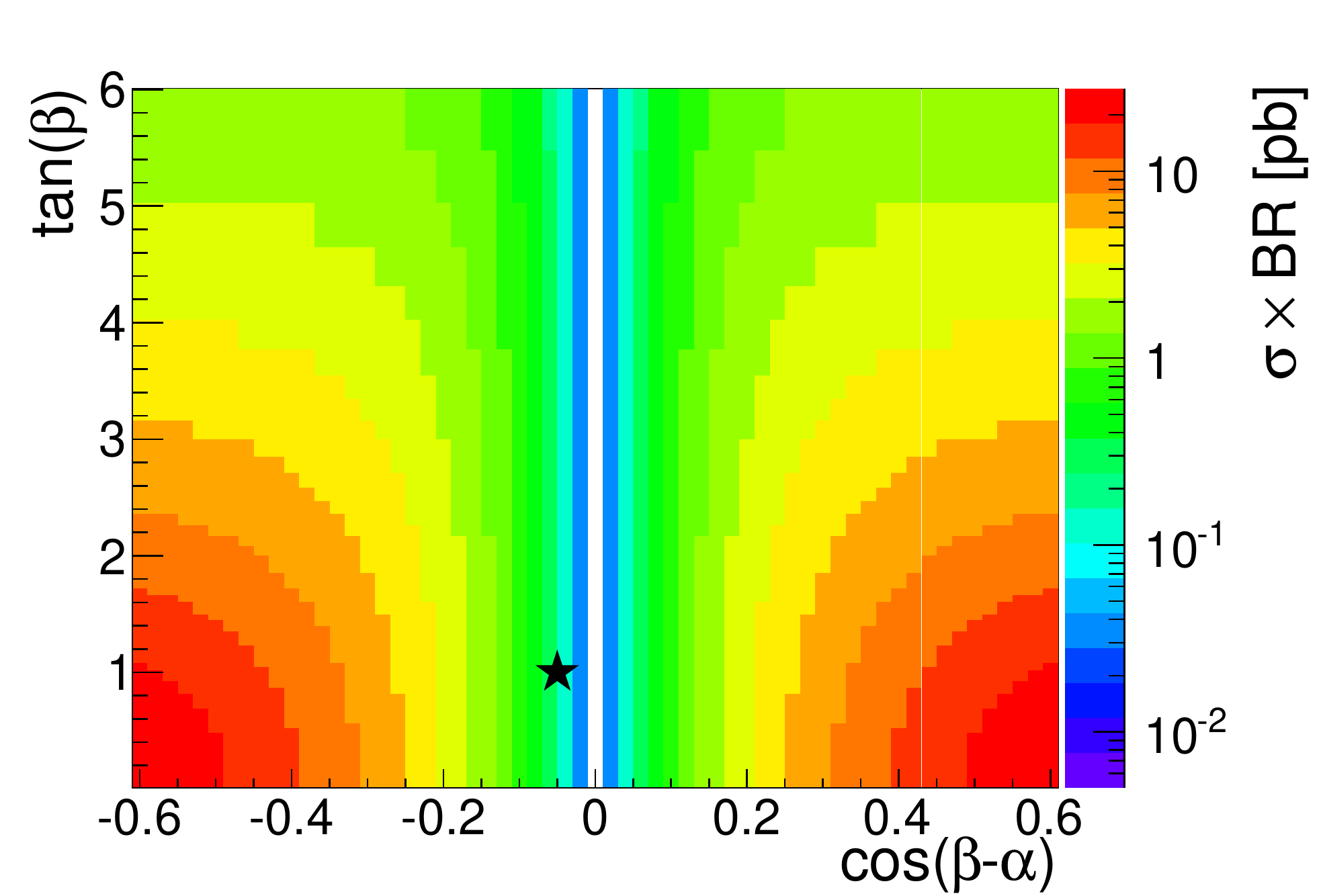}
\includegraphics[width=0.4\columnwidth,height=0.4\textheight,keepaspectratio=true]{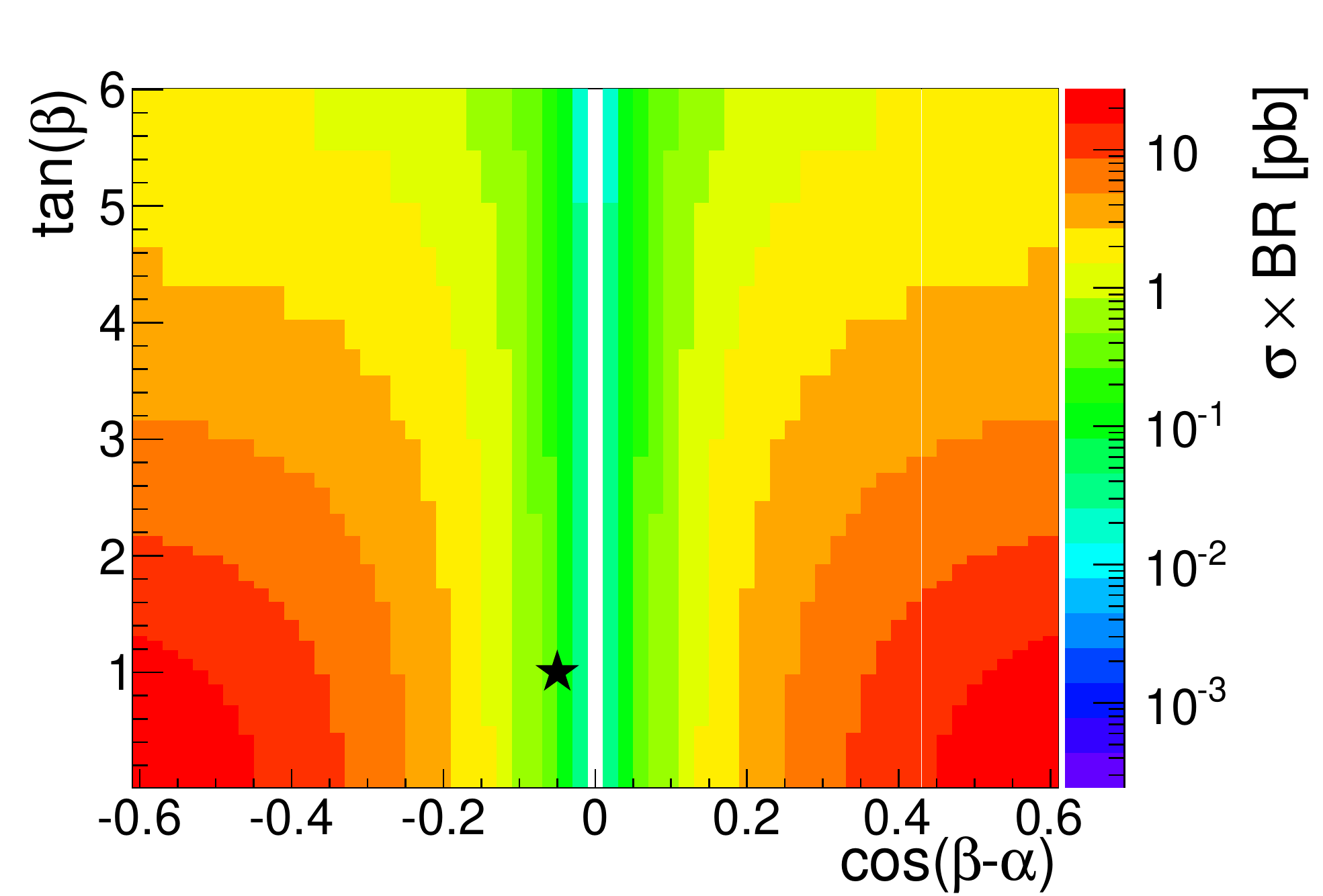}
\caption{Cross section times branching ratio for $pp\rightarrow H\rightarrow ZZ$~(top) and $pp\rightarrow A\rightarrow Zh$~(bottom) at $\sqrt{s}=33$~TeV in the tan$(\beta)$ versus cos($\beta - \alpha$) plane, for type I 2HDM (left) and type II 2HDM (right).  All plots assume $m(H/A)=500$ GeV.}
\label{fig:sigmaBR33tanBeta}
\end{center}
\end{figure}

Signal MC is generated using the same processing chain as described for background MC. To efficiently cover the 2HDM parameter space, for each signal mass hypothesis the signal MC is generated for a single benchmark point in $\alpha$ and $\beta$, while the signal cross section and branching ratios are calculated analytically as a function of $\alpha$ and $\beta$ for signal mass hypotheses from 200~GeV to 1~TeV in both type I and II 2HDMs. The gluon fusion production cross sections at $\sqrt{s}= $14 and 33 TeV for $H$ and $A$ are computed using \texttt{ggh@nnlo} \cite{gghnnlo} with Standard Model Yukawa couplings and analytically re-weighted using the leading-order dependence on $\alpha$ and $\beta$ in each 2HDM type. The branching ratios are computed analogously using the procedure outlined in~\cite{constraints} with $\lambda_5 = \lambda_6 = \lambda_7 = 0$. Note that as a result the scalar widths are held fixed in the signal MC for a given signal mass hypothesis, although the physical widths vary across the parameter space of $\alpha$ and $\beta$. This approximation is reasonable near the alignment limit where the neutral scalars $A$ and $H$ are narrow and finite width effects play a small role in experimental resolution.  The cross section times branching ratio in the tan$(\beta)$ versus cos($\beta-\alpha$) plane for a 500~GeV $H$ ($A$) decaying to $ZZ$ ($Zh$) is shown in Figures~\ref{fig:sigmaBR14tanBeta} and \ref{fig:sigmaBR33tanBeta}.  

Unless otherwise noted, all tables and plots in this paper assume a signal cross section for a type II 2HDM with cos$(\beta - \alpha) = -0.06$ and tan$(\beta) = 1$.  This point in parameter space is allowed by current limits from Higgs coupling measurements.  The signal cross section times branching ratio for this benchmark point are given in Tables~\ref{tab:HZZSigma14TeV}, \ref{tab:AZhSigma14TeV}, \ref{tab:HZZSigma33TeV}, and \ref{tab:AZhSigma33TeV}.  All intermediate results assume branching ratios for $h\rightarrow b \bar b$ and $\tauptaum$ of a 125 GeV SM Higgs, but the final results allow these branching ratios to vary across parameter space.


\section{Analysis}

Three scenarios are considered, corresponding to the expectations for LHC Phase I, LHC Phase II (HL-LHC), and HE-LHC running.  For LHC Phase I, we assume a total integrated luminosity of $\int Ldt=300$~\ifb~at $\sqrt{s}=14$~TeV, with an average of 50 pileup interactions per bunch crossing.  For the LHC Phase II, we assume a total integrated luminosity of $\int Ldt=3000$~\ifb~at $\sqrt{s}=14$~TeV, with an average of 140 pileup interactions per bunch crossing.  For the HE-LHC, we assume a total integrated luminosity of $\int Ldt=3000$~\ifb~at $\sqrt{s}=33$~TeV, with an average of 140 pileup interactions per bunch crossing.

For each of these three scenarios, the sensitivity of direct searches for the heavy, neutral Higgses is determined.  The sensitivity for $H$ discovery and exclusion are evaluated in the four lepton ($\ell=e,\mu$) final state.  The sensitivity for $A$ discovery and exclusion are evaluated in $\ell^+\ell^- bb$ and  $\ell^+\ell^-\tautau$ final states.


\boldmath \subsection[H to ZZ]{$H\rightarrow ZZ$} \unboldmath
\label{sec:HZZ}


The four lepton final state provides the best sensitivity in a search for the CP-even scalar $H$, due to the low background rate and fully-reconstructible, high-resolution final state.  We therefore focus on this decay mode.


The following criteria are applied to leptons:
\begin{itemize}
\item $p_T \geq 5$~GeV
\item $\left| \eta\right| \leq$~2.5
\item Relative isolation $\leq$ 0.1 (after FastJet $\rho\times A$ correction)
\end{itemize}

$Z$ boson candidates are constructed from pairs of opposite-sign (OS), same-flavor (SF) lepton pairs, with a dilepton invariant mass between 60~GeV and 100~GeV.  $H$ candidates are constructed from $Z$ pairs with an invariant mass above 150~GeV.

In order to fire a leptonic trigger, it is assumed that events must contain either a leading lepton with $p_T\geq$~30~GeV, or a leading lepton with $p_T\geq$~20~GeV and a sub-leading lepton with $p_T\geq$~10~GeV.  Events are further required to contain exactly four leptons and exactly two $Z$ candidates, one of which must have an invariant mass between 80~GeV and 100~GeV.  The signal and background yields at various stages of selection for the 300~\ifb~analysis are shown in Tables~\ref{tab:HZZSignal_14_300} and \ref{tab:HZZBackground_14_300}.  Analogous tables for the 3000~\ifb~analysis at $\sqrt{s}=14$~TeV ($\sqrt{s}=33$~TeV) are shown in Tables~\ref{tab:HZZSignal_14_3000} and \ref{tab:HZZBackground_14_3000} (\ref{tab:HZZSignal_33_3000} and \ref{tab:HZZBackground_33_3000}).  Various kinematic distributions for selected events are shown in Figures~\ref{fig:HZZplots_14_300} (300~\ifb at $\sqrt{s}=14$~TeV), \ref{fig:HZZplots_14_3000} (3000~\ifb at $\sqrt{s}=14$~TeV), and \ref{fig:HZZplots_33_3000} (3000~\ifb at $\sqrt{s}=33$~TeV).

\begin{table}[htbp]
\begin{center}
\begin{tabular}{|l|c|c|c|c|c|c|}
\hline
Signal Mass [GeV]                                                   & $N_{lepton} = 4$ & Lepton Trigger & $N_{Z} \geq 1$  & $N_{Z} = 2$     & $N_{H} = 1$     \\ \hline
200                 & 1.09e+3        & 1.09e+3        & 1.09e+3        & 1.07e+3        & 1.07e+3        \\
250                 & 1.24e+3        & 1.24e+3        & 1.23e+3        & 1.21e+3        & 1.21e+3        \\
300                 & 497             & 497             & 496             & 486             & 486             \\
350                 & 251             & 251             & 251             & 246             & 246             \\
400                 & 33              & 33              & 33              & 32.3            & 32.3            \\
450                 & 18.7            & 18.7            & 18.7            & 18.3            & 18.3            \\
500                 & 13              & 13              & 13              & 12.8            & 12.8            \\
600                 & 7.28            & 7.28            & 7.27            & 7.13            & 7.13            \\
700                 & 4.5             & 4.5             & 4.5             & 4.41            & 4.41            \\
800                 & 2.97            & 2.97            & 2.97            & 2.91            & 2.91            \\
900                 & 2               & 2               & 2               & 1.96            & 1.96            \\
1000                & 1.34            & 1.34            & 1.34            & 1.32            & 1.32            \\ \hline
\end{tabular}
\end{center}
\caption{Expected number of events for the $H\rightarrow ZZ$ signal in the four lepton final state for $\int Ldt=$ 300~\ifb~at $\sqrt{s}=14$ TeV with $<N_{PU}>=50$.}
\label{tab:HZZSignal_14_300}
\end{table}

\begin{table}[htbp]
\begin{center}
\begin{tabular}{|l|c|c|c|c|c|c|}
\hline
Background                                                   & $N_{lepton} = 4$ & Lepton Trigger & $N_{Z} \geq 1$  & $N_{Z} = 2$     & $N_{H} = 1$     \\ \hline
B, Bj, Bjj-vbf, BB, BBB                            & 5.35e+3        & 5.35e+3        & 5.3e+3         & 4.53e+3        & 4.53e+3        \\
tj, tB, tt, ttB                                    & 475             & 475             & 231             & 29.5            & 29.5            \\
H                                                  & 74.1            & 73.3            & 69.6            & 0               & 0               \\ \hline
Total Background                                   & 5.9e+3         & 5.9e+3         & 5.6e+3         & 4.56e+3        & 4.56e+3        \\ \hline
\end{tabular}
\end{center}
\caption{Expected number of events for the SM backgrounds in the four lepton final state for $\int Ldt=$ 300~\ifb~at $\sqrt{s}=14$ TeV with $<N_{PU}>=50$.}
\label{tab:HZZBackground_14_300}
\end{table}


\begin{table}[htbp]
\begin{center}
\begin{tabular}{|l|c|c|c|c|c|c|}
\hline
Signal Mass [GeV]                                                   & $N_{lepton} = 4$ & Lepton Trigger & $N_{Z} \geq 1$  & $N_{Z} = 2$     & $N_{H} = 1$     \\ \hline
200                & 1.17e+4        & 1.17e+4        & 1.17e+4        & 1.14e+4        & 1.14e+4        \\
250                & 1.33e+4        & 1.33e+4        & 1.33e+4        & 1.3e+4         & 1.3e+4         \\
300                & 5.29e+3        & 5.29e+3        & 5.29e+3        & 5.17e+3        & 5.17e+3        \\
350                & 2.63e+3        & 2.63e+3        & 2.62e+3        & 2.57e+3        & 2.57e+3        \\
400                & 343             & 343             & 342             & 335             & 335             \\
450                & 196             & 196             & 195             & 192             & 192             \\
500                & 135             & 135             & 135             & 132             & 132             \\
600                & 74.2            & 74.2            & 74.1            & 72.7            & 72.7            \\
700                & 46.4            & 46.4            & 46.3            & 45.4            & 45.4            \\
800                & 30.4            & 30.4            & 30.4            & 29.8            & 29.8            \\
900                & 20.4            & 20.4            & 20.4            & 20              & 20              \\
1000               & 13.8            & 13.8            & 13.8            & 13.6            & 13.6            \\ \hline
\end{tabular}
\end{center}
\caption{Expected number of events for the $H\rightarrow ZZ$ signal in the four lepton final state for $\int Ldt=$ 3000~\ifb~at $\sqrt{s}=14$ TeV with $<N_{PU}>=$~140.}
\label{tab:HZZSignal_14_3000}
\end{table}

\begin{table}[htbp]
\begin{center}
\begin{tabular}{|l|c|c|c|c|c|c|}
\hline
Background                                                   & $N_{lepton} = 4$ & Lepton Trigger & $N_{Z} \geq 1$  & $N_{Z} = 2$     & $N_{H} = 1$     \\ \hline
B, Bj, Bjj-vbf, BB, BBB                            & 5.65e+4        & 5.64e+4        & 5.58e+4        & 4.77e+4        & 4.77e+4        \\
tj, tB, tt, ttB                                    & 1.54e+4        & 1.53e+4        & 5.47e+3        & 336             & 336             \\
H                                                  & 874             & 865             & 813             & 0               & 0               \\ \hline
Total Background                                   & 7.27e+4        & 7.26e+4        & 6.21e+4        & 4.8e+4         & 4.8e+4         \\ \hline
\end{tabular}
\end{center}
\caption{Expected number of events for the SM backgrounds in the four lepton final state for $\int Ldt=$ 3000~\ifb~at $\sqrt{s}=14$ TeV with $<N_{PU}>=$~140.}
\label{tab:HZZBackground_14_3000}
\end{table}


\begin{table}[htbp]
\begin{center}
\begin{tabular}{|l|c|c|c|c|c|c|}
\hline
Signal Mass [GeV]                                                   & $N_{lepton} = 4$ & Lepton Trigger & $N_{Z} \geq 1$  & $N_{Z} = 2$     & $N_{H} = 1$     \\ \hline
200                & 4.17e+4        & 4.17e+4        & 4.16e+4        & 4.07e+4        & 4.07e+4        \\
250                & 5.14e+4        & 5.14e+4        & 5.13e+4        & 5.03e+4        & 5.03e+4        \\
300                & 2.14e+4        & 2.14e+4        & 2.14e+4        & 2.09e+4        & 2.09e+4        \\
350                & 1.15e+4        & 1.15e+4        & 1.14e+4        & 1.12e+4        & 1.12e+4        \\
400                & 1.6e+3         & 1.6e+3         & 1.6e+3         & 1.57e+3        & 1.57e+3        \\
450                & 973             & 973             & 971             & 950             & 950             \\
500                & 699             & 699             & 698             & 683             & 683             \\
600                & 435             & 435             & 435             & 426             & 426             \\
700                & 303             & 303             & 302             & 296             & 296             \\
800                & 217             & 217             & 217             & 212             & 212             \\
900                & 159             & 159             & 159             & 155             & 155             \\
1000               & 118             & 118             & 118             & 115             & 115             \\ \hline
\end{tabular}
\end{center}
\caption{Expected number of events for the $H\rightarrow ZZ$ signal in the four lepton final state for $\int Ldt=$ 3000~\ifb~at $\sqrt{s}=33$ TeV with $<N_{PU}>=$~140.}
\label{tab:HZZSignal_33_3000}
\end{table}

\begin{table}[htbp]
\begin{center}
\begin{tabular}{|l|c|c|c|c|c|c|}
\hline
Background                                                   & $N_{lepton} = 4$ & Lepton Trigger & $N_{Z} \geq 1$  & $N_{Z} = 2$     & $N_{H} = 1$     \\ \hline
B, Bj, Bjj-vbf, BB, BBB                            & 1.45e+5        & 1.45e+5        & 1.41e+5        & 1.14e+5        & 1.14e+5        \\
tj, tB, tt, ttB                                    & 1.63e+5        & 1.61e+5        & 4.99e+4        & 2.97e+3        & 2.97e+3        \\
H                                                  & 3.58e+3        & 3.55e+3        & 3.32e+3        & 0               & 0               \\
Total Background                                   & 3.11e+5        & 3.1e+5         & 1.94e+5        & 1.17e+5        & 1.17e+5        \\ \hline
\end{tabular}
\end{center}
\caption{Expected number of events for the SM backgrounds in the four lepton final state for $\int Ldt=$ 3000~\ifb~at $\sqrt{s}=33$ TeV with $<N_{PU}>=$~140.}
\label{tab:HZZBackground_33_3000}
\end{table}


\begin{figure}[htbp]
\begin{center}
\includegraphics[width=0.4\columnwidth,height=0.4\textheight,keepaspectratio=true]{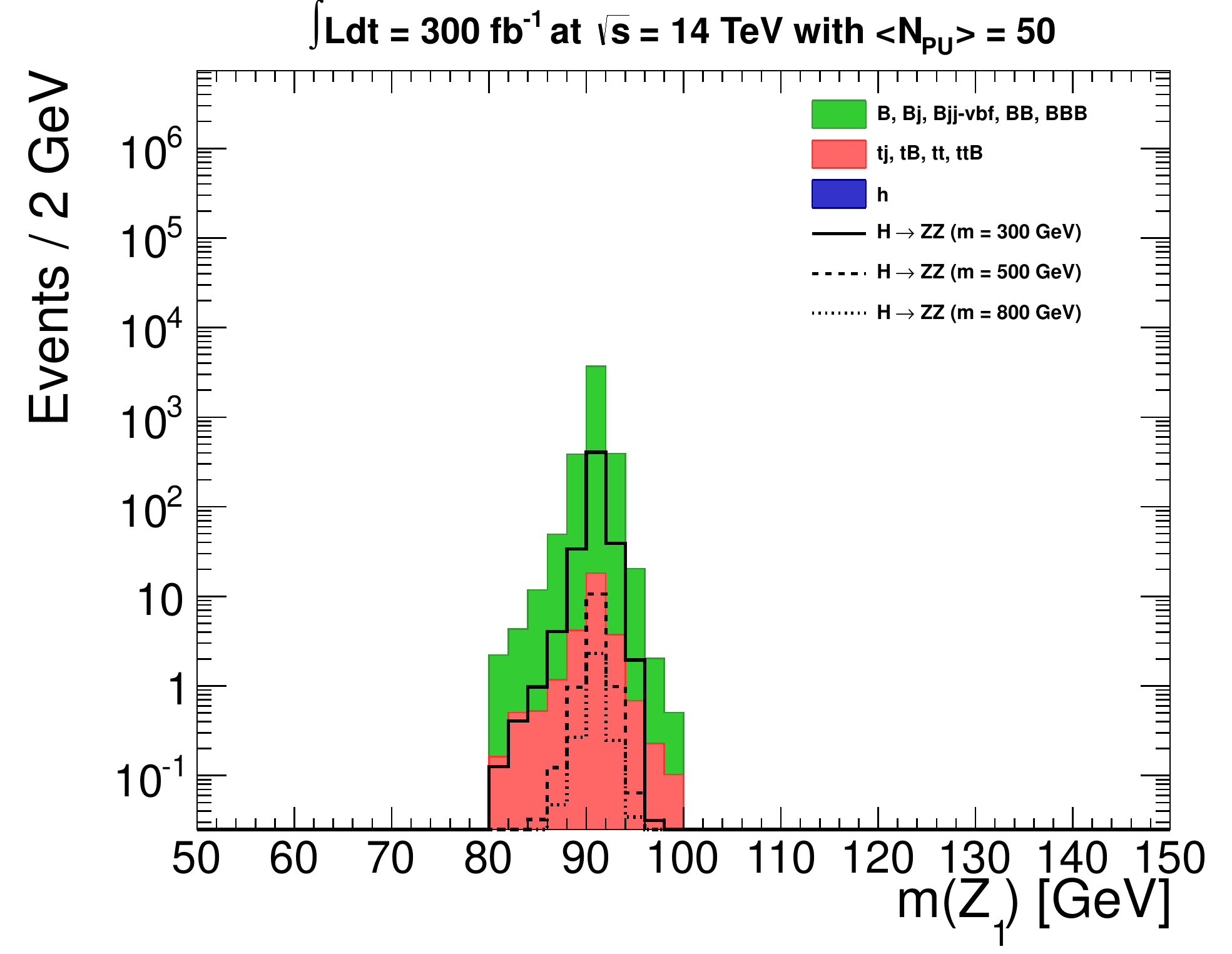}
\includegraphics[width=0.4\columnwidth,height=0.4\textheight,keepaspectratio=true]{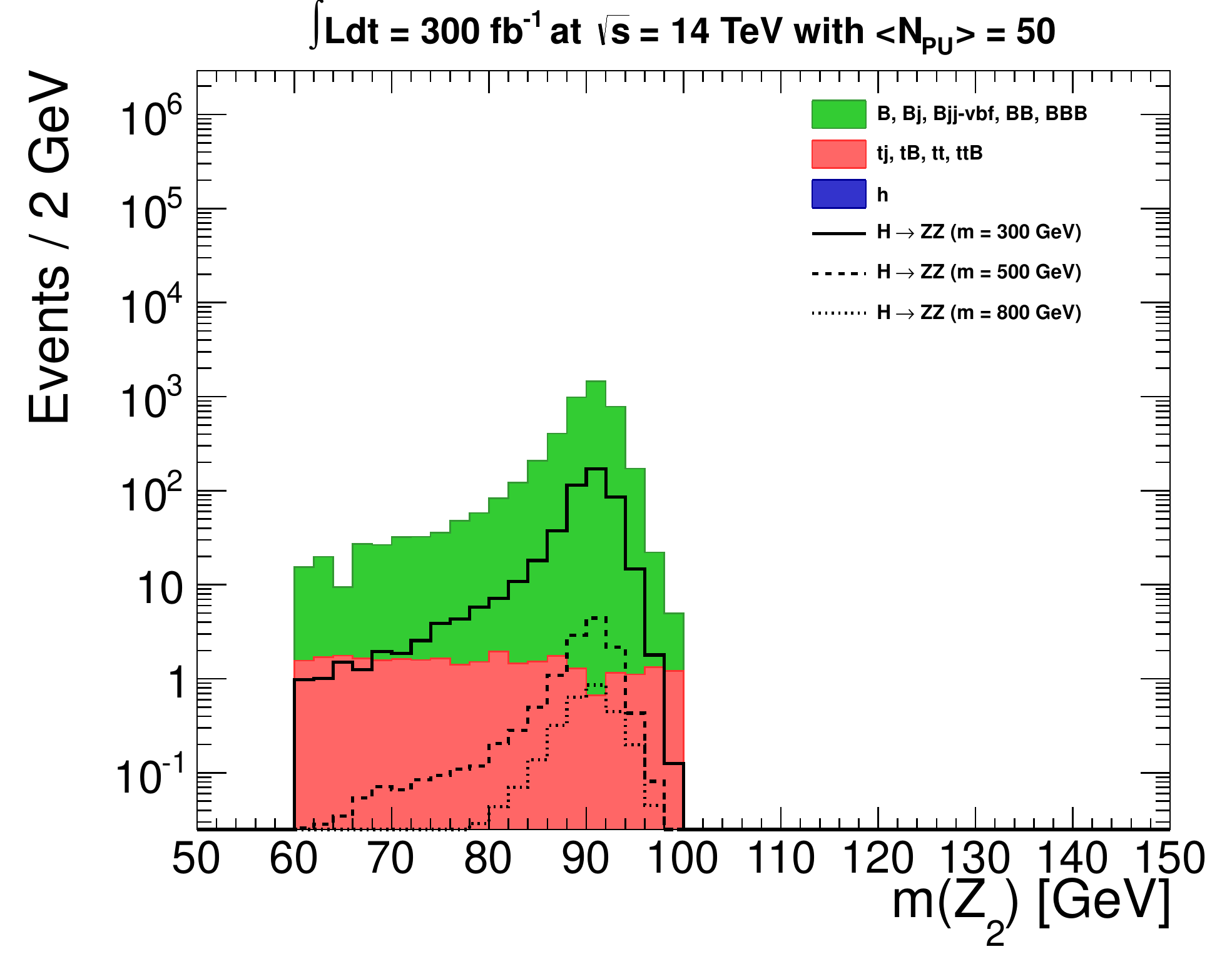}
\includegraphics[width=0.4\columnwidth,height=0.4\textheight,keepaspectratio=true]{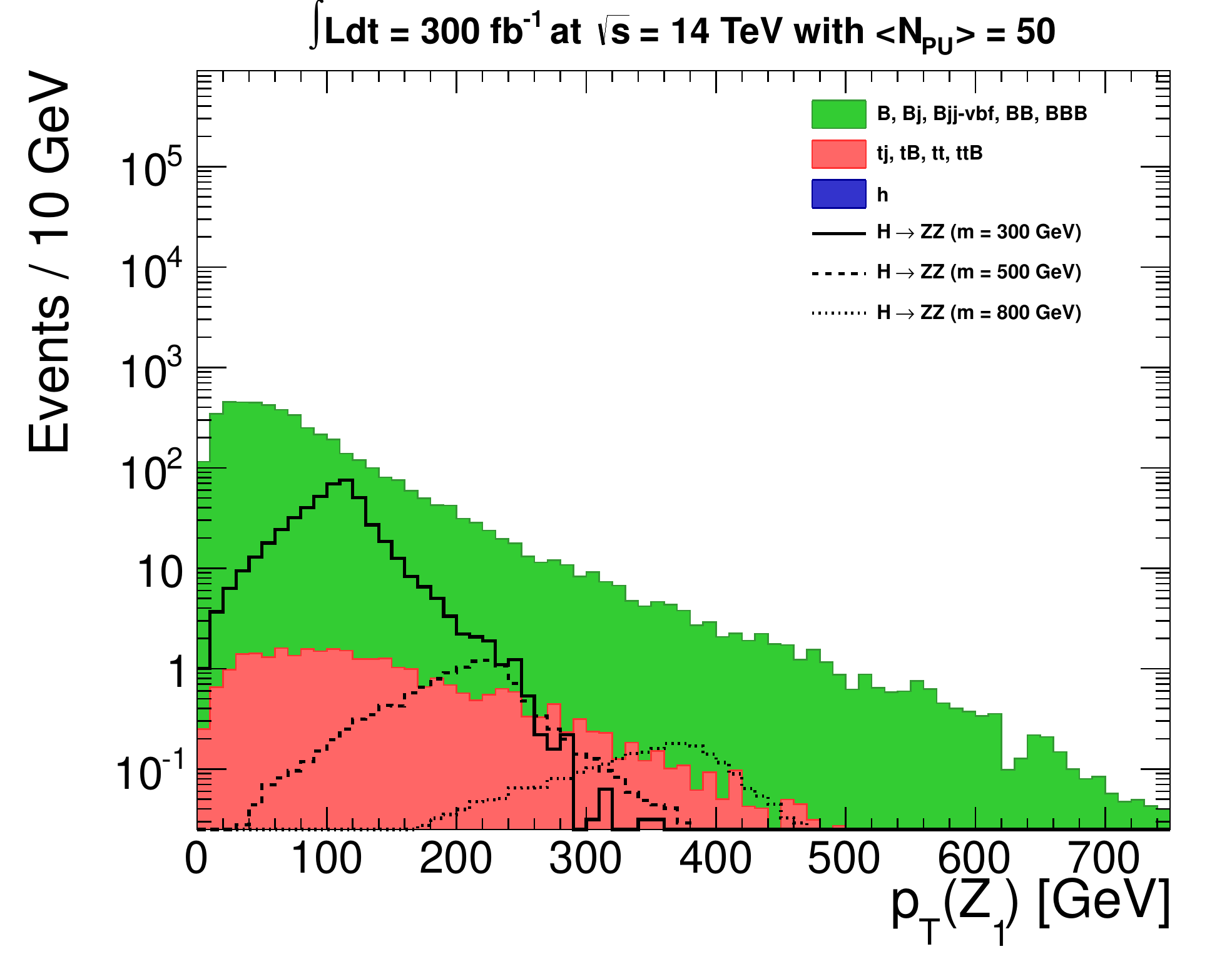}
\includegraphics[width=0.4\columnwidth,height=0.4\textheight,keepaspectratio=true]{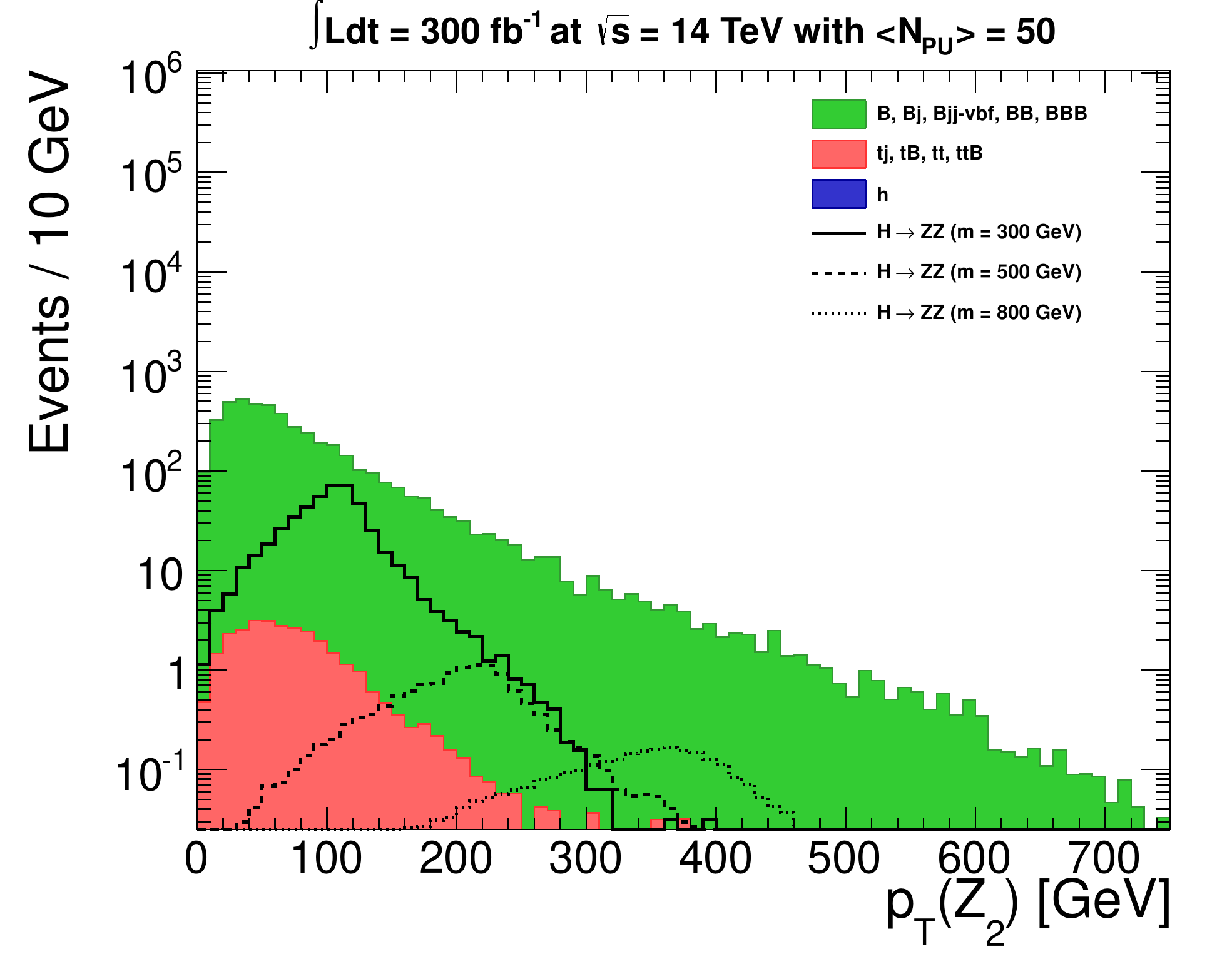}
\includegraphics[width=0.4\columnwidth,height=0.4\textheight,keepaspectratio=true]{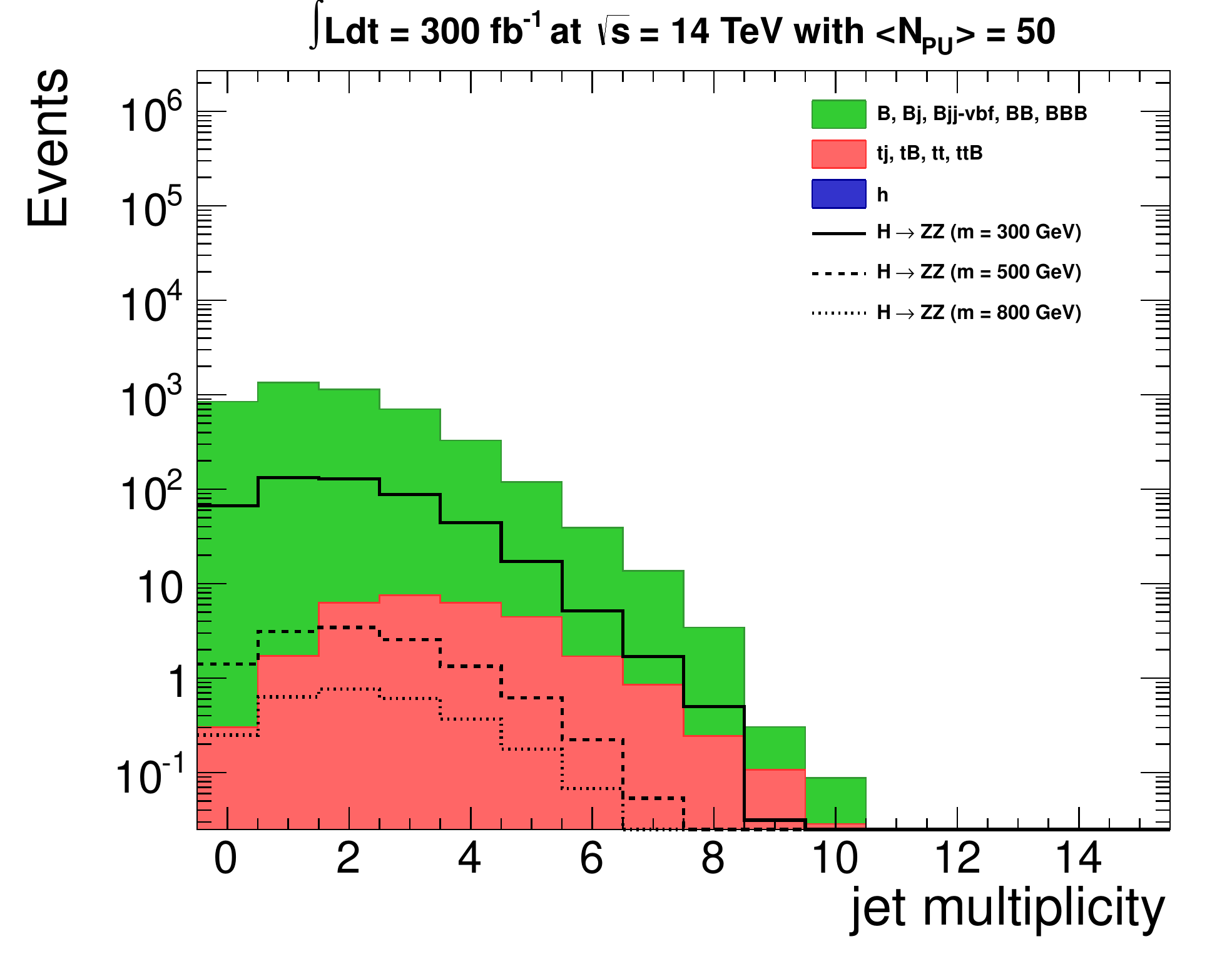}
\includegraphics[width=0.4\columnwidth,height=0.4\textheight,keepaspectratio=true]{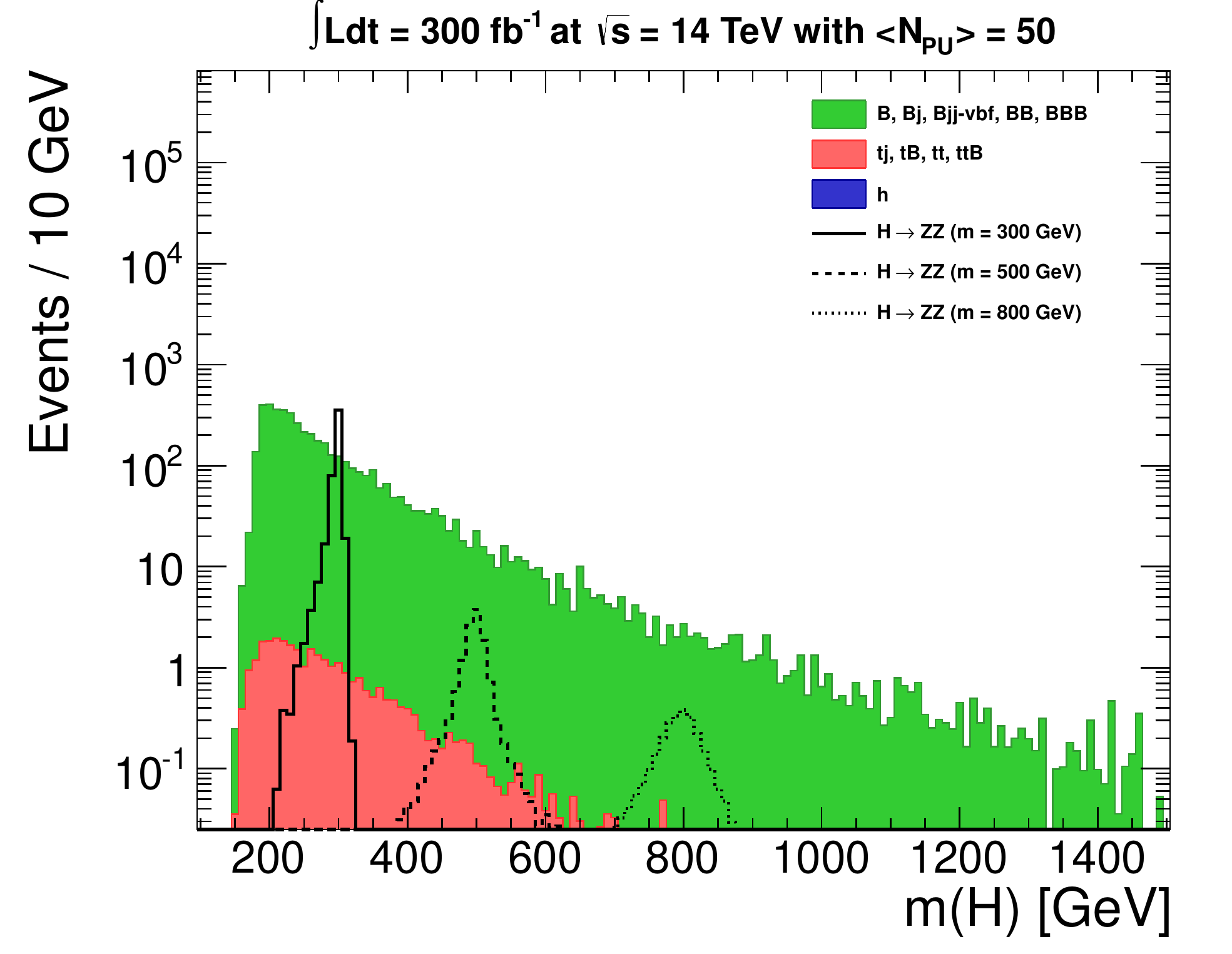}
\caption{Kinematic distributions for selected events in $\int Ldt=$ 300~\ifb~at $\sqrt{s}=14$ TeV with $<N_{PU}>=50$.}
\label{fig:HZZplots_14_300}
\end{center}
\end{figure}


\begin{figure}[htbp]
\begin{center}
\includegraphics[width=0.4\columnwidth,height=0.4\textheight,keepaspectratio=true]{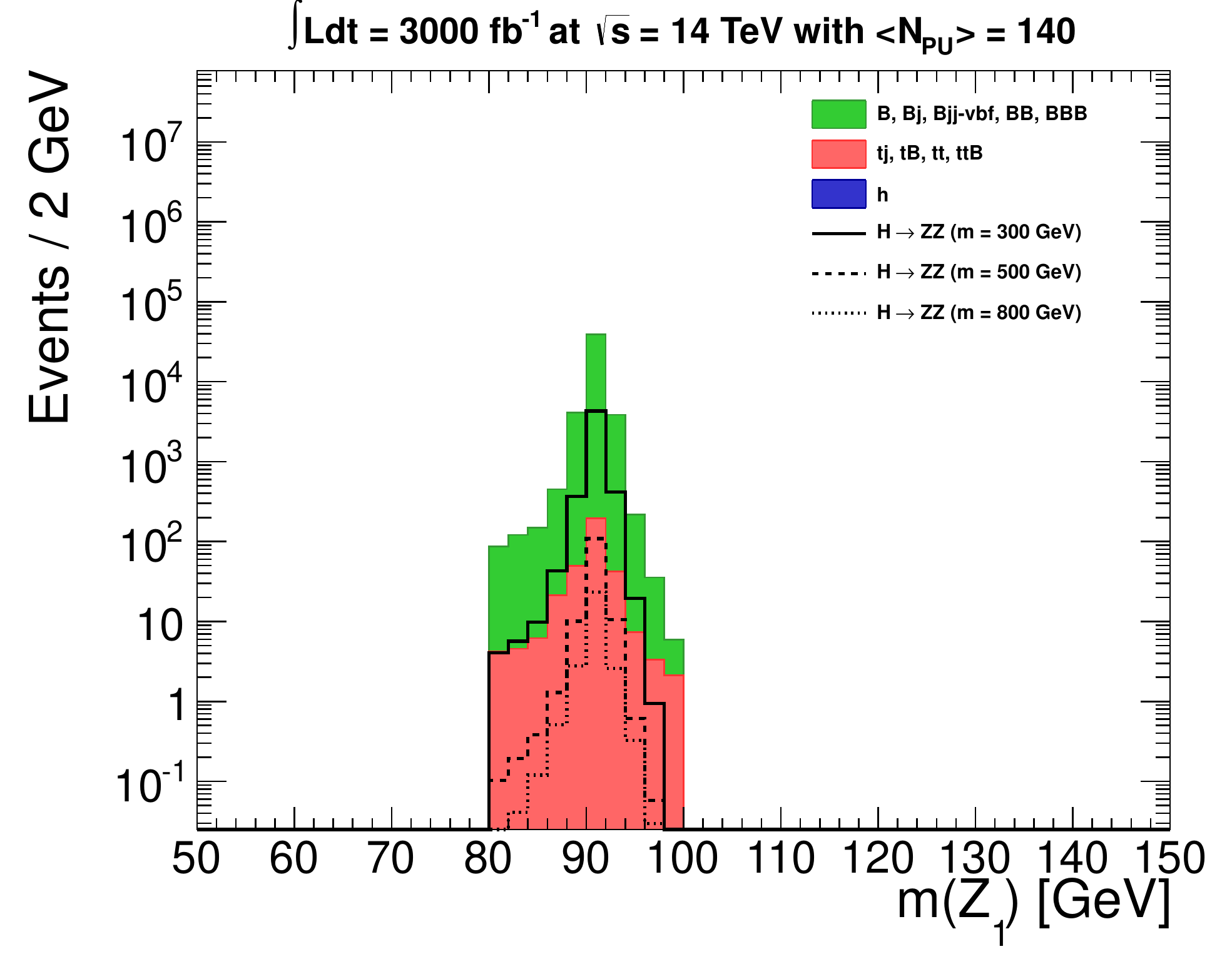}
\includegraphics[width=0.4\columnwidth,height=0.4\textheight,keepaspectratio=true]{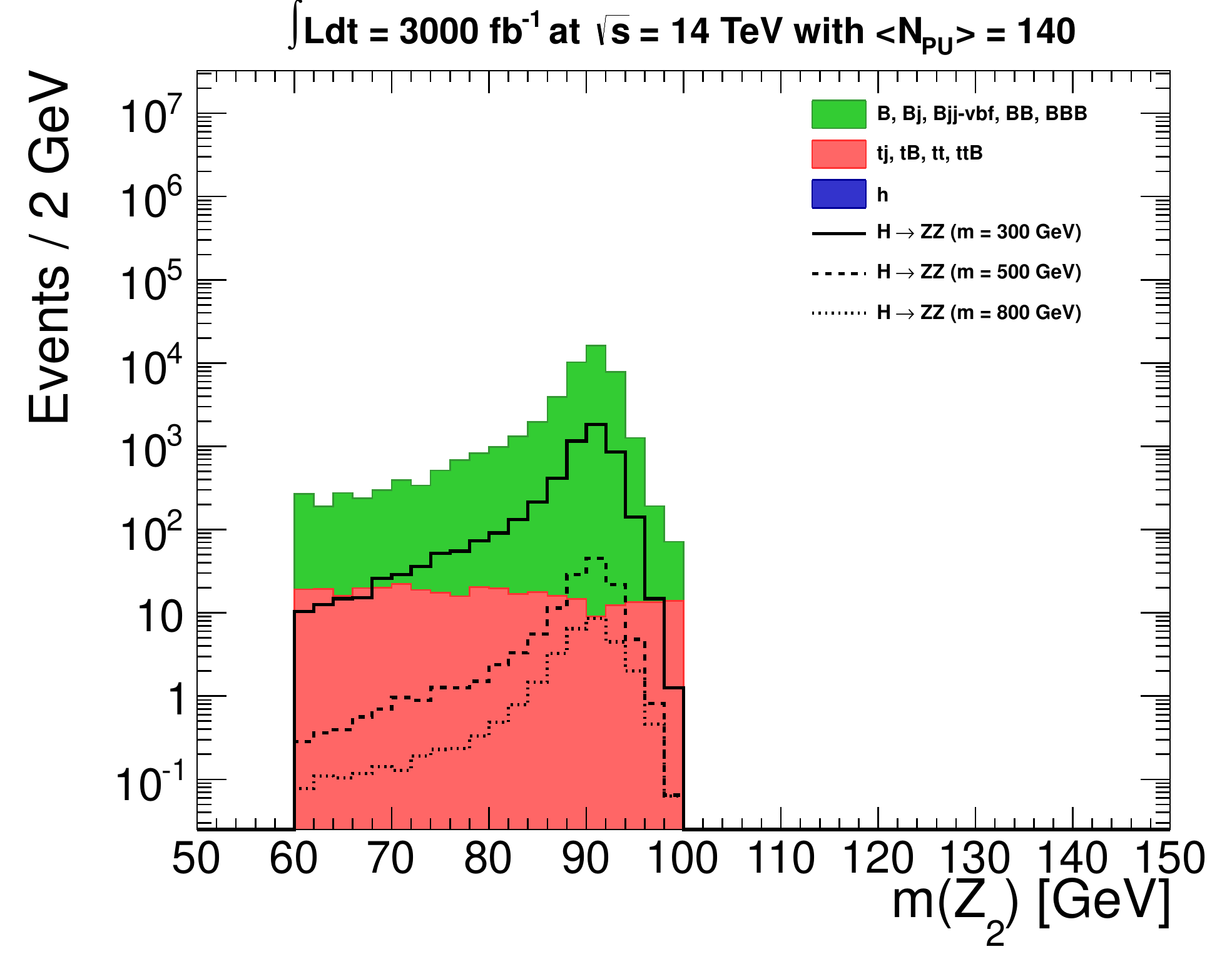}
\includegraphics[width=0.4\columnwidth,height=0.4\textheight,keepaspectratio=true]{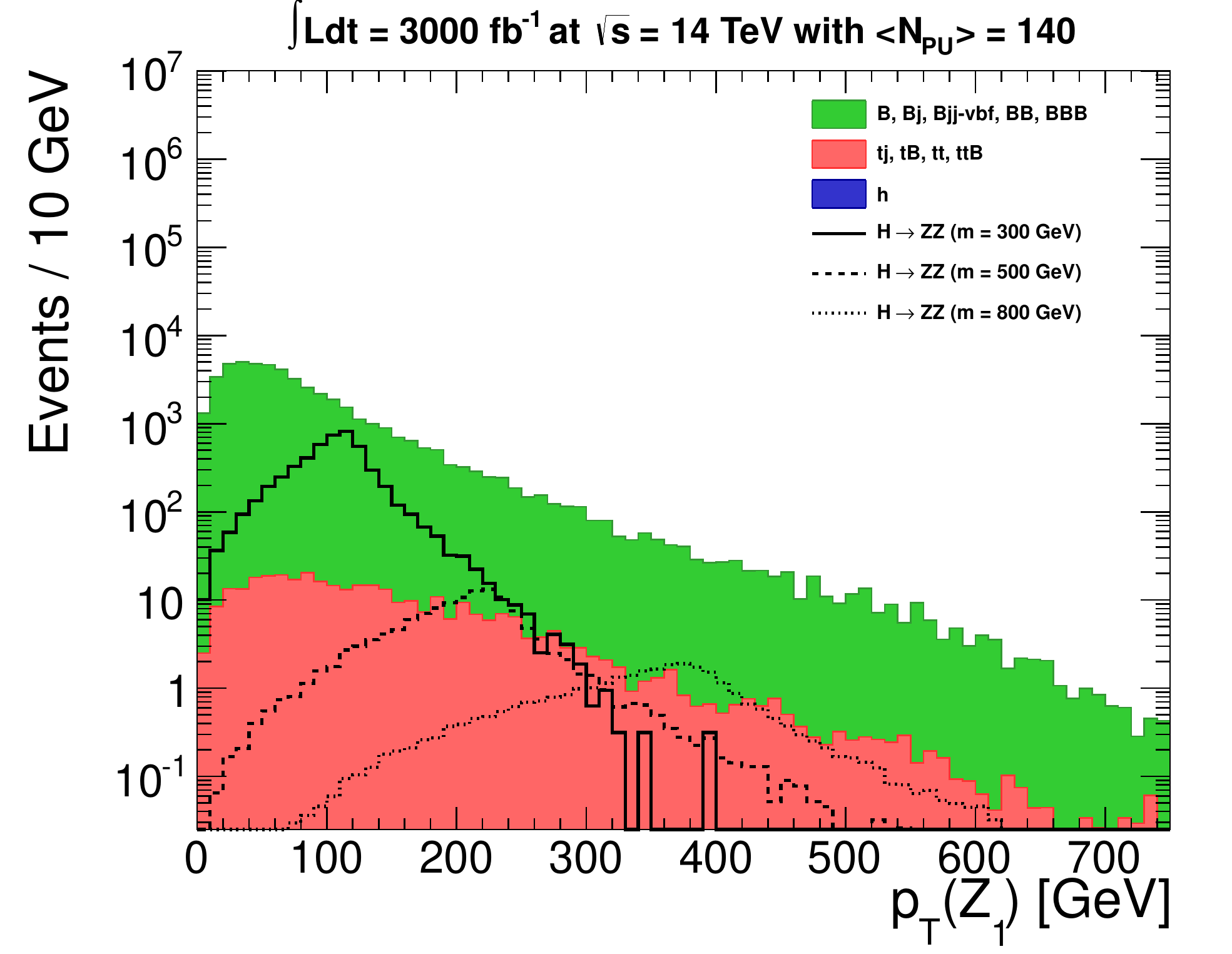}
\includegraphics[width=0.4\columnwidth,height=0.4\textheight,keepaspectratio=true]{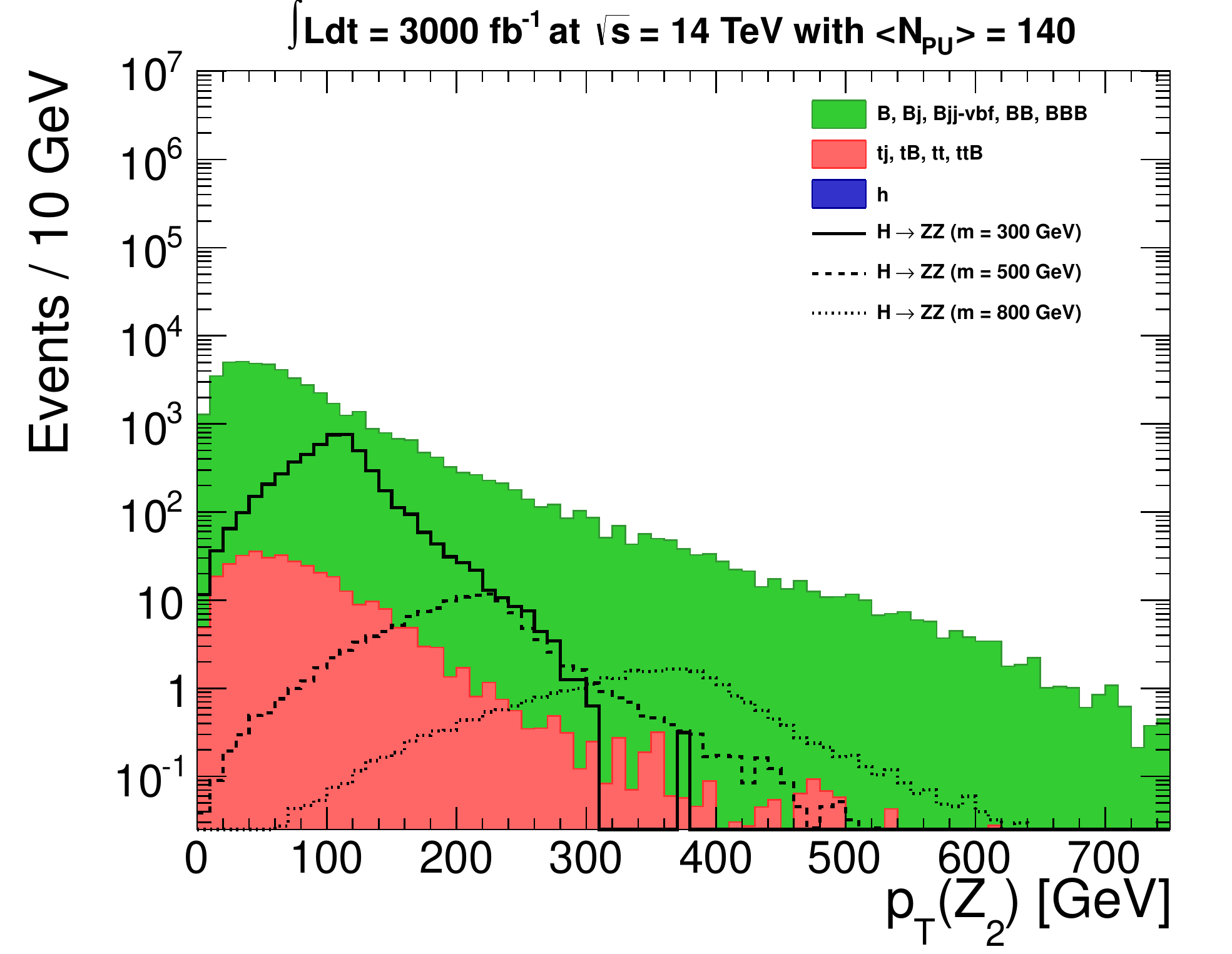}
\includegraphics[width=0.4\columnwidth,height=0.4\textheight,keepaspectratio=true]{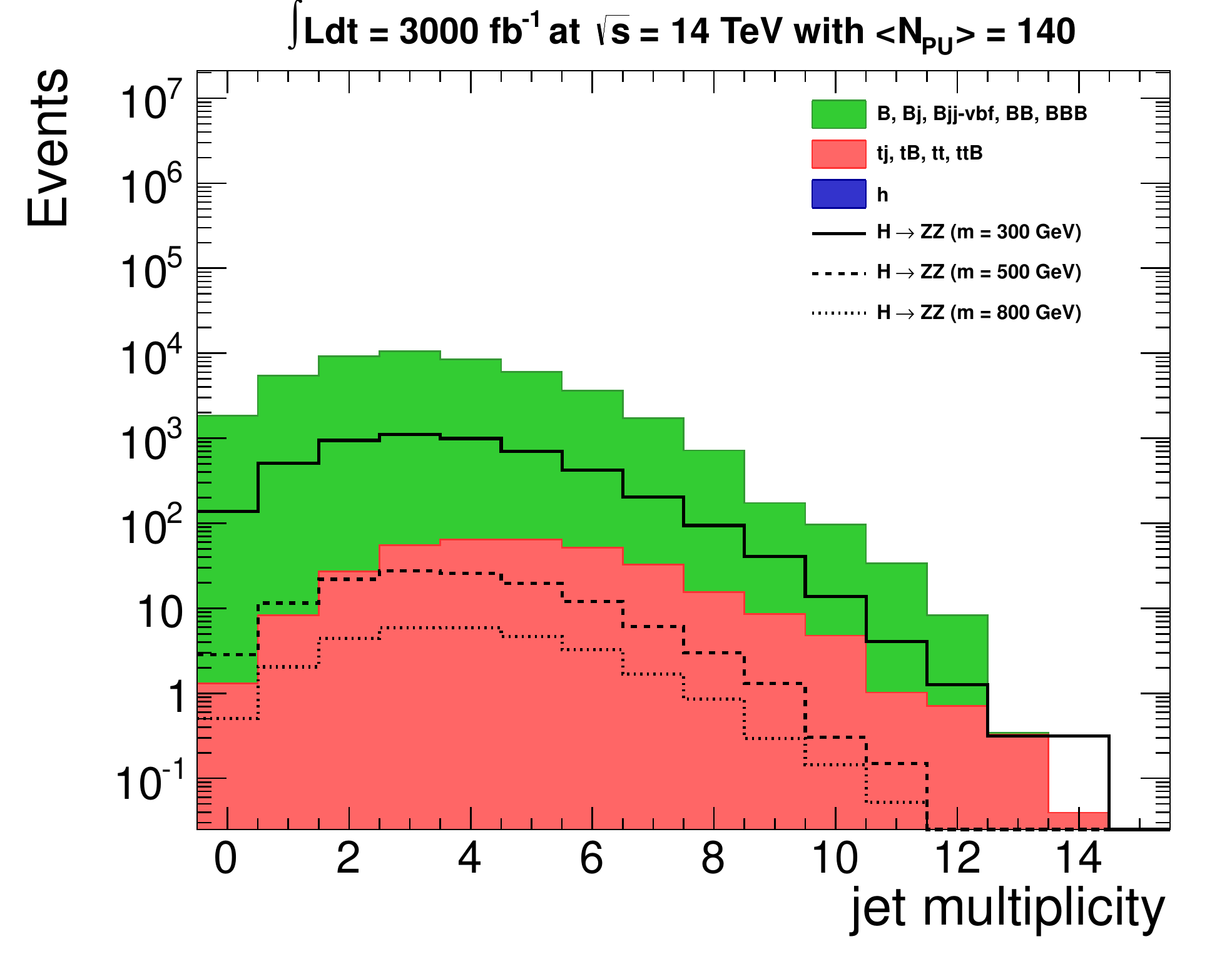}
\includegraphics[width=0.4\columnwidth,height=0.4\textheight,keepaspectratio=true]{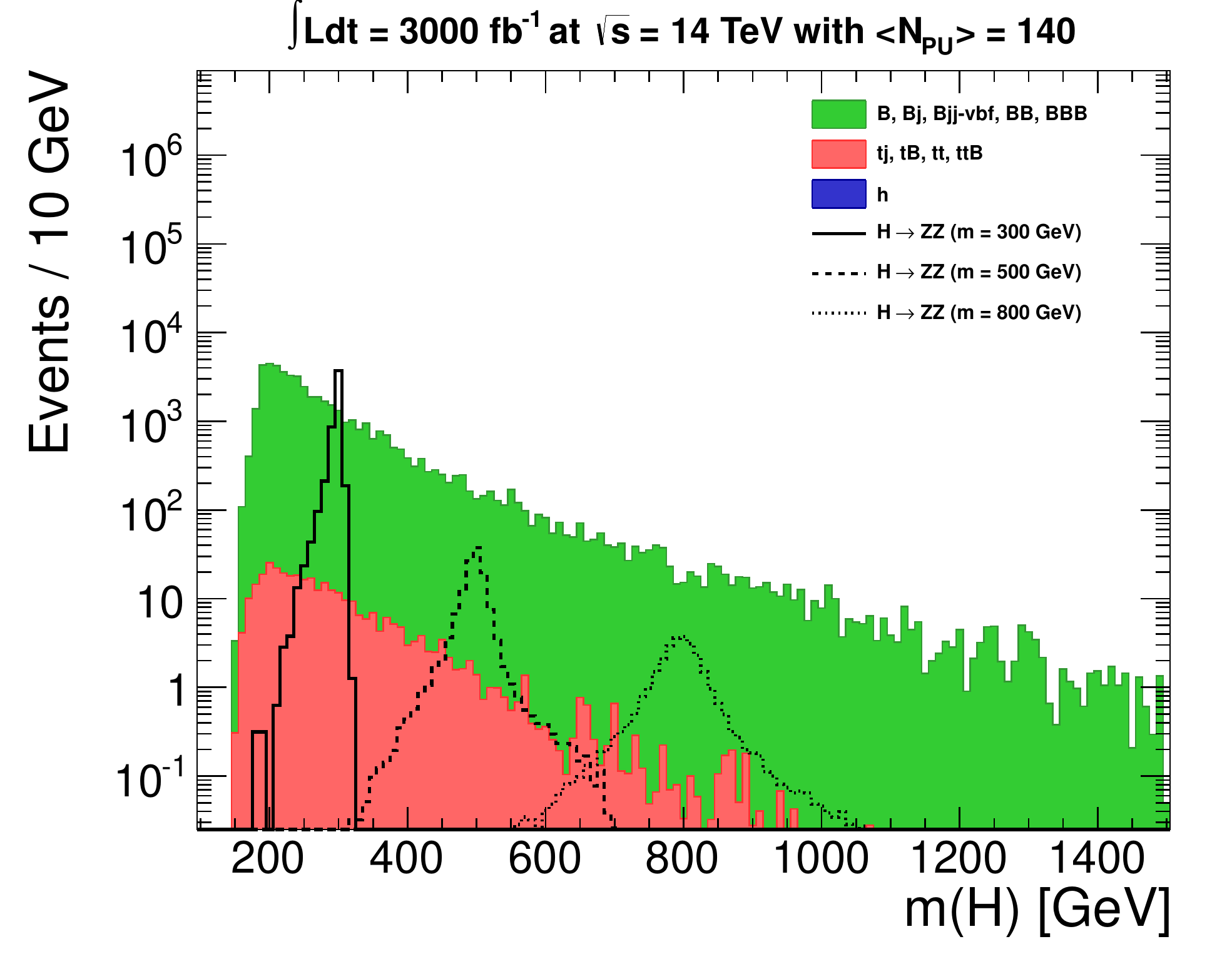}
\caption{Kinematic distributions for selected events in $\int Ldt=$ 3000~\ifb~at $\sqrt{s}=14$ TeV with $<N_{PU}>=$~140.}
\label{fig:HZZplots_14_3000}
\end{center}
\end{figure}

\begin{figure}[htbp]
\begin{center}
\includegraphics[width=0.4\columnwidth,height=0.4\textheight,keepaspectratio=true]{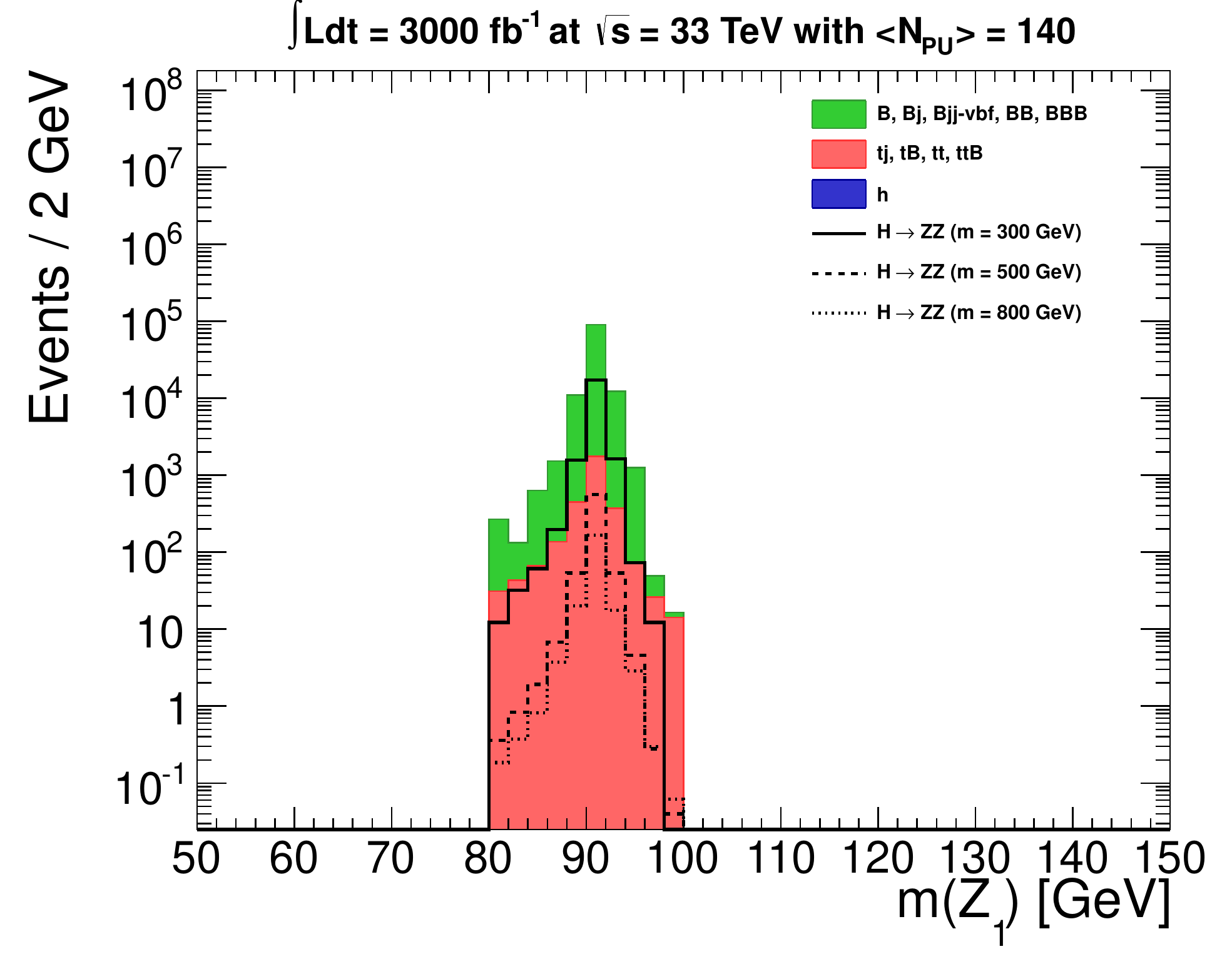}
\includegraphics[width=0.4\columnwidth,height=0.4\textheight,keepaspectratio=true]{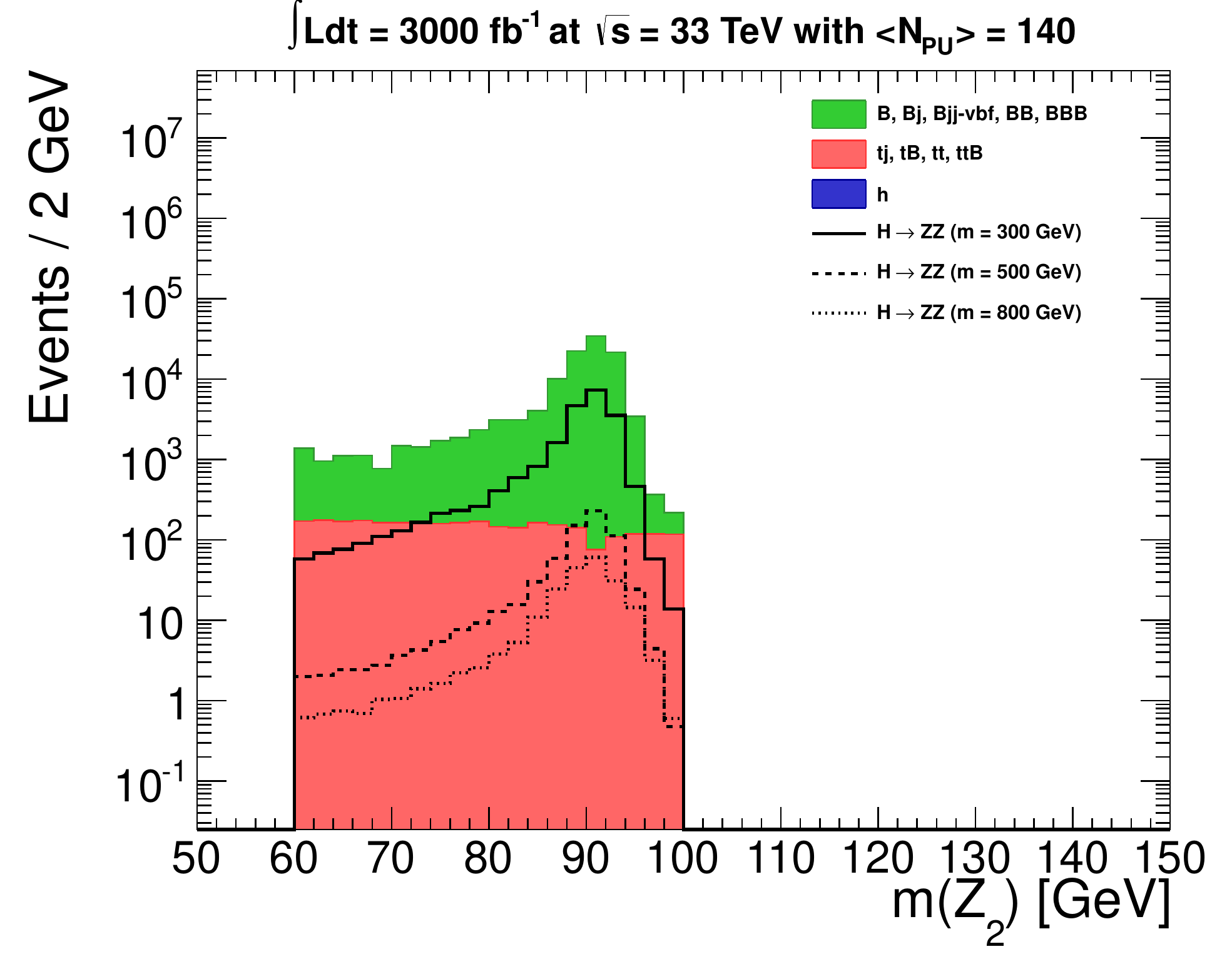}
\includegraphics[width=0.4\columnwidth,height=0.4\textheight,keepaspectratio=true]{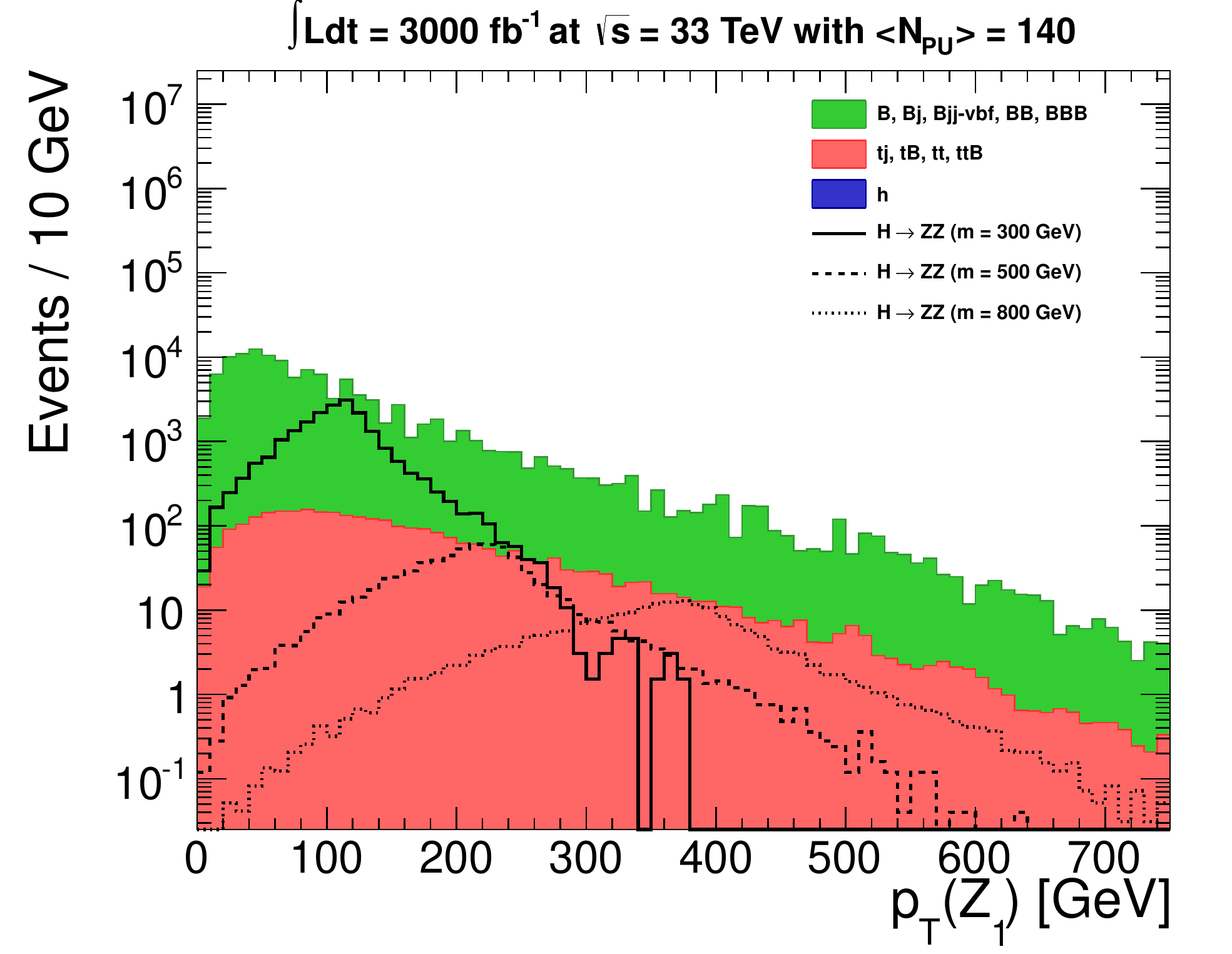}
\includegraphics[width=0.4\columnwidth,height=0.4\textheight,keepaspectratio=true]{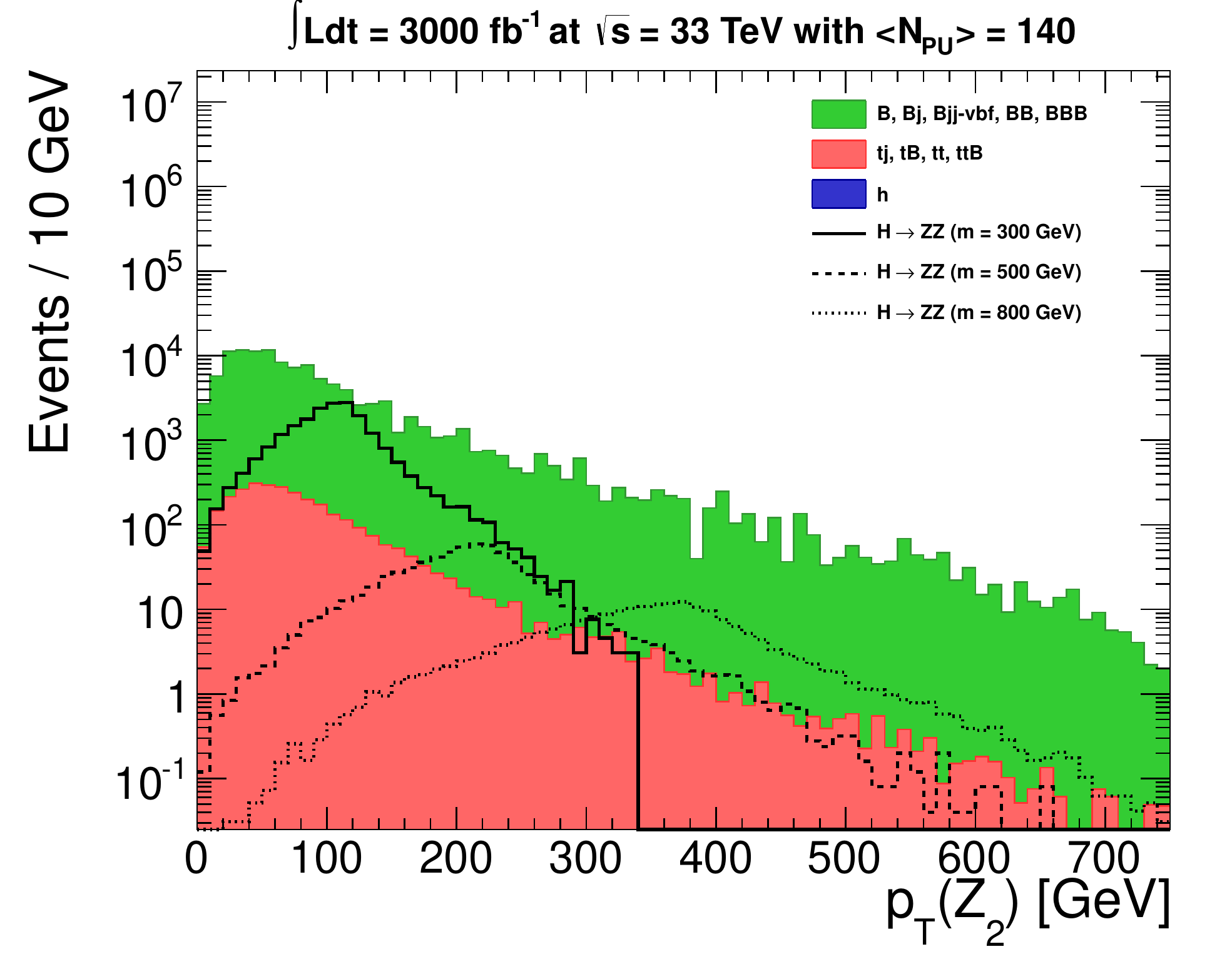}
\includegraphics[width=0.4\columnwidth,height=0.4\textheight,keepaspectratio=true]{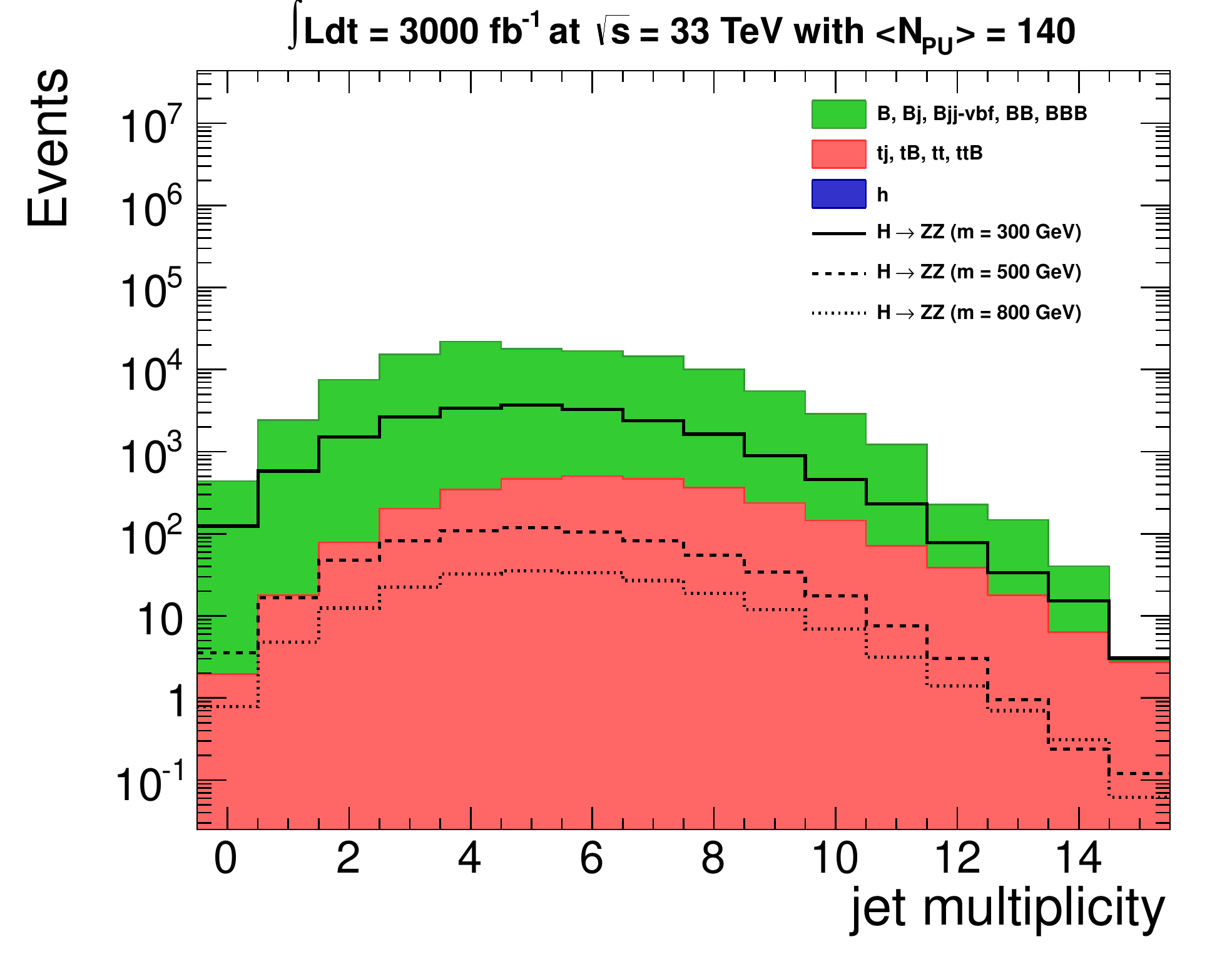}
\includegraphics[width=0.4\columnwidth,height=0.4\textheight,keepaspectratio=true]{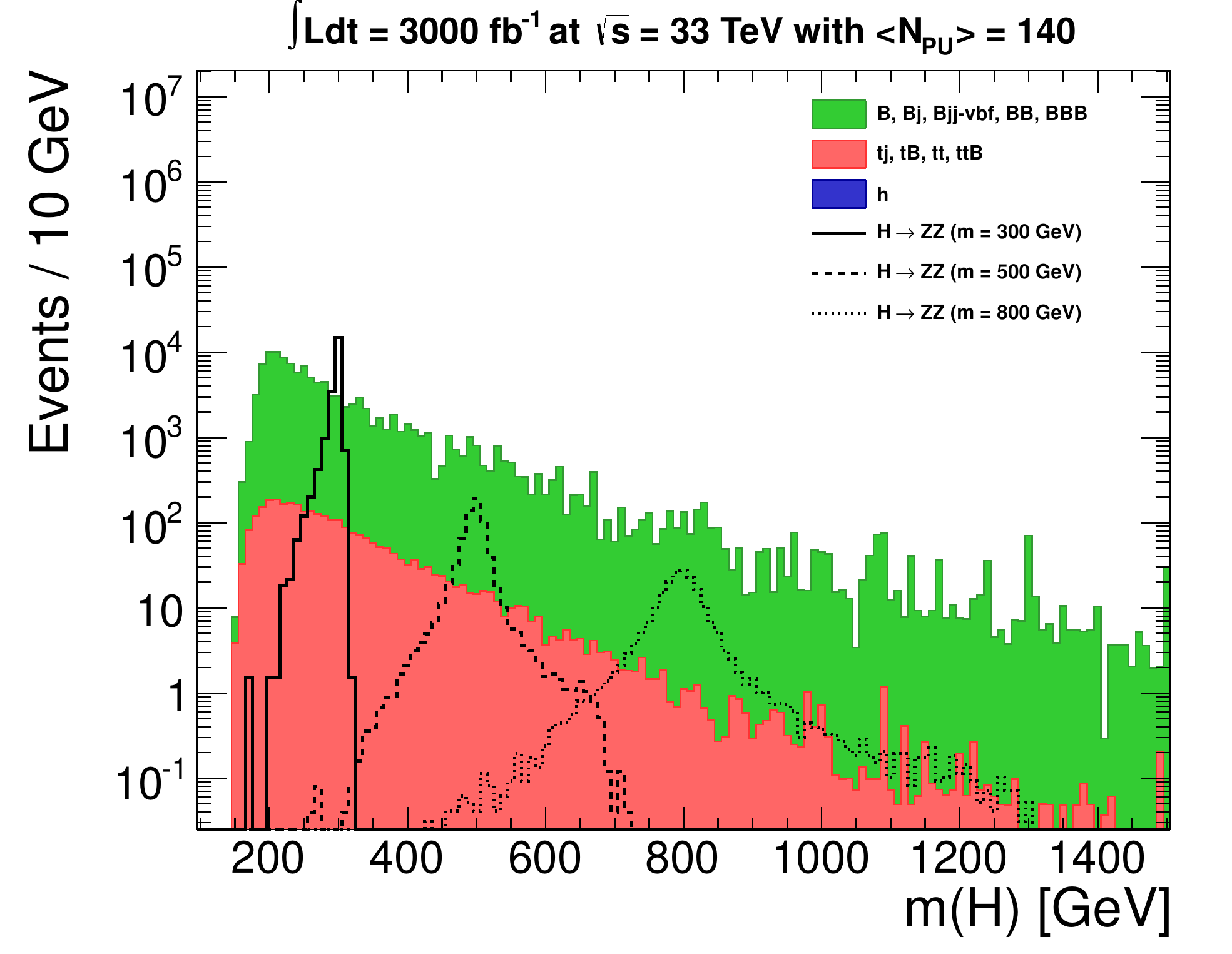}
\caption{Kinematic distributions for selected events in $\int Ldt=$ 3000~\ifb~at $\sqrt{s}=33$ TeV with $<N_{PU}>=$~140.}
\label{fig:HZZplots_33_3000}
\end{center}
\end{figure}


\subsubsection{Results}

The distribution of the $H$ candidate invariant mass is used to assess discovery and exclusion potential.  A 20\% rate uncertainty in the backgrounds is assumed.  Figure~\ref{fig:HZZ1DLim} shows the cross section which can be excluded at 95\% CL for each $H$ mass hypothesis, as well as the cross section required for $3\sigma$ and $5\sigma$ signal significance, at both $\sqrt{s}=14$ and 33~TeV.  Also shown is the cross section for a type II 2HDM with cos$(\beta - \alpha) = -0.06$ and tan$(\beta) = 1$.

The cross sections required for exclusion, observation, and discovery are then interpreted in the parameter space of Type I and II 2HDMs.  Figure~\ref{fig:HZZExc} shows the regions in the $\tan \beta$ versus cos($\beta - \alpha$) plane which can be excluded at 95\% CL for the three scenarios under consideration.  Figures~\ref{fig:HZZObs} and \ref{fig:HZZDisco} show the regions for which $3\sigma$ and $5\sigma$ signal significance can be obtained.  These plots also show the complementarity between direct searches for an extended Higgs sector and precision Higgs coupling measurements using the LHC Higgs coupling projections of~\cite{snowCouplings}.    There is a considerable region that can only be excluded through direct search, and also regions where coupling measurements are stronger. In particular, the direct search limits weaken only at large $\tan \beta$ where the production cross section for $H$ falls due to diminishing coupling to the top quark, as well as close to the alignment limit $|\cos(\beta - \alpha)| \to  0$ where the branching ratio for $H \to ZZ$ vanishes.


\begin{figure}[htbp]
\begin{center}
\includegraphics[width=0.4\columnwidth,height=0.4\textheight,keepaspectratio=true]{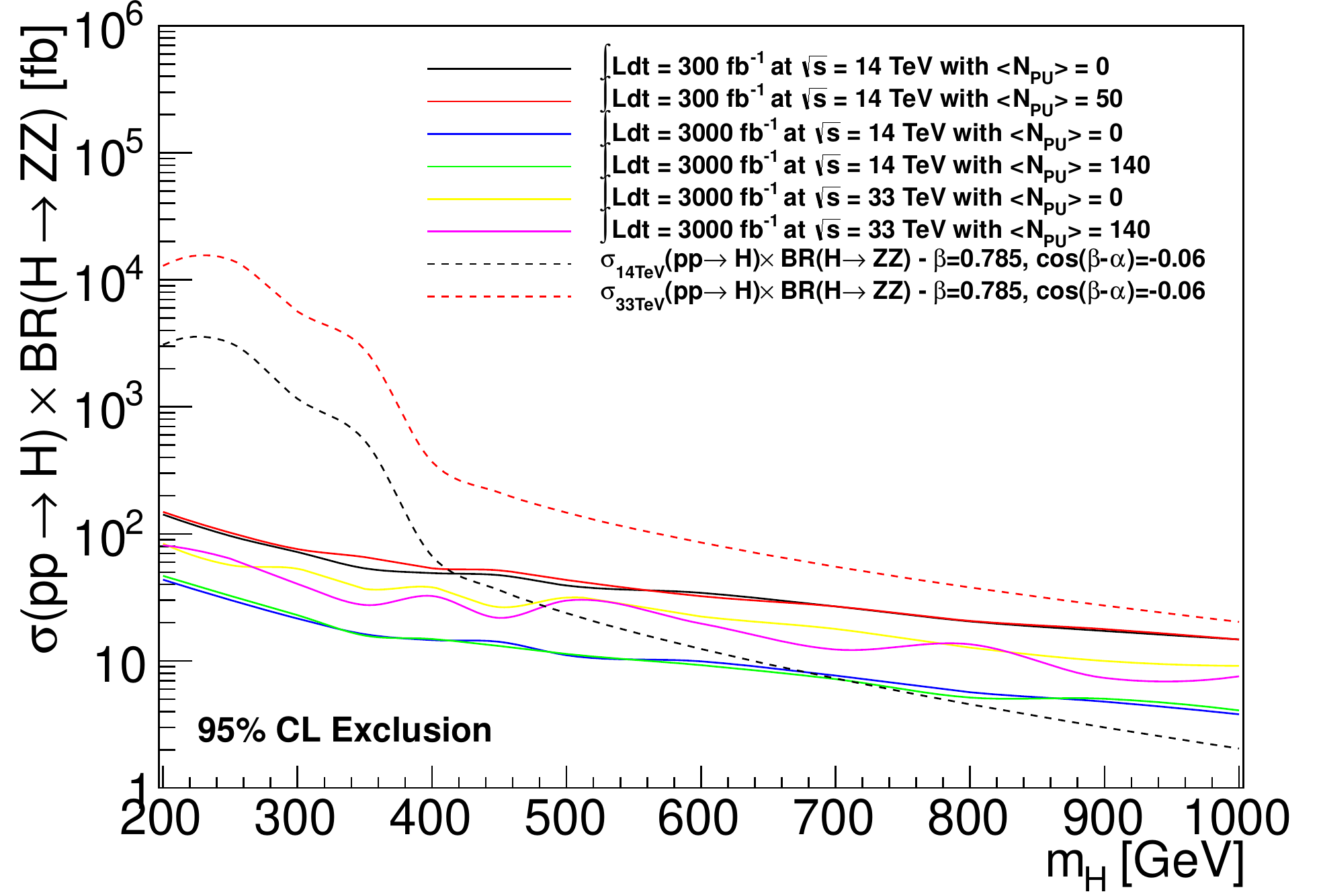}
\includegraphics[width=0.4\columnwidth,height=0.4\textheight,keepaspectratio=true]{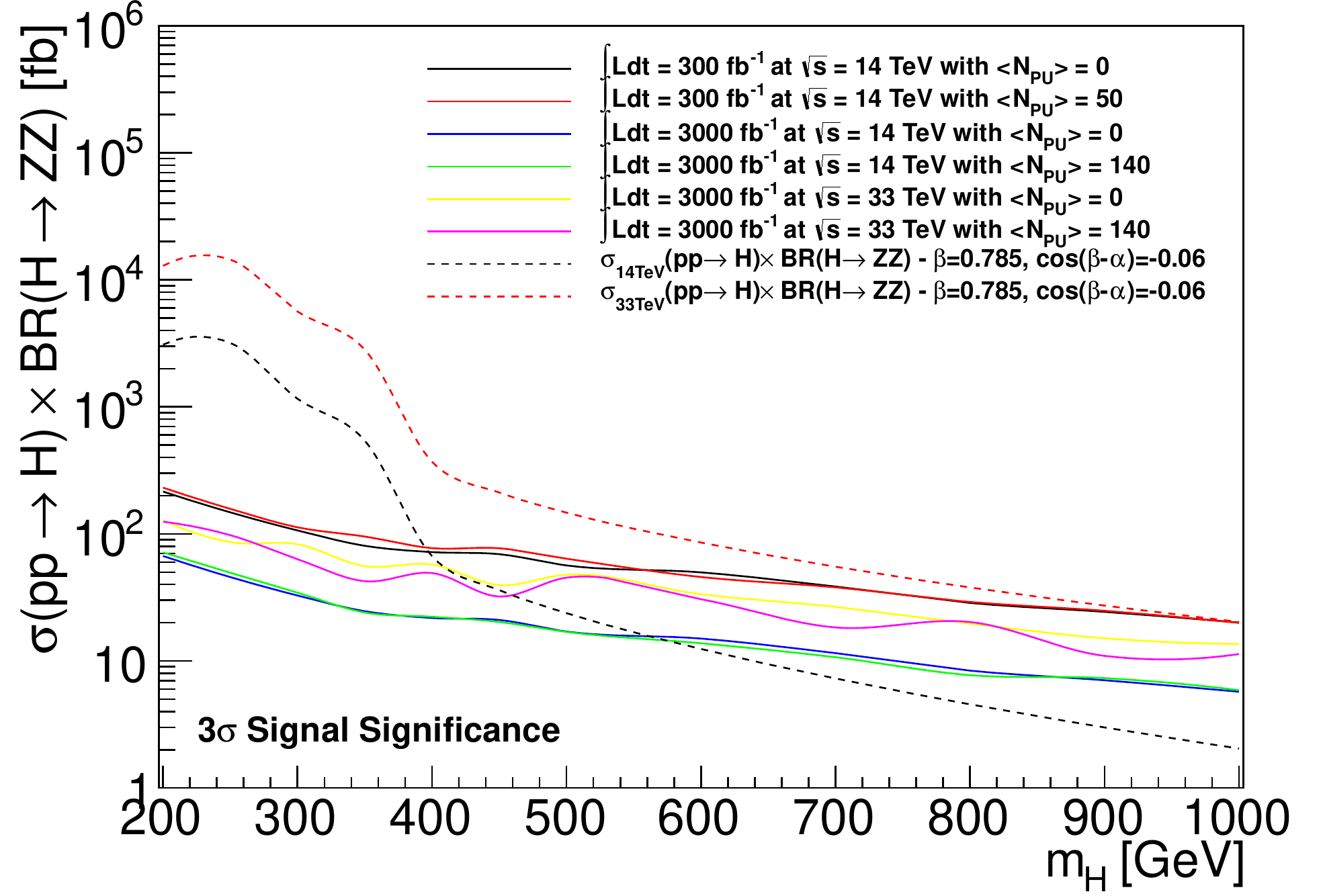}
\includegraphics[width=0.4\columnwidth,height=0.4\textheight,keepaspectratio=true]{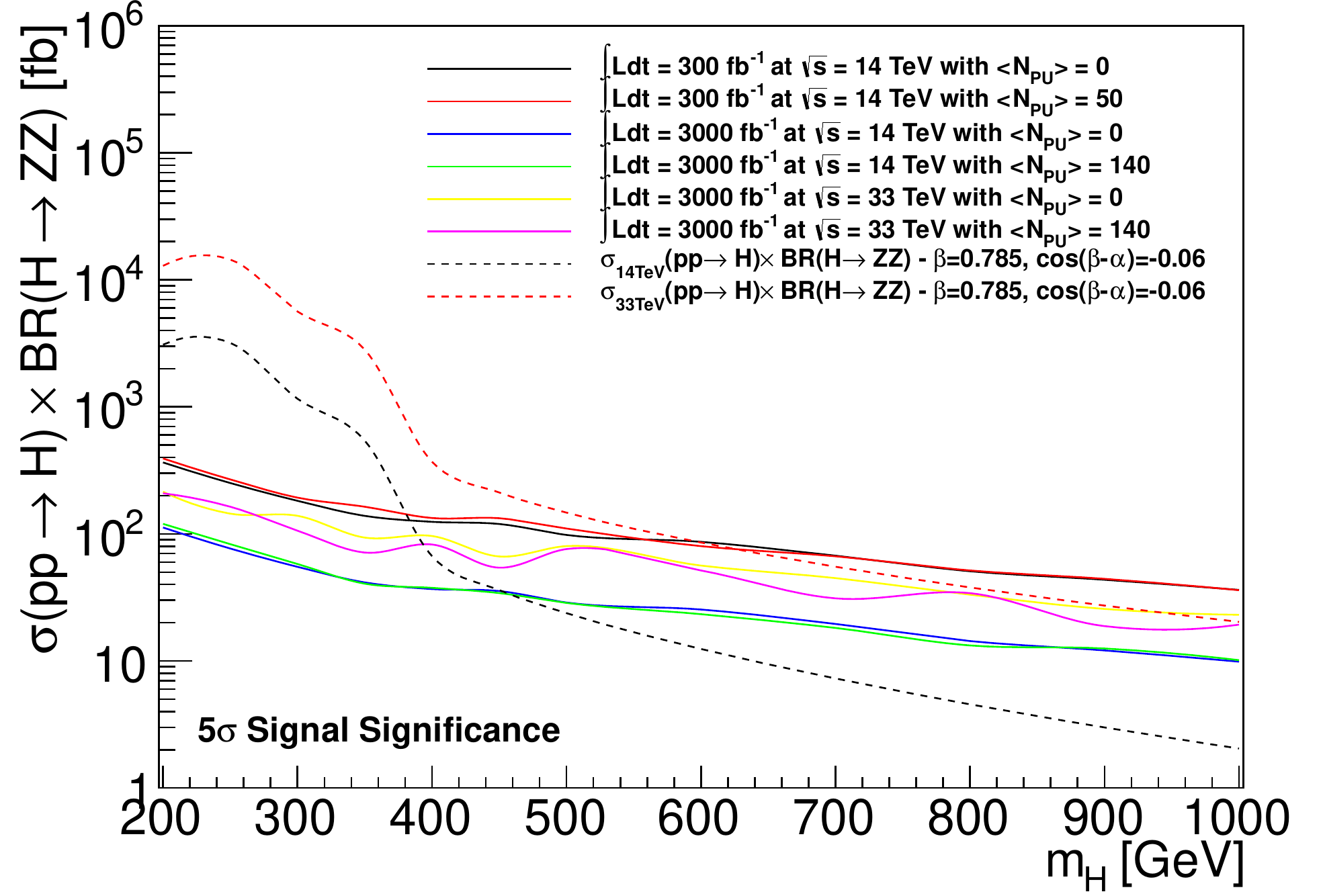}
\caption{The cross section which can be excluded at 95\% CL, and the cross section required for a $3\sigma$ and $5\sigma$ signal significance, for each $H$ mass hypothesis.}
\label{fig:HZZ1DLim}
\end{center}
\end{figure}


\begin{figure}[htbp]
\begin{center}
\includegraphics[width=0.4\columnwidth,height=0.4\textheight,keepaspectratio=true]{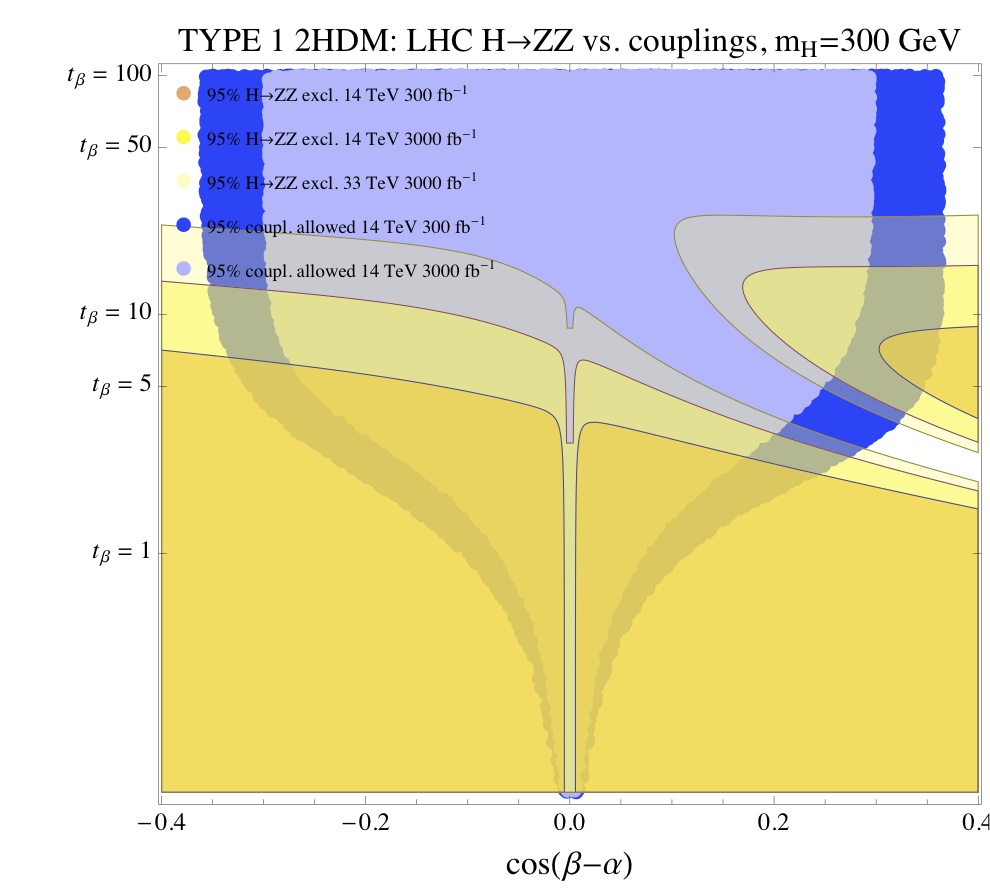}
\includegraphics[width=0.4\columnwidth,height=0.4\textheight,keepaspectratio=true]{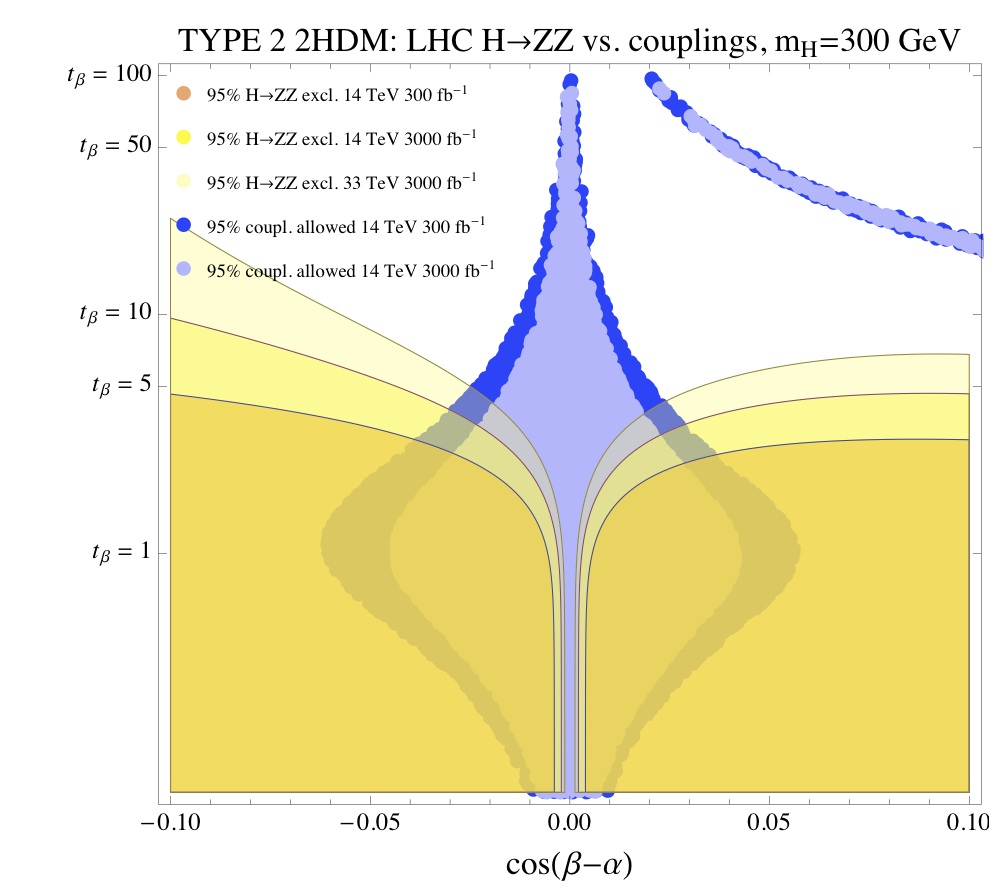}
\includegraphics[width=0.4\columnwidth,height=0.4\textheight,keepaspectratio=true]{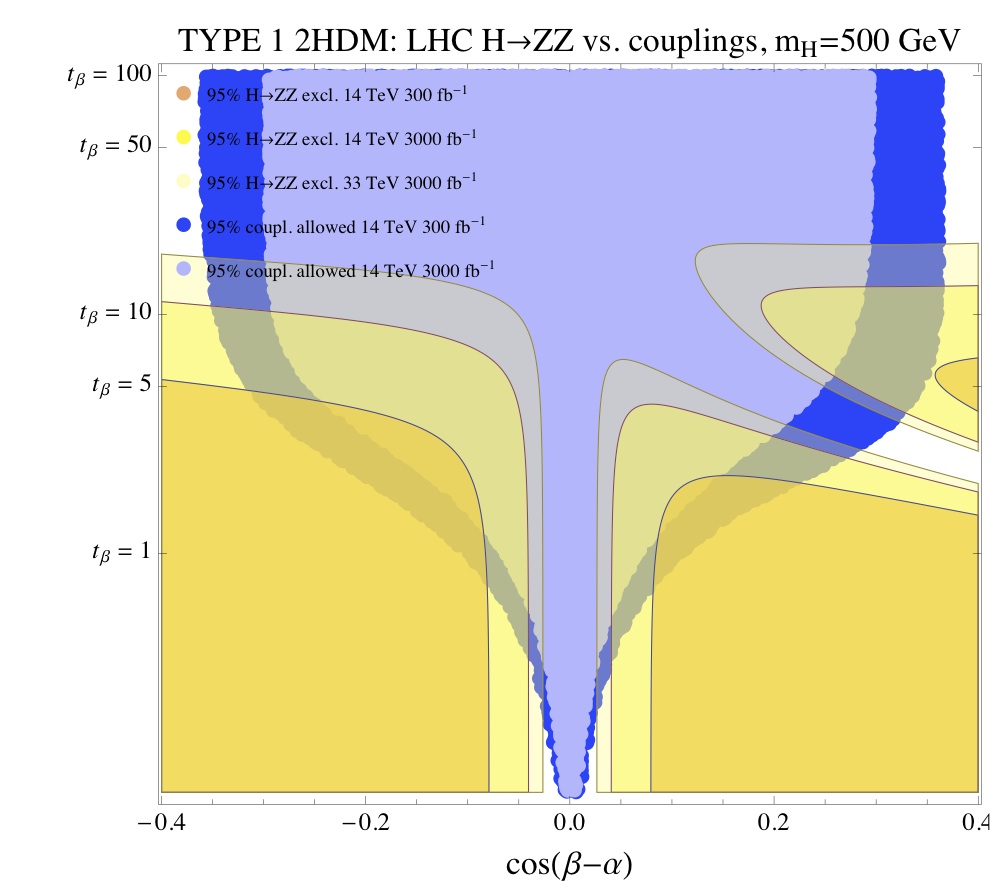}
\includegraphics[width=0.4\columnwidth,height=0.4\textheight,keepaspectratio=true]{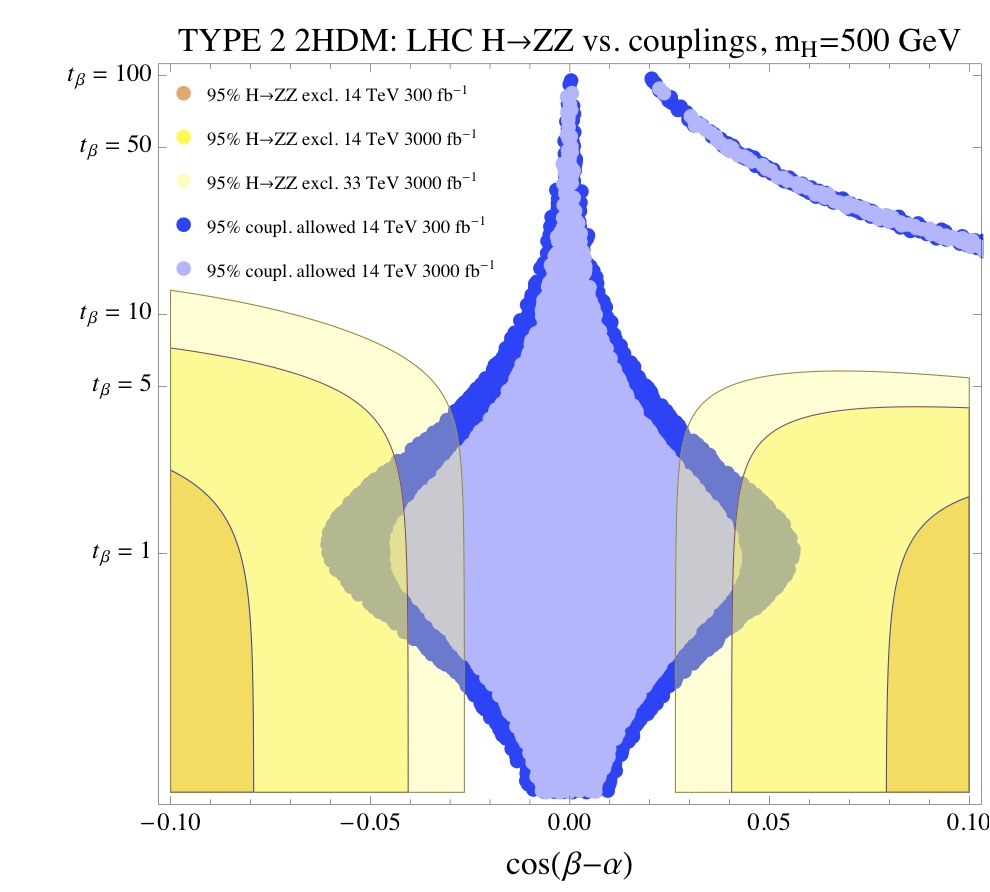}
\includegraphics[width=0.4\columnwidth,height=0.4\textheight,keepaspectratio=true]{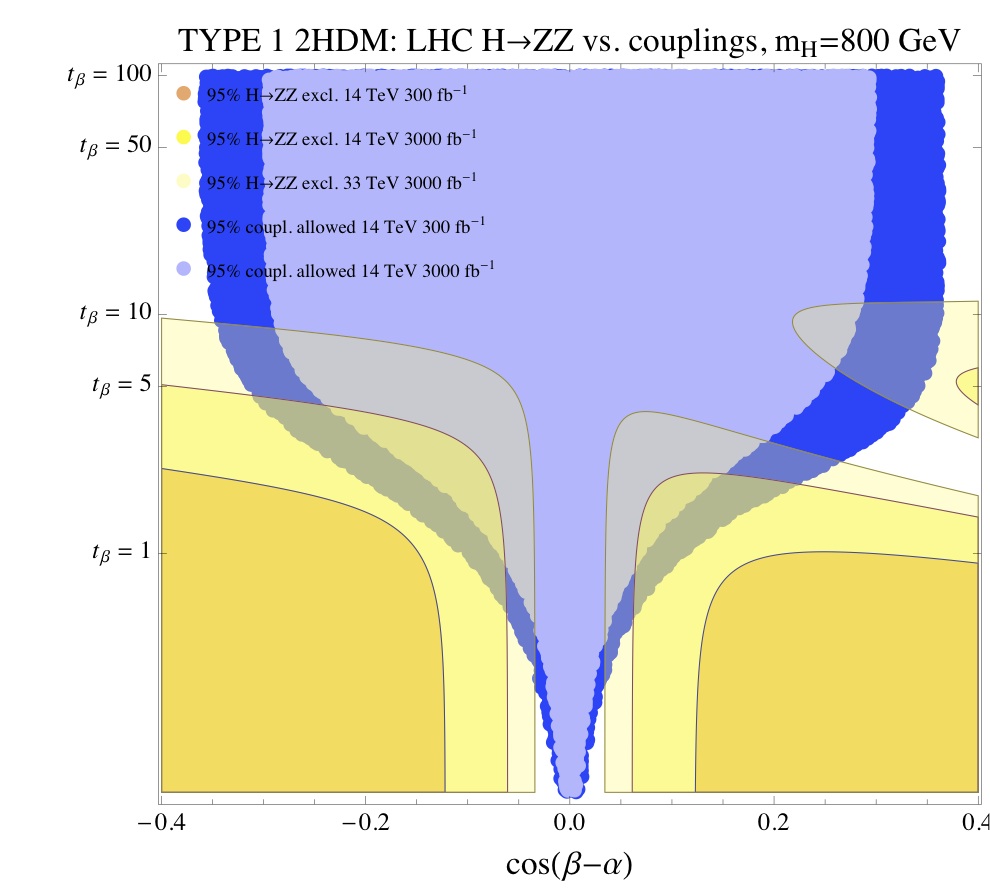}
\includegraphics[width=0.4\columnwidth,height=0.4\textheight,keepaspectratio=true]{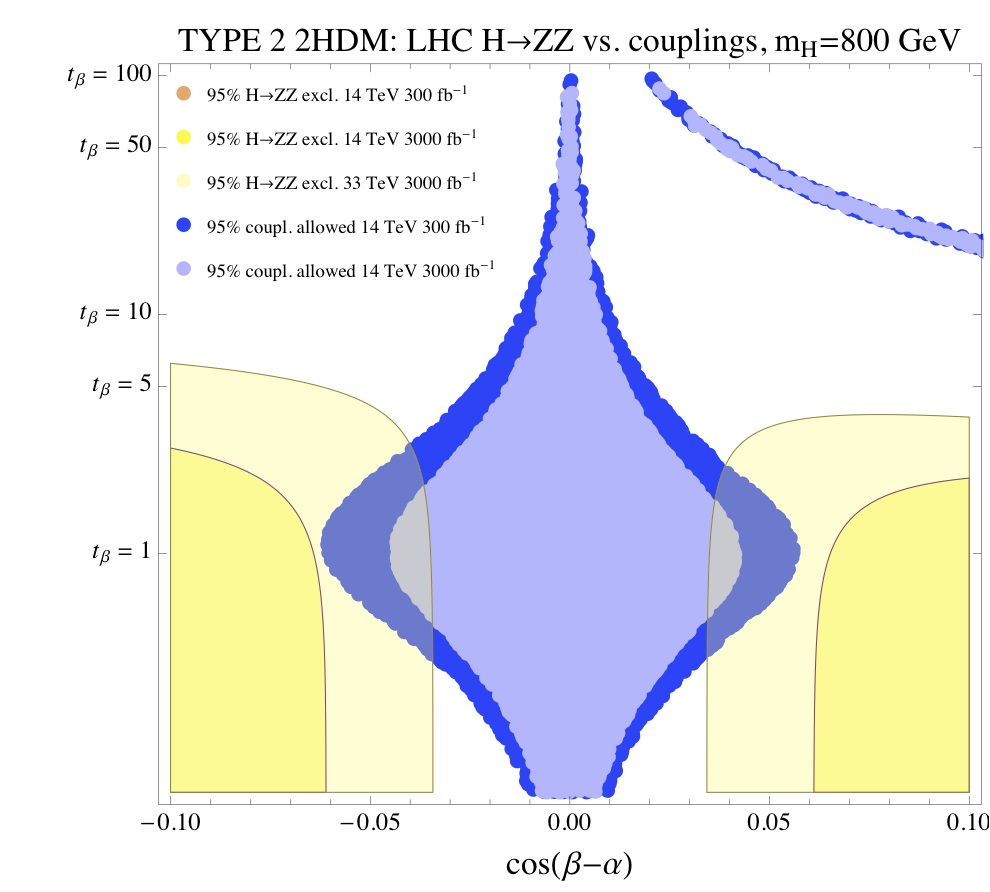}
\caption{The region of parameter space which could be excluded at 95\% CL for various $H$ mass hypotheses in a type I (left) and type II (right) 2HDM.  The dark yellow region corresponds to 300~\ifb~at $\sqrt{s}=14$~TeV, the yellow region to 3000~\ifb~at $\sqrt{s}=14$~TeV, and the light yellow region to 3000~\ifb~at $\sqrt{s}=33$~TeV.  The region which would remain allowed at 95\% CL based on non-observation of deviations from the SM in precision Higgs coupling measurements is shown in dark (light) blue for 300~\ifb~(3000~\ifb)~\cite{snowCouplings}.}
\label{fig:HZZExc}
\end{center}
\end{figure}


\begin{figure}[htbp]
\begin{center}
\includegraphics[width=0.4\columnwidth,height=0.4\textheight,keepaspectratio=true]{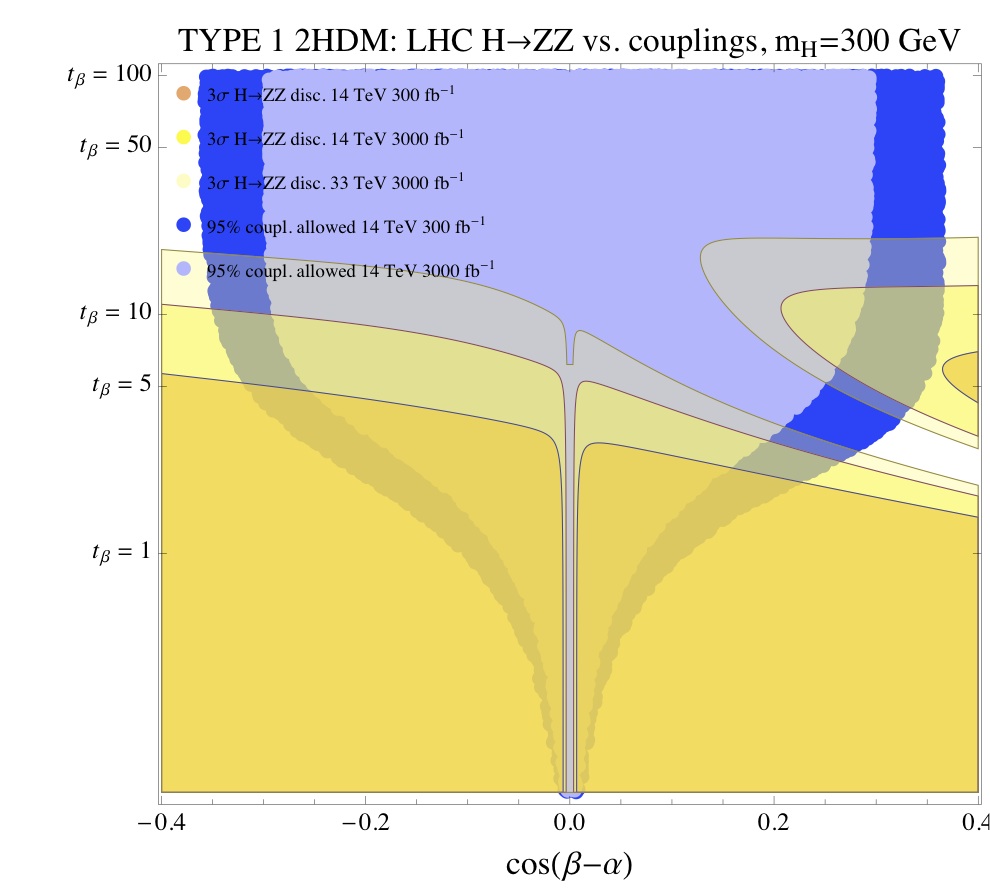}
\includegraphics[width=0.4\columnwidth,height=0.4\textheight,keepaspectratio=true]{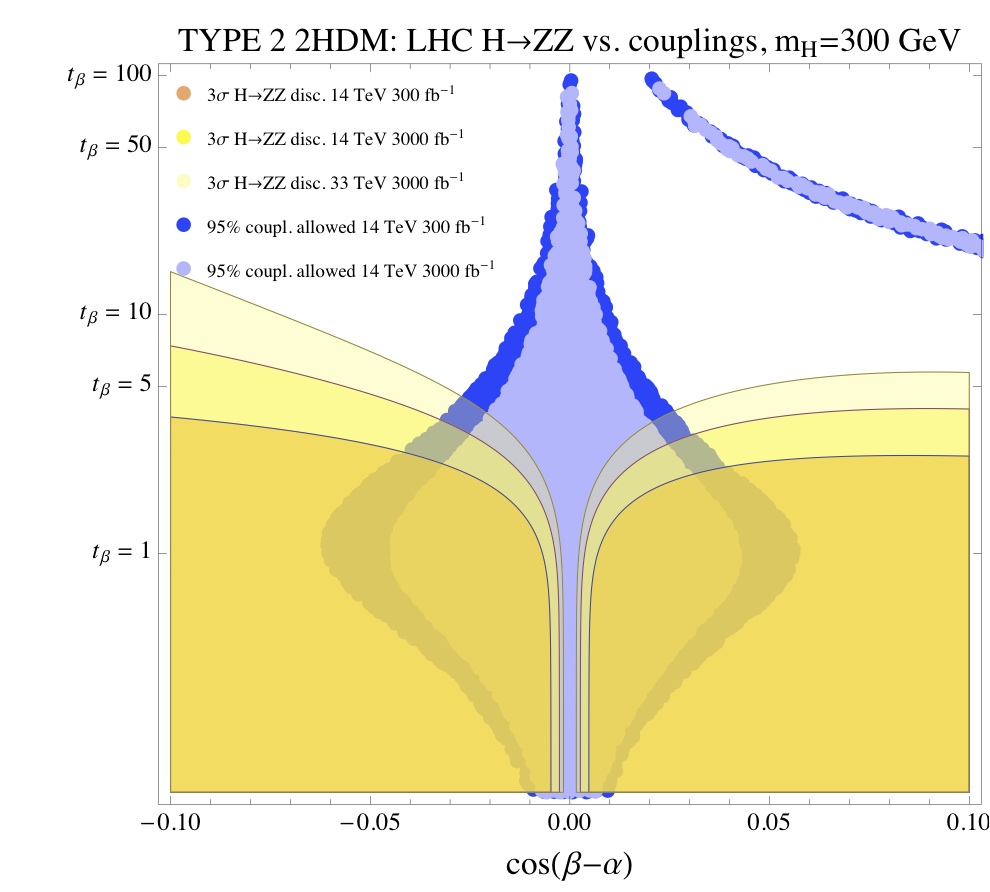}
\includegraphics[width=0.4\columnwidth,height=0.4\textheight,keepaspectratio=true]{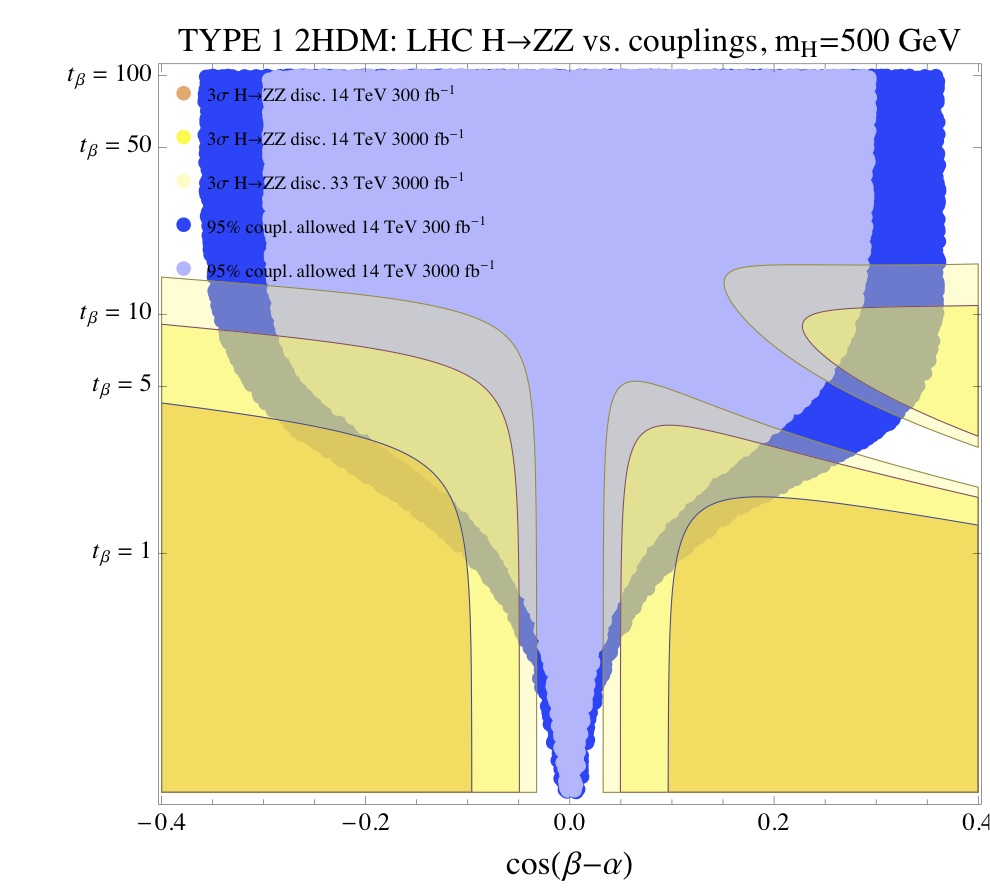}
\includegraphics[width=0.4\columnwidth,height=0.4\textheight,keepaspectratio=true]{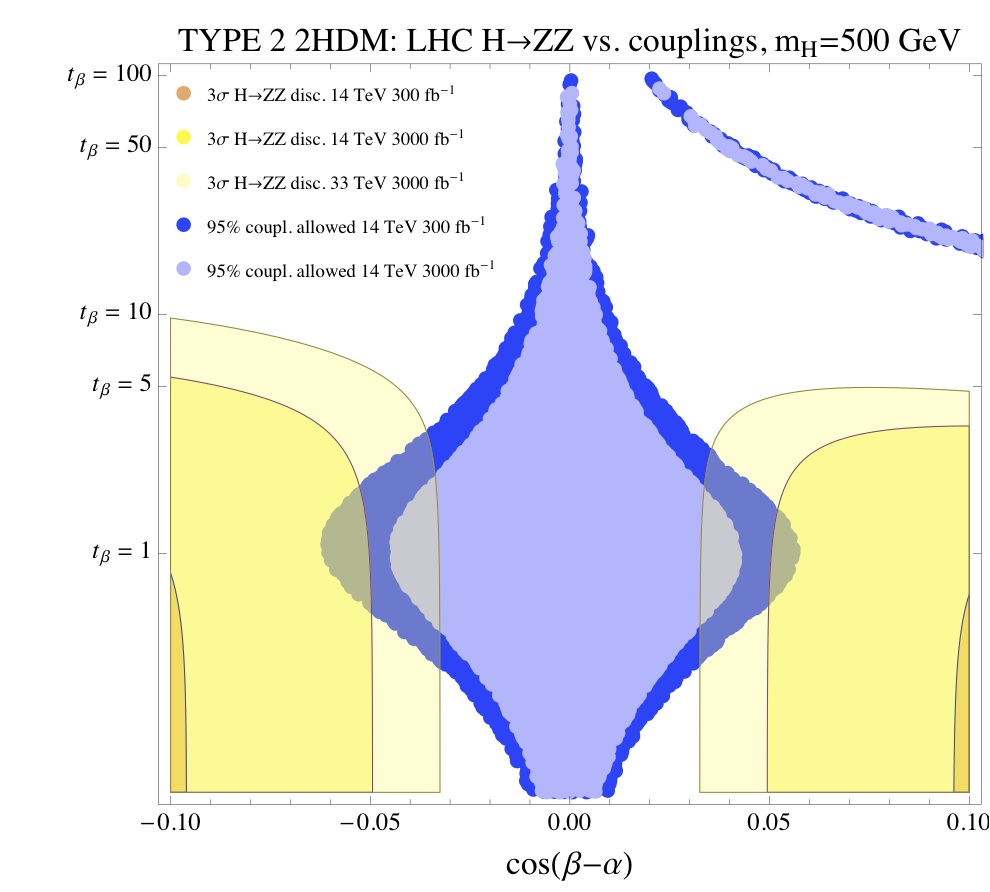}
\includegraphics[width=0.4\columnwidth,height=0.4\textheight,keepaspectratio=true]{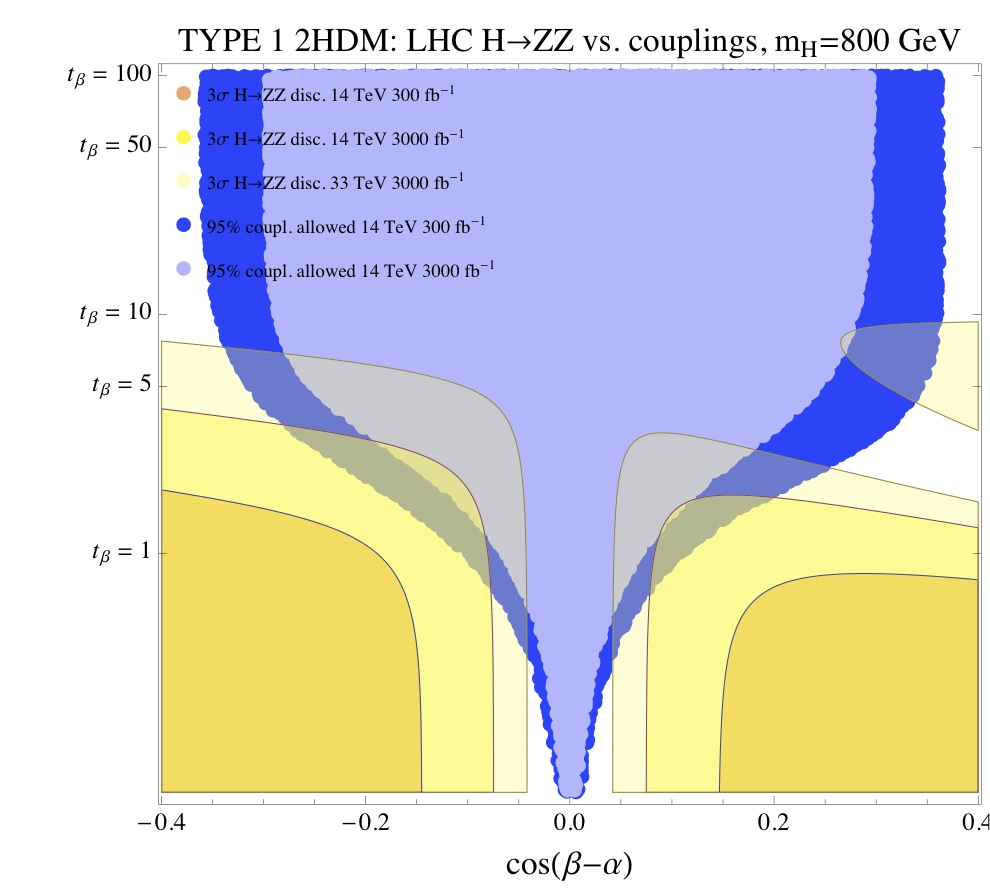}
\includegraphics[width=0.4\columnwidth,height=0.4\textheight,keepaspectratio=true]{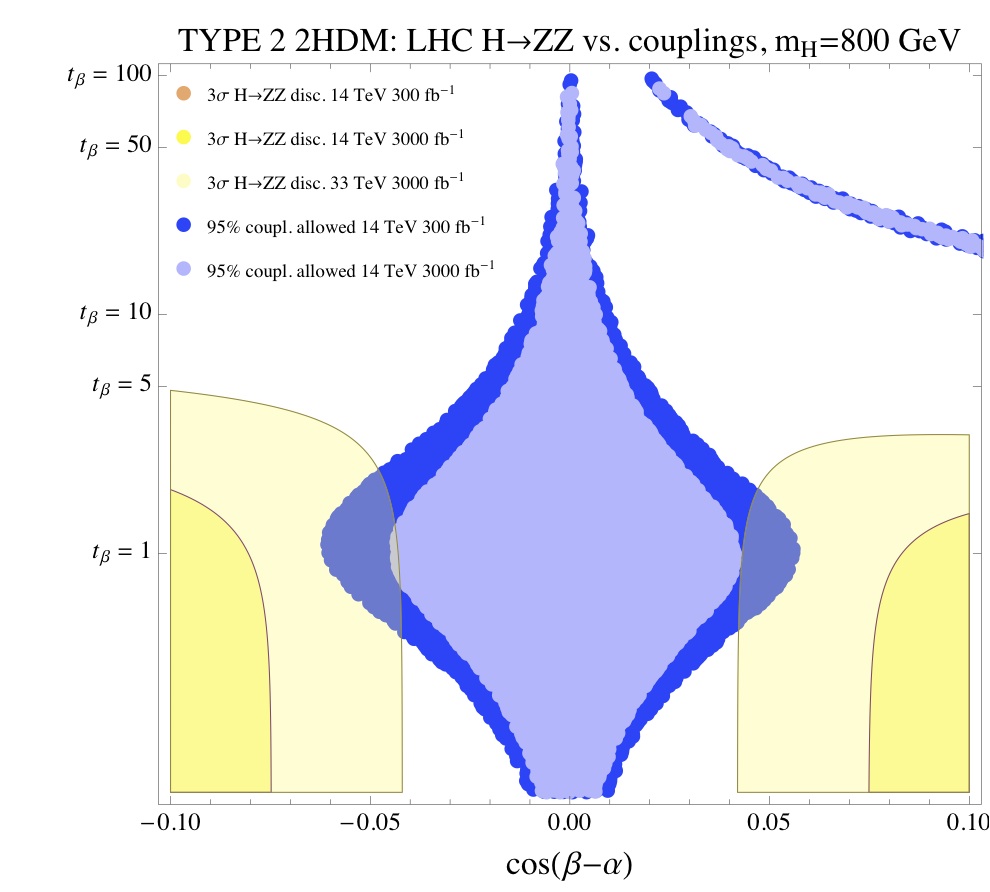}
\caption{The region of parameter space for which a 3$\sigma$ signal significance could be obtained for various $H$ mass hypotheses in a type I (left) and type II (right) 2HDM.  The dark yellow region corresponds to 300~\ifb~at $\sqrt{s}=14$~TeV, the yellow region to 3000~\ifb~at $\sqrt{s}=14$~TeV, and the light yellow region to 3000~\ifb~at $\sqrt{s}=33$~TeV.  The region which would remain allowed at 95\% CL based on non-observation of deviations from the SM in precision Higgs coupling measurements is shown in dark (light) blue for 300~\ifb~(3000~\ifb)~\cite{snowCouplings}.}
\label{fig:HZZObs}
\end{center}
\end{figure}


\begin{figure}[htbp]
\begin{center}
\includegraphics[width=0.4\columnwidth,height=0.4\textheight,keepaspectratio=true]{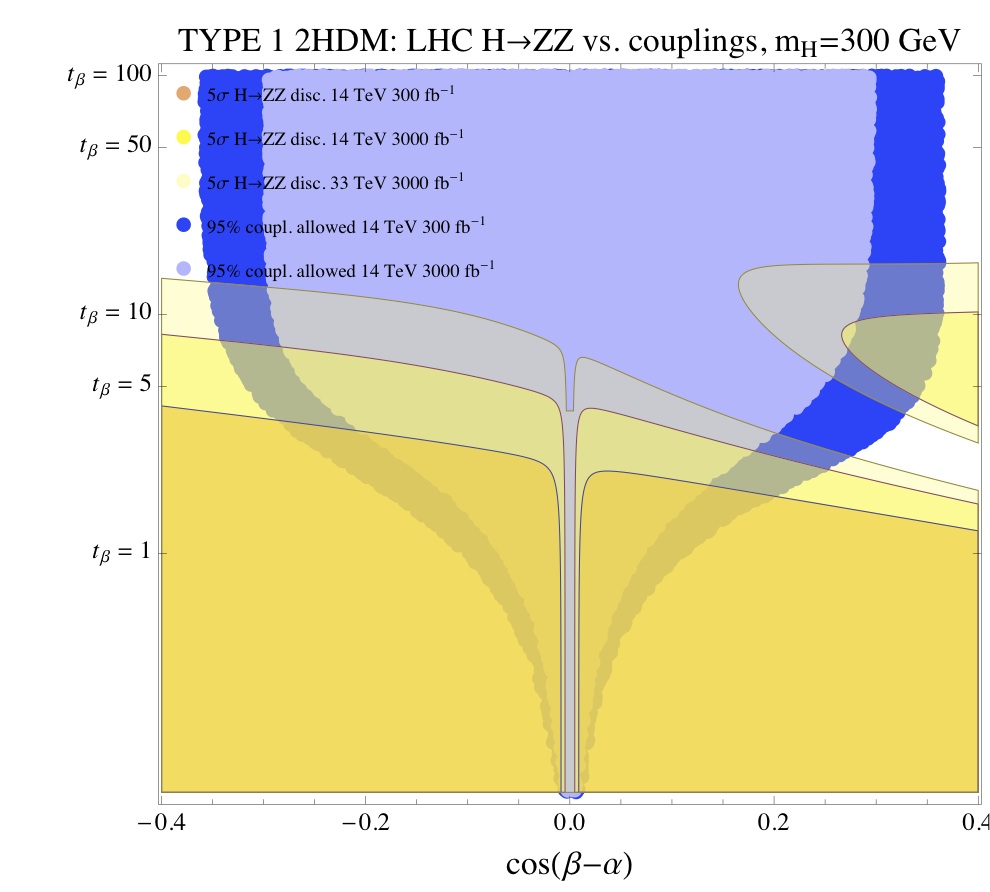}
\includegraphics[width=0.4\columnwidth,height=0.4\textheight,keepaspectratio=true]{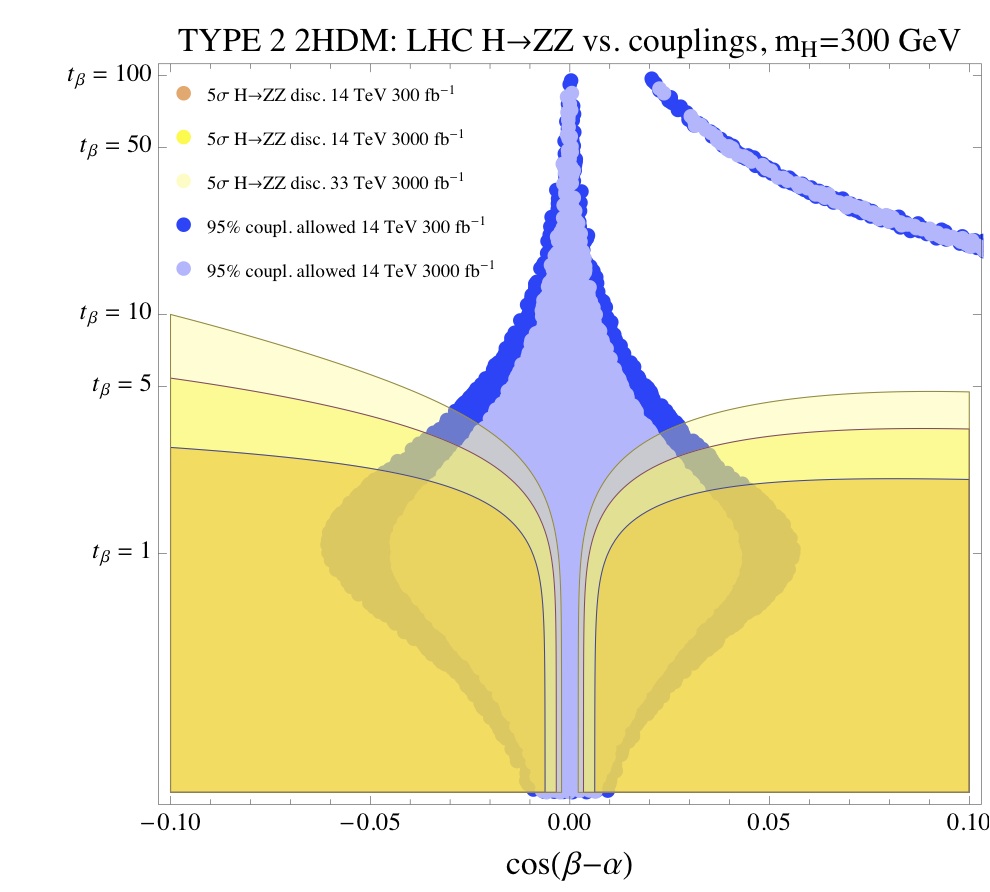}
\includegraphics[width=0.4\columnwidth,height=0.4\textheight,keepaspectratio=true]{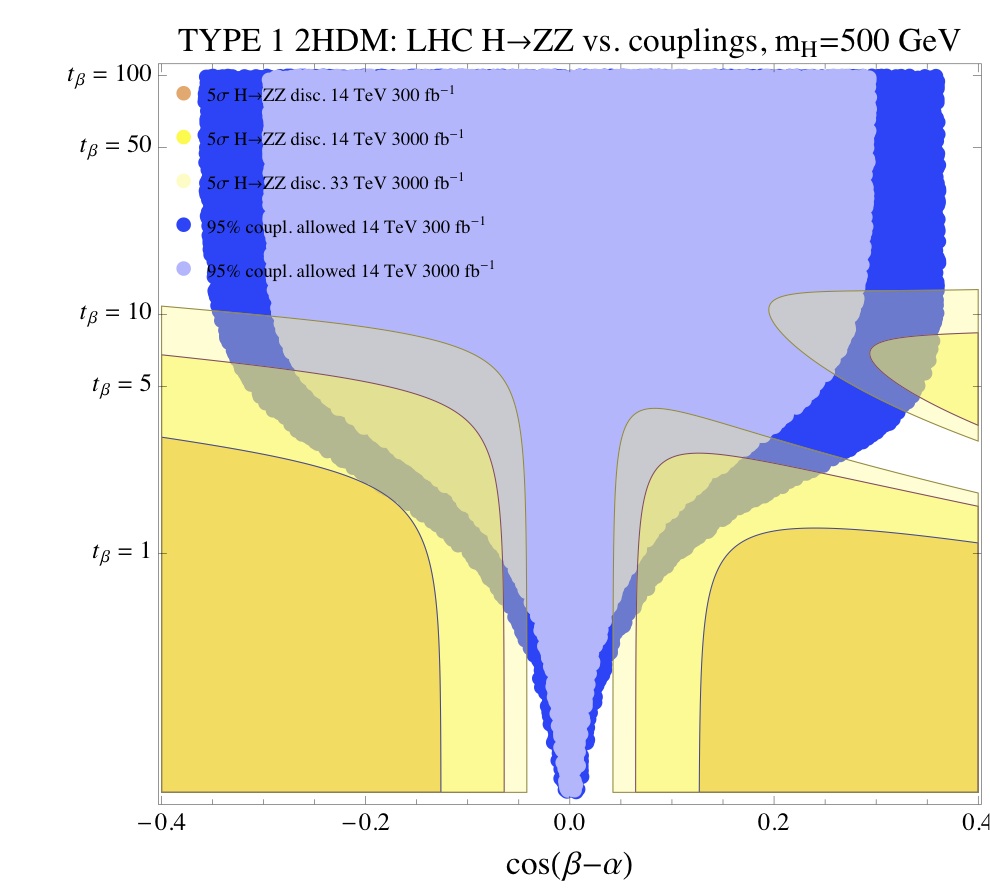}
\includegraphics[width=0.4\columnwidth,height=0.4\textheight,keepaspectratio=true]{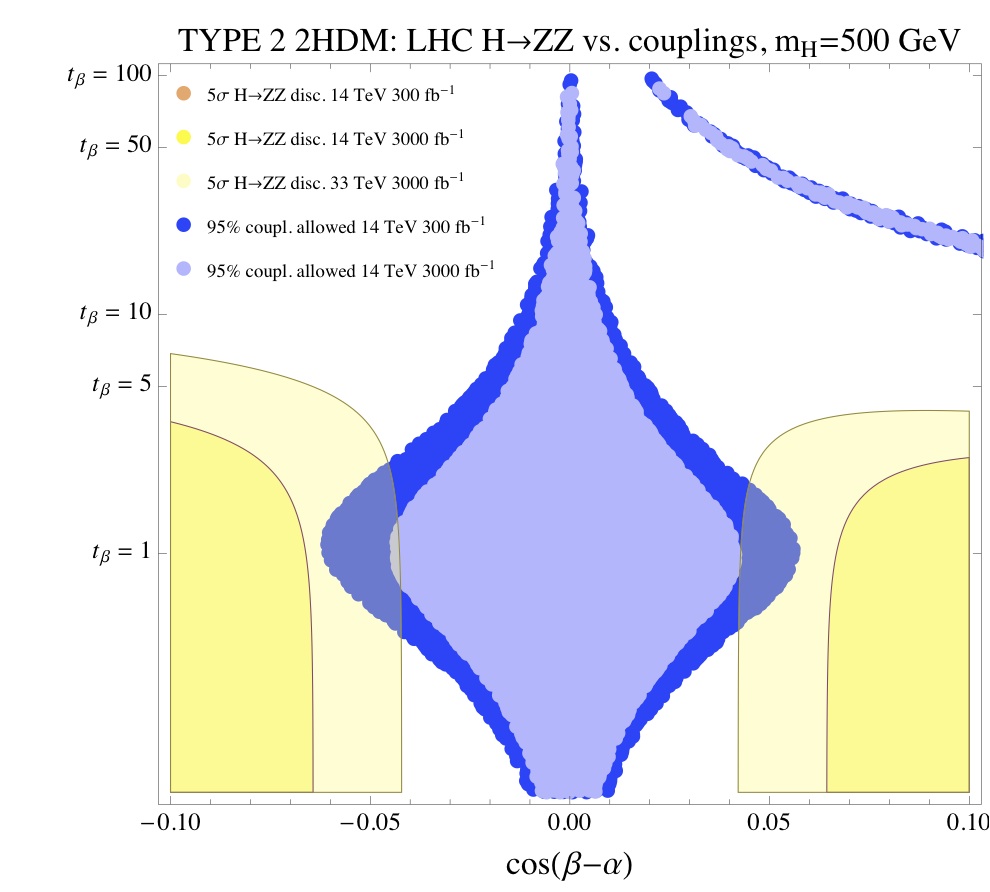}
\includegraphics[width=0.4\columnwidth,height=0.4\textheight,keepaspectratio=true]{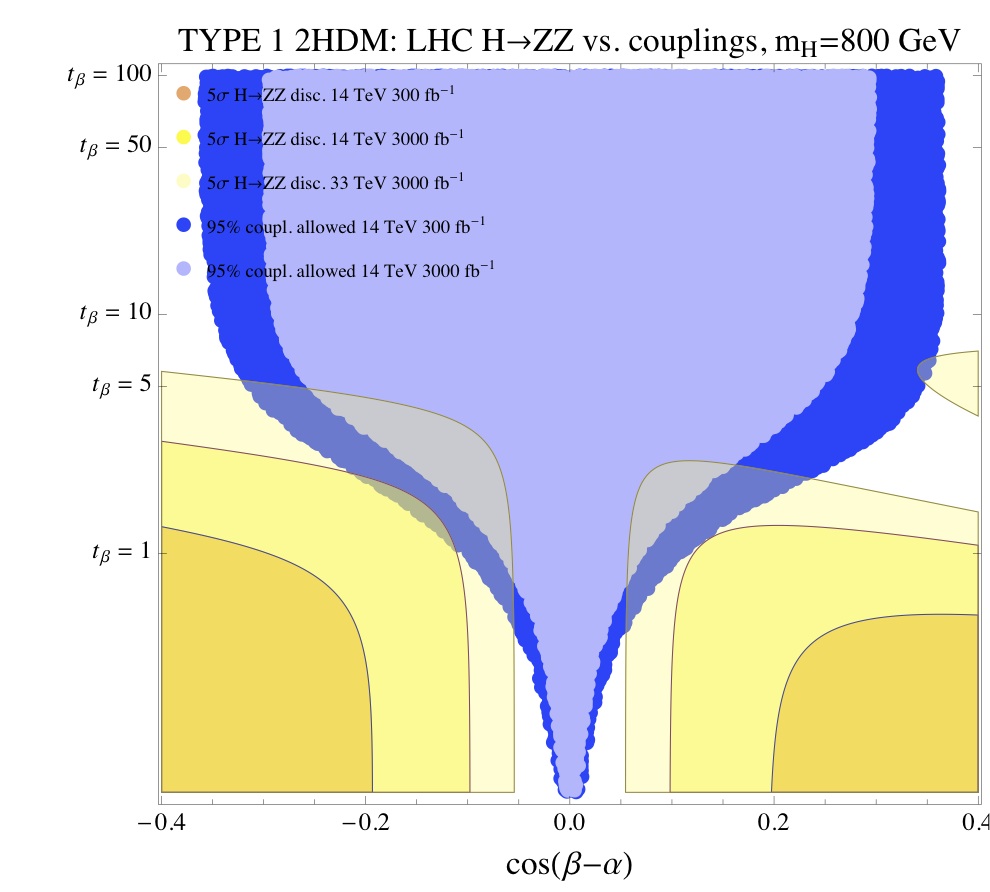}
\includegraphics[width=0.4\columnwidth,height=0.4\textheight,keepaspectratio=true]{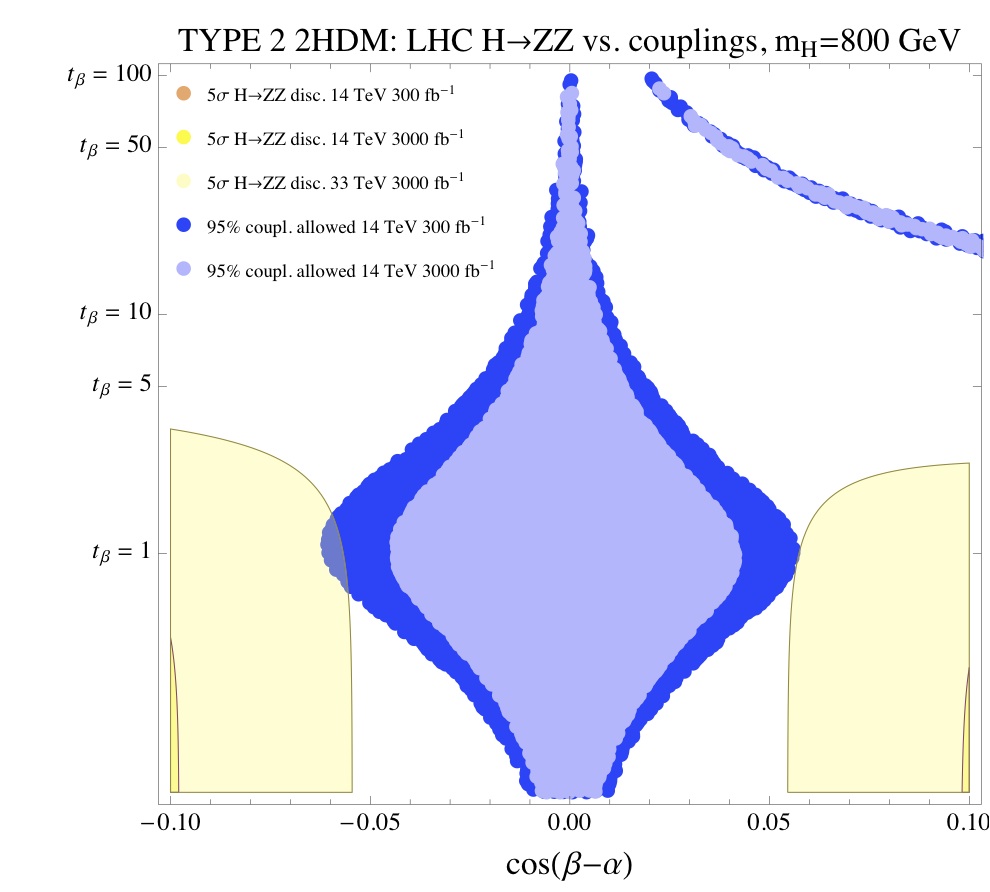}
\caption{The region of parameter space for which a 5$\sigma$ signal significance could be obtained for various $H$ mass hypotheses in a type I (left) and type II (right) 2HDM.  The dark yellow region corresponds to 300~\ifb~at $\sqrt{s}=14$~TeV, the yellow region to 3000~\ifb~at $\sqrt{s}=14$~TeV, and the light yellow region to 3000~\ifb~at $\sqrt{s}=33$~TeV.  The region which would remain allowed at 95\% CL based on non-observation of deviations from the SM in precision Higgs coupling measurements is shown in dark (light) blue for 300~\ifb~(3000~\ifb)~\cite{snowCouplings}.}
\label{fig:HZZDisco}
\end{center}
\end{figure}


\newpage 
\boldmath \subsection[A to Zh]{$A\rightarrow Zh$} \unboldmath
\label{sec:AZh}

A heavy CP-odd pseudo-scalar $A$ cannot decay to $ZZ$ pairs, but often possesses a large width for decay to the distinctive $Zh$ final state.  Therefore, the $A\rightarrow Zh$ decay mode is the most promising channel for exclusion or discovery in a broad region of parameter space, and we choose to focus on this decay mode with subsequent $Z\rightarrow\ell^+\ell^-$ and $h \to b \bar b$ or $h\to\tauptaum$ decays.


The following criteria are applied to leptons:
\begin{itemize}
\item $p_T \geq 5$~GeV
\item $\left| \eta\right| \leq$~2.5
\item Relative isolation $\leq$ 0.1 (after FastJet $\rho\times A$ correction)
\end{itemize}

The following criteria are applied to b-jets and $\tau$ leptons:
\begin{itemize}
\item $p_T \geq 20$~GeV
\item $\left| \eta\right| \leq$~2.5
\end{itemize}

The tight b-tagging working point~\cite{snowmassDet} is used to identify jets originating from $b$ quarks.  This working point assumes a $p_T$- and $\eta$-dependent b-tagging efficiency, which peaks at 70\% (60\%) for high-$p_T$ jets with $\left|\eta\right|\leq 1.2$ ($\left|\eta\right|>$1.2).  The mistag rate for $c$ quarks is similarly $p_T$- and $\eta$-dependent, peaking at $\sim$19\%.  The mistag rate for light quarks and gluons is a flat 0.01\%.  The $\tau$-tagging working point assumes a flat 65\% (0.4\%) efficiency (mis-tag rate).

A pre-selection is applied to select events with a topology consistent with signal events.  Events are required to contain exactly two leptons.  In order to fire a leptonic trigger, it is assumed that the leading lepton must satisfy $p_T\geq$~30~GeV, or the leading lepton must satisfy $p_T\geq$~20~GeV and the sub-leading lepton must satisfy $p_T\geq$~10~GeV.  In the $bb$ channel, events are required to contain exactly two b-jets and fewer than two $\tau$ leptons (to maintain orthogonality with the \tautau~channel).  Similarly, in the \tautau~channel, events are required to contain exactly two $\tau$ leptons and fewer than two b-jets.  In both channels, events are required to contain exactly one $Z$ boson candidate, constructed from an OS SF lepton pair with an invariant mass satisfying 80 $\leq m_{\ell\ell}\leq$ 100.  Events must also contain exactly one SM Higgs candidate.  In the $bb$ channel, Higgs candidates are constructed from a $bb$ pair with an invariant mass satisfying $90\leq m_{bb}\leq 150$.  In the \tautau~channel, Higgs candidates are constructed from a $\tau\tau$ pair with a visible mass satisfying $55\leq m_{\tautau}\leq 125$.  Finally, events must contain an $A$ candidate, constructed from the $Z$ and $h$ candidates, with an invariant mass $m_A\geq$~150~GeV.  Pre-selected event yields for 300~\ifb at $\sqrt{s}=14$~TeV (3000~\ifb at $\sqrt{s}=33$~TeV) in the $bb$ channel are shown in Tables~\ref{tab:AZhbbPreselCutflowSignal} and \ref{tab:AZhbbPreselCutflowBackground} (\ref{tab:AZhbbPreselCutflowSignal_33} and \ref{tab:AZhbbPreselCutflowBackground_33}).  Analogous tables for the \tautau~channel are shown in Tables~\ref{tab:AZhtautauPreselCutflowSignal} and \ref{tab:AZhtautauPreselCutflowBackground} (\ref{tab:AZhtautauPreselCutflowSignal_33} and \ref{tab:AZhtautauPreselCutflowBackground_33}).  Several kinematic distributions of interest for the 300~\ifb analysis at $\sqrt{s}=14$~TeV are shown in Figures~\ref{fig:AZhbbPresel} and \ref{fig:AZhtautauPresel} for the $bb$ and \tautau~channels, respectively.  Analogous plots for the 3000~\ifb analysis at $\sqrt{s}=33$~TeV are shown in Figures~\ref{fig:AZhbbPresel_33} and \ref{fig:AZhtautauPresel_33}.


\begin{table}[htbp]
\begin{center}
\begin{footnotesize}
\begin{tabular}{|l|c|c|c|c|c|c|c|}
\hline
Signal Mass [GeV] & $N_{lepton}=2$ & Lepton Trigger & $N_{b}=2$ & $N_{\tau}<2$ & $N_Z=1$ & $N_h=1$ & $N_A=1$ \\ \hline
250        & 3.15e+4        & 3.14e+4        & 2.47e+3        & 2.47e+3        & 2.38e+3        & 1.73e+3        & 1.73e+3        \\
300        & 6.6e+4         & 6.59e+4        & 6.24e+3        & 6.24e+3        & 6e+3           & 4.3e+3         & 4.3e+3         \\
350        & 1.2e+3         & 1.2e+3         & 136             & 136             & 131             & 96.2            & 96.2            \\
400        & 604             & 604             & 86.6            & 86.6            & 83.8            & 61.9            & 61.9            \\
450        & 433             & 433             & 68.4            & 68.4            & 65.9            & 50.3            & 50.3            \\
500        & 325             & 325             & 56.5            & 56.5            & 54.7            & 42.5            & 42.5            \\
600        & 195             & 195             & 40.9            & 40.9            & 39.3            & 31.8            & 31.8            \\
700        & 121             & 121             & 28              & 28              & 26.9            & 22              & 22              \\
800        & 77.5            & 77.5            & 18.7            & 18.7            & 18              & 14.8            & 14.8            \\
900        & 50.5            & 50.5            & 12.3            & 12.3            & 11.9            & 9.96            & 9.96            \\
1000       & 33.9            & 33.8            & 7.55            & 7.55            & 7.31            & 5.98            & 5.98            \\ \hline
\end{tabular}
\end{footnotesize}
\end{center}
\caption{Expected number of pre-selected events for the $A\rightarrow Zh\rightarrow \ell \ell bb$ signal for $\int Ldt=$ 300~\ifb~at $\sqrt{s}=14$ TeV with $<N_{PU}>=50$.}
\label{tab:AZhbbPreselCutflowSignal}
\end{table}


\begin{table}[htbp]
\begin{center}
\begin{footnotesize}
\begin{tabular}{|l|c|c|c|c|c|c|c|}
\hline
Background & $N_{lepton}=2$ & Lepton Trigger & $N_{b}=2$ & $N_{\tau}<2$ & $N_Z=1$ & $N_h=1$ & $N_A=1$ \\ \hline
B, Bj, Bjj-vbf, BB, BBB                            & 4.58e+8        & 4.05e+8        & 1.15e+5        & 1.15e+5        & 9.91e+4        & 2.66e+4        & 2.66e+4        \\
tj, tB, tt, ttB                                    & 7.53e+6        & 7.36e+6        & 8.31e+5        & 8.31e+5        & 5.69e+4        & 1.51e+4        & 1.51e+4        \\
H                                                  & 4.99e+4        & 4.48e+4        & 44.5            & 44.5            & 6.37            & 1.58            & 1.58            \\ \hline
Total Background                                   & 4.65e+8        & 4.13e+8        & 9.47e+5        & 9.47e+5        & 1.56e+5        & 4.16e+4        & 4.16e+4        \\ \hline
\end{tabular}
\end{footnotesize}
\end{center}
\caption{Expected number of pre-selected events for the SM backgrounds to $A\rightarrow Zh\rightarrow \ell \ell bb$ for $\int Ldt=$ 300~\ifb~at $\sqrt{s}=14$ TeV with $<N_{PU}>=50$.}
\label{tab:AZhbbPreselCutflowBackground}
\end{table}


\begin{table}[htbp]
\begin{center}
\begin{footnotesize}
\begin{tabular}{|l|c|c|c|c|c|c|c|}
\hline
Signal Mass [GeV] & $N_{lepton}=2$ & Lepton Trigger & $N_{b}=2$ & $N_{\tau}<2$ & $N_Z=1$ & $N_h=1$ & $N_A=1$ \\ \hline
250       & 1.29e+6        & 1.29e+6        & 5.44e+4        & 5.43e+4        & 5.07e+4        & 2.5e+4         & 2.5e+4         \\
300       & 2.89e+6        & 2.89e+6        & 1.74e+5        & 1.74e+5        & 1.65e+5        & 8.47e+4        & 8.47e+4        \\
350       & 5.64e+4        & 5.63e+4        & 4.31e+3        & 4.31e+3        & 4.1e+3         & 2.04e+3        & 2.04e+3        \\
400       & 3.02e+4        & 3.02e+4        & 2.84e+3        & 2.84e+3        & 2.7e+3         & 1.38e+3        & 1.38e+3        \\
450       & 2.3e+4         & 2.3e+4         & 2.56e+3        & 2.56e+3        & 2.46e+3        & 1.28e+3        & 1.28e+3        \\
500       & 1.85e+4        & 1.85e+4        & 2.39e+3        & 2.39e+3        & 2.29e+3        & 1.25e+3        & 1.25e+3        \\
600       & 1.23e+4        & 1.23e+4        & 1.98e+3        & 1.98e+3        & 1.9e+3         & 1.07e+3        & 1.07e+3        \\
700       & 8.53e+3        & 8.52e+3        & 1.57e+3        & 1.57e+3        & 1.5e+3         & 893             & 893             \\
800       & 5.98e+3        & 5.98e+3        & 1.14e+3        & 1.14e+3        & 1.1e+3         & 662             & 662             \\
900       & 4.31e+3        & 4.31e+3        & 850             & 850             & 816             & 502             & 502             \\
1000      & 3.11e+3        & 3.11e+3        & 569             & 568             & 546             & 325             & 325             \\ \hline
\end{tabular}
\end{footnotesize}
\end{center}
\caption{Expected number of pre-selected events for the $A\rightarrow Zh\rightarrow \ell \ell bb$ signal for $\int Ldt=$ 3000~\ifb~at $\sqrt{s}=33$ TeV with $<N_{PU}>=$~140.}
\label{tab:AZhbbPreselCutflowSignal_33}
\end{table}

\begin{table}[htbp]
\begin{center}
\begin{footnotesize}
\begin{tabular}{|l|c|c|c|c|c|c|c|}
\hline
Background & $N_{lepton}=2$ & Lepton Trigger & $N_{b}=2$ & $N_{\tau}<2$ & $N_Z=1$ & $N_h=1$ & $N_A=1$ \\ \hline
B, Bj, Bjj-vbf, BB, BBB                            & 8.66e+9        & 7.68e+9        & 4.13e+6        & 4.13e+6        & 3.34e+6        & 6.68e+5        & 6.68e+5         \\
tj, tB, tt, ttB                                    & 5.06e+8        & 4.86e+8        & 3.61e+7        & 3.61e+7        & 2.39e+6        & 5.81e+5        & 5.81e+5         \\
H                                                  & 2.21e+6        & 1.97e+6        & 2.54e+3        & 2.54e+3        & 370             & 103             & 103              \\ \hline
Total Background                                   & 9.17e+9        & 8.17e+9        & 4.03e+7        & 4.03e+7        & 5.73e+6        & 1.25e+6        & 1.25e+6         \\ \hline
\end{tabular}
\end{footnotesize}
\end{center}
\caption{Expected number of pre-selected events for the SM backgrounds to $A\rightarrow Zh\rightarrow \ell \ell bb$ for $\int Ldt=$ 3000~\ifb~at $\sqrt{s}=33$ TeV with $<N_{PU}>=$~140.}
\label{tab:AZhbbPreselCutflowBackground_33}
\end{table}


\begin{figure}[htbp]
\begin{center}
\includegraphics[width=0.4\columnwidth,height=0.4\textheight,keepaspectratio=true]{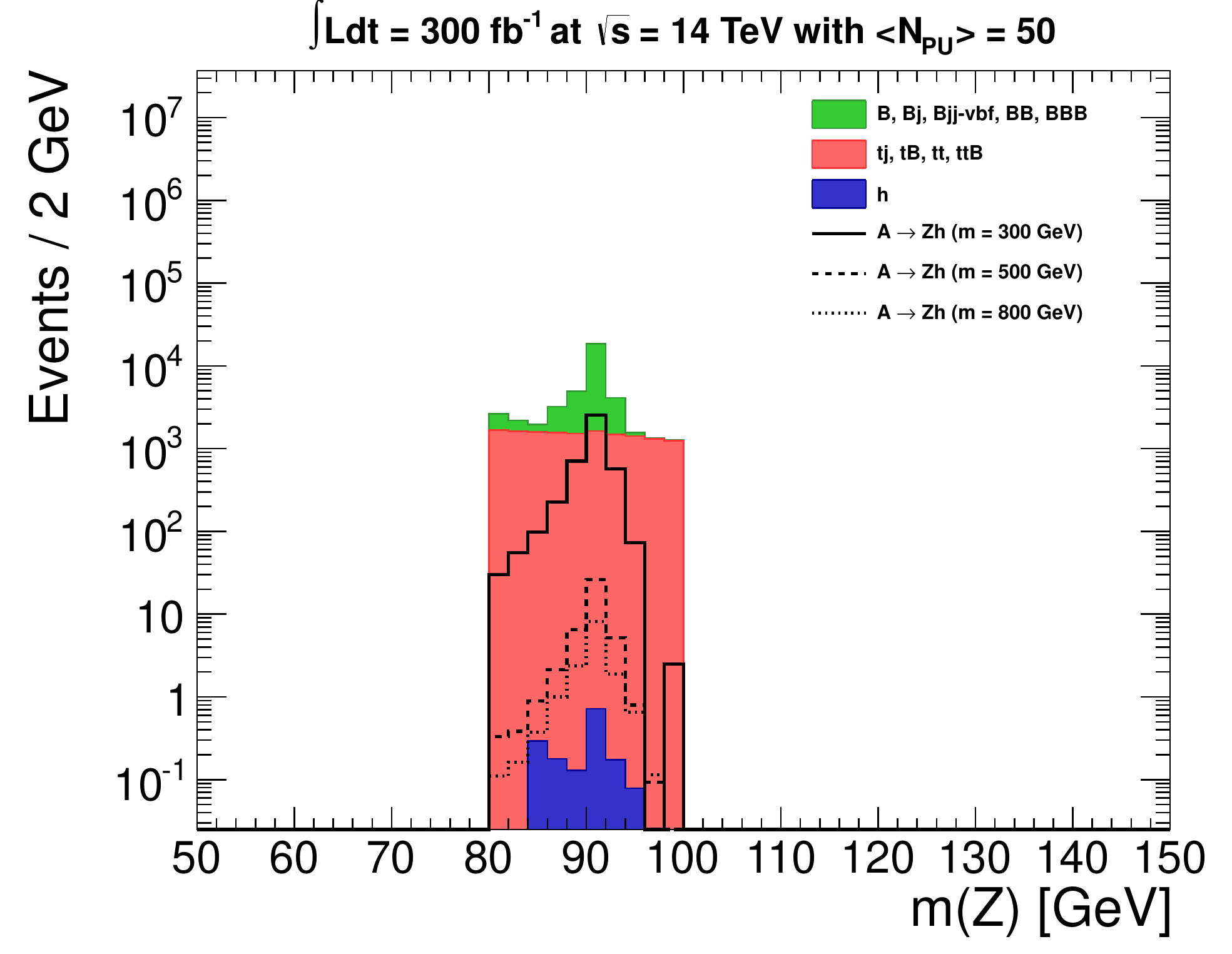}
\includegraphics[width=0.4\columnwidth,height=0.4\textheight,keepaspectratio=true]{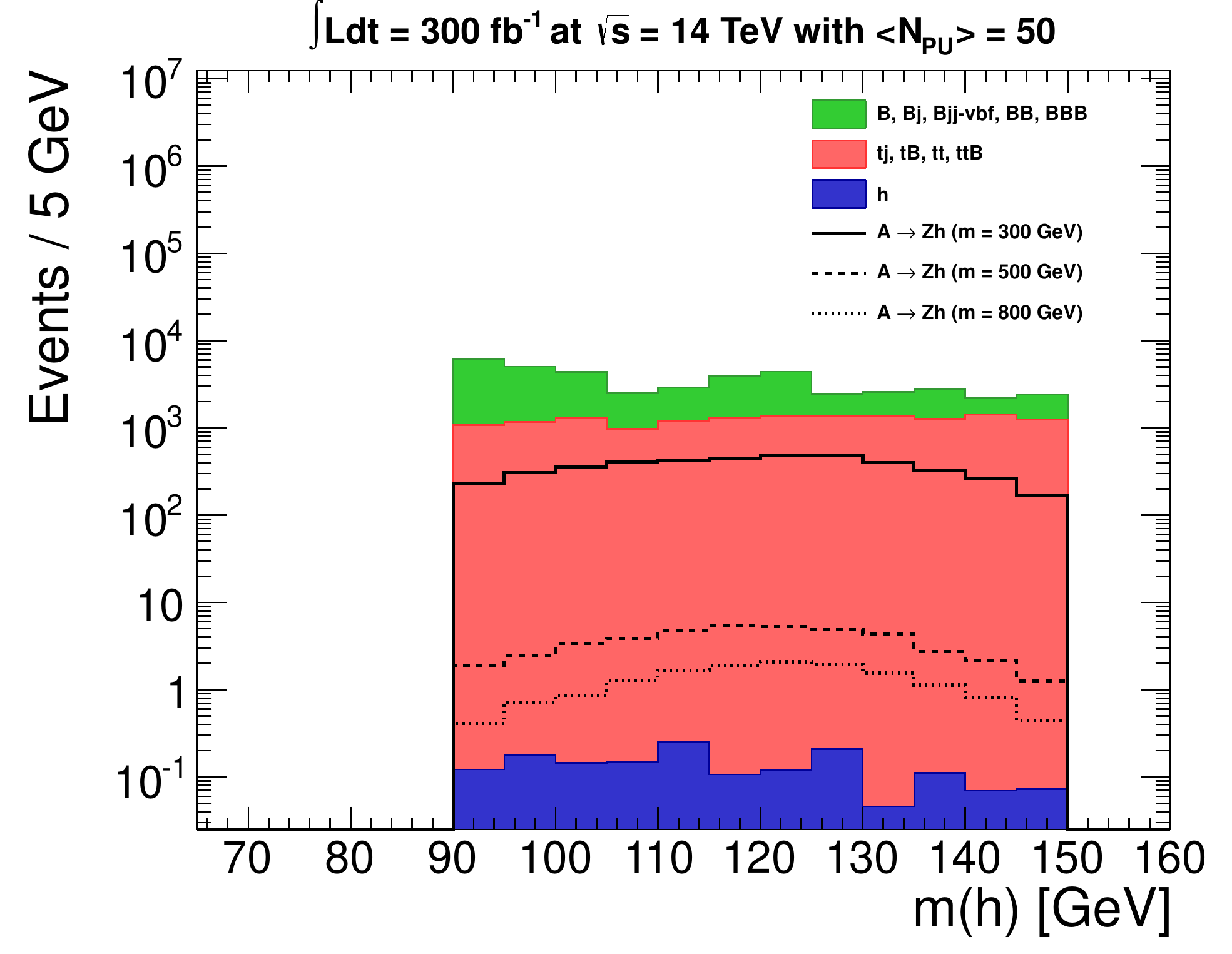}
\includegraphics[width=0.4\columnwidth,height=0.4\textheight,keepaspectratio=true]{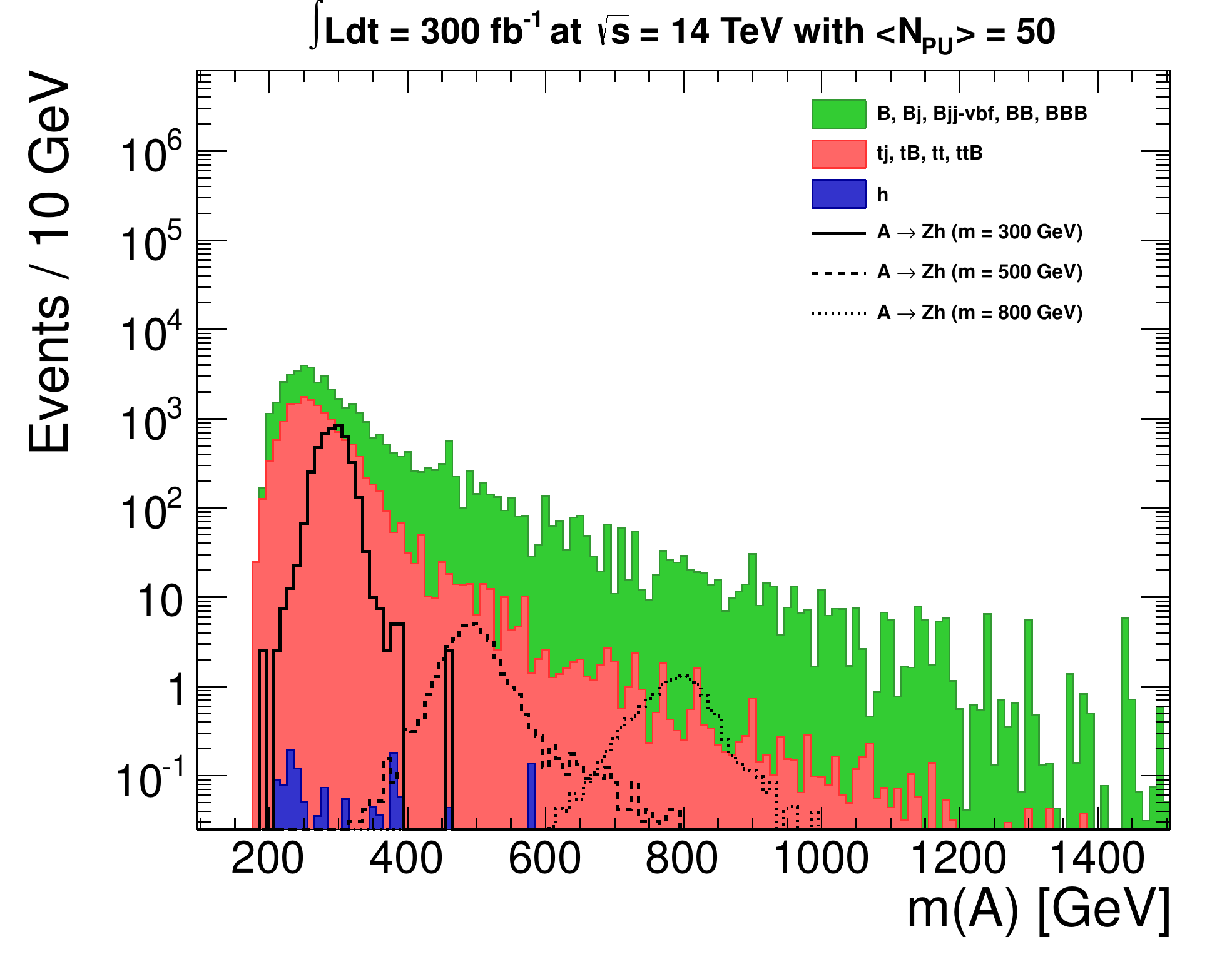}
\includegraphics[width=0.4\columnwidth,height=0.4\textheight,keepaspectratio=true]{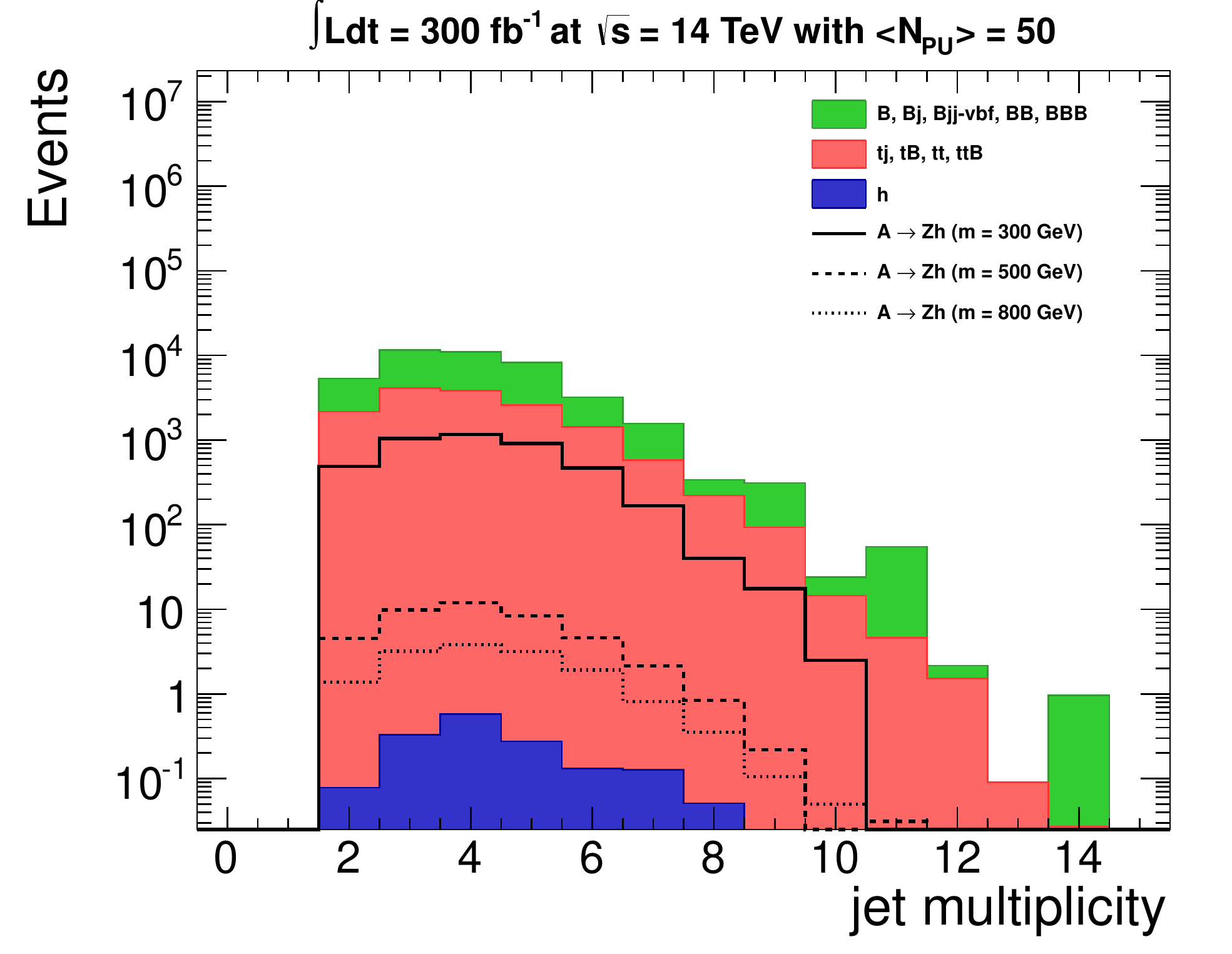}
\includegraphics[width=0.4\columnwidth,height=0.4\textheight,keepaspectratio=true]{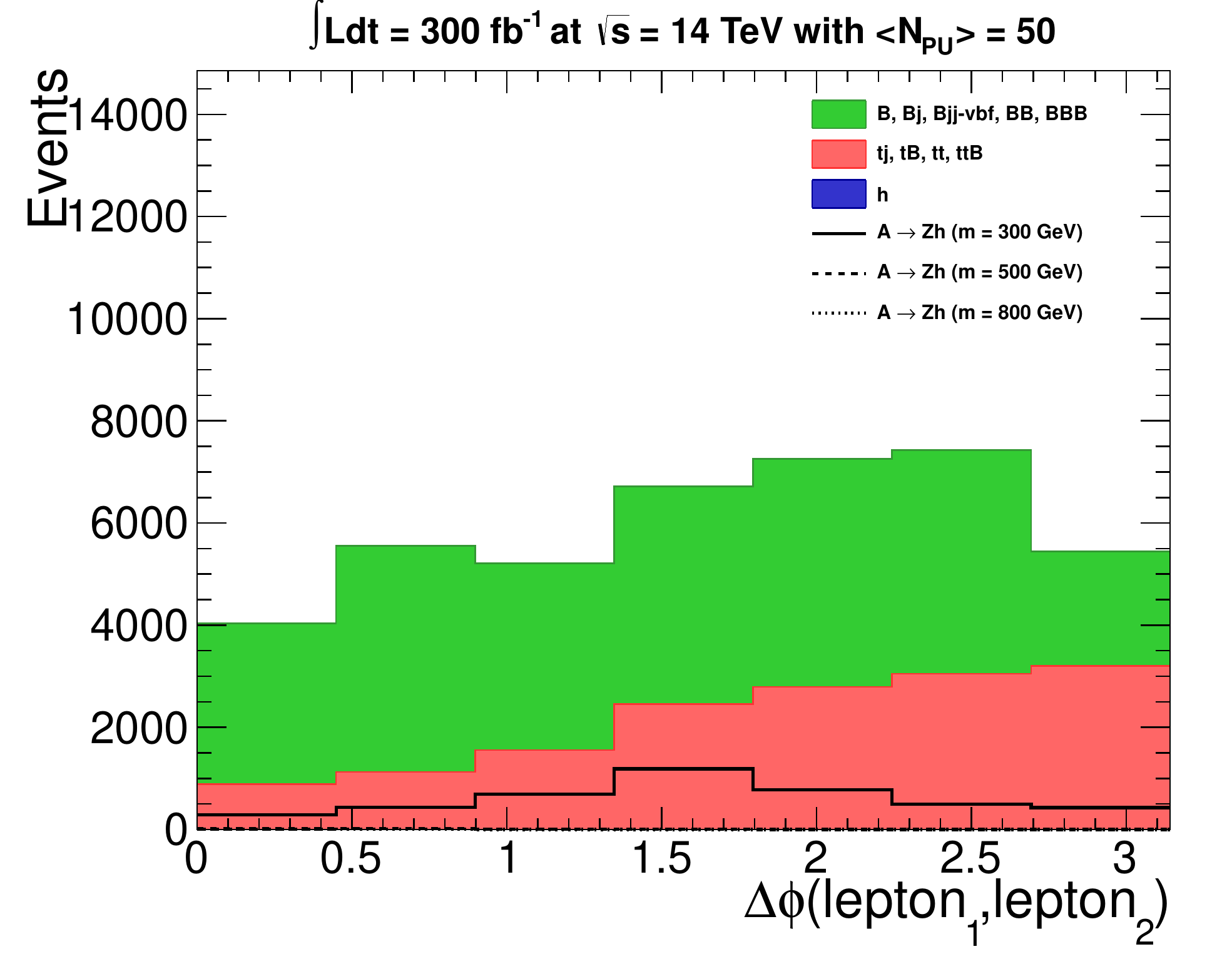}
\includegraphics[width=0.4\columnwidth,height=0.4\textheight,keepaspectratio=true]{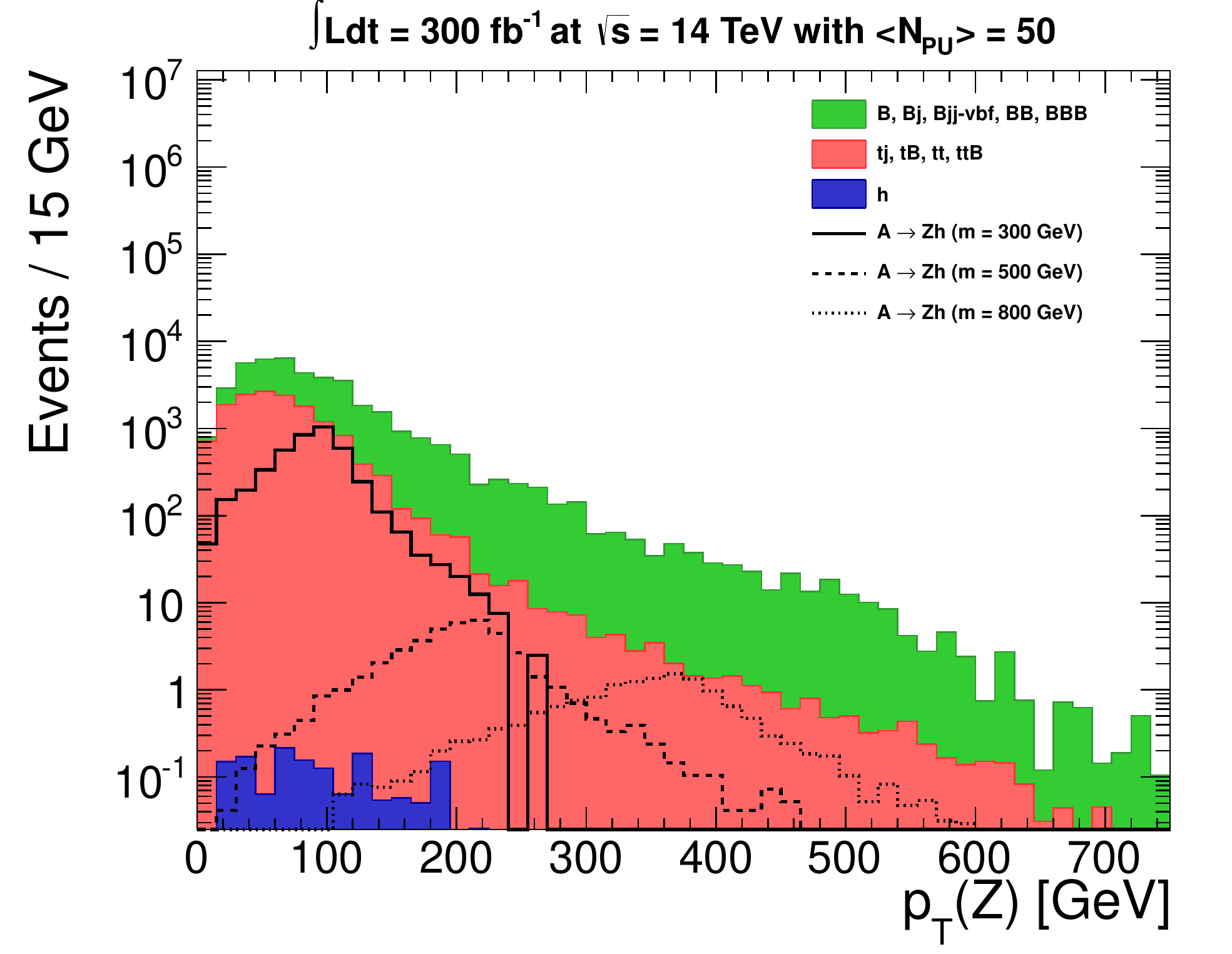}
\includegraphics[width=0.4\columnwidth,height=0.4\textheight,keepaspectratio=true]{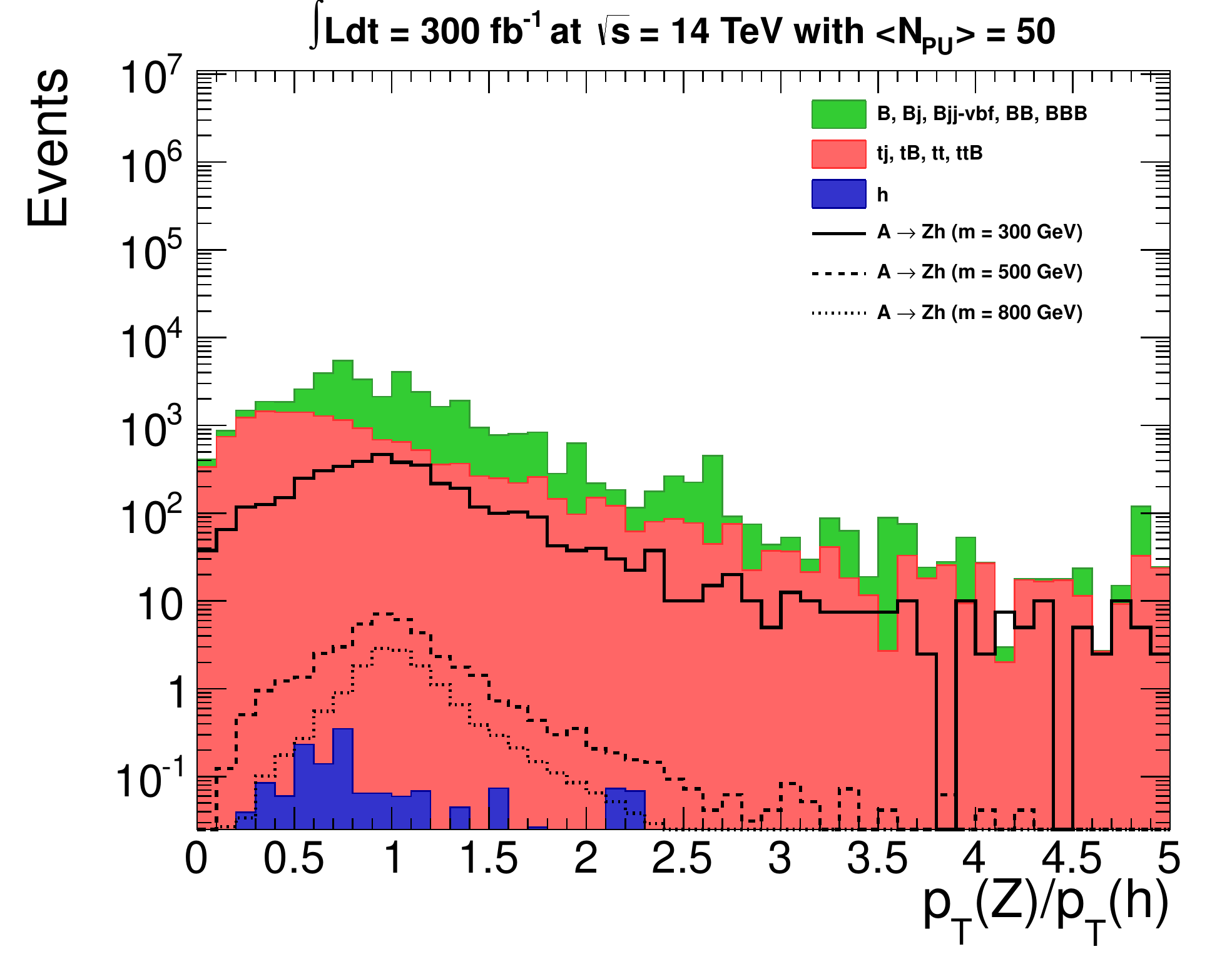}
\includegraphics[width=0.4\columnwidth,height=0.4\textheight,keepaspectratio=true]{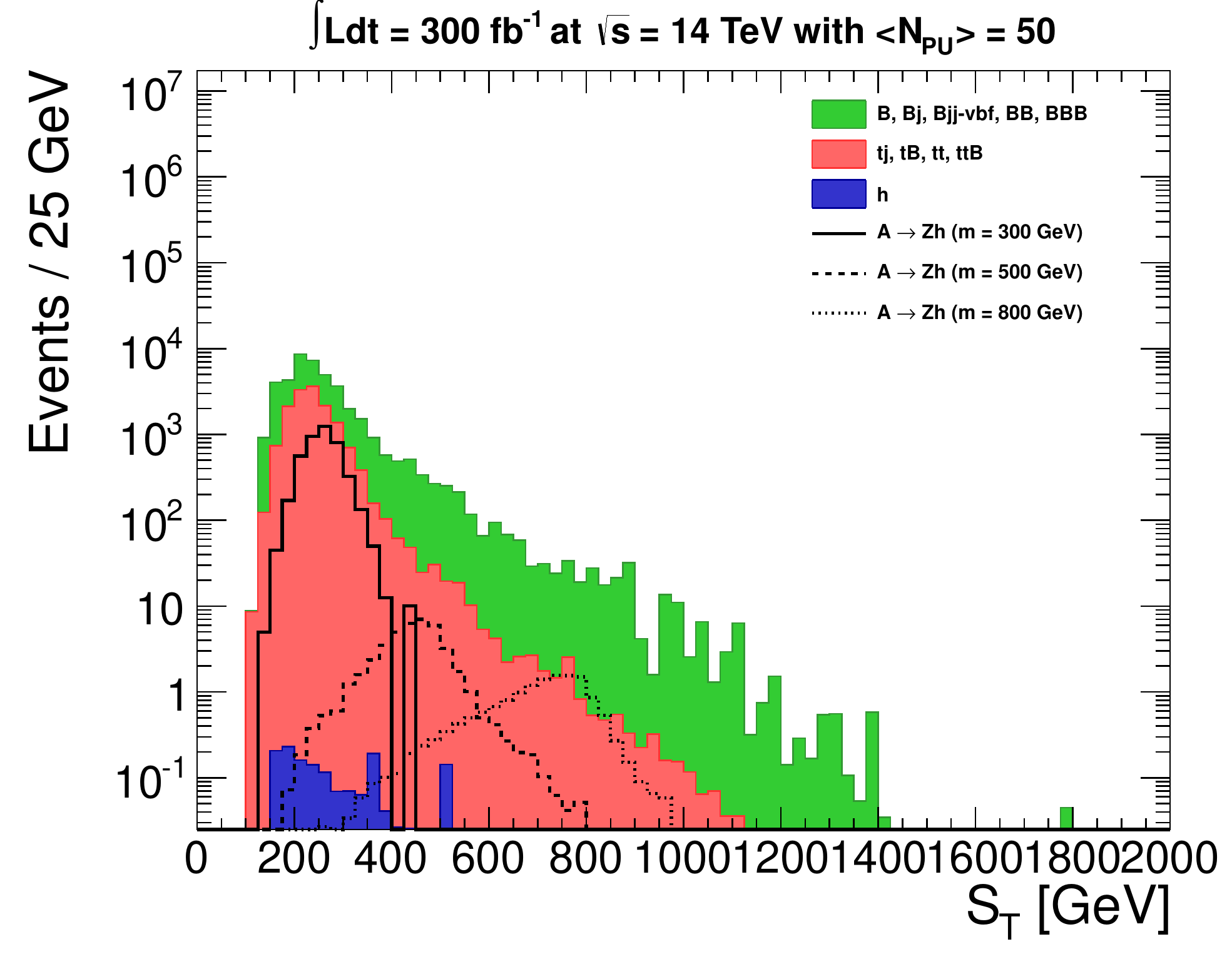}
\caption{Kinematic distributions for pre-selected events in the $bb$ channel, for $\int Ldt=$ 300~\ifb~at $\sqrt{s}=14$ TeV with $<N_{PU}>=50$.}
\label{fig:AZhbbPresel}
\end{center}
\end{figure}


\begin{figure}[htbp]
\begin{center}
\includegraphics[width=0.4\columnwidth,height=0.4\textheight,keepaspectratio=true]{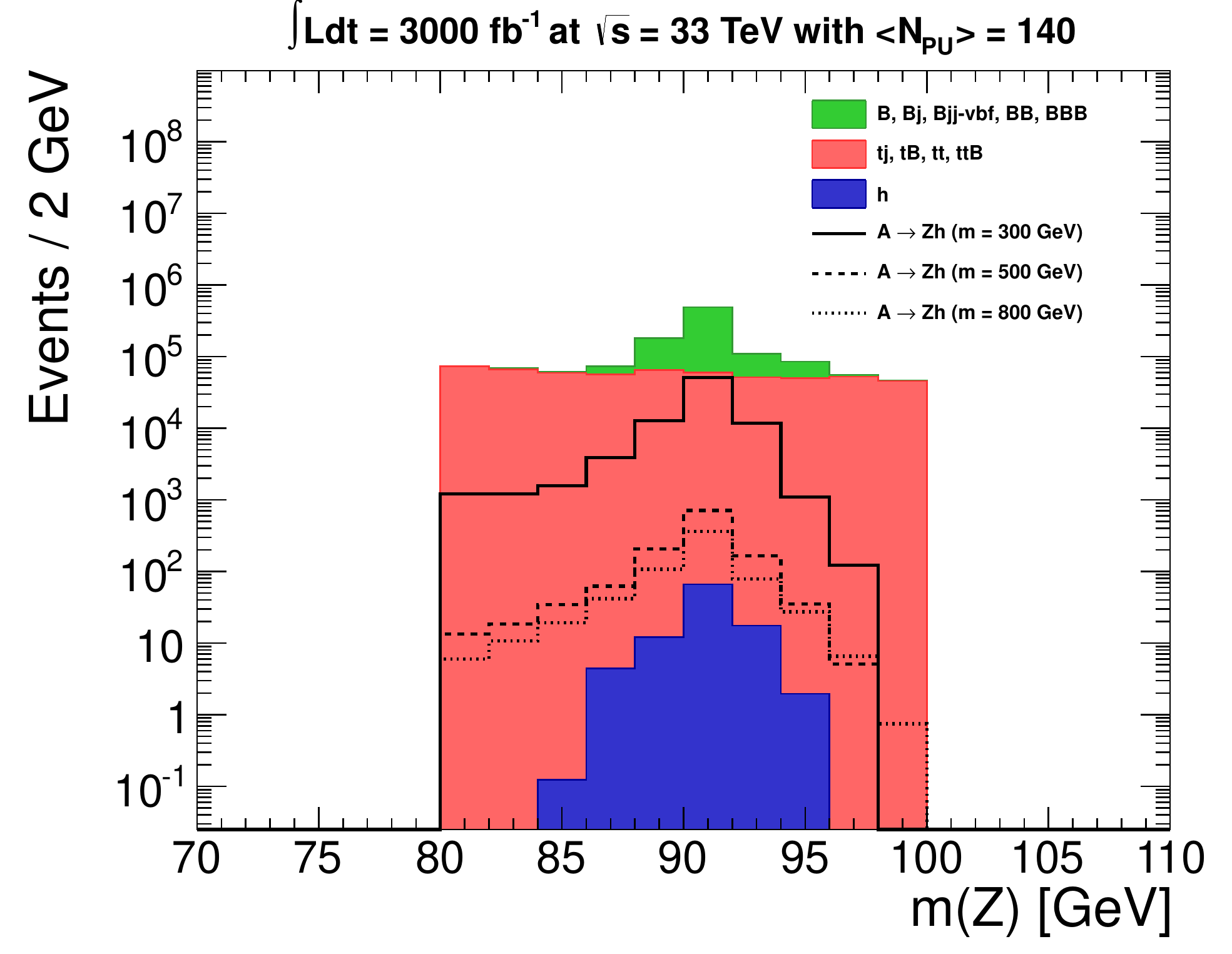}
\includegraphics[width=0.4\columnwidth,height=0.4\textheight,keepaspectratio=true]{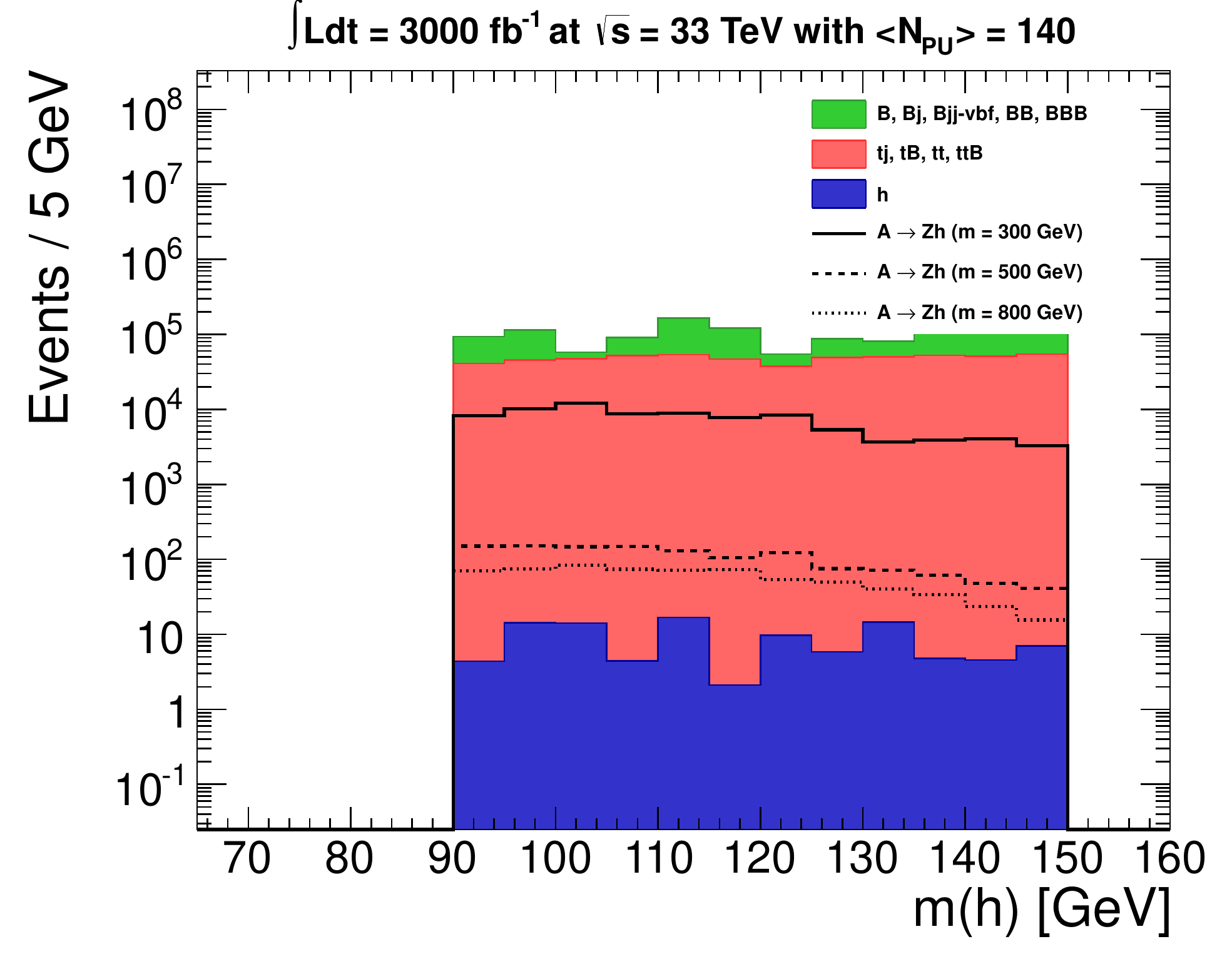}
\includegraphics[width=0.4\columnwidth,height=0.4\textheight,keepaspectratio=true]{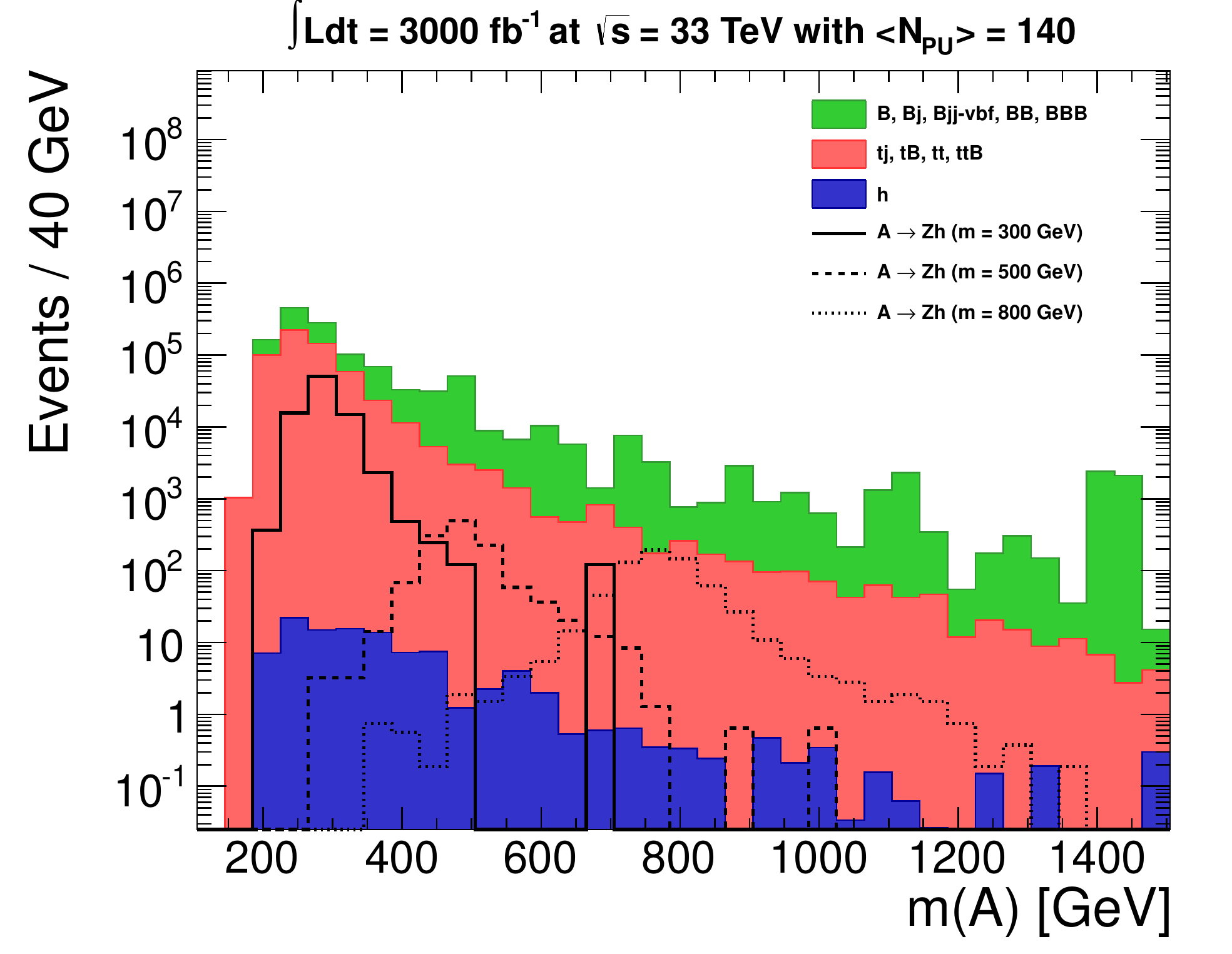}
\includegraphics[width=0.4\columnwidth,height=0.4\textheight,keepaspectratio=true]{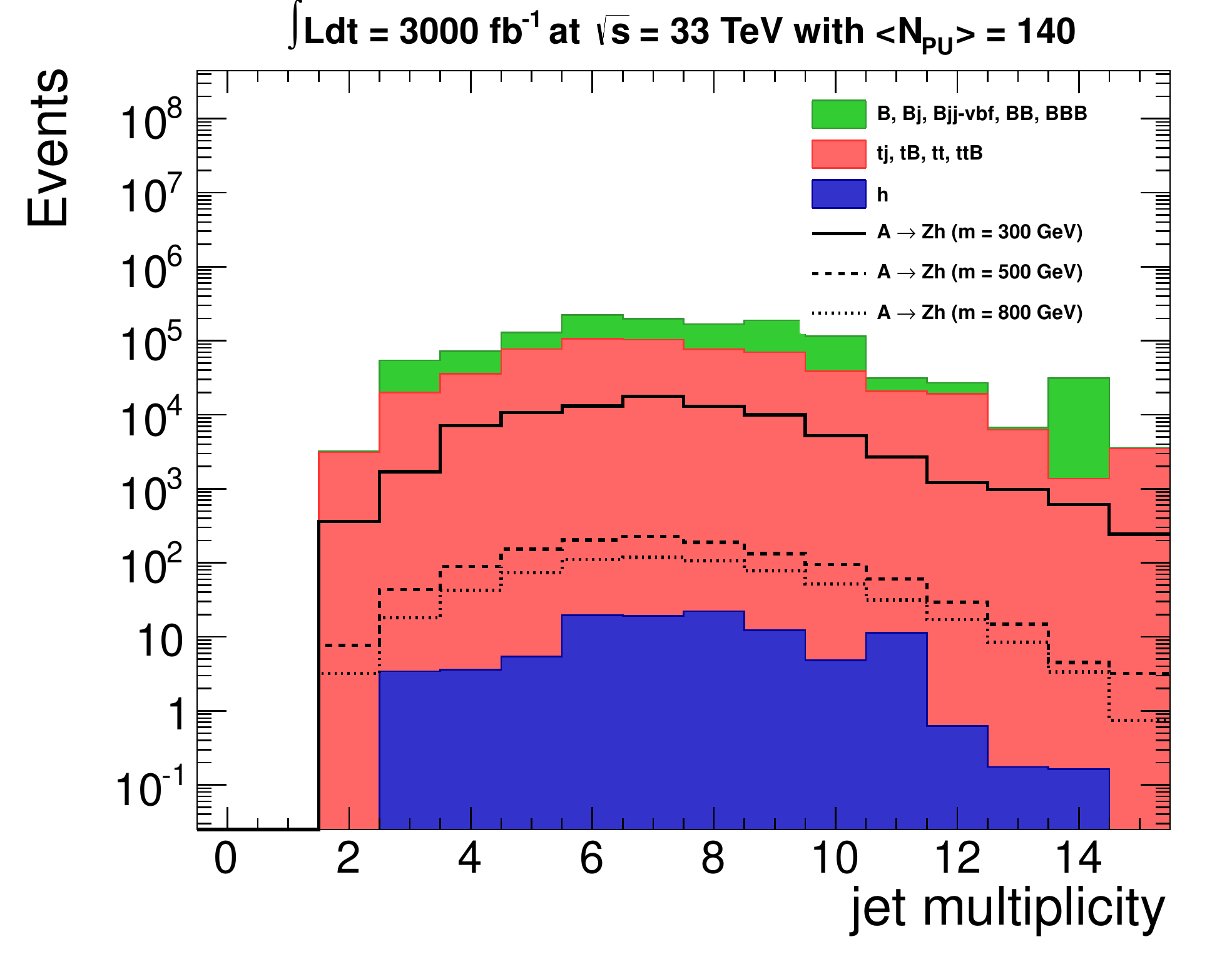}
\includegraphics[width=0.4\columnwidth,height=0.4\textheight,keepaspectratio=true]{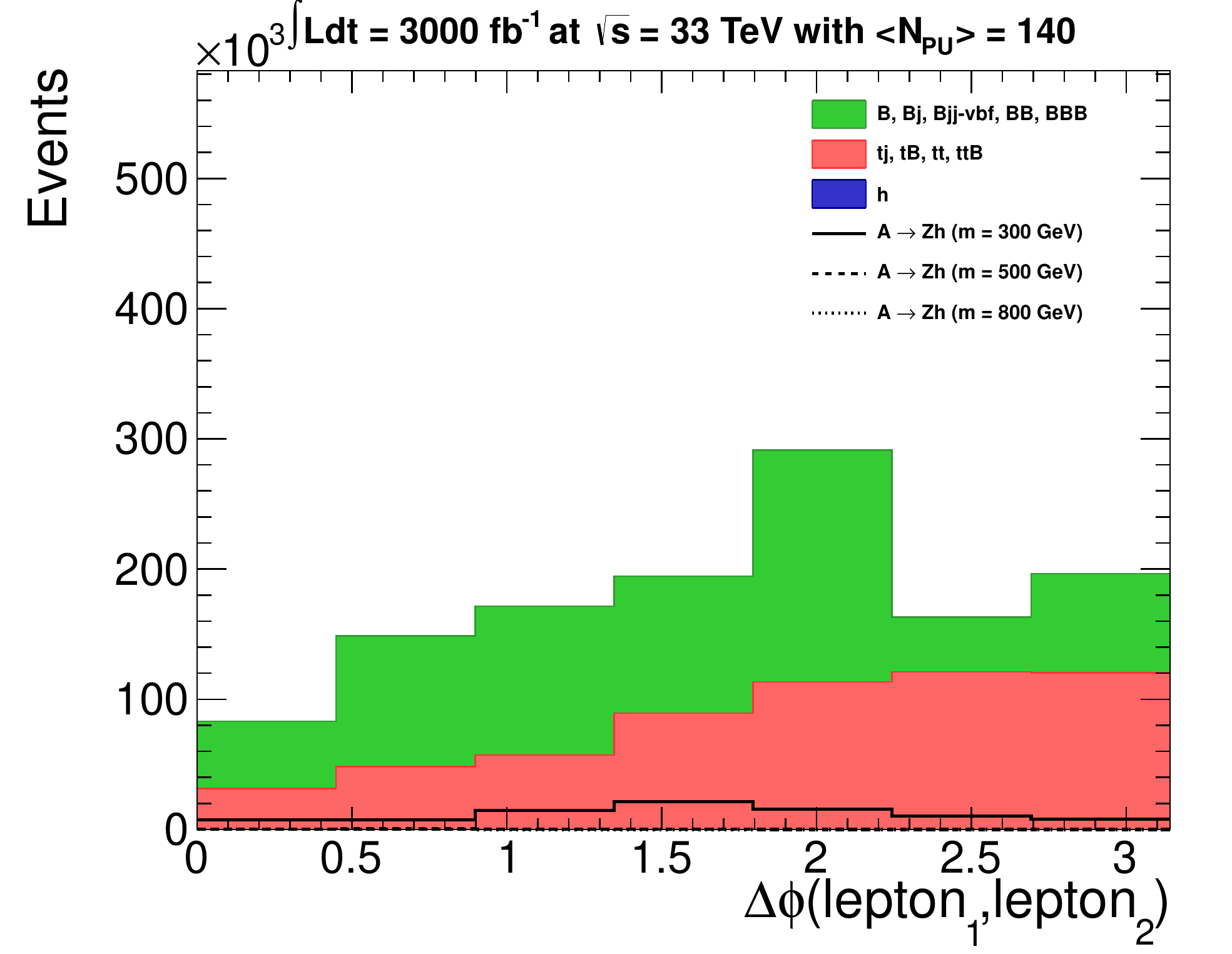}
\includegraphics[width=0.4\columnwidth,height=0.4\textheight,keepaspectratio=true]{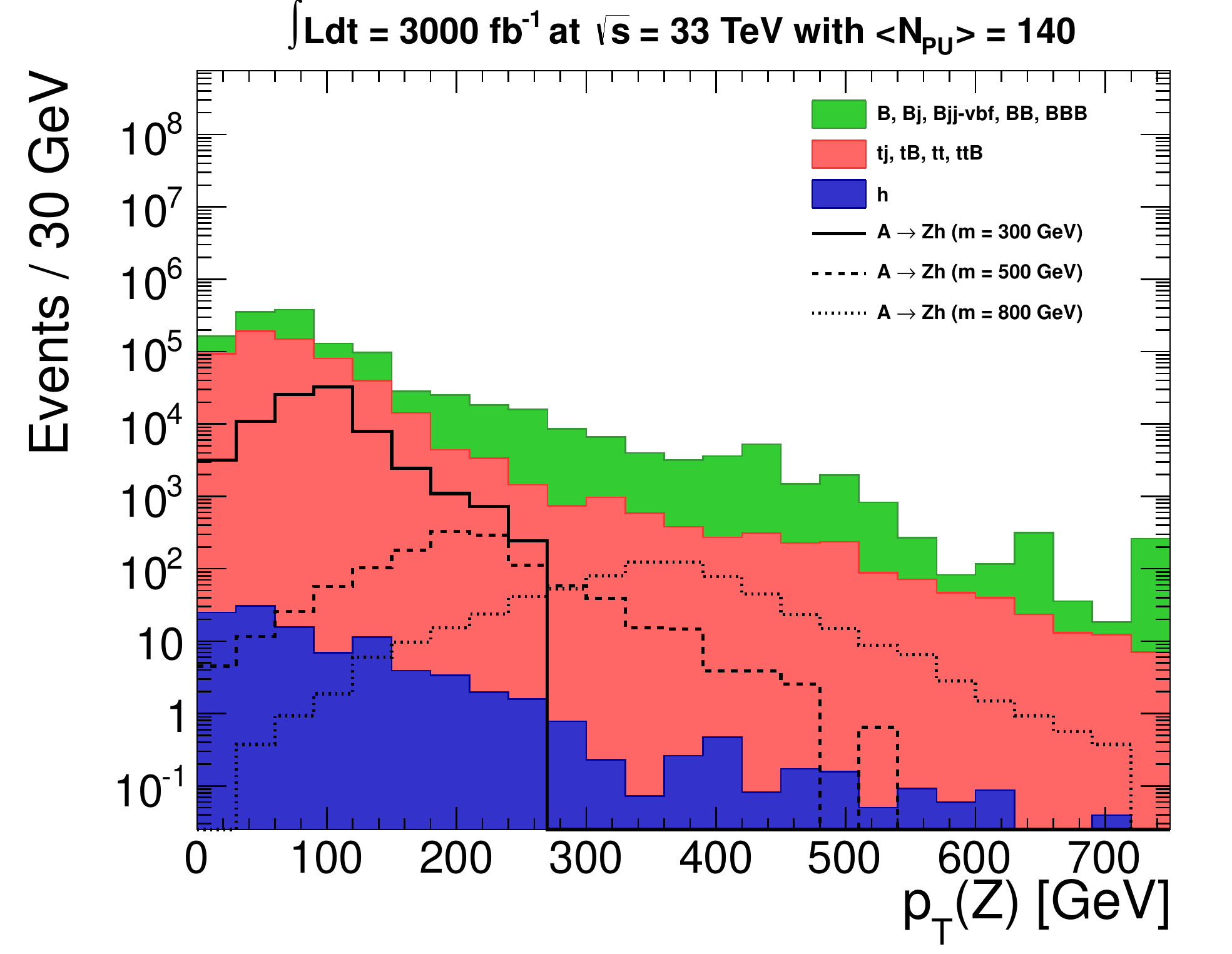}
\includegraphics[width=0.4\columnwidth,height=0.4\textheight,keepaspectratio=true]{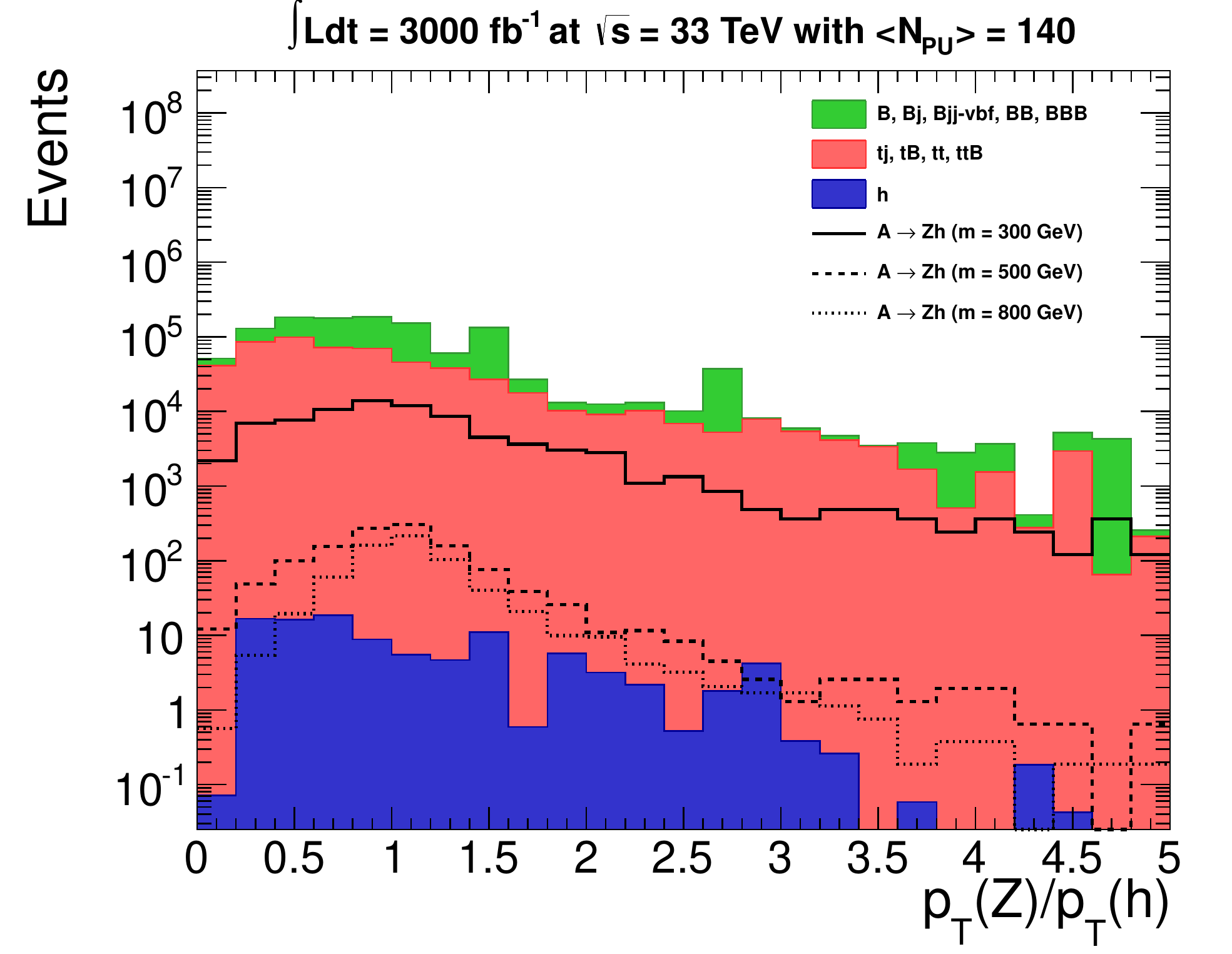}
\includegraphics[width=0.4\columnwidth,height=0.4\textheight,keepaspectratio=true]{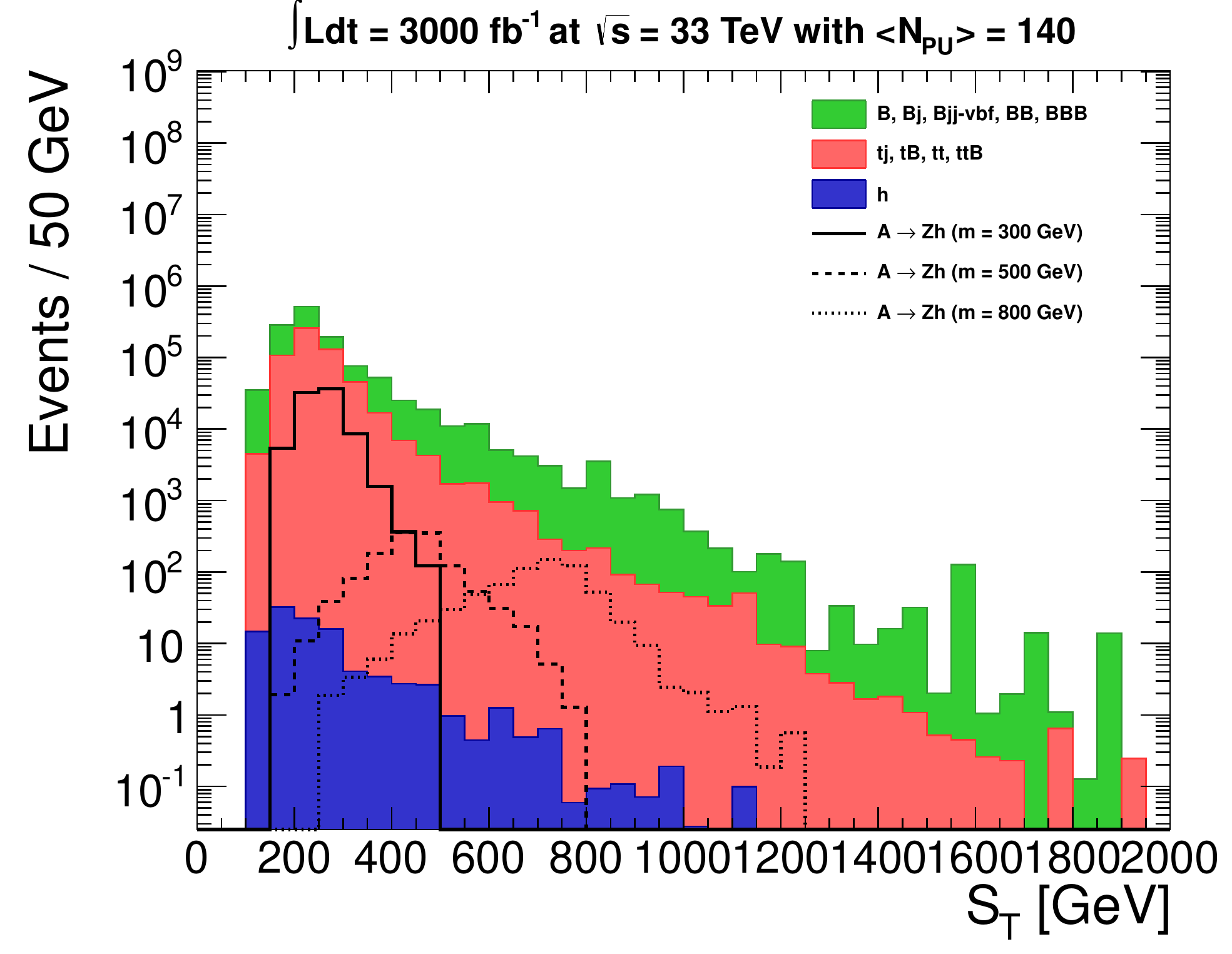}
\caption{Kinematic distributions for pre-selected events in the $bb$ channel, for $\int Ldt=$ 3000~\ifb~at $\sqrt{s}=33$ TeV with $<N_{PU}>=$~140.}
\label{fig:AZhbbPresel_33}
\end{center}
\end{figure}


Several variables are found to discriminate between pre-selected signal and background events.  The three variables which discriminate most strongly are chosen, and cut thresholds are varied simultaneously to determine the set which maximizes signal sensitivity.  As a result, we apply the following selection cuts:
\begin{itemize}
\item Azimuthal angle between the two leptons $|\Delta\phi(\ell_1,\ell_2)|\leq$~1.9
\item Transverse momentum of the $Z$ candidate $p_T(Z)\geq$~40~GeV
\item Ratio of the $p_T$ of the $Z$ and $h$ candidates 0.4$\leq p_T(Z)/p_T(h)\leq$ 2.75
\end{itemize}

The selected signal and background yields for the $bb$ channel are shown in Tables~\ref{tab:AZhbbSelCutflowSignal} and \ref{tab:AZhbbSelCutflowBackground} for the 300~\ifb analysis at $\sqrt{s}=14$~TeV, and Tables~\ref{tab:AZhbbSelCutflowSignal33} and \ref{tab:AZhbbSelCutflowBackground33} for the 3000~\ifb analysis at $\sqrt{s}=33$~TeV.  Analogous tables for the \tautau~channel are shown in Tables~\ref{tab:AZhtautauSelCutflowSignal}, \ref{tab:AZhtautauSelCutflowBackground}, \ref{tab:AZhtautauSelCutflowSignal33}, \ref{tab:AZhtautauSelCutflowBackground33}.  The $m_A$ distributions for both channels are shown in Figure~\ref{fig:AZhplots}, for the analyses based on 300~\ifb~at $\sqrt{s}=14$~TeV and 3000~\ifb~at $\sqrt{s}=33$~TeV.


\begin{table}[htbp]
\begin{center}
\begin{footnotesize}
\begin{tabular}{|l|c|c|c|c|}
\hline
Signal Mass [GeV] & Pre-selection & $\left|\Delta\phi(\ell_1\ell_2)\right|\leq$~1.9 & $p_T(Z)\geq$~40~GeV & 0.4$\leq\frac{p_T(Z)}{p_T(h)}\leq$ 2.75 \\ \hline
250        & 1.73e+3        & 544             & 529             & 447             \\
300        & 4.3e+3         & 2.83e+3        & 2.82e+3        & 2.57e+3        \\
350        & 96.2            & 77              & 76.9            & 72.4            \\
400        & 61.9            & 54.6            & 54.6            & 52.1            \\
450        & 50.3            & 46              & 46              & 43.6            \\
500        & 42.5            & 40.3            & 40.3            & 38.4            \\
600        & 31.8            & 31              & 31              & 30.1            \\
700        & 22              & 21.7            & 21.7            & 21.3            \\
800        & 14.8            & 14.7            & 14.7            & 14.4            \\
900        & 9.96            & 9.91            & 9.91            & 9.76            \\
1000       & 5.98            & 5.97            & 5.97            & 5.9             \\ \hline
\end{tabular}
\end{footnotesize}
\end{center}
\caption{Expected number of selected events for the $A\rightarrow Zh\rightarrow \ell \ell bb$ signal for $\int Ldt=$ 300~\ifb~at $\sqrt{s}=14$ TeV with $<N_{PU}>=50$.}
\label{tab:AZhbbSelCutflowSignal}
\end{table}

\begin{table}[htbp]
\begin{center}
\begin{footnotesize}
\begin{tabular}{|l|c|c|c|c|}
\hline
Background & Pre-selection & $\left|\Delta\phi(\ell_1\ell_2)\right|\leq$~1.9 & $p_T(Z)\geq$~40~GeV & 0.4$\leq\frac{p_T(Z)}{p_T(h)}\leq$ 2.75 \\ \hline
B, Bj, Bjj-vbf, BB, BBB                            & 2.66e+4        & 1.73e+4        & 7.92e+3        & 7.06e+3        \\
tj, tB, tt, ttB                                    & 1.51e+4        & 6.68e+3        & 6.48e+3        & 5.78e+3        \\
H                                                  & 1.58            & 0.765           & 0.765           & 0.718           \\ \hline
Total Background                                   & 4.16e+4        & 2.4e+4         & 1.44e+4        & 1.28e+4        \\ \hline
\end{tabular}
\end{footnotesize}
\end{center}
\caption{Expected number of selected events for the SM backgrounds to $A\rightarrow Zh\rightarrow \ell \ell bb$ for $\int Ldt=$ 300~\ifb~at $\sqrt{s}=14$ TeV with $<N_{PU}>=50$.}
\label{tab:AZhbbSelCutflowBackground}
\end{table}


\begin{table}[htbp]
\begin{center}
\begin{footnotesize}
\begin{tabular}{|l|c|c|c|c|}
\hline
Signal Mass [GeV] & Pre-selection & $\left|\Delta\phi(\ell_1\ell_2)\right|\leq$~1.9 & $p_T(Z)\geq$~40~GeV & 0.4$\leq\frac{p_T(Z)}{p_T(h)}\leq$ 2.75 \\ \hline
250       & 2.5e+4         & 8.04e+3        & 7.98e+3        & 6.72e+3        \\
300       & 8.47e+4        & 5.55e+4        & 5.53e+4        & 4.88e+4        \\
350       & 2.04e+3        & 1.62e+3        & 1.62e+3        & 1.45e+3        \\
400       & 1.38e+3        & 1.21e+3        & 1.21e+3        & 1.14e+3        \\
450       & 1.28e+3        & 1.19e+3        & 1.19e+3        & 1.1e+3         \\
500       & 1.25e+3        & 1.16e+3        & 1.16e+3        & 1.1e+3         \\
600       & 1.07e+3        & 1.04e+3        & 1.04e+3        & 1e+3           \\
700       & 893             & 881             & 881             & 851             \\
800       & 662             & 659             & 659             & 646             \\
900       & 502             & 499             & 499             & 491             \\
1000      & 325             & 324             & 324             & 319             \\ \hline
\end{tabular}
\end{footnotesize}
\end{center}
\caption{Expected number of selected events for the $A\rightarrow Zh\rightarrow \ell \ell bb$ signal for $\int Ldt=$ 3000~\ifb~at $\sqrt{s}=33$ TeV with $<N_{PU}>=$~140.}
\label{tab:AZhbbSelCutflowSignal33}
\end{table}

\begin{table}[htbp]
\begin{center}
\begin{footnotesize}
\begin{tabular}{|l|c|c|c|c|}
\hline
Background & Pre-selection & $\left|\Delta\phi(\ell_1\ell_2)\right|\leq$~1.9 & $p_T(Z)\geq$~40~GeV & 0.4$\leq\frac{p_T(Z)}{p_T(h)}\leq$ 2.75 \\ \hline
B, Bj, Bjj-vbf, BB, BBB                            & 6.68e+5        & 4.38e+5        & 4.38e+5        & 4.16e+5         \\
tj, tB, tt, ttB                                    & 5.81e+5        & 2.51e+5        & 2.44e+5        & 2.1e+5          \\
H                                                  & 103             & 46.8            & 46.8            & 17.5             \\ \hline
Total Background                                   & 1.25e+6        & 6.89e+5        & 6.82e+5        & 6.26e+5         \\ \hline
\end{tabular}
\end{footnotesize}
\end{center}
\caption{Expected number of selected events for the SM backgrounds to $A\rightarrow Zh\rightarrow \ell \ell bb$ for $\int Ldt=$ 3000~\ifb~at $\sqrt{s}=33$ TeV with $<N_{PU}>=$~140.}
\label{tab:AZhbbSelCutflowBackground33}
\end{table}


\begin{figure}[htbp]
\begin{center}
\includegraphics[width=0.4\columnwidth,height=0.4\textheight,keepaspectratio=true]{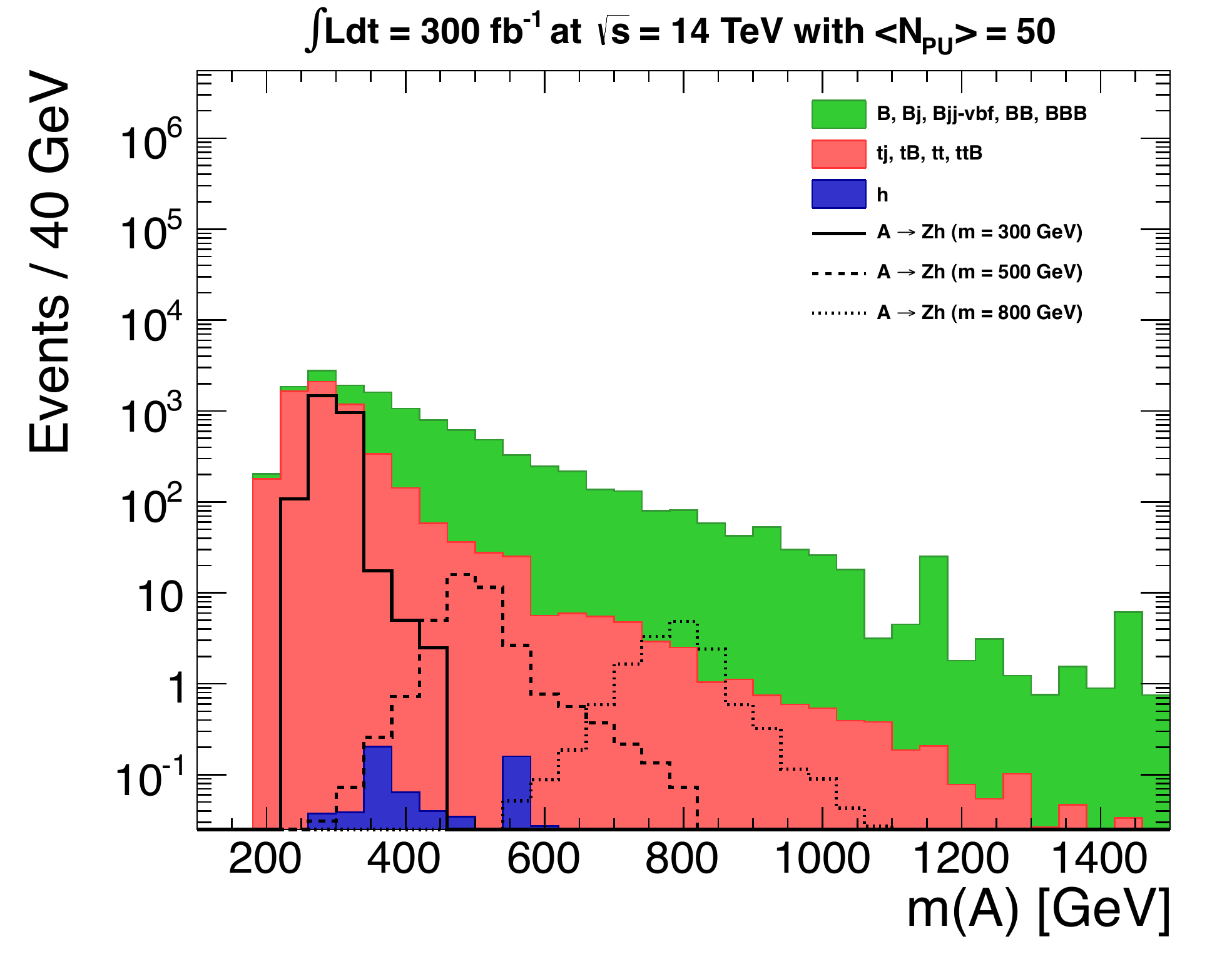}
\includegraphics[width=0.4\columnwidth,height=0.4\textheight,keepaspectratio=true]{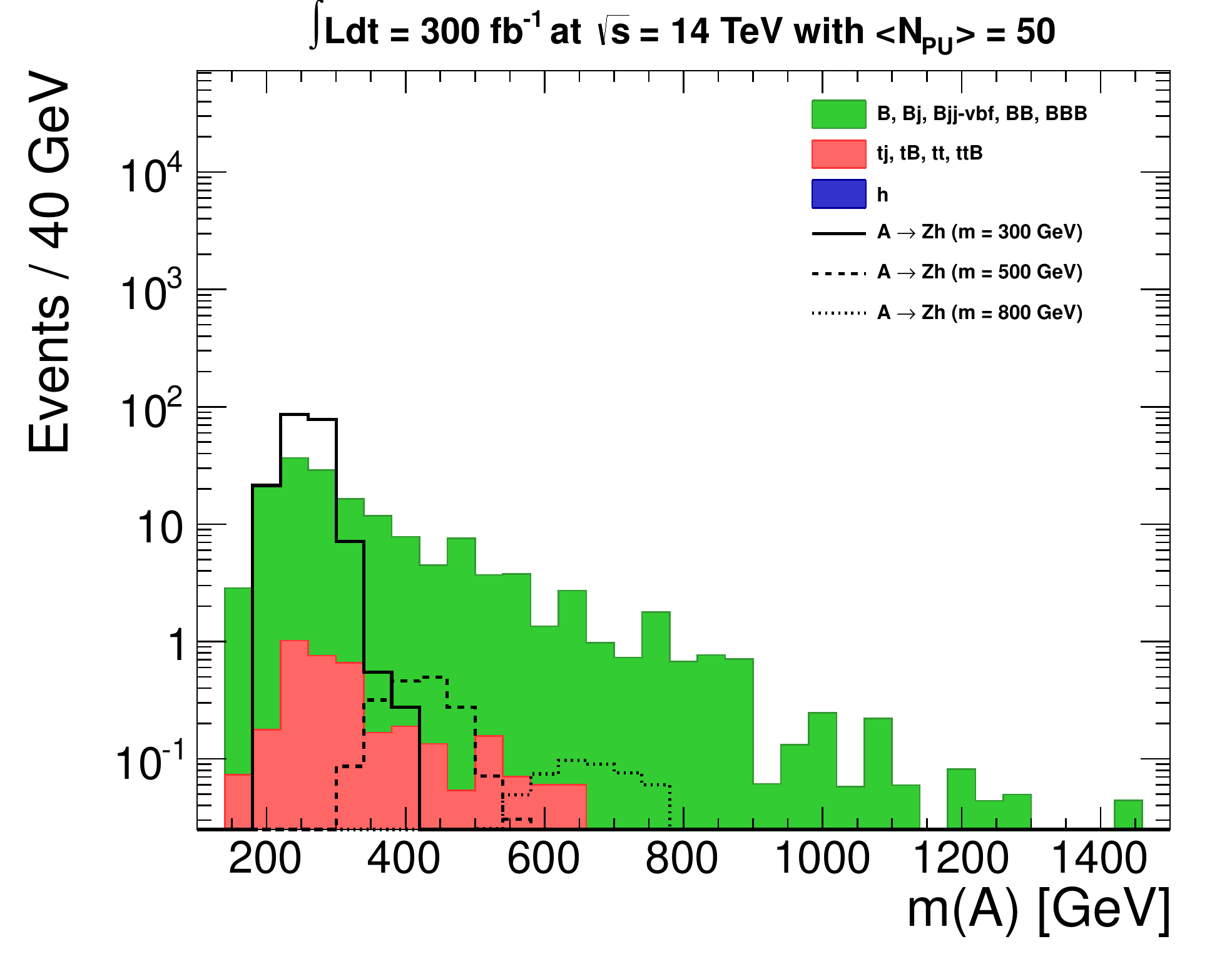}
\includegraphics[width=0.4\columnwidth,height=0.4\textheight,keepaspectratio=true]{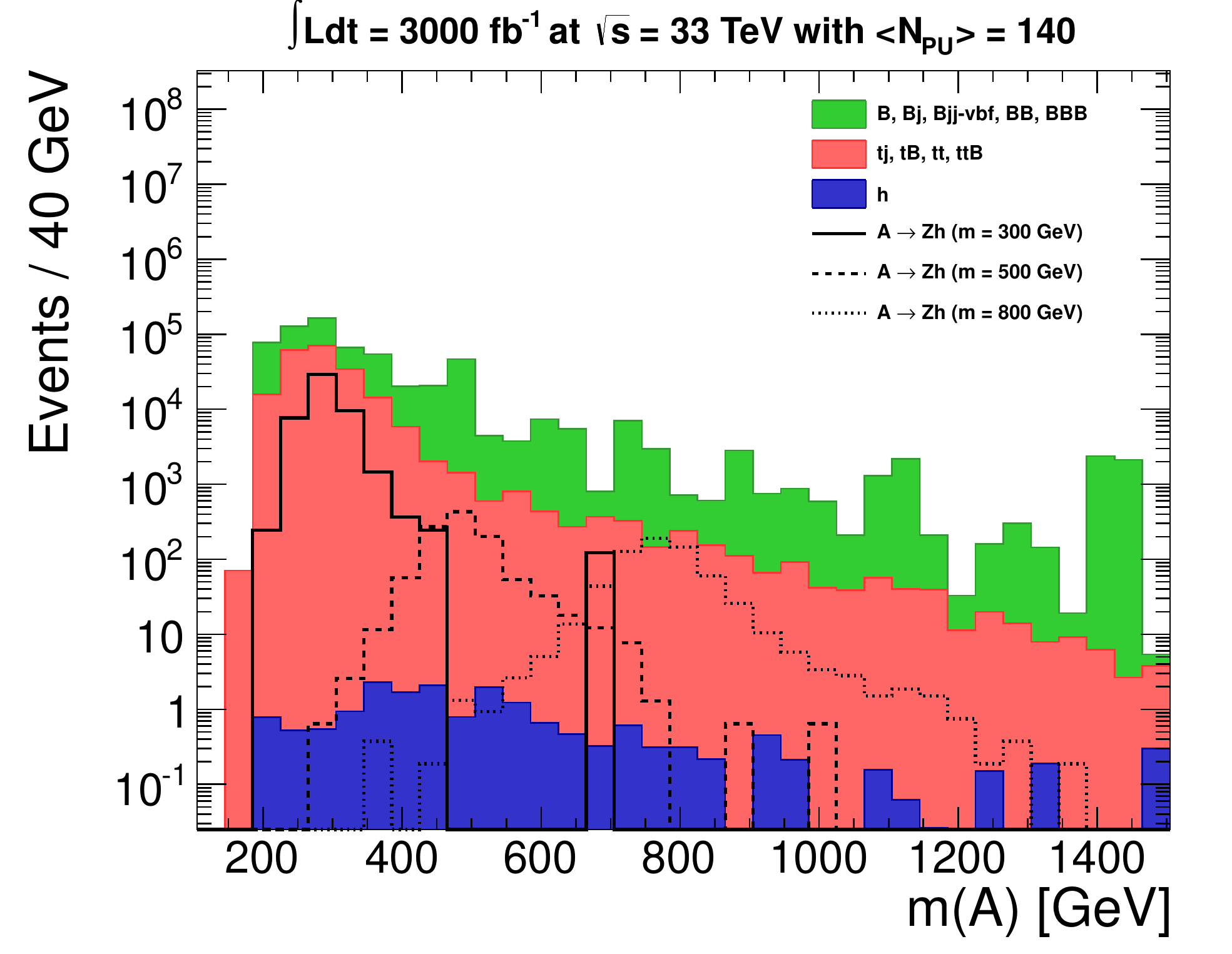}
\includegraphics[width=0.4\columnwidth,height=0.4\textheight,keepaspectratio=true]{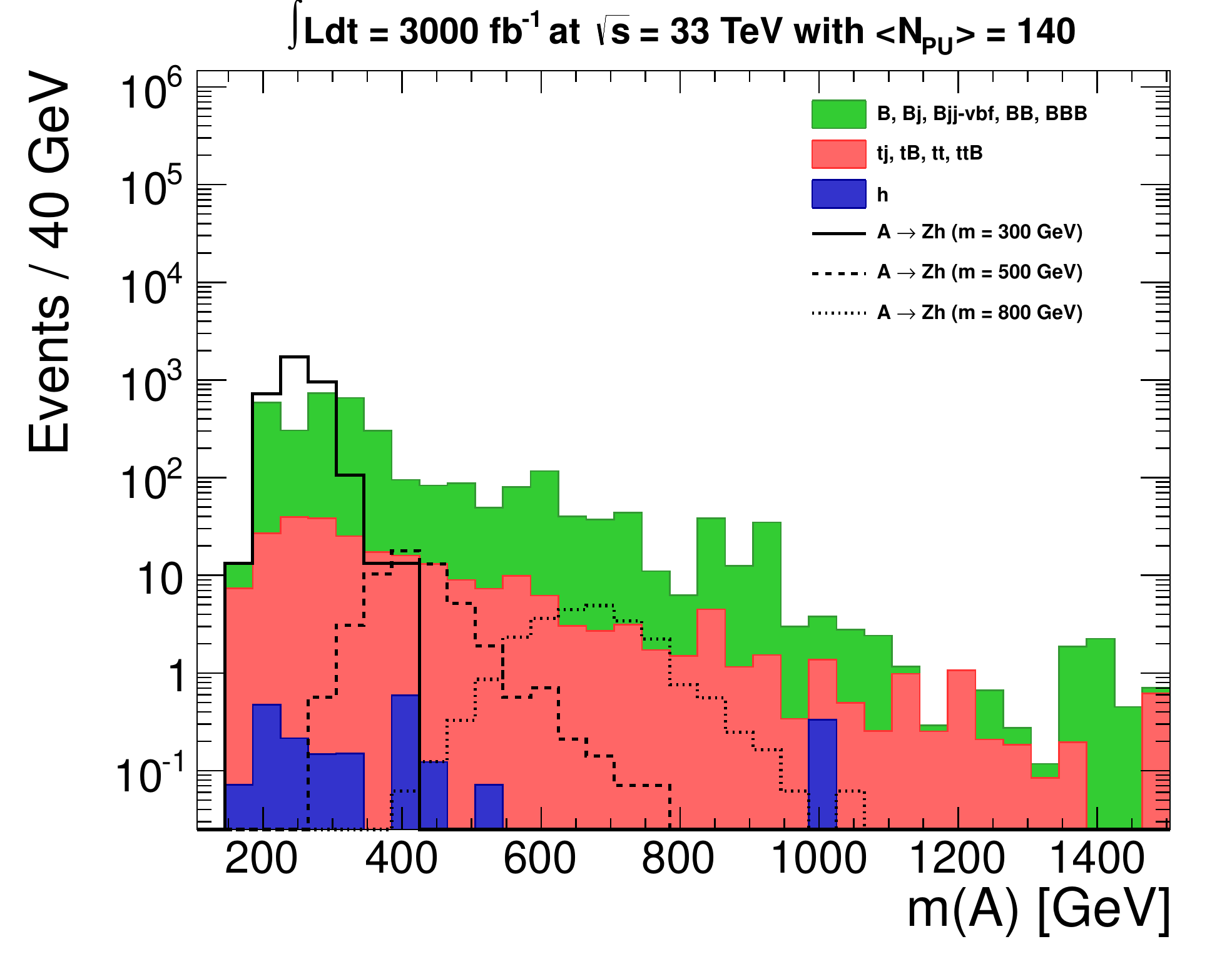}
\caption{Distribution of the $A$ candidate invariant mass in selected events for $\int Ldt=$ 300~\ifb~at $\sqrt{s}=14$ TeV with $<N_{PU}>=50$ (top) and $\int Ldt=$ 3000~\ifb~at $\sqrt{s}=33$ TeV with $<N_{PU}>=$~140 (bottom).  The $bb$ (\tautau) channel is on the left (right).}
\label{fig:AZhplots}
\end{center}
\end{figure}





\subsubsection{Results}

The distribution of the $A$ candidate invariant mass is used to assess discovery and exclusion potential.  A 20\% rate uncertainty in the backgrounds is assumed.  

Figure~\ref{AZhCrossSectionLimitsByChannel} shows the signal cross section required to exclude the $A\rightarrow Zh$ signal at 95\% CL in both channels separately, and the combination, for $\int Ldt=$ 300~\ifb~and 3000~\ifb~at $\sqrt{s}=14$ TeV.  Also shown is the cross section required for a 3$\sigma$ and 5$\sigma$ signal significance.  Figure~\ref{AZhCrossSectionLimits} overlays the signal cross sections required for exclusion or discovery based on the combination of the two channels, for various run conditions. These cross section limits are then interpreted in terms of the actual signal cross section for each point in parameter space, in order to determine discovery and exclusion potential for each mass hypothesis.  Figure~\ref{fig:AZhExc} shows the regions in the $\tan \beta$ versus cos($\beta - \alpha$) plane which can be excluded at 95\% CL for the three scenarios under consideration.  Figures~\ref{fig:AZhObs} and \ref{fig:AZhDisco} show the masses for which $3\sigma$ and $5\sigma$ significance can be obtained in the tan$(\beta)$ versus cos($\beta - \alpha$) plane. Once again, direct search limits provide considerable complementarity to Higgs coupling measurements~\cite{snowCouplings}. In this case, the direct search limits weaken only at large tan$(\beta)$ where the production cross section for $A$ falls due to diminishing coupling to the top quark, as well as close to the alignment limit $|\cos(\beta - \alpha)| \to  0$ where the branching ratio for $A \to Zh$ vanishes.


\begin{figure}[htbp]
\begin{center}
\includegraphics[width=0.4\columnwidth,height=0.4\textheight,keepaspectratio=true]{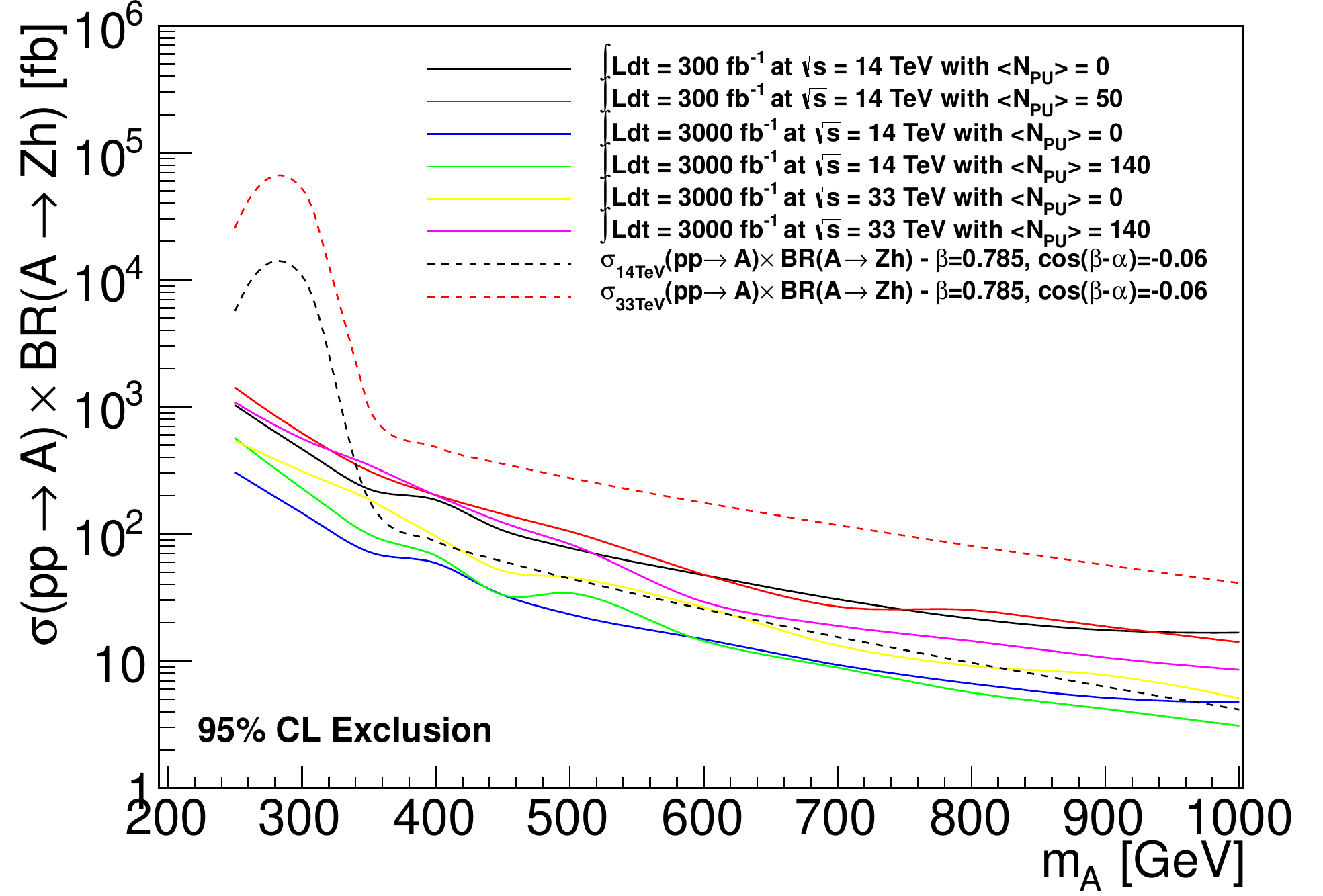}
\includegraphics[width=0.4\columnwidth,height=0.4\textheight,keepaspectratio=true]{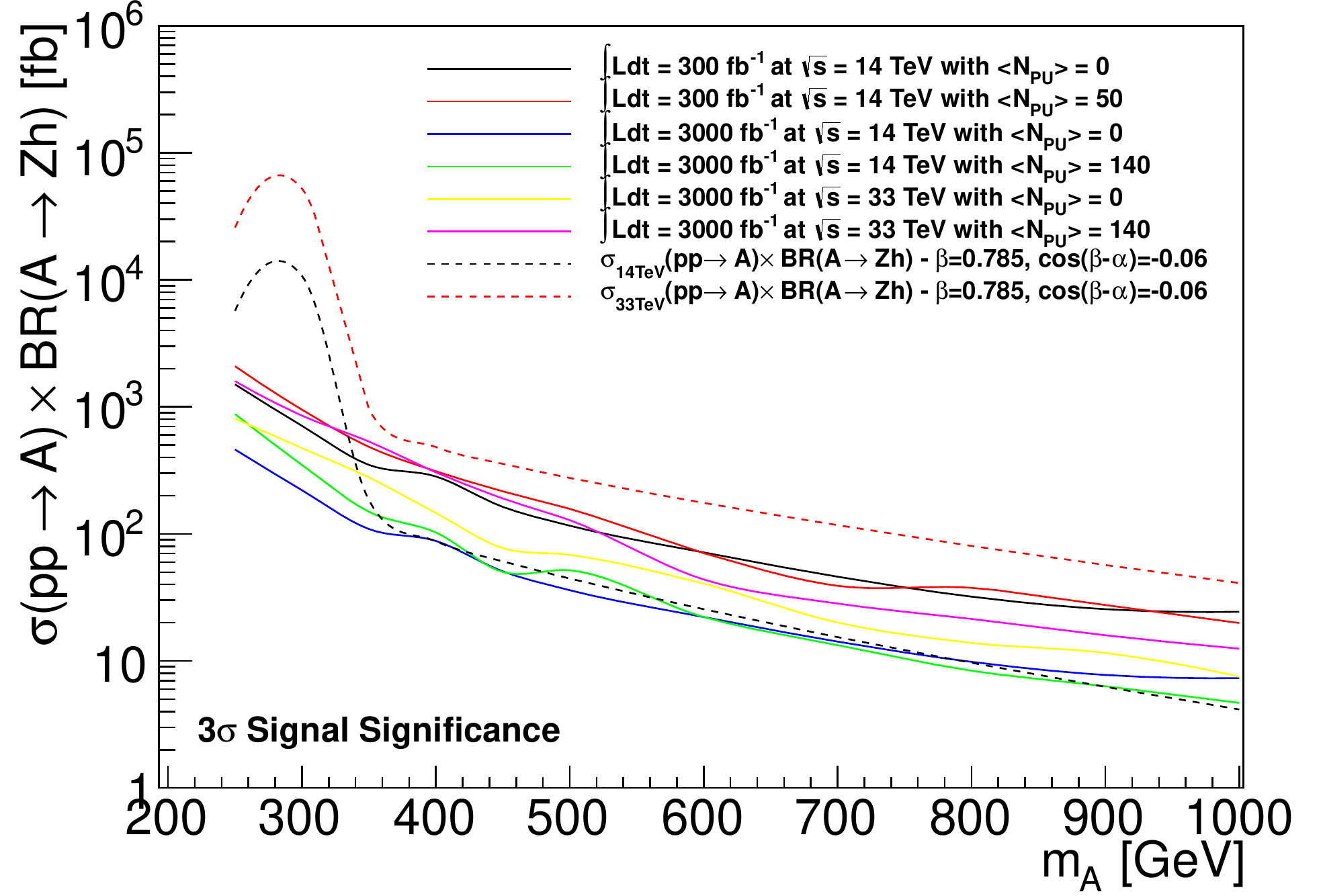}
\includegraphics[width=0.4\columnwidth,height=0.4\textheight,keepaspectratio=true]{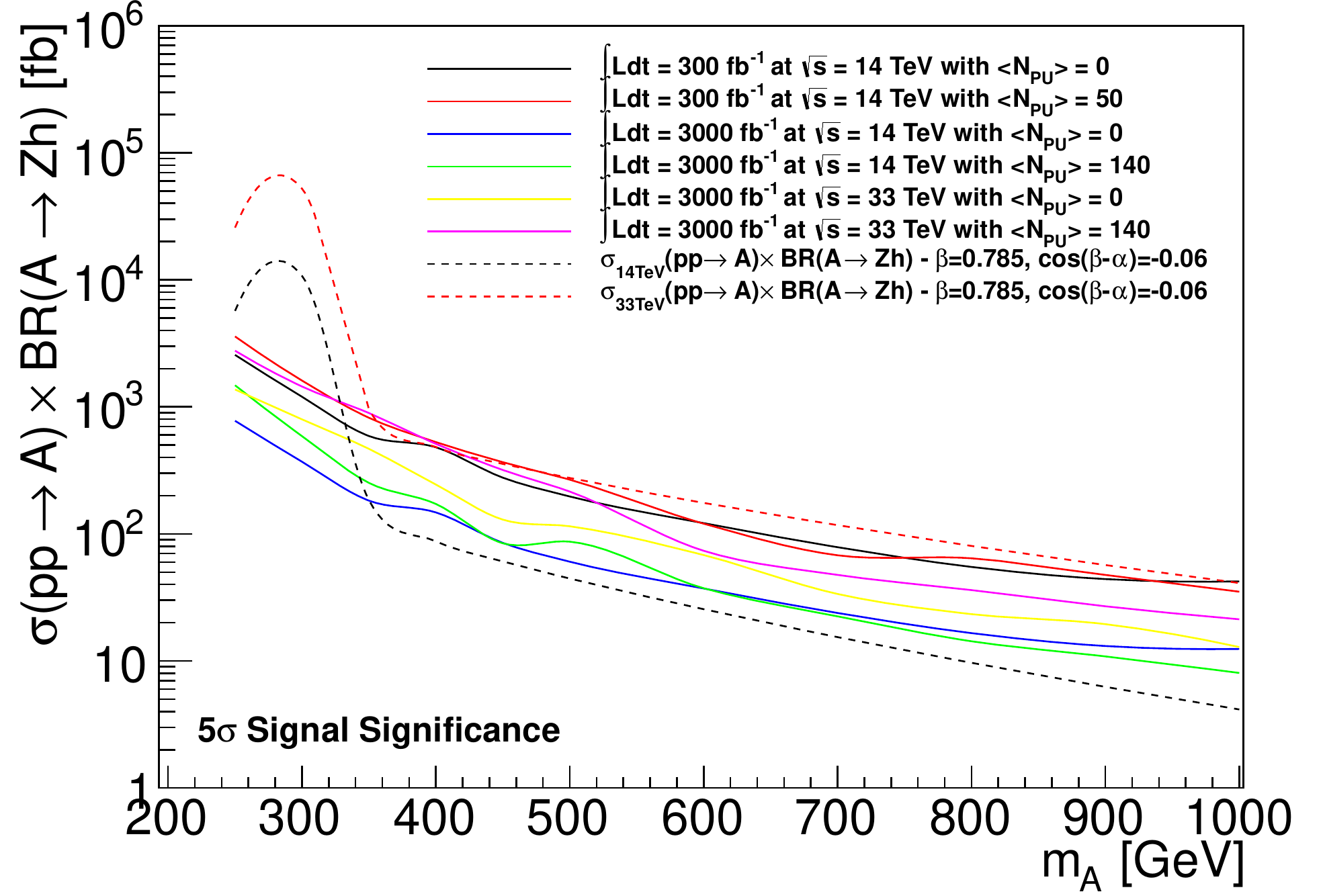}
\caption{The cross section which can be excluded at 95\% CL, and the cross section required for a 3$\sigma$ and 5$\sigma$ signal significance, as a function of the $A$ mass.}
\label{AZhCrossSectionLimits}
\end{center}
\end{figure}


\begin{figure}[htbp]
\begin{center}
\includegraphics[width=0.4\columnwidth,height=0.4\textheight,keepaspectratio=true]{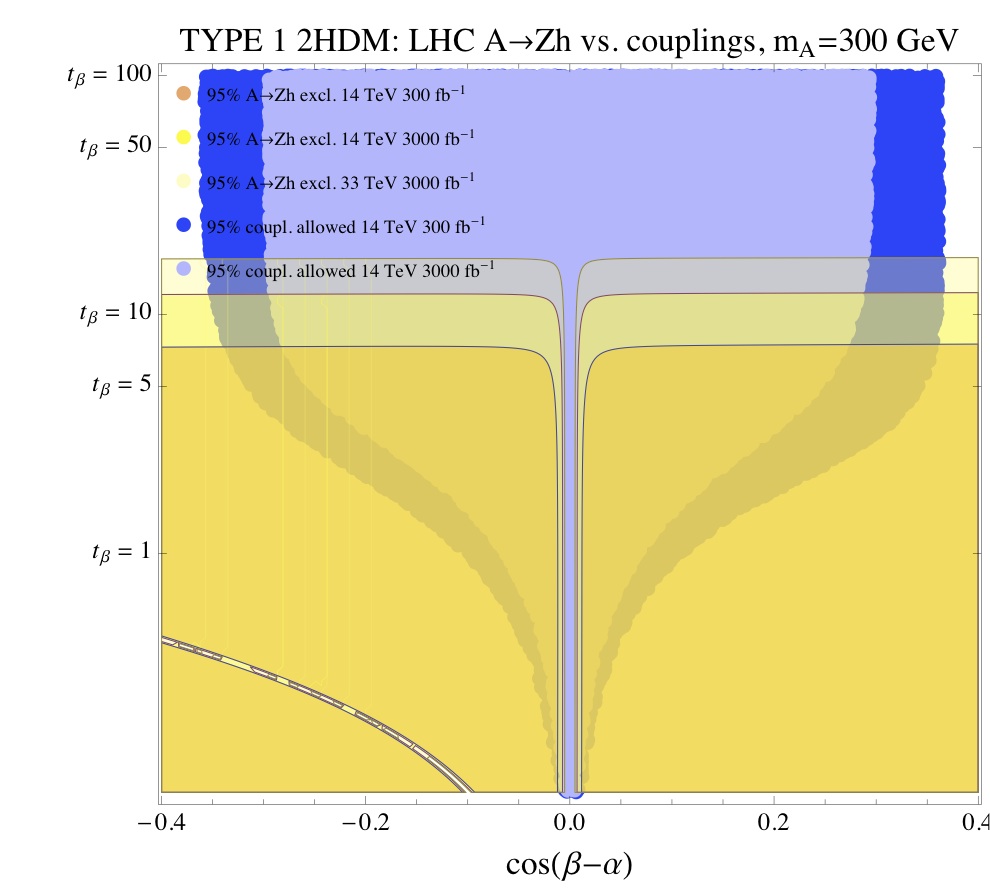}
\includegraphics[width=0.4\columnwidth,height=0.4\textheight,keepaspectratio=true]{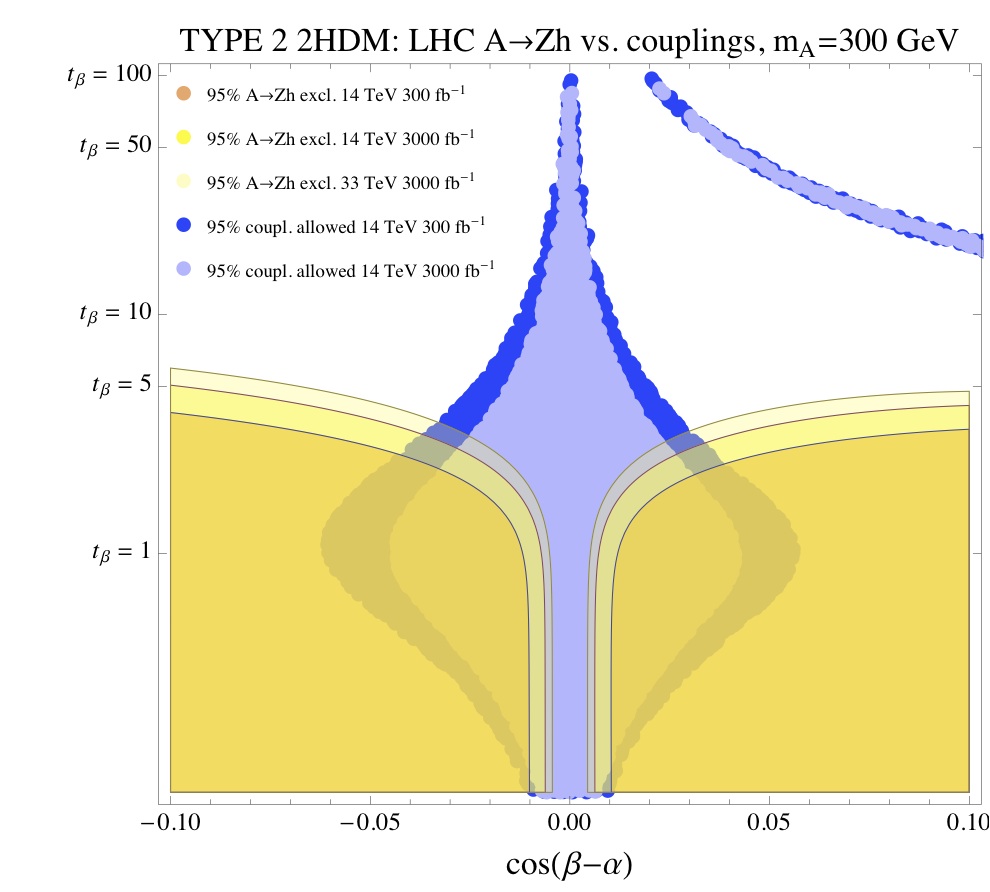}
\includegraphics[width=0.4\columnwidth,height=0.4\textheight,keepaspectratio=true]{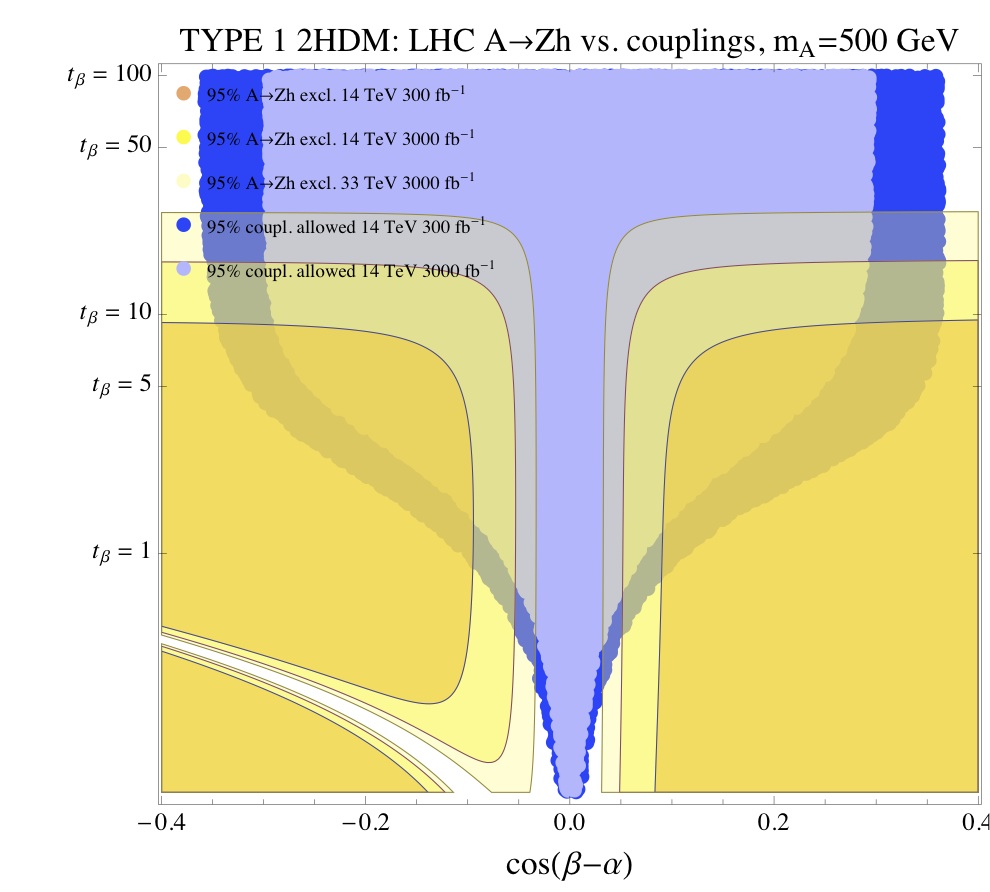}
\includegraphics[width=0.4\columnwidth,height=0.4\textheight,keepaspectratio=true]{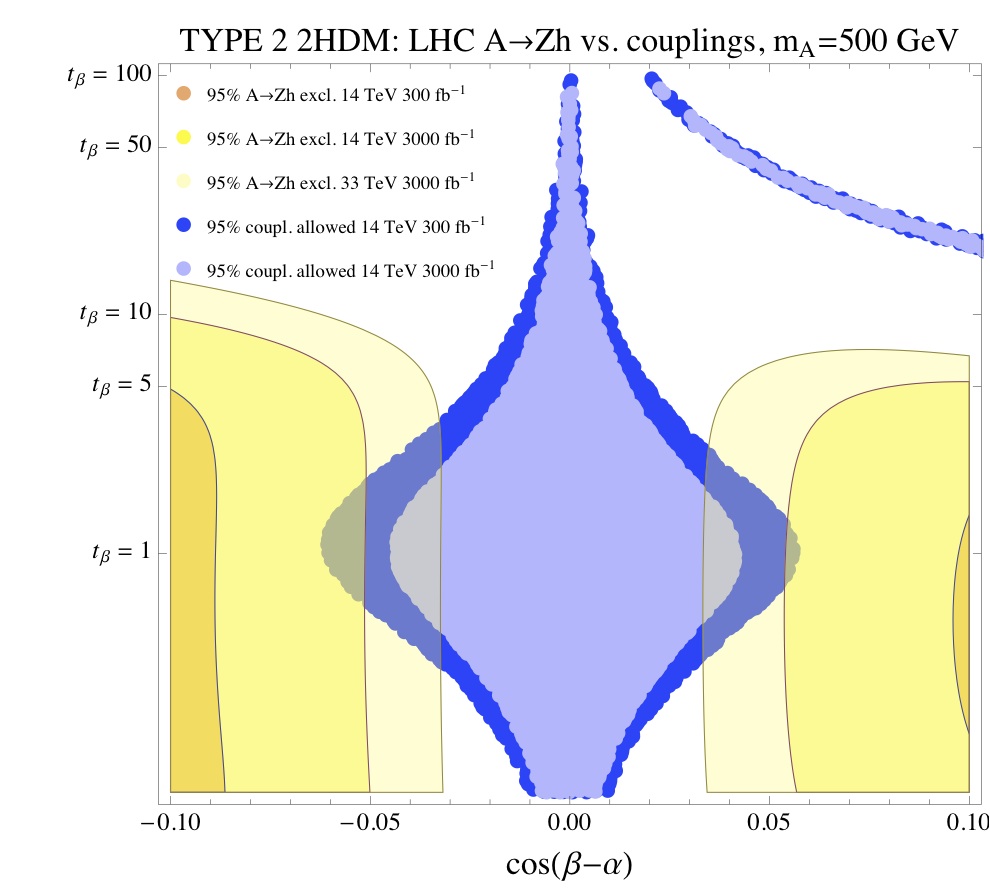}
\includegraphics[width=0.4\columnwidth,height=0.4\textheight,keepaspectratio=true]{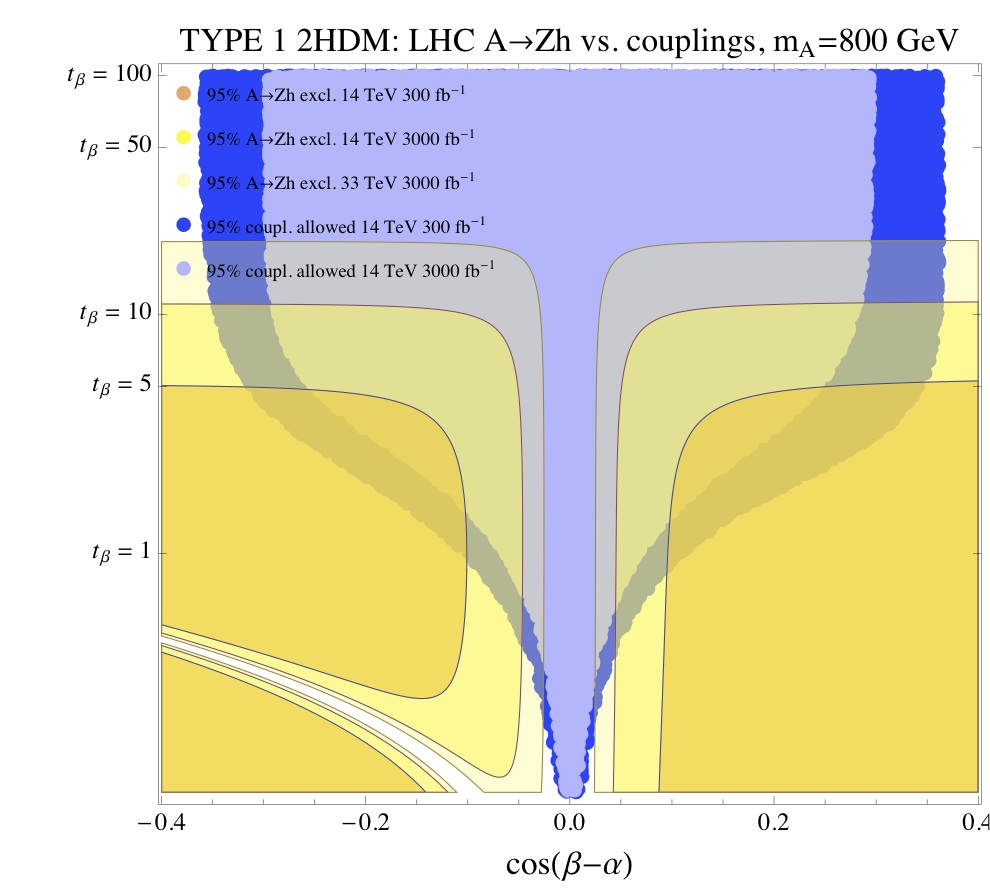}
\includegraphics[width=0.4\columnwidth,height=0.4\textheight,keepaspectratio=true]{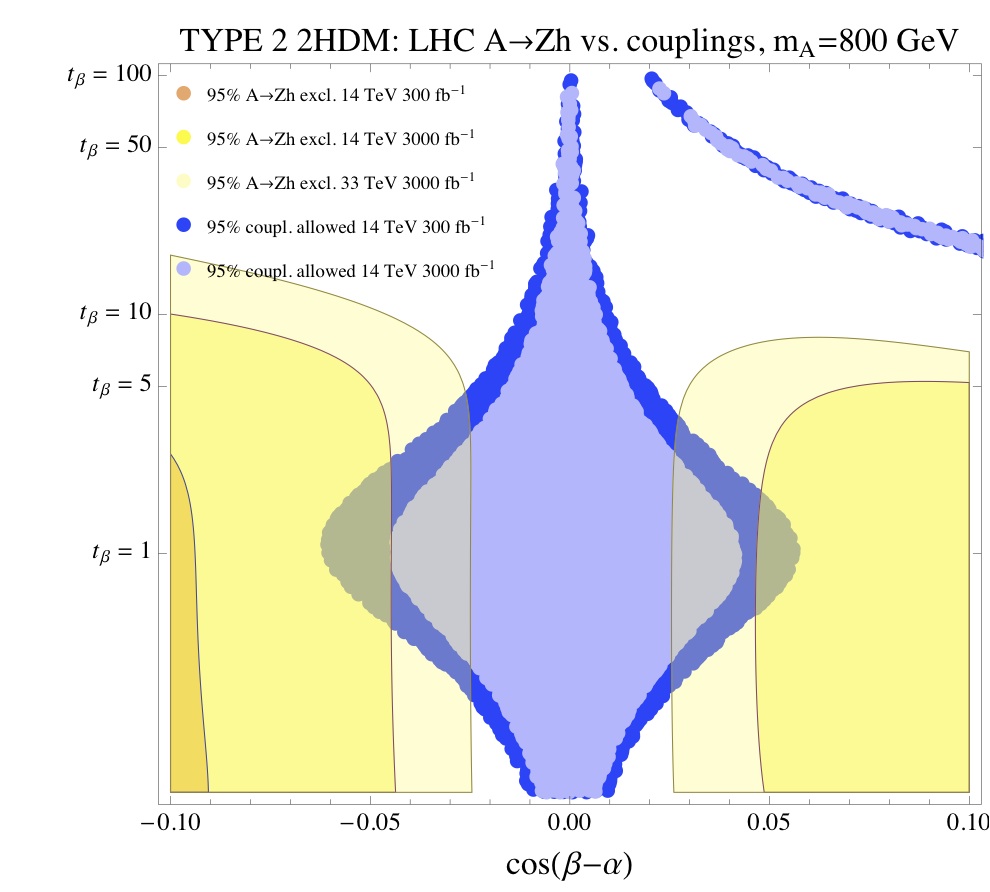}
\caption{The region of parameter space which could be excluded at 95\% CL for various $A$ mass hypotheses in a type I (left) and type II (right) 2HDM.  The dark yellow region corresponds to 300~\ifb~at $\sqrt{s}=14$~TeV, the yellow region to 3000~\ifb~at $\sqrt{s}=14$~TeV, and the light yellow region to 3000~\ifb~at $\sqrt{s}=33$~TeV.  The region which would remain allowed at 95\% CL based on non-observation of deviations from the SM in precision Higgs coupling measurements is shown in dark (light) blue for 300~\ifb~(3000~\ifb)~\cite{snowCouplings}.}
\label{fig:AZhExc}
\end{center}
\end{figure}


\begin{figure}[htbp]
\begin{center}
\includegraphics[width=0.4\columnwidth,height=0.4\textheight,keepaspectratio=true]{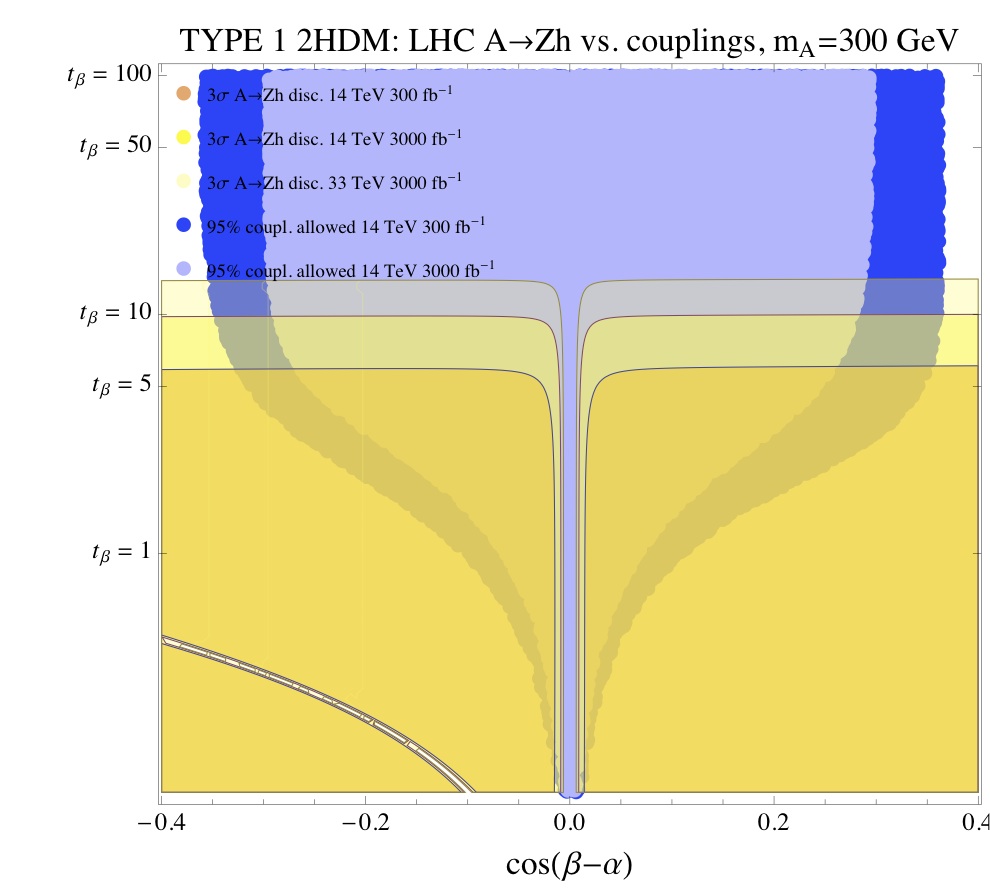}
\includegraphics[width=0.4\columnwidth,height=0.4\textheight,keepaspectratio=true]{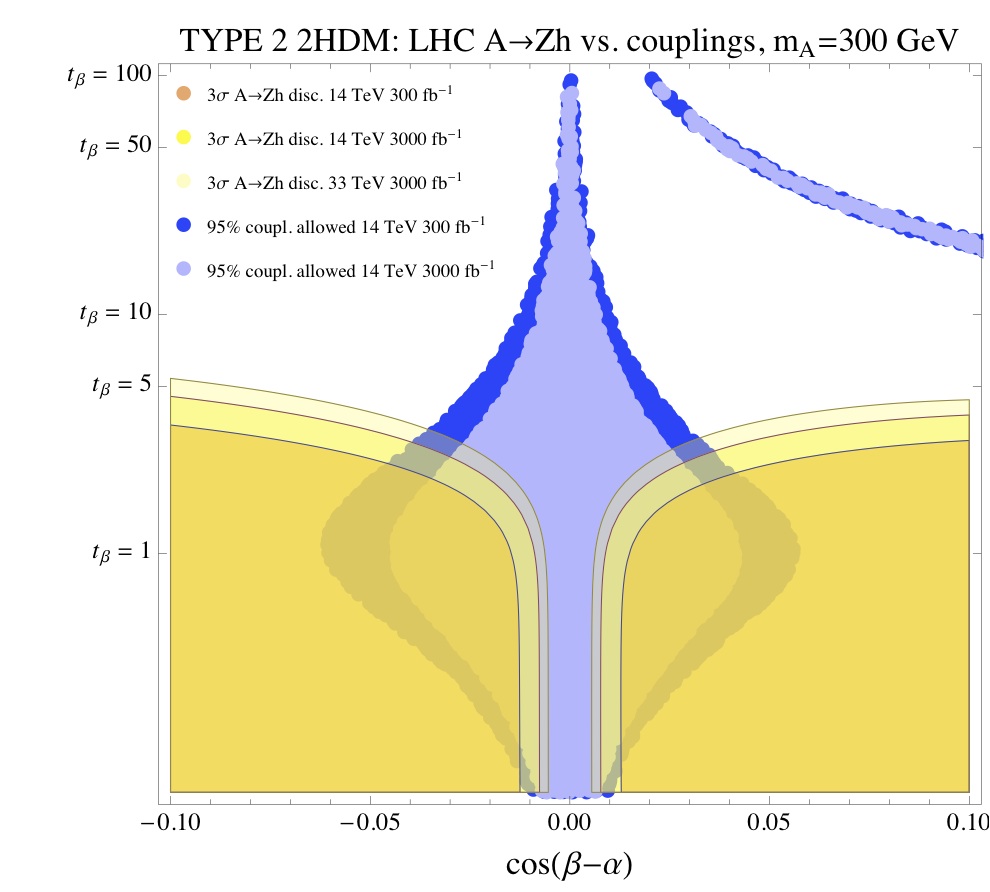}
\includegraphics[width=0.4\columnwidth,height=0.4\textheight,keepaspectratio=true]{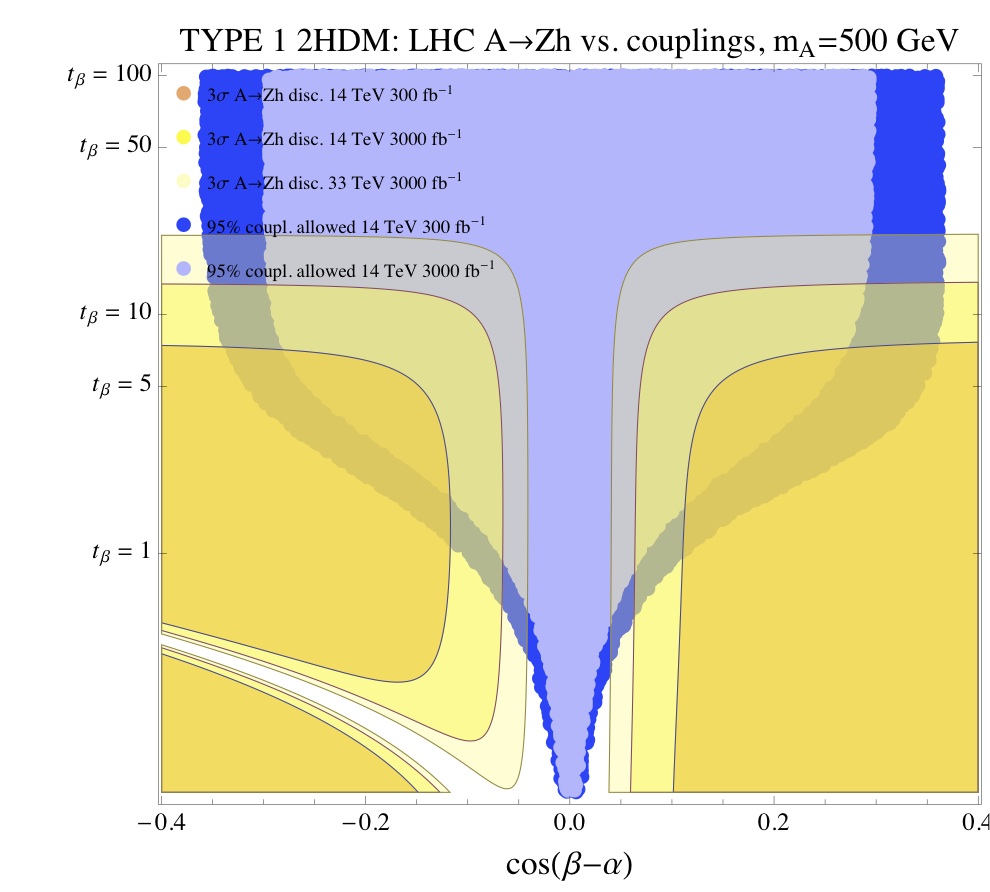}
\includegraphics[width=0.4\columnwidth,height=0.4\textheight,keepaspectratio=true]{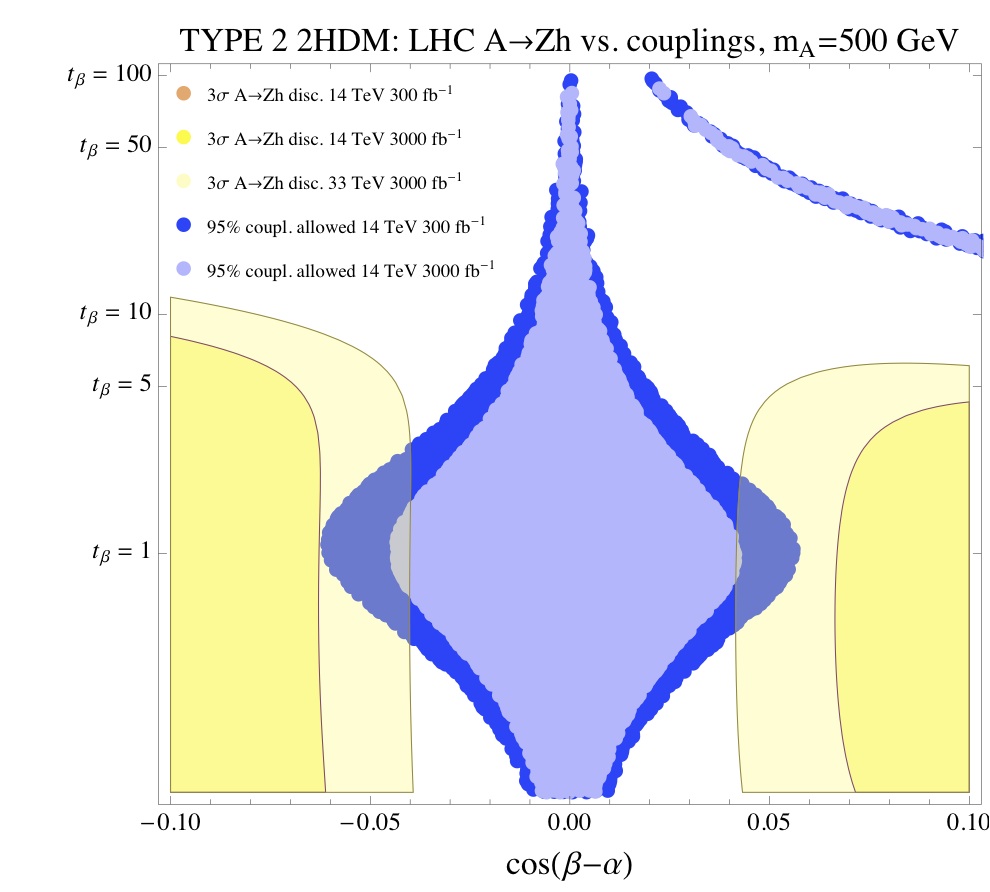}
\includegraphics[width=0.4\columnwidth,height=0.4\textheight,keepaspectratio=true]{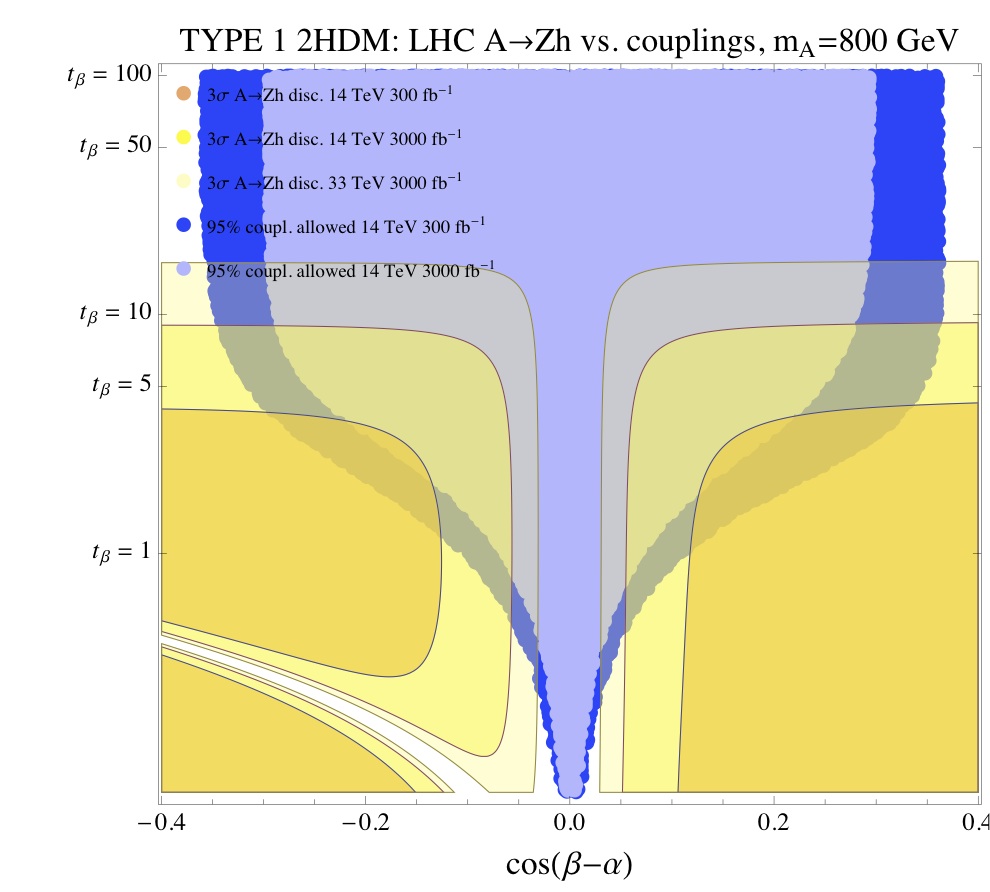}
\includegraphics[width=0.4\columnwidth,height=0.4\textheight,keepaspectratio=true]{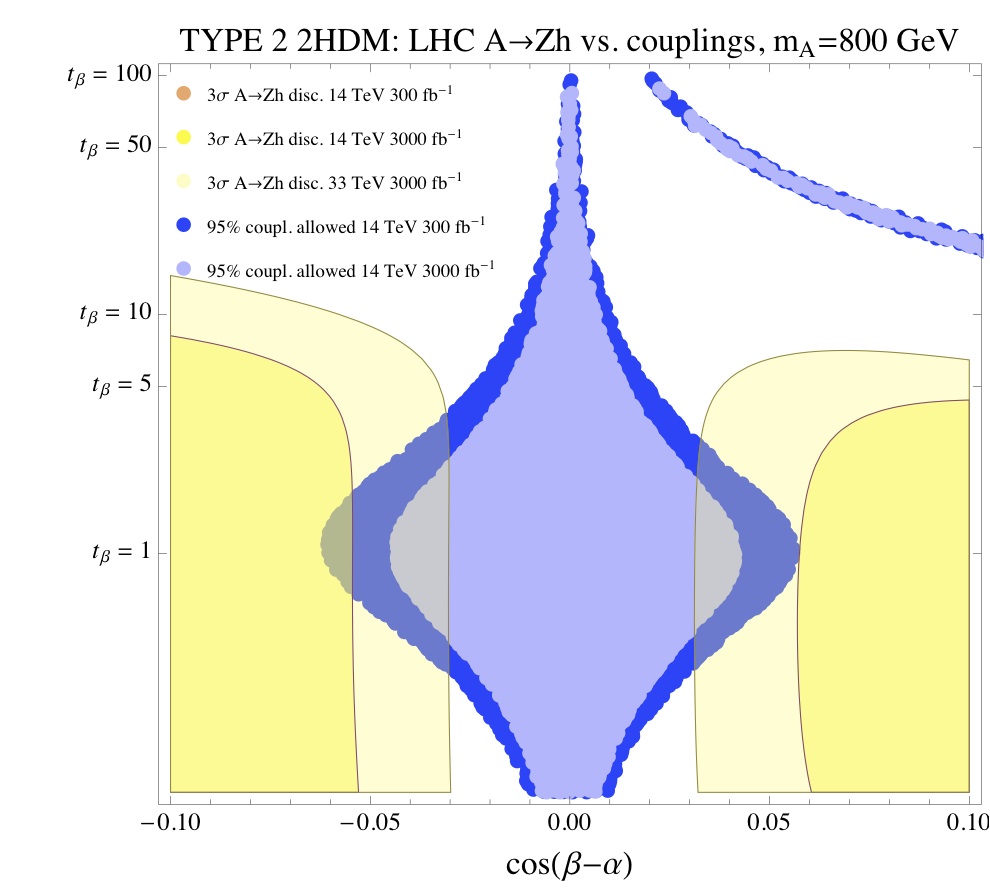}
\caption{The region of parameter space for which a 3$\sigma$ signal significance could be obtained for various $A$ mass hypotheses in a type I (left) and type II (right) 2HDM.  The dark yellow region corresponds to 300~\ifb~at $\sqrt{s}=14$~TeV, the yellow region to 3000~\ifb~at $\sqrt{s}=14$~TeV, and the light yellow region to 3000~\ifb~at $\sqrt{s}=33$~TeV.  The region which would remain allowed at 95\% CL based on non-observation of deviations from the SM in precision Higgs coupling measurements is shown in dark (light) blue for 300~\ifb~(3000~\ifb)~\cite{snowCouplings}.}
\label{fig:AZhObs}
\end{center}
\end{figure}


\begin{figure}[htbp]
\begin{center}
\includegraphics[width=0.4\columnwidth,height=0.4\textheight,keepaspectratio=true]{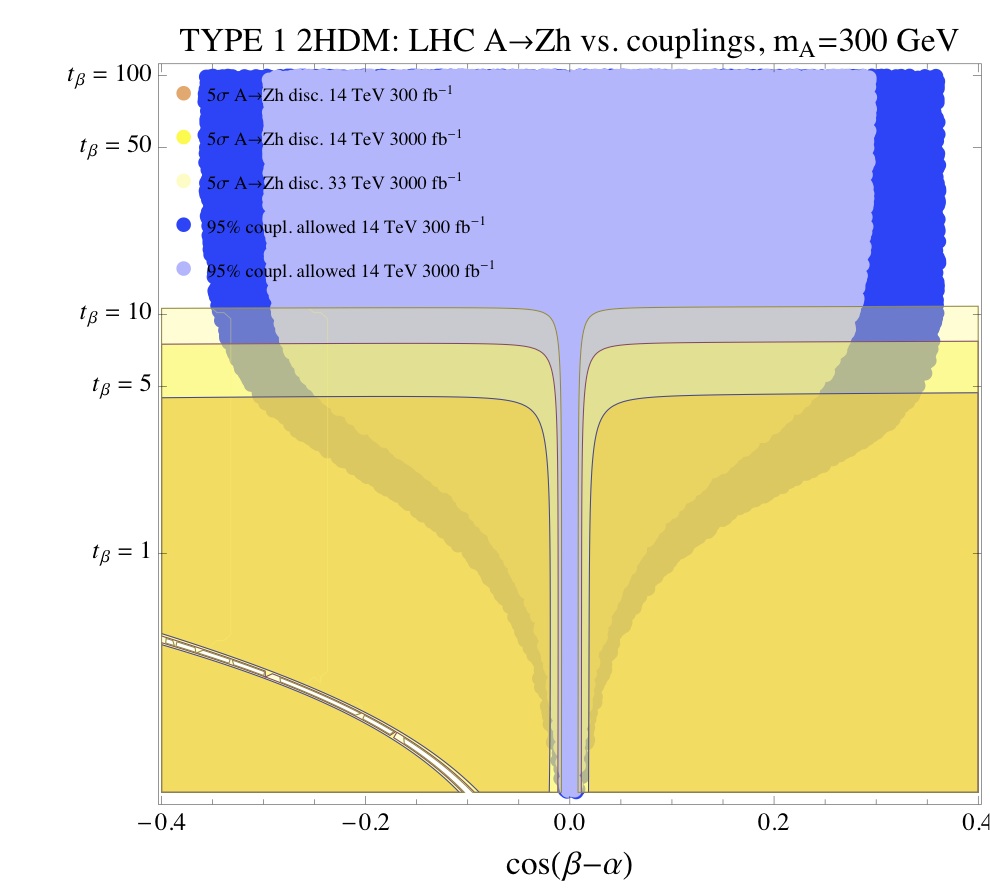}
\includegraphics[width=0.4\columnwidth,height=0.4\textheight,keepaspectratio=true]{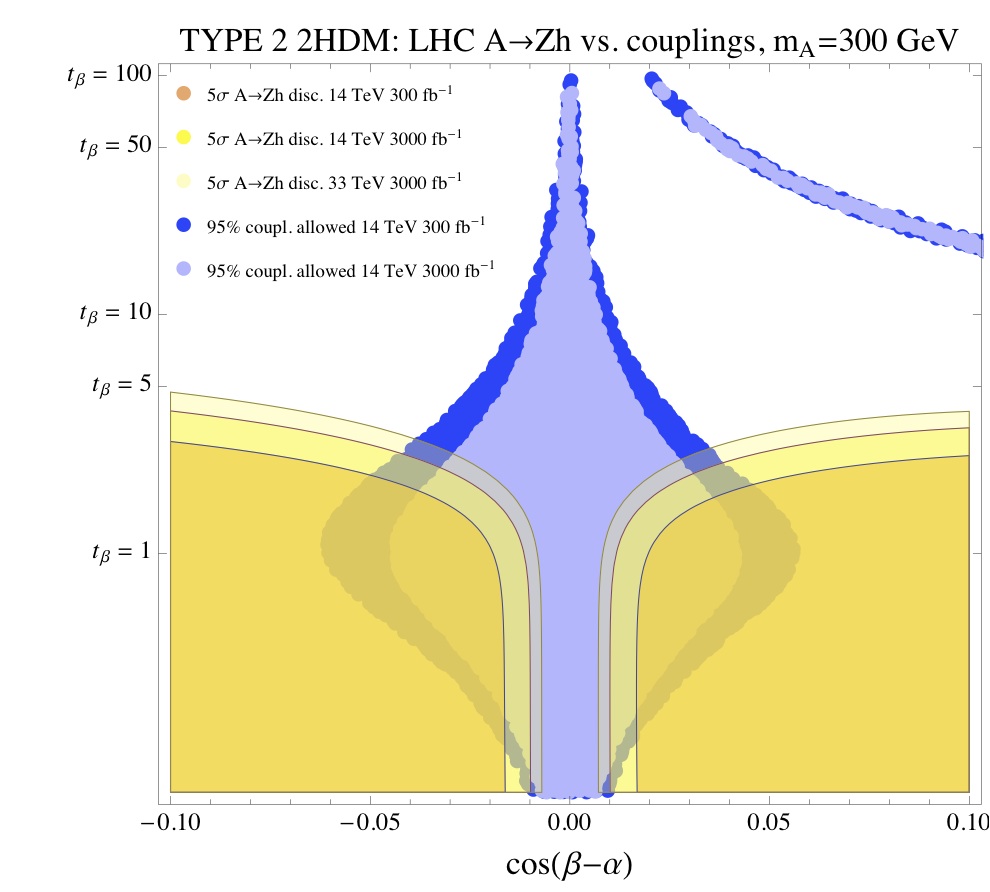}
\includegraphics[width=0.4\columnwidth,height=0.4\textheight,keepaspectratio=true]{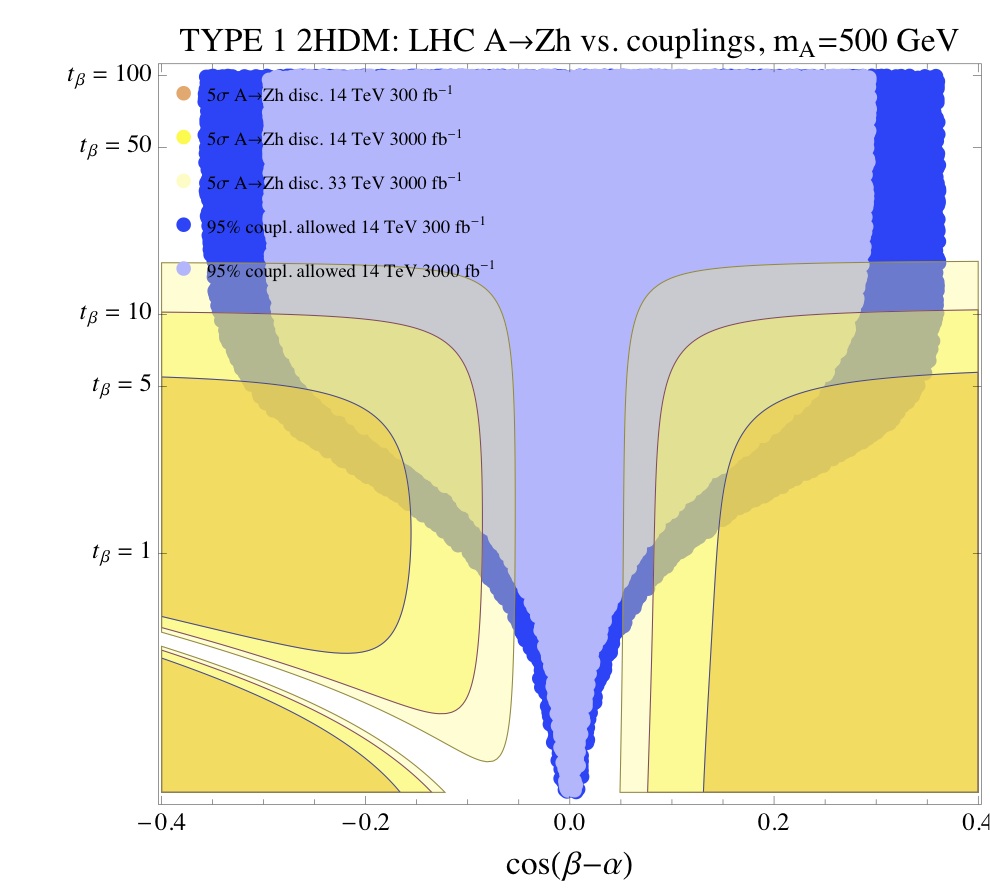}
\includegraphics[width=0.4\columnwidth,height=0.4\textheight,keepaspectratio=true]{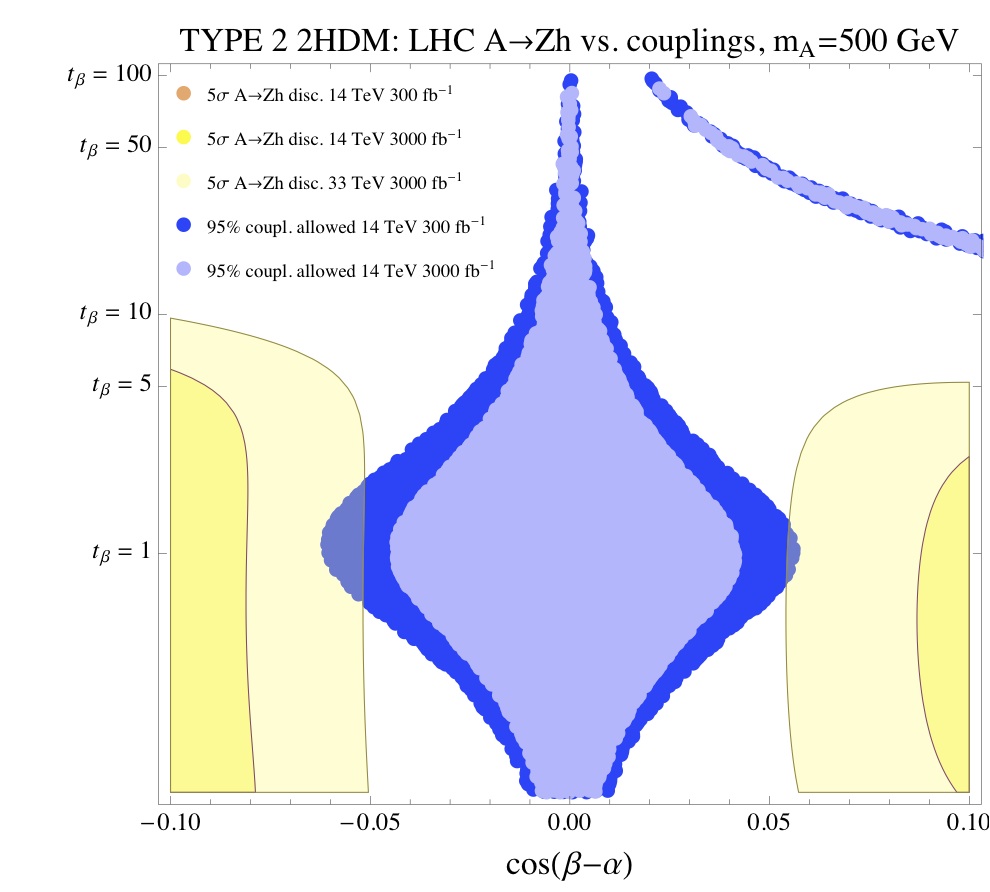}
\includegraphics[width=0.4\columnwidth,height=0.4\textheight,keepaspectratio=true]{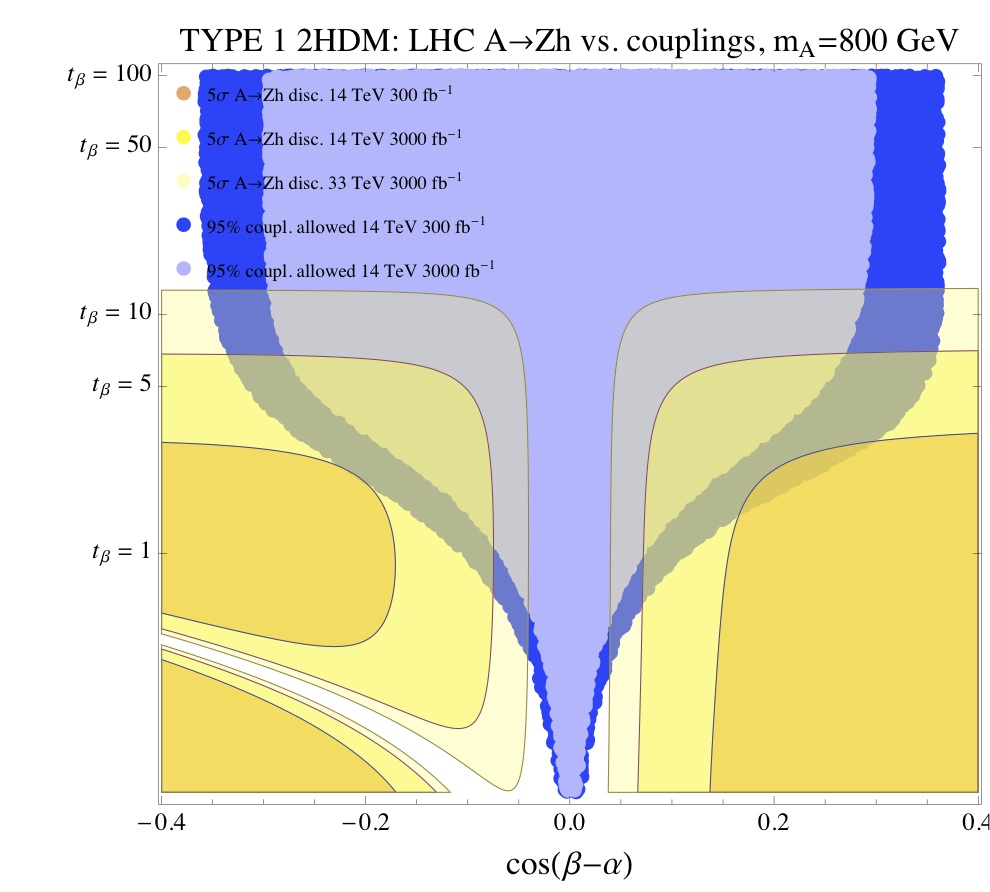}
\includegraphics[width=0.4\columnwidth,height=0.4\textheight,keepaspectratio=true]{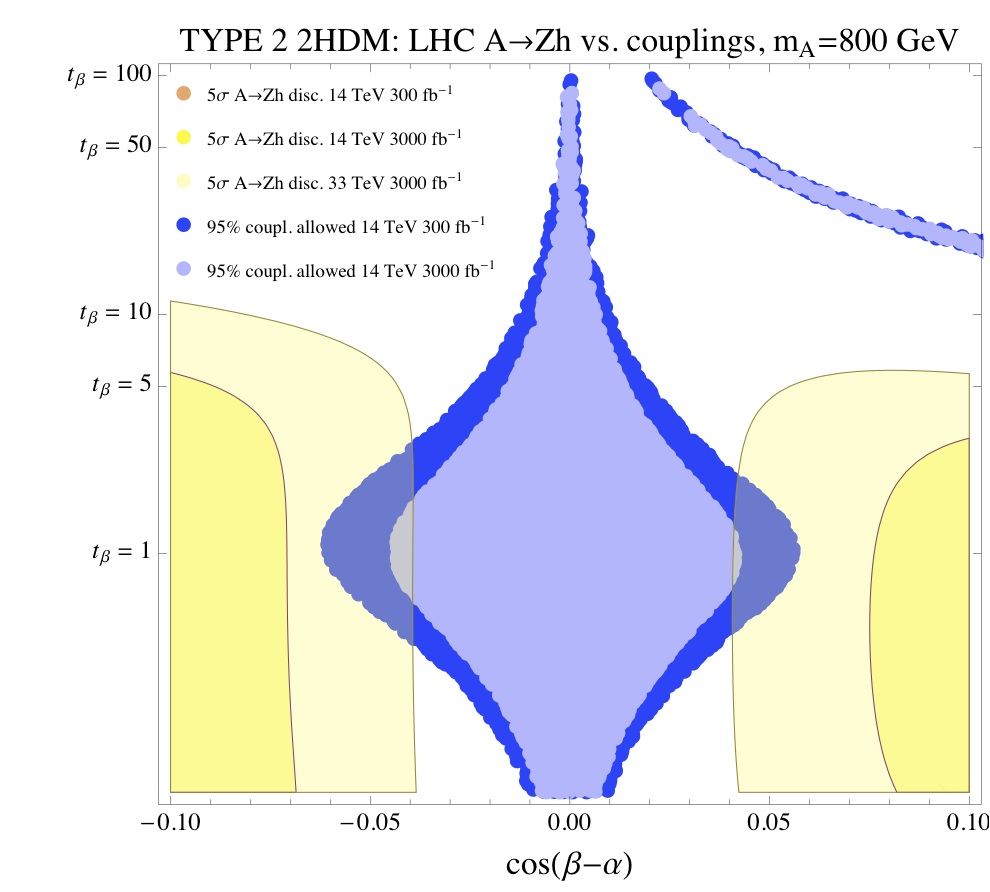}
\caption{The region of parameter space for which a 5$\sigma$ signal significance could be obtained for various $A$ mass hypotheses in a type I (left) and type II (right) 2HDM.  The dark yellow region corresponds to 300~\ifb~at $\sqrt{s}=14$~TeV, the yellow region to 3000~\ifb~at $\sqrt{s}=14$~TeV, and the light yellow region to 3000~\ifb~at $\sqrt{s}=33$~TeV.  The region which would remain allowed at 95\% CL based on non-observation of deviations from the SM in precision Higgs coupling measurements is shown in dark (light) blue for 300~\ifb~(3000~\ifb)~\cite{snowCouplings}.}
\label{fig:AZhDisco}
\end{center}
\end{figure}


\section{Conclusion}
\label{sec:Conclusion}

Searches for extended Higgs sectors at future colliders are extremely well motivated.  We have analyzed the sensitivity of  direct searches for heavy, neutral Higgs bosons at future hadron colliders with $\sqrt{s}=14$~TeV and $\sqrt{s}=33$~TeV.  The potential to either exclude or discover an extended Higgs sector is found to significantly exceed existing exclusion limits, and probe regions of parameter space which can't be constrained by precision measurements of the SM-like Higgs boson couplings.  The complementarity between direct search and precision measurement is a key finding of this analysis; the exploitation of which will allow, at future hadron colliders, exclusion or discovery of additional neutral Higgs scalars or pseudoscalars with a mass of 300~GeV, for values of $\left| \mathrm{cos}\left(\beta - \alpha\right)\right|$ as small as $\sim 0.01$ with tan$\left(\beta\right)\lesssim 1$, in both type I and II 2HDMs.  The results are compared with and without the addition of additional minimum bias events, and found to be robust against pileup.


\bibliography{2HDM/2HDM.bib}


\newpage 
\renewcommand\thefigure{A.\arabic{figure}}    
\setcounter{figure}{0} 
\renewcommand\thetable{A.\arabic{table}}    
\setcounter{table}{0} 

\section{Appendix}


\begin{table}[htbp]
\begin{center}
\begin{tabular}{|l|c|c|}
\hline
Mass [GeV]  & Cross Section [fb]    & Branching Ratio      \\ \hline
200 & 2.41e+4 & 0.128 \\
250 & 1.62e+4 & 0.196 \\
300 & 1.27e+4 & 0.0915 \\
350 & 1.32e+4 & 0.0411 \\
400 & 1.2e+4 & 0.00558 \\
450 & 8.53e+3 & 0.00423 \\
500 & 5.74e+3 & 0.00414 \\
600 & 2.55e+3 & 0.00486 \\
700 & 1.19e+3 & 0.00614 \\
800 & 580 & 0.00785 \\
900 & 299 & 0.01 \\
1000 & 161 & 0.0127 \\ \hline
\end{tabular}
\end{center}
\caption{Cross section and branching ratio for $pp\rightarrow H\rightarrow ZZ$ at $\sqrt{s}=14$~TeV in a type II 2HDM for the benchmark point with cos$(\beta - \alpha) = -0.06$ and tan$(\beta) = 1$.}
\label{tab:HZZSigma14TeV}
\end{table}

\begin{table}[htbp]
\begin{center}
\begin{tabular}{|l|c|c|}
\hline
Mass [GeV]  & Cross Section [fb]    & Branching Ratio      \\ \hline
250 & 3.7e+4 & 0.154 \\
300 & 3.37e+4 & 0.318 \\
350 & 5.98e+4 & 0.00308 \\
400 & 3.08e+4 & 0.00284 \\
450 & 1.69e+4 & 0.00359 \\
500 & 9.77e+3 & 0.00456 \\
600 & 3.67e+3 & 0.00696 \\
700 & 1.55e+3 & 0.00993 \\
800 & 718 & 0.0135 \\
900 & 356 & 0.0176 \\
1000 & 187 & 0.0222 \\ \hline
\end{tabular}
\end{center}
\caption{Cross section and branching ratio for $pp\rightarrow A\rightarrow Zh$ at $\sqrt{s}=14$~TeV in a type II 2HDM for the benchmark point with cos$(\beta - \alpha) = -0.06$ and tan($\beta) = 1$.}
\label{tab:AZhSigma14TeV}
\end{table}

\begin{table}[htbp]
\begin{center}
\begin{tabular}{|l|c|c|}
\hline
Mass [GeV]  & Cross Section [fb]    & Branching Ratio      \\ \hline
200 & 1.01e+5 & 0.128 \\
250 & 7.35e+4 & 0.196 \\
300 & 6.15e+4 & 0.0915 \\
350 & 6.87e+4 & 0.0411 \\
400 & 6.63e+4 & 0.00558 \\
450 & 5e+4 & 0.00423 \\
500 & 3.55e+4 & 0.00414 \\
600 & 1.76e+4 & 0.00486 \\
700 & 8.98e+3 & 0.00614 \\
800 & 4.83e+3 & 0.00785 \\
900 & 2.73e+3 & 0.01 \\
1000 & 1.6e+3 & 0.0127 \\ \hline
\end{tabular}
\end{center}
\caption{Cross section and branching ratio for $pp\rightarrow H\rightarrow ZZ$ at $\sqrt{s}=33$~TeV in a type II 2HDM for the benchmark point with cos$(\beta - \alpha) = -0.06$ and tan($\beta) = 1$.}
\label{tab:HZZSigma33TeV}
\end{table}

\begin{table}[htbp]
\begin{center}
\begin{tabular}{|l|c|c|}
\hline
Mass [GeV]  & Cross Section [fb]    & Branching Ratio      \\ \hline
250 & 1.68e+5 & 0.154 \\
300 & 1.64e+5 & 0.318 \\
350 & 3.11e+5 & 0.00308 \\
400 & 1.7e+5 & 0.00284 \\
450 & 9.91e+4 & 0.00359 \\
500 & 6.05e+4 & 0.00456 \\
600 & 2.52e+4 & 0.00696 \\
700 & 1.18e+4 & 0.00993 \\
800 & 5.97e+3 & 0.0135 \\
900 & 3.24e+3 & 0.0176 \\
1000 & 1.85e+3 & 0.0222 \\ \hline
\end{tabular}
\end{center}
\caption{Cross section and branching ratio for $pp\rightarrow A\rightarrow Zh$ at $\sqrt{s}=33$~TeV in a type II 2HDM for the benchmark point with cos$(\beta - \alpha) = -0.06$ and tan$(\beta) = 1$.}
\label{tab:AZhSigma33TeV}
\end{table}


\begin{table}[htbp]
\begin{center}
\begin{footnotesize}
\begin{tabular}{|l|c|c|c|c|c|c|c|}
\hline
Signal Mass [GeV] & $N_{lepton}=2$ & Lepton Trigger & $N_{\tau}=2$ & $N_{b}<2$ & $N_Z=1$ & $N_h=1$ & $N_A=1$ \\ \hline
250      & 3.28e+3        & 3.25e+3        & 221             & 221             & 210             & 176             & 176             \\
300      & 6.47e+3        & 6.44e+3        & 471             & 471             & 453             & 377             & 377             \\
350      & 114             & 114             & 9.02            & 9               & 8.52            & 6.9             & 6.9             \\
400      & 55.1            & 54.9            & 4.67            & 4.67            & 4.45            & 3.64            & 3.64            \\
450      & 39              & 38.9            & 3.67            & 3.66            & 3.51            & 2.81            & 2.81            \\
500      & 28.5            & 28.5            & 2.79            & 2.78            & 2.66            & 2.13            & 2.13            \\
600      & 16.6            & 16.6            & 1.77            & 1.77            & 1.68            & 1.28            & 1.28            \\
700      & 9.9             & 9.89            & 1.18            & 1.18            & 1.12            & 0.874           & 0.874           \\
800      & 6.21            & 6.21            & 0.775           & 0.774           & 0.74            & 0.568           & 0.568           \\
900      & 4               & 4               & 0.521           & 0.521           & 0.498           & 0.388           & 0.388           \\
1000     & 2.63            & 2.62            & 0.332           & 0.332           & 0.317           & 0.243           & 0.243           \\ \hline
\end{tabular}
\end{footnotesize}
\end{center}
\caption{Expected number of pre-selected events for the $A\rightarrow Zh\rightarrow \ell \ell \tautau$ signal for $\int Ldt=$ 300~\ifb~at $\sqrt{s}=14$ TeV with $<N_{PU}>=50$.}
\label{tab:AZhtautauPreselCutflowSignal}
\end{table}

\begin{table}[htbp]
\begin{center}
\begin{footnotesize}
\begin{tabular}{|l|c|c|c|c|c|c|c|}
\hline
Background & $N_{lepton}=2$ & Lepton Trigger & $N_{\tau}=2$ & $N_{b}<2$ & $N_Z=1$ & $N_h=1$ & $N_A=1$ \\ \hline
B, Bj, Bjj-vbf, BB, BBB                            & 4.58e+8        & 4.05e+8        & 1.9e+3         & 1.9e+3         & 701             & 399             & 399              \\
tj, tB, tt, ttB                                    & 7.53e+6        & 7.36e+6        & 2.49e+3        & 2.43e+3        & 30.2            & 8.92            & 8.92             \\
H                                                  & 4.99e+4        & 4.48e+4        & 5.13            & 5.13            & 1.33            & 0.384           & 0.384            \\ \hline
Total Background                                   & 4.65e+8        & 4.13e+8        & 4.4e+3         & 4.33e+3        & 732             & 408             & 408              \\ \hline
\end{tabular}
\end{footnotesize}
\end{center}
\caption{Expected number of pre-selected events for the SM backgrounds to $A\rightarrow Zh\rightarrow \ell \ell \tautau$ for $\int Ldt=$ 300~\ifb~at $\sqrt{s}=14$ TeV with $<N_{PU}>=50$.}
\label{tab:AZhtautauPreselCutflowBackground}
\end{table}


\begin{table}[htbp]
\begin{center}
\begin{footnotesize}
\begin{tabular}{|l|c|c|c|c|c|c|c|}
\hline
Signal Mass [GeV] & $N_{lepton}=2$ & Lepton Trigger & $N_{\tau}=2$ & $N_{b}<2$ & $N_Z=1$ & $N_h=1$ & $N_A=1$ \\ \hline
250     & 1.3e+5         & 1.29e+5        & 5.14e+3        & 5.14e+3        & 4.65e+3        & 3.16e+3        & 3.16e+3        \\
300     & 2.82e+5        & 2.8e+5         & 1.22e+4        & 1.22e+4        & 1.13e+4        & 7.25e+3        & 7.25e+3        \\
350     & 5.38e+3        & 5.36e+3        & 290             & 290             & 266             & 173             & 173             \\
400     & 2.81e+3        & 2.8e+3         & 156             & 156             & 143             & 93.6            & 93.6            \\
450     & 2.08e+3        & 2.07e+3        & 133             & 132             & 122             & 72              & 72              \\
500     & 1.64e+3        & 1.64e+3        & 118             & 118             & 111             & 67              & 67              \\
600     & 1.07e+3        & 1.07e+3        & 87              & 86.9            & 82              & 46.7            & 46.7            \\
700     & 714             & 713             & 65.3            & 65.2            & 61.6            & 36.4            & 36.4            \\
800     & 494             & 493             & 50.6            & 50.5            & 47.7            & 27.1            & 27.1            \\
900     & 348             & 348             & 35.8            & 35.7            & 33.7            & 18.6            & 18.6            \\
1000    & 249             & 248             & 26.3            & 26.2            & 24.9            & 14.1            & 14.1            \\ \hline
\end{tabular}
\end{footnotesize}
\end{center}
\caption{Expected number of pre-selected events for the $A\rightarrow Zh\rightarrow \ell \ell \tautau$ signal for $\int Ldt=$ 3000~\ifb~at $\sqrt{s}=33$ TeV with $<N_{PU}>=$~140.}
\label{tab:AZhtautauPreselCutflowSignal_33}
\end{table}

\begin{table}[htbp]
\begin{center}
\begin{footnotesize}
\begin{tabular}{|l|c|c|c|c|c|c|c|}
\hline
Background & $N_{lepton}=2$ & Lepton Trigger & $N_{\tau}=2$ & $N_{b}<2$ & $N_Z=1$ & $N_h=1$ & $N_A=1$ \\ \hline
B, Bj, Bjj-vbf, BB, BBB                            & 8.66e+9        & 7.68e+9        & 1.6e+6         & 1.6e+6         & 1.43e+6        & 9.75e+3        & 9.75e+3         \\
tj, tB, tt, ttB                                    & 5.06e+8        & 4.86e+8        & 3.9e+5         & 3.78e+5        & 1.97e+4        & 610             & 609              \\
H                                                  & 2.21e+6        & 1.97e+6        & 791             & 790             & 102             & 37.4            & 37.4             \\ \hline
Total Background                                   & 9.17e+9        & 8.17e+9        & 1.99e+6        & 1.98e+6        & 1.45e+6        & 1.04e+4        & 1.04e+4         \\ \hline
\end{tabular}
\end{footnotesize}
\end{center}
\caption{Expected number of pre-selected events for the SM backgrounds to $A\rightarrow Zh\rightarrow \ell \ell \tautau$ for $\int Ldt=$ 3000~\ifb~at $\sqrt{s}=33$ TeV with $<N_{PU}>=$~140.}
\label{tab:AZhtautauPreselCutflowBackground_33}
\end{table}


\begin{figure}[htbp]
\begin{center}
\includegraphics[width=0.4\columnwidth,height=0.4\textheight,keepaspectratio=true]{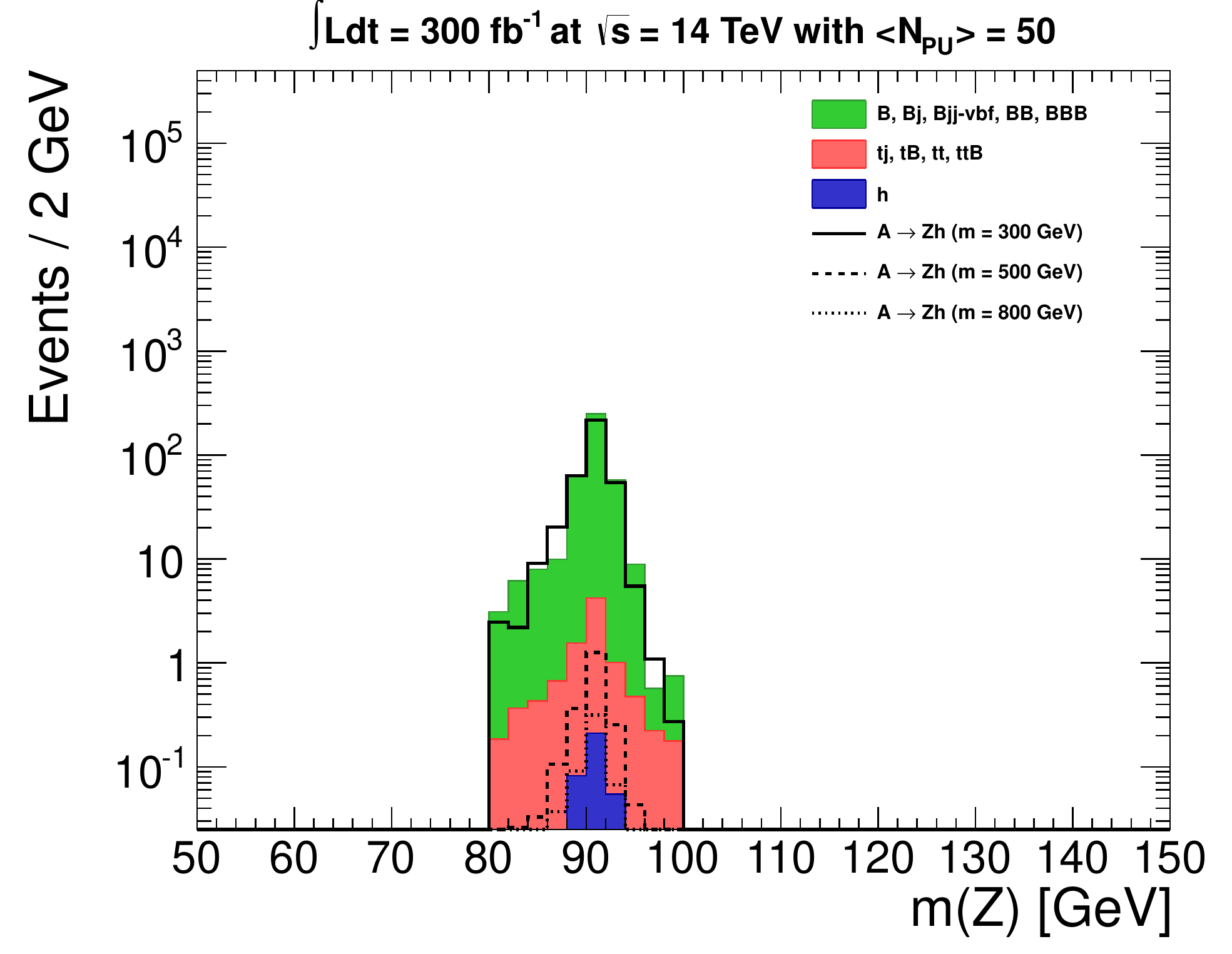}
\includegraphics[width=0.4\columnwidth,height=0.4\textheight,keepaspectratio=true]{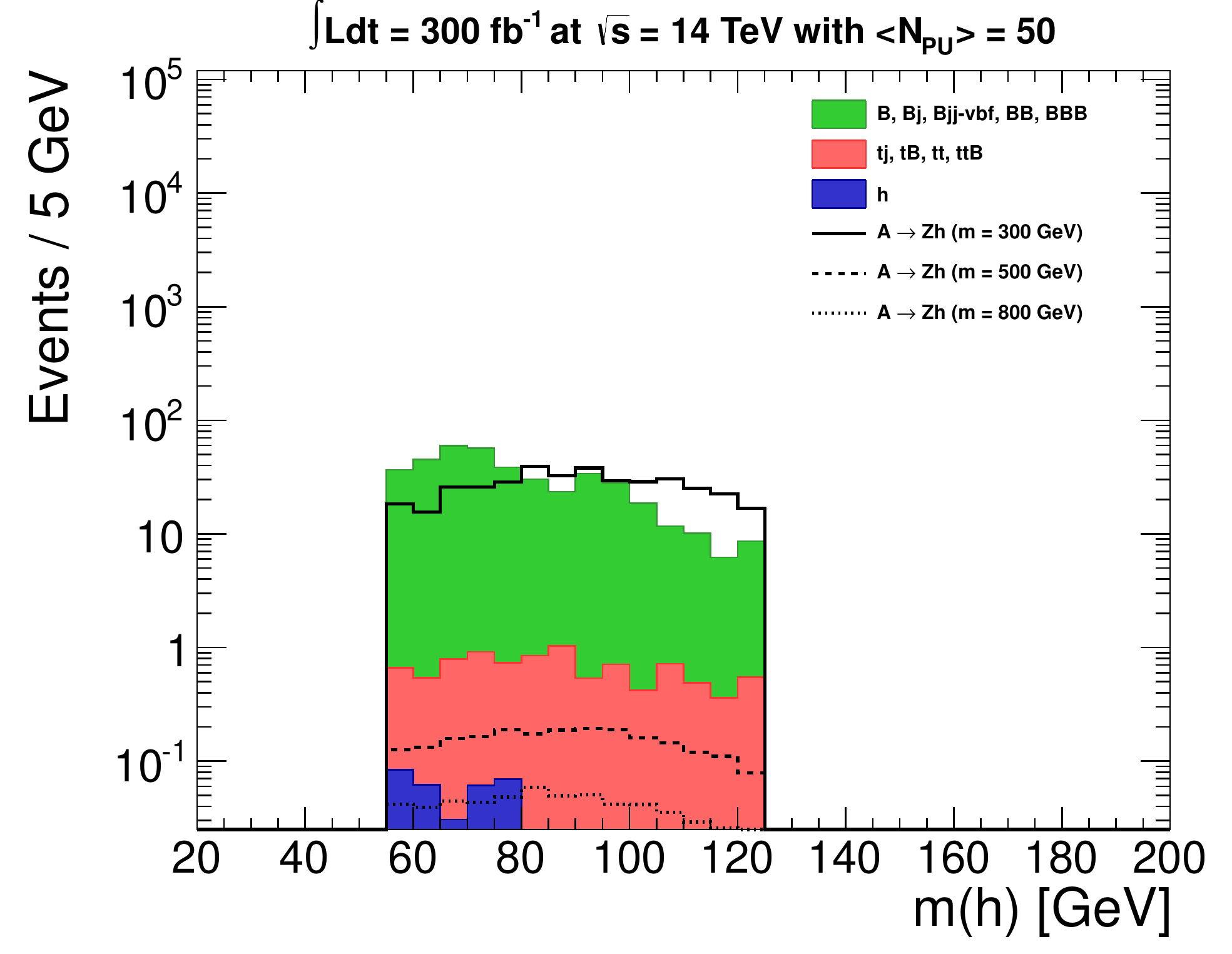}
\includegraphics[width=0.4\columnwidth,height=0.4\textheight,keepaspectratio=true]{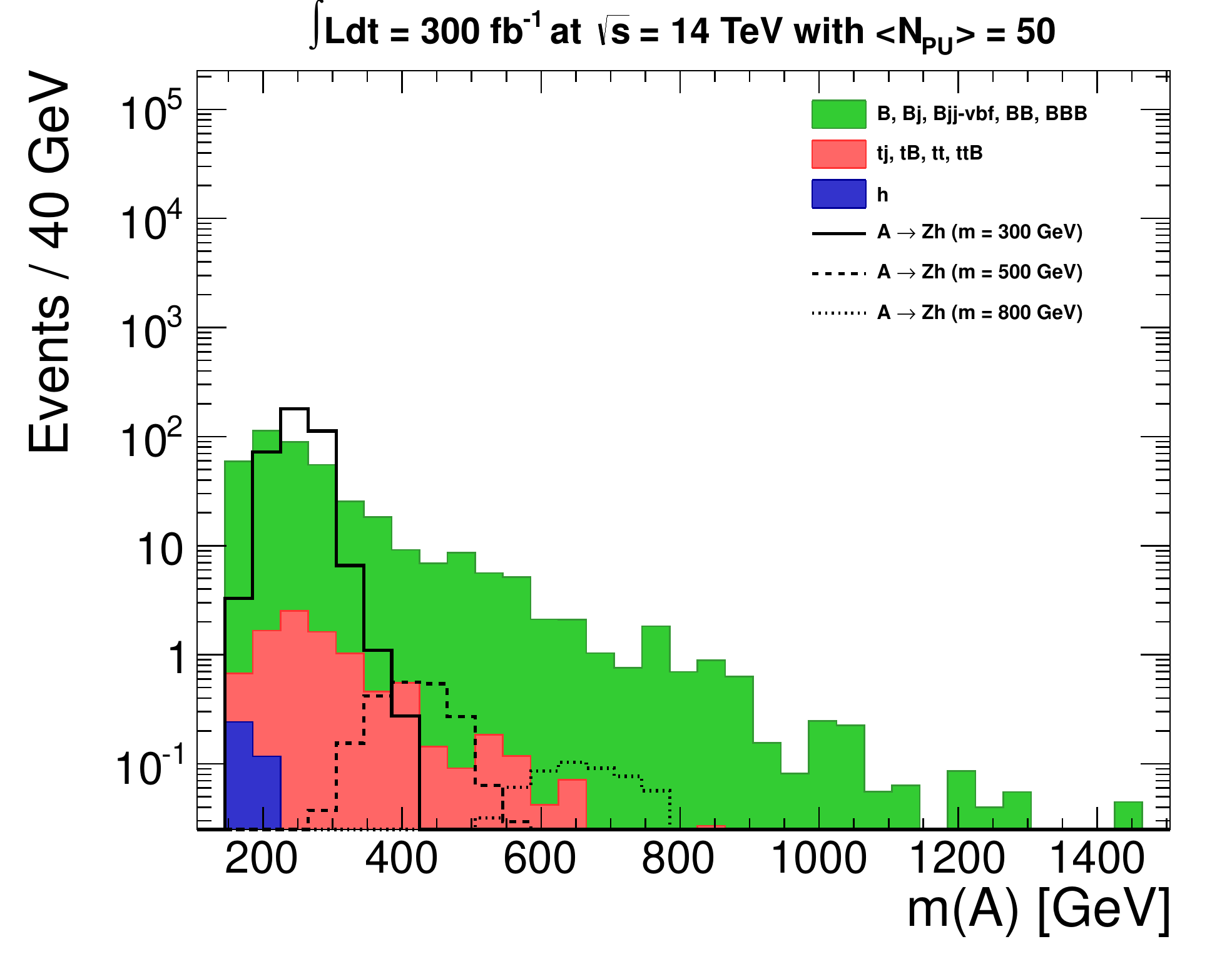}
\includegraphics[width=0.4\columnwidth,height=0.4\textheight,keepaspectratio=true]{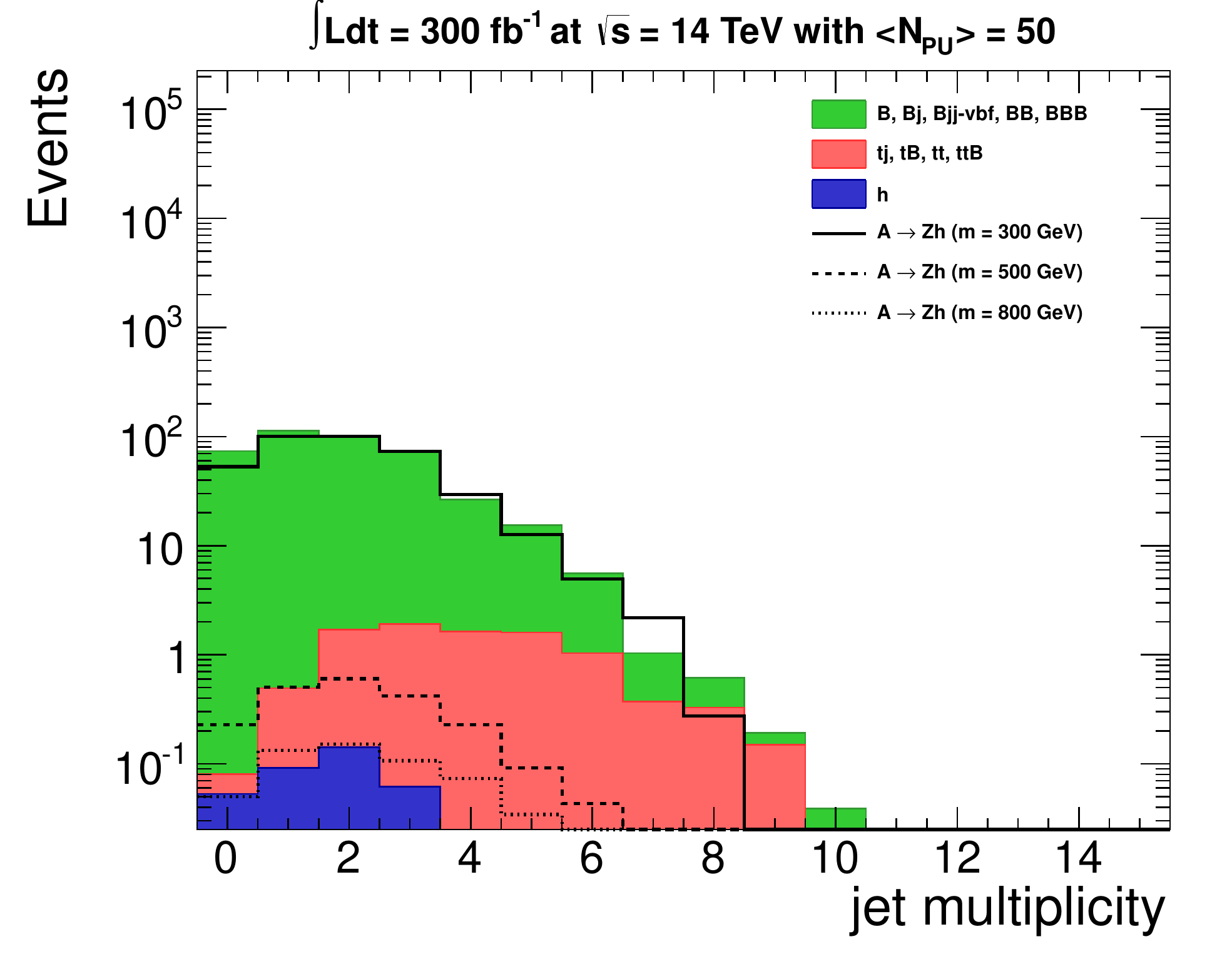}
\includegraphics[width=0.4\columnwidth,height=0.4\textheight,keepaspectratio=true]{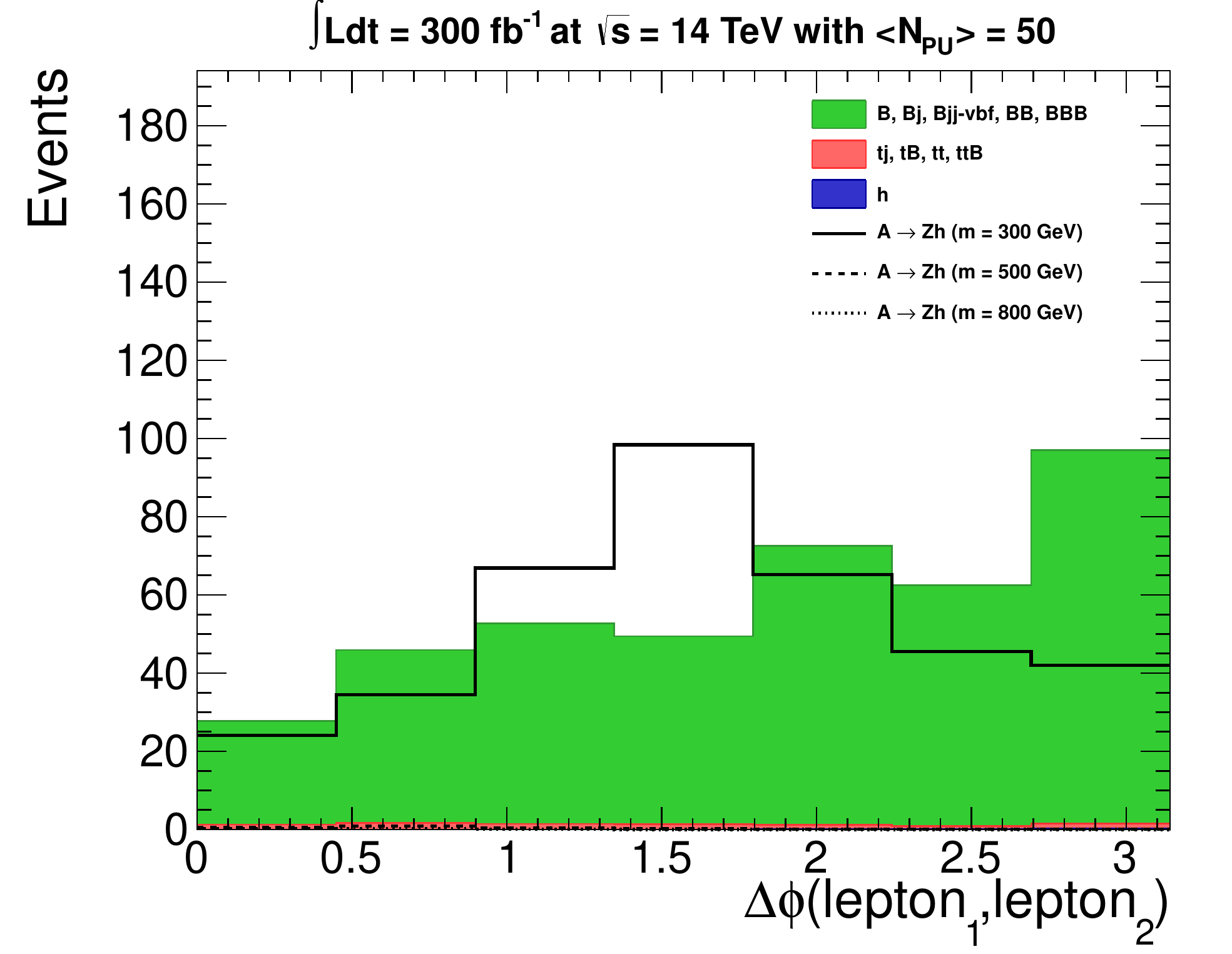}
\includegraphics[width=0.4\columnwidth,height=0.4\textheight,keepaspectratio=true]{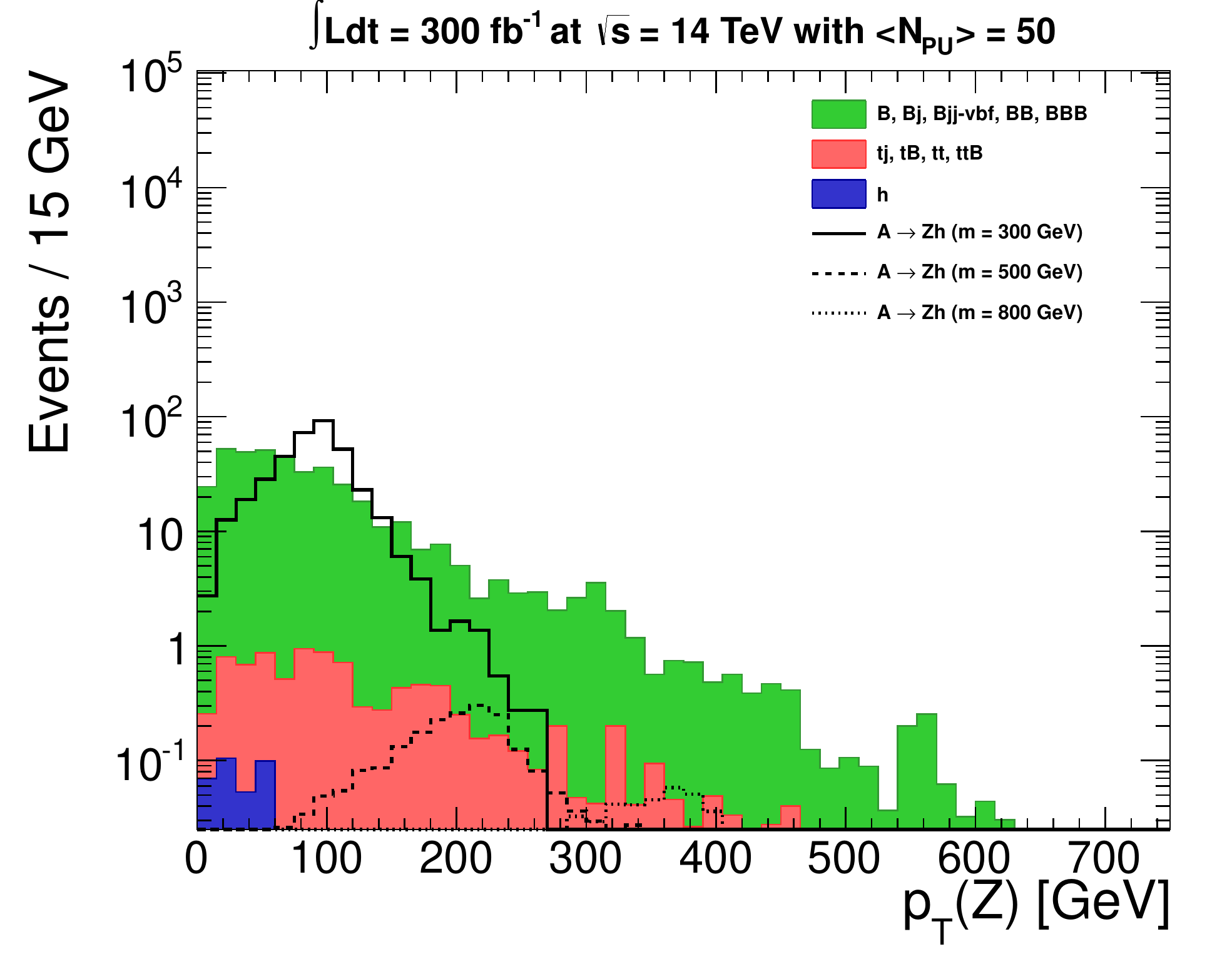}
\includegraphics[width=0.4\columnwidth,height=0.4\textheight,keepaspectratio=true]{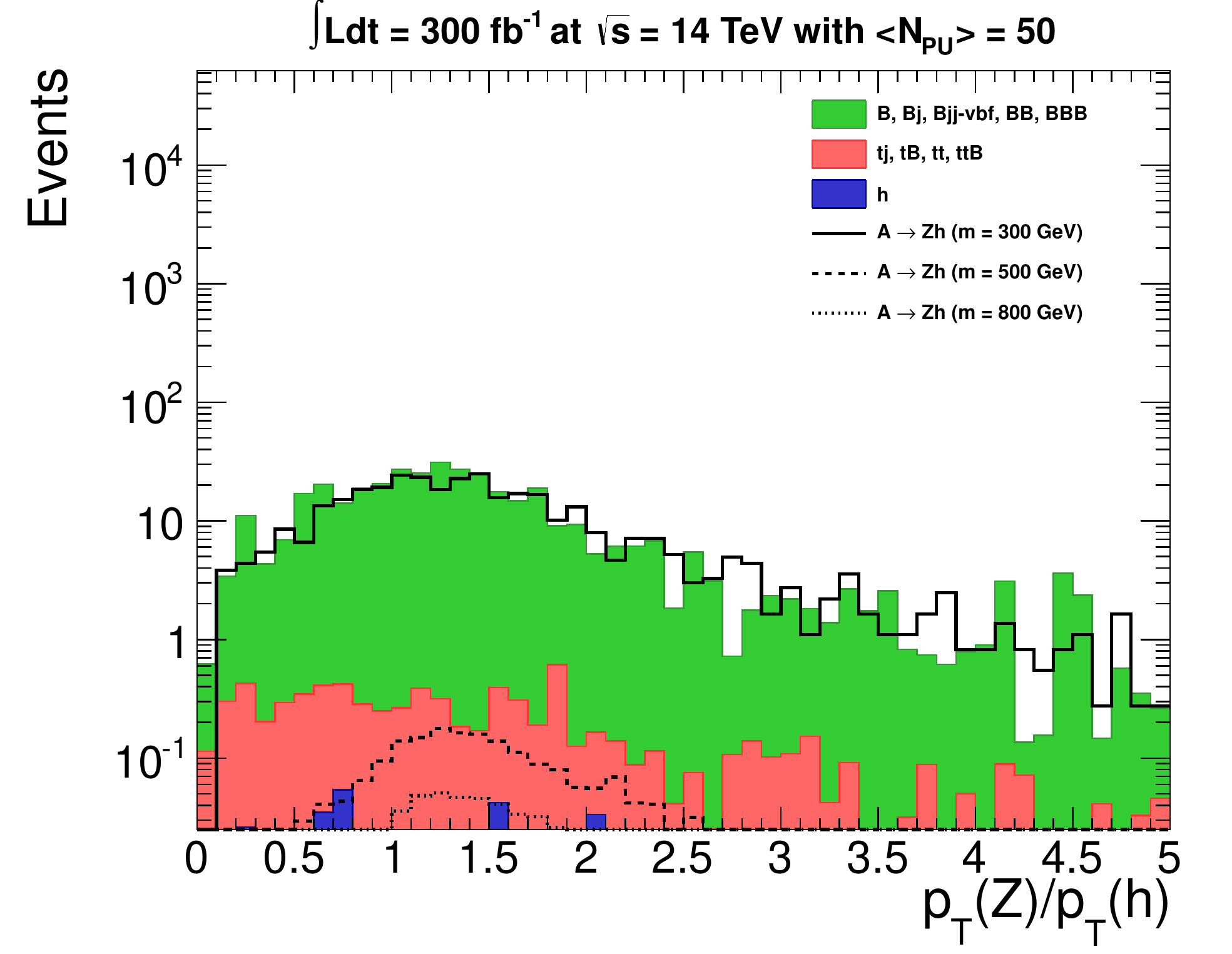}
\includegraphics[width=0.4\columnwidth,height=0.4\textheight,keepaspectratio=true]{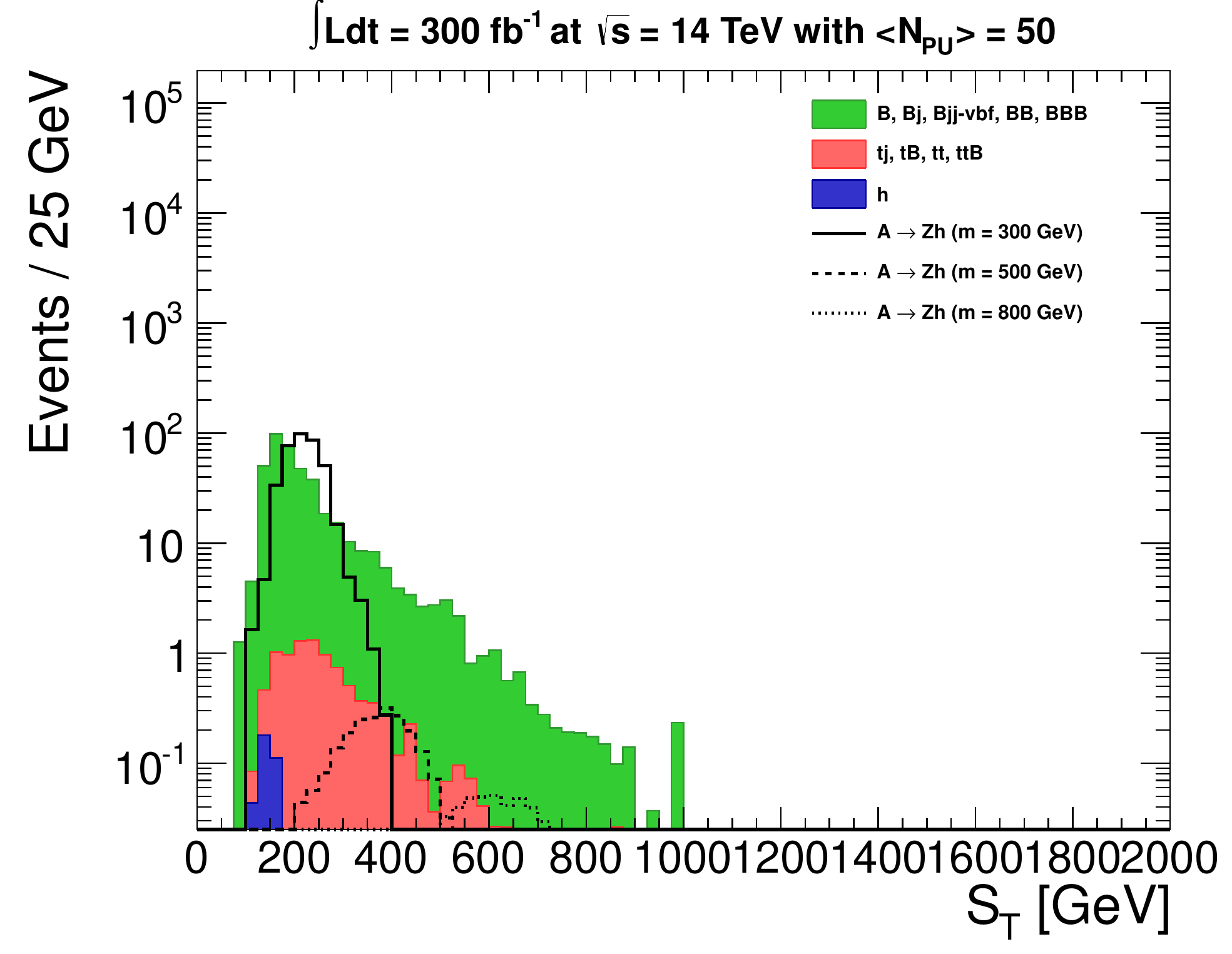}
\caption{Kinematic distributions for pre-selected events in the \tautau~channel, for $\int Ldt=$ 300~\ifb~at $\sqrt{s}=14$ TeV with $<N_{PU}>=50$.}
\label{fig:AZhtautauPresel}
\end{center}
\end{figure}


\begin{figure}[htbp]
\begin{center}
\includegraphics[width=0.4\columnwidth,height=0.4\textheight,keepaspectratio=true]{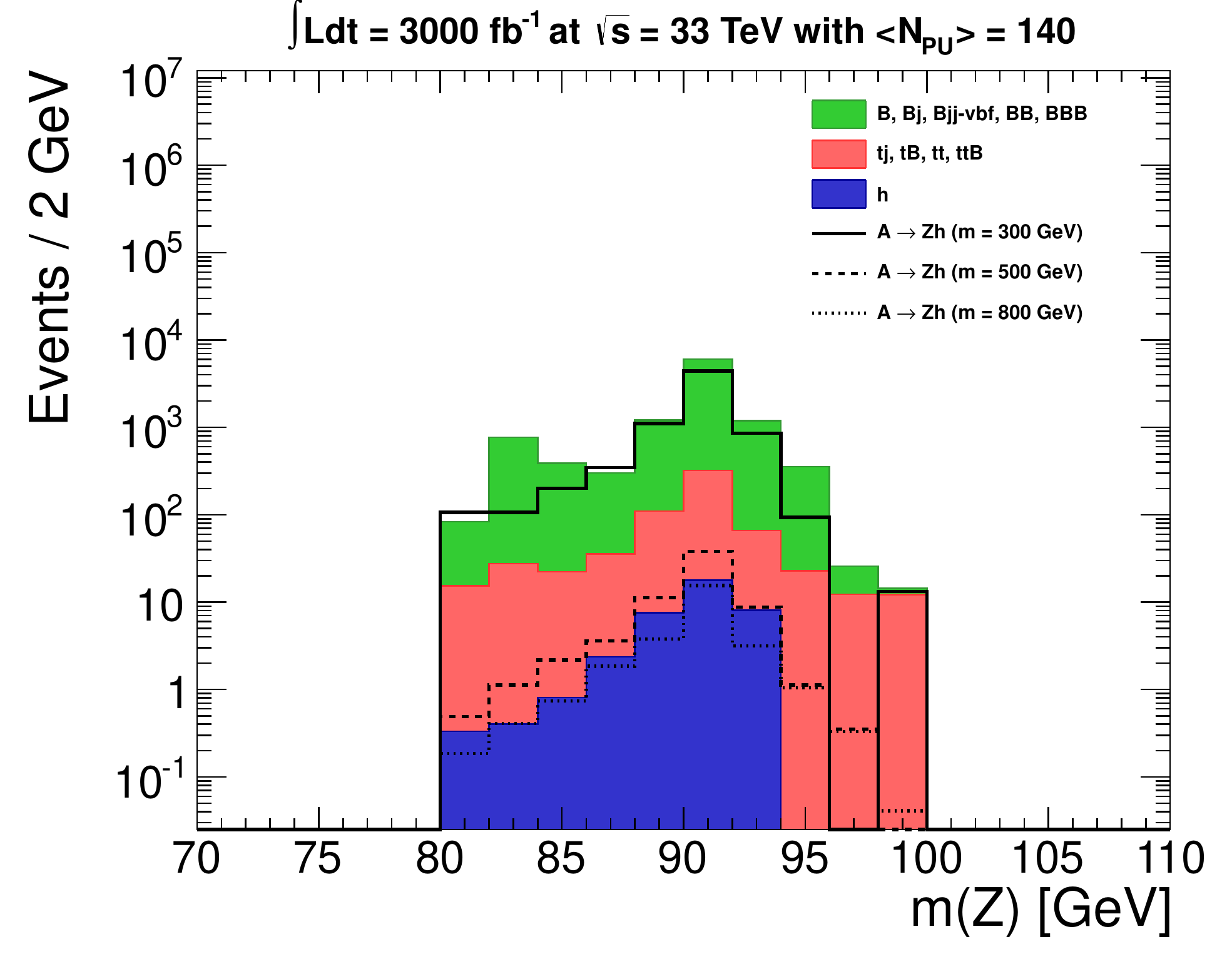}
\includegraphics[width=0.4\columnwidth,height=0.4\textheight,keepaspectratio=true]{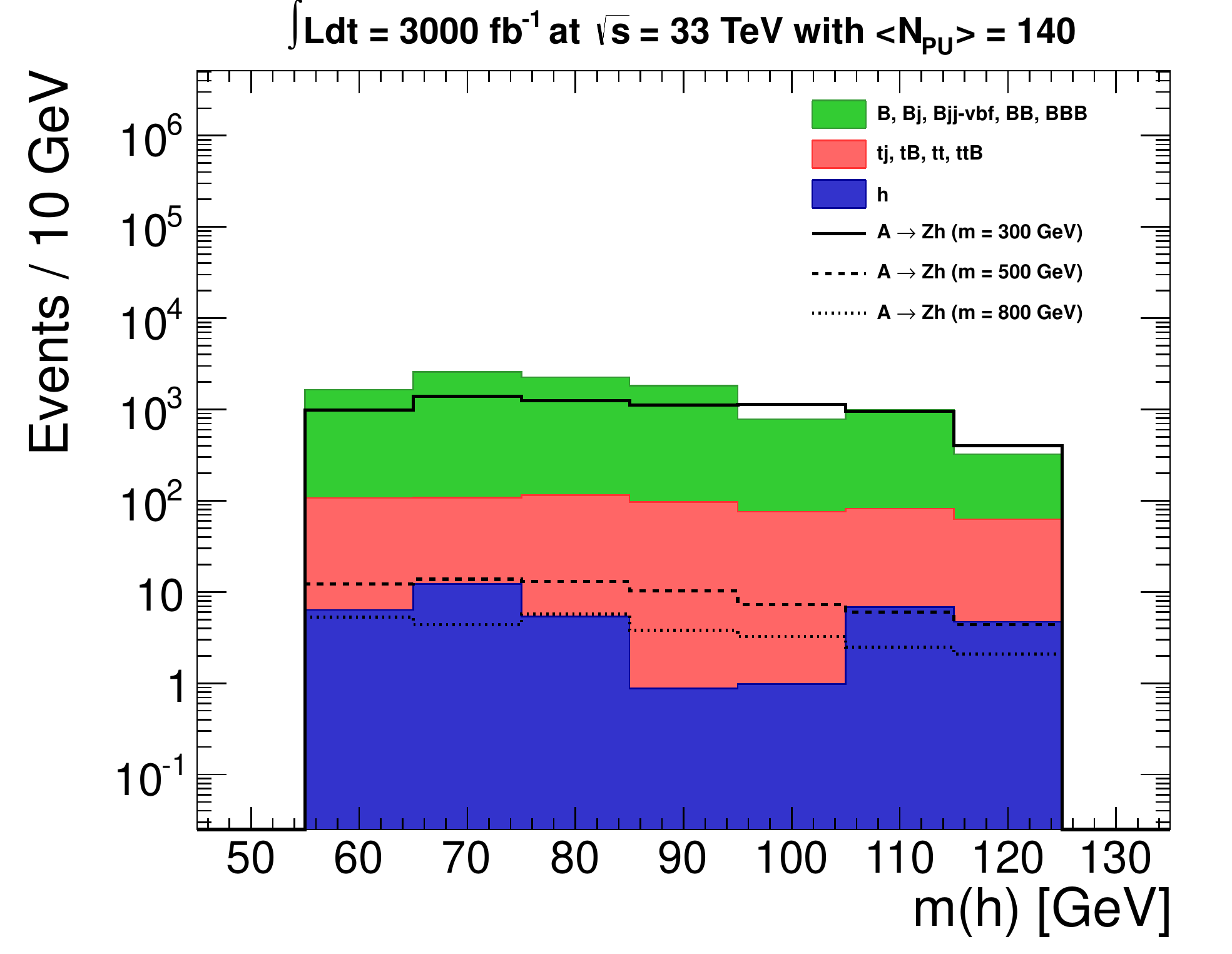}
\includegraphics[width=0.4\columnwidth,height=0.4\textheight,keepaspectratio=true]{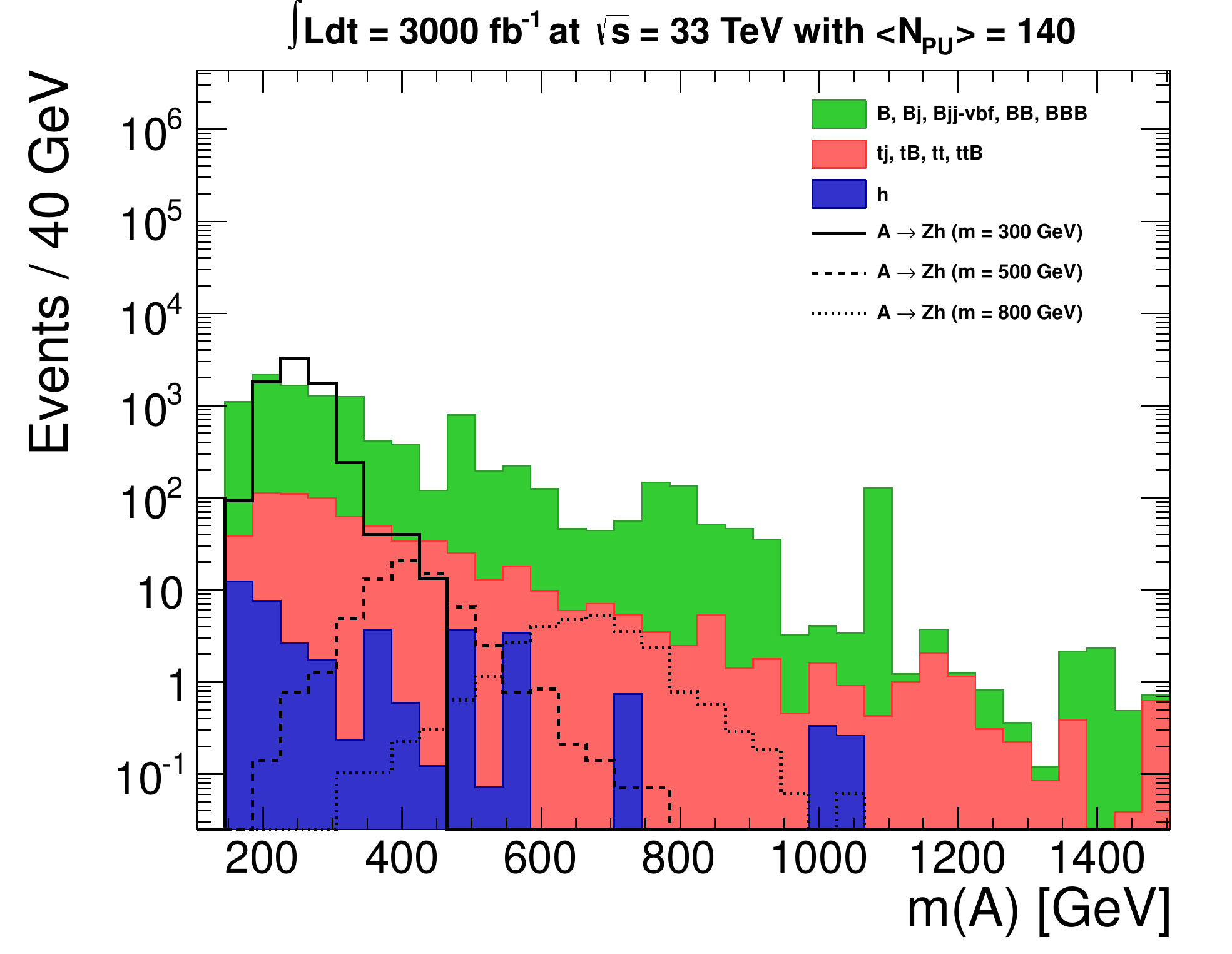}
\includegraphics[width=0.4\columnwidth,height=0.4\textheight,keepaspectratio=true]{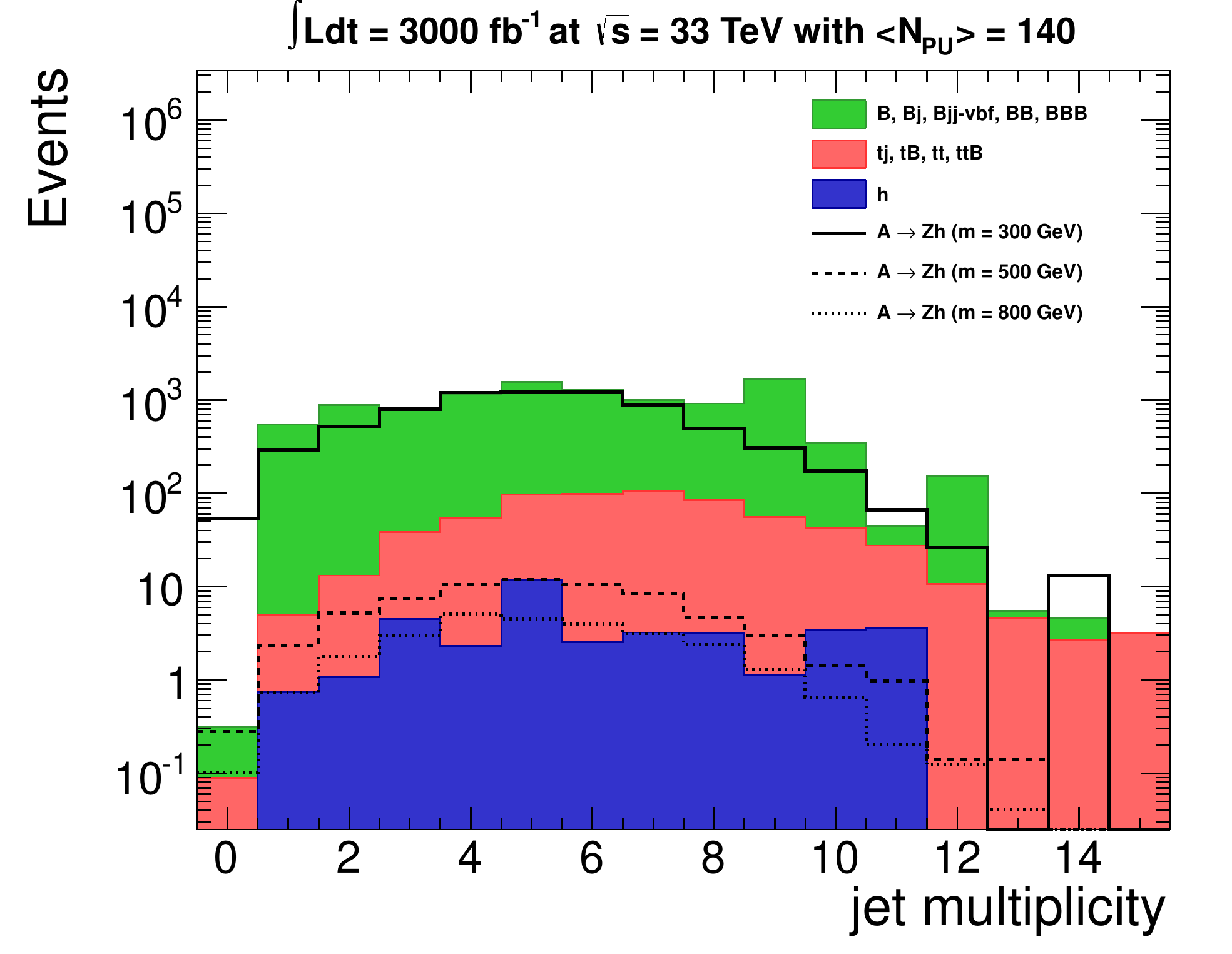}
\includegraphics[width=0.4\columnwidth,height=0.4\textheight,keepaspectratio=true]{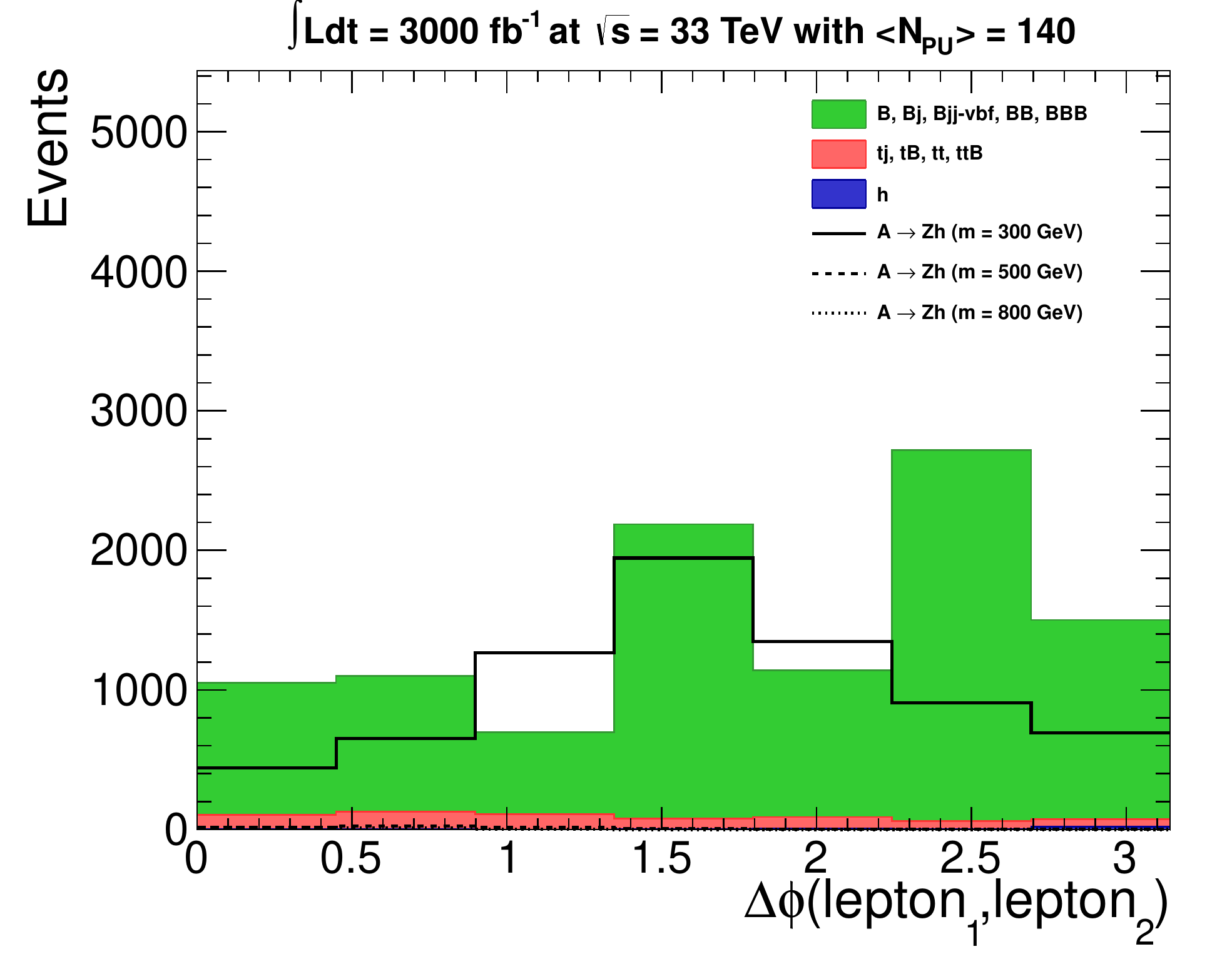}
\includegraphics[width=0.4\columnwidth,height=0.4\textheight,keepaspectratio=true]{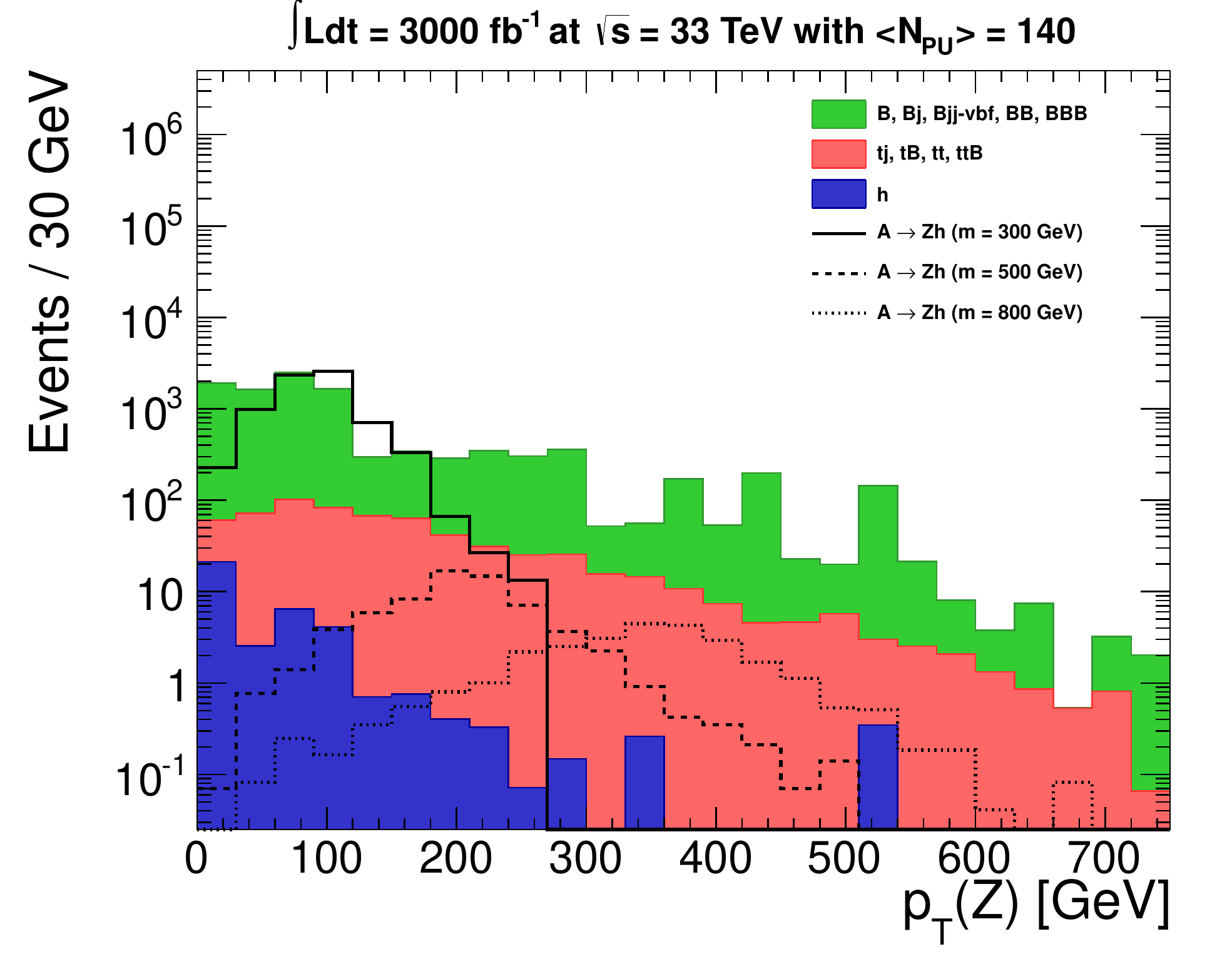}
\includegraphics[width=0.4\columnwidth,height=0.4\textheight,keepaspectratio=true]{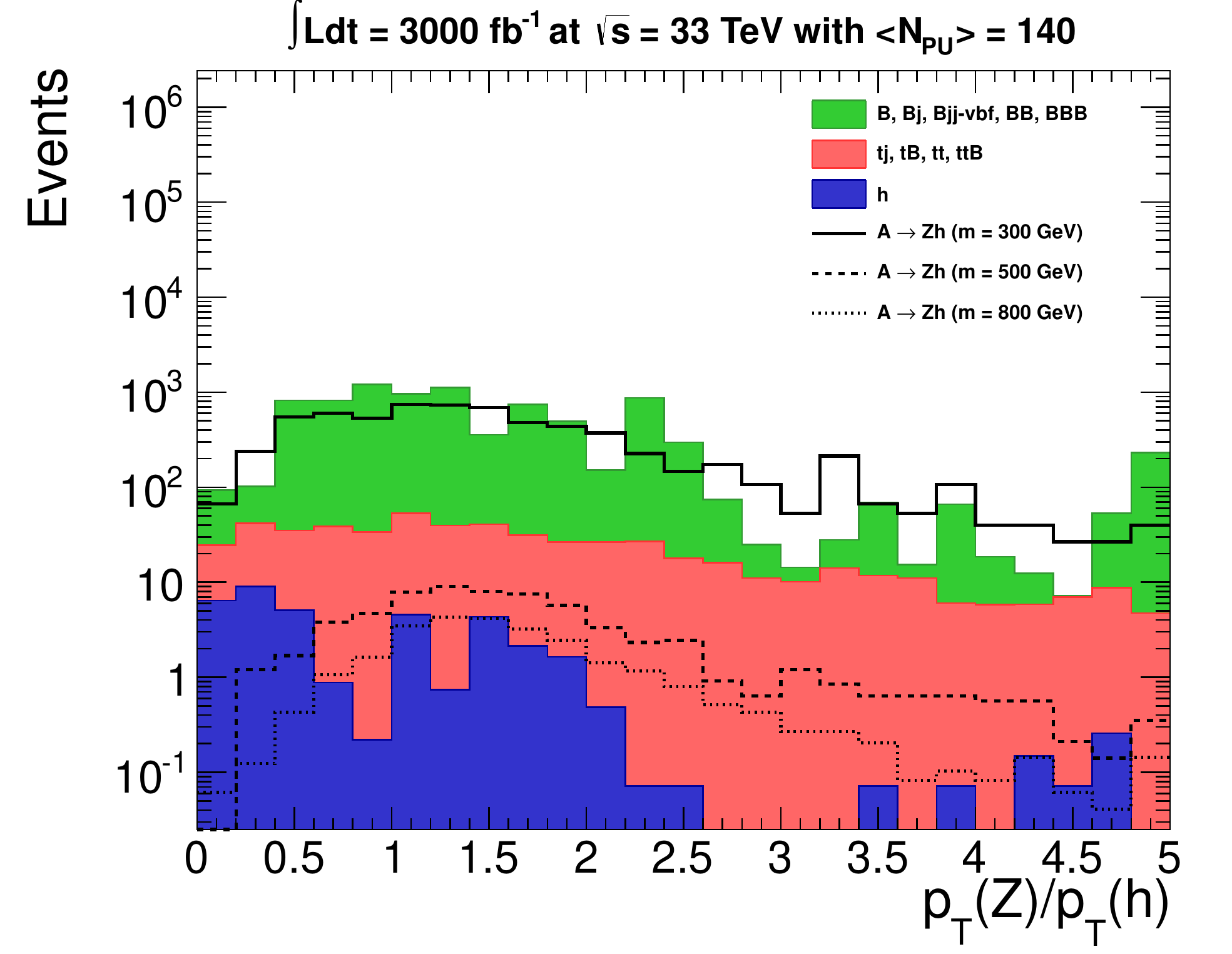}
\includegraphics[width=0.4\columnwidth,height=0.4\textheight,keepaspectratio=true]{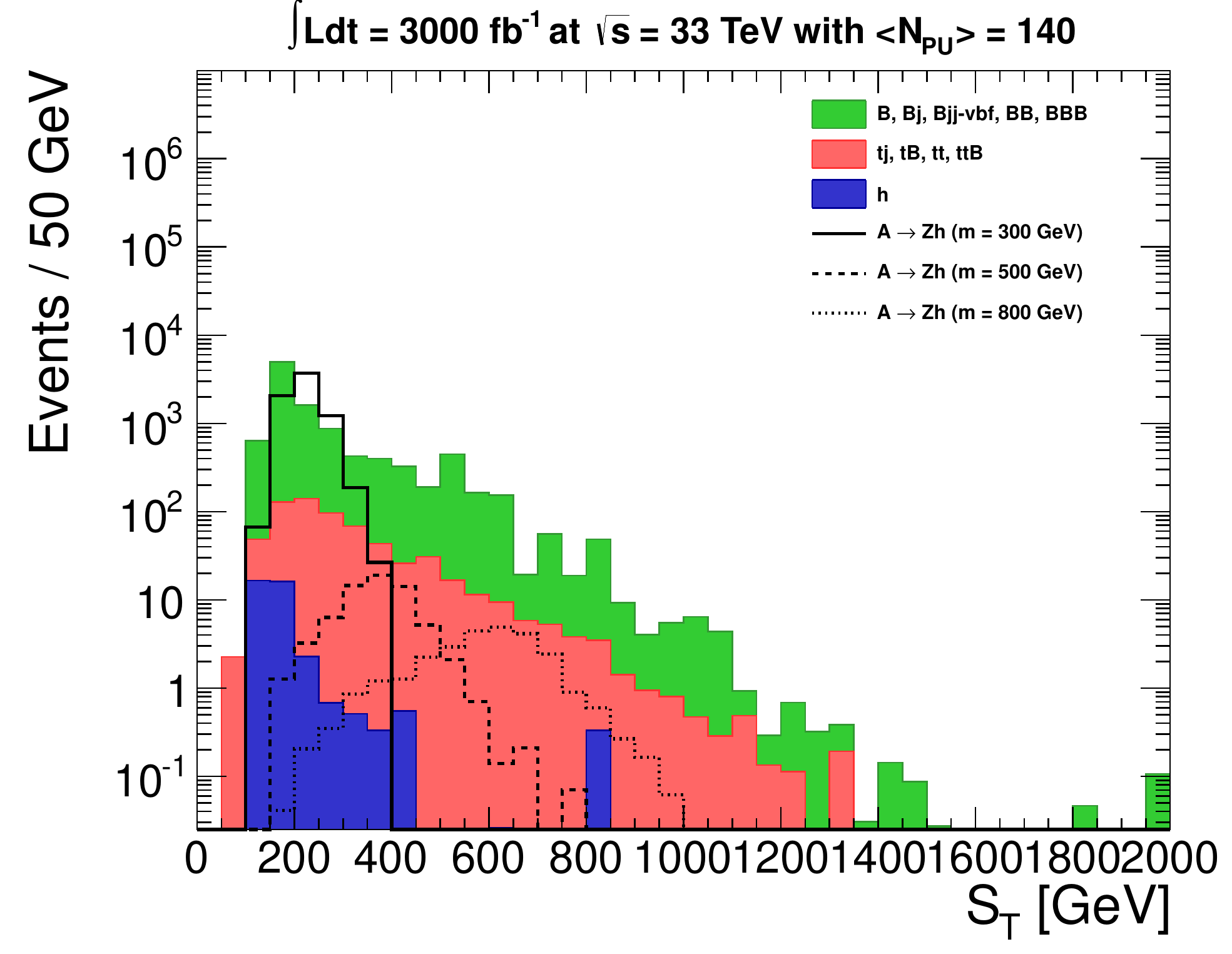}
\caption{Kinematic distributions for pre-selected events in the \tautau~channel, for $\int Ldt=$ 3000~\ifb~at $\sqrt{s}=33$ TeV with $<N_{PU}>=$~140.}
\label{fig:AZhtautauPresel_33}
\end{center}
\end{figure}


\begin{table}[htbp]
\begin{center}
\begin{footnotesize}
\begin{tabular}{|l|c|c|c|c|}
\hline
Signal Mass [GeV] & Pre-selection & $\left|\Delta\phi(\ell_1\ell_2)\right|\leq$~1.9 & $p_T(Z)\geq$~40~GeV & 0.4$\leq\frac{p_T(Z)}{p_T(h)}\leq$ 2.75 \\ \hline
250      & 176             & 54.8            & 52.2            & 38.7            \\
300      & 377             & 240             & 239             & 194             \\
350      & 6.9             & 5.42            & 5.42            & 4.57            \\
400      & 3.64            & 3.13            & 3.13            & 2.74            \\
450      & 2.81            & 2.55            & 2.55            & 2.22            \\
500      & 2.13            & 2.02            & 2.01            & 1.77            \\
600      & 1.28            & 1.24            & 1.24            & 1.13            \\
700      & 0.874           & 0.855           & 0.855           & 0.787           \\
800      & 0.568           & 0.564           & 0.564           & 0.52            \\
900      & 0.388           & 0.385           & 0.385           & 0.358           \\
1000     & 0.243           & 0.242           & 0.242           & 0.228           \\ \hline
\end{tabular}
\end{footnotesize}
\end{center}
\caption{Expected number of selected events for the $A\rightarrow Zh\rightarrow \ell \ell \tautau$ signal for $\int Ldt=$ 300~\ifb~at $\sqrt{s}=14$ TeV with $<N_{PU}>=50$.}
\label{tab:AZhtautauSelCutflowSignal}
\end{table}

\begin{table}[htbp]
\begin{center}
\begin{footnotesize}
\begin{tabular}{|l|c|c|c|c|}
\hline
Background & Pre-selection & $\left|\Delta\phi(\ell_1\ell_2)\right|\leq$~1.9 & $p_T(Z)\geq$~40~GeV & 0.4$\leq\frac{p_T(Z)}{p_T(h)}\leq$ 2.75 \\ \hline
B, Bj, Bjj-vbf, BB, BBB                            & 399             & 190             & 185             & 153              \\
tj, tB, tt, ttB                                    & 8.92            & 6.02            & 6               & 3.66             \\
H                                                  & 0.384           & 0.0233          & 0.0233          & 0.0163           \\ \hline
Total Background                                   & 408             & 196             & 191             & 157              \\ \hline
\end{tabular}
\end{footnotesize}
\end{center}
\caption{Expected number of selected events for the SM backgrounds to $A\rightarrow Zh\rightarrow \ell \ell \tautau$ for $\int Ldt=$ 300~\ifb~at $\sqrt{s}=14$ TeV with $<N_{PU}>=50$.}
\label{tab:AZhtautauSelCutflowBackground}
\end{table}


\begin{table}[htbp]
\begin{center}
\begin{footnotesize}
\begin{tabular}{|l|c|c|c|c|}
\hline
Signal Mass [GeV] & Pre-selection & $\left|\Delta\phi(\ell_1\ell_2)\right|\leq$~1.9 & $p_T(Z)\geq$~40~GeV & 0.4$\leq\frac{p_T(Z)}{p_T(h)}\leq$ 2.75 \\ \hline
250     & 3.16e+3        & 1.01e+3        & 980             & 671             \\
300     & 7.25e+3        & 4.58e+3        & 4.58e+3        & 3.54e+3        \\
350     & 173             & 135             & 135             & 110             \\
400     & 93.6            & 81.7            & 81.6            & 68.5            \\
450     & 72              & 65.8            & 65.8            & 56.6            \\
500     & 67              & 62.6            & 62.6            & 53.5            \\
600     & 46.7            & 45              & 45              & 39              \\
700     & 36.4            & 35.5            & 35.5            & 31.6            \\
800     & 27.1            & 26.6            & 26.6            & 24.2            \\
900     & 18.6            & 18.5            & 18.5            & 16.7            \\
1000    & 14.1            & 14              & 14              & 12.7            \\ \hline
\end{tabular}
\end{footnotesize}
\end{center}
\caption{Expected number of selected events for the $A\rightarrow Zh\rightarrow \ell \ell \tautau$ signal for $\int Ldt=$ 3000~\ifb~at $\sqrt{s}=33$ TeV with $<N_{PU}>=$~140.}
\label{tab:AZhtautauSelCutflowSignal33}
\end{table}

\begin{table}[htbp]
\begin{center}
\begin{footnotesize}
\begin{tabular}{|l|c|c|c|c|}
\hline
Background & Pre-selection & $\left|\Delta\phi(\ell_1\ell_2)\right|\leq$~1.9 & $p_T(Z)\geq$~40~GeV & 0.4$\leq\frac{p_T(Z)}{p_T(h)}\leq$ 2.75 \\ \hline
B, Bj, Bjj-vbf, BB, BBB                            & 9.75e+3        & 4.7e+3         & 4.7e+3         & 3.12e+3         \\
tj, tB, tt, ttB                                    & 609             & 431             & 429             & 241              \\
H                                                  & 37.4            & 3.54            & 3.54            & 2.25             \\ \hline
Total Background                                   & 1.04e+4        & 5.14e+3        & 5.13e+3        & 3.36e+3         \\ \hline
\end{tabular}
\end{footnotesize}
\end{center}
\caption{Expected number of selected events for the SM backgrounds to $A\rightarrow Zh\rightarrow \ell \ell \tautau$ for $\int Ldt=$ 3000~\ifb~at $\sqrt{s}=33$ TeV with $<N_{PU}>=$~140.}
\label{tab:AZhtautauSelCutflowBackground33}
\end{table}

\begin{figure}[htbp]
\begin{center}
\includegraphics[width=0.4\columnwidth,height=0.4\textheight,keepaspectratio=true]{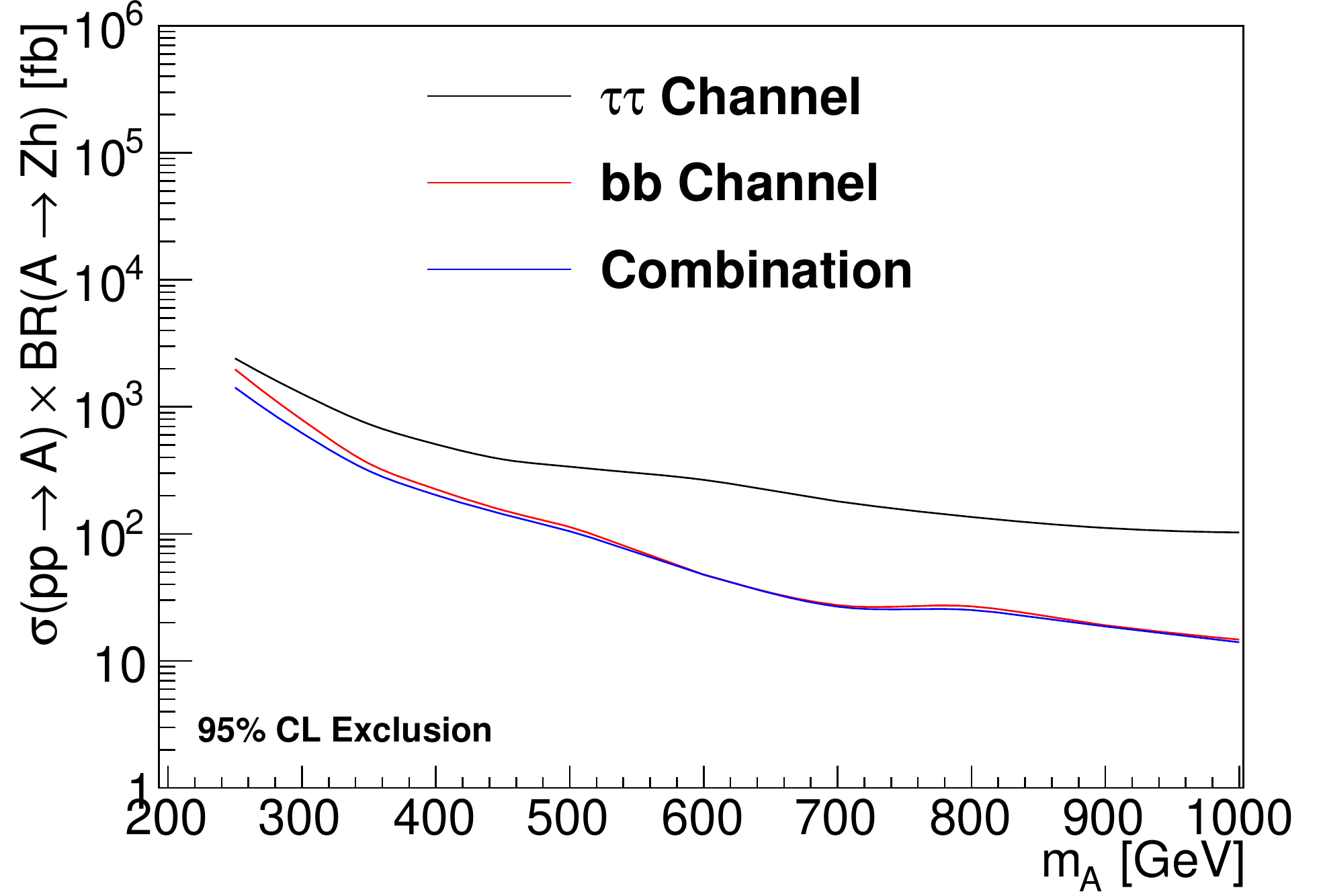}
\includegraphics[width=0.4\columnwidth,height=0.4\textheight,keepaspectratio=true]{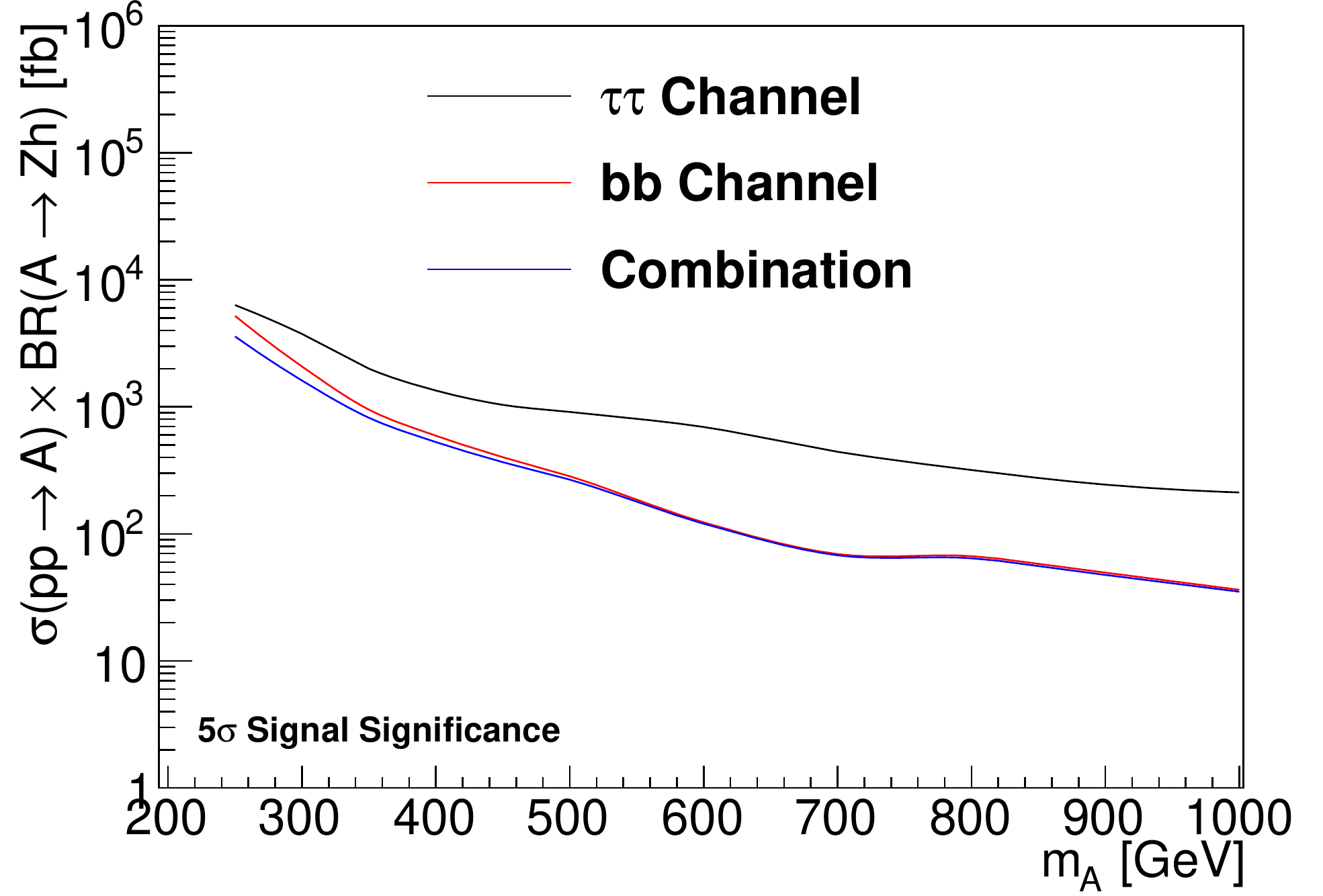}
\includegraphics[width=0.4\columnwidth,height=0.4\textheight,keepaspectratio=true]{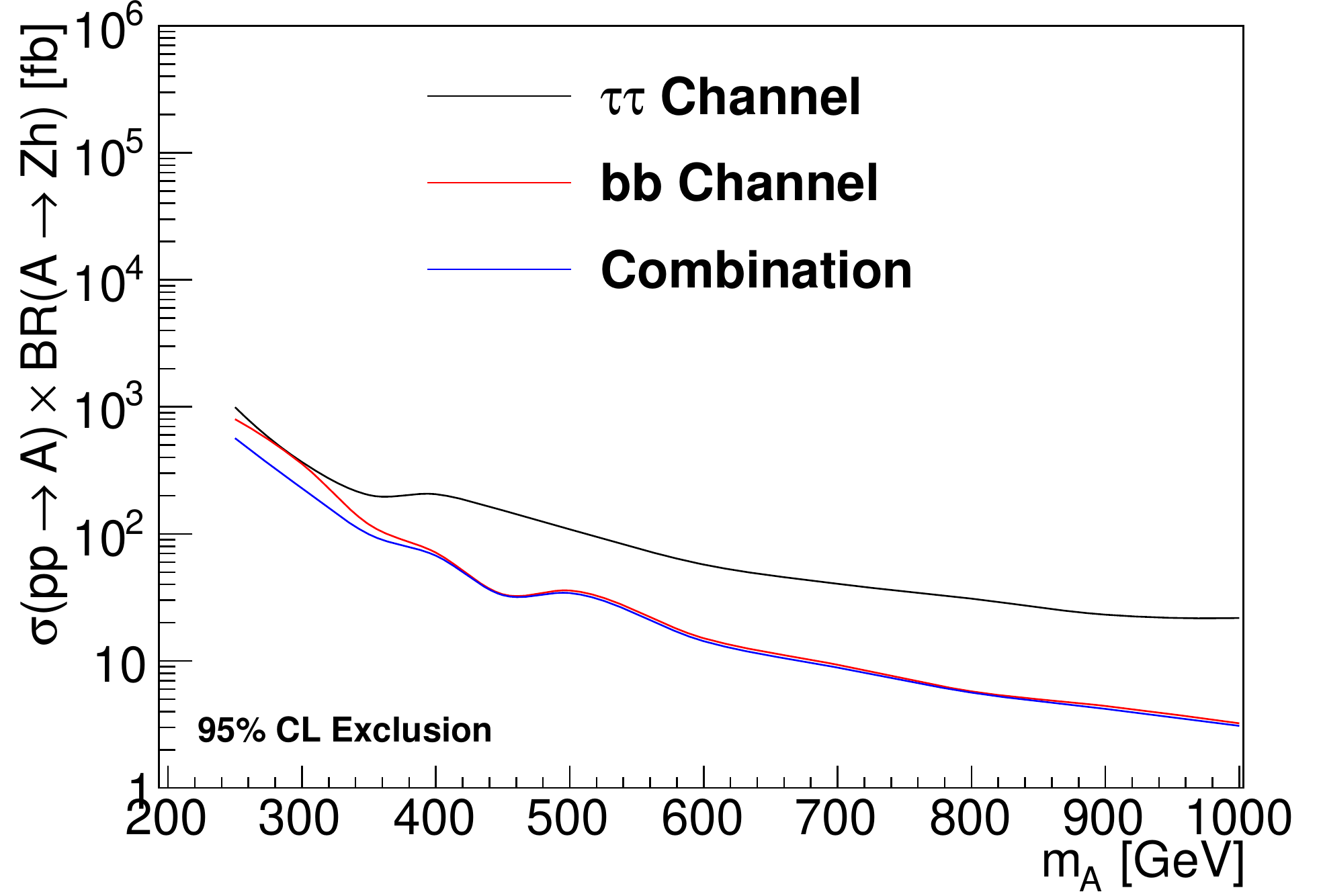}
\includegraphics[width=0.4\columnwidth,height=0.4\textheight,keepaspectratio=true]{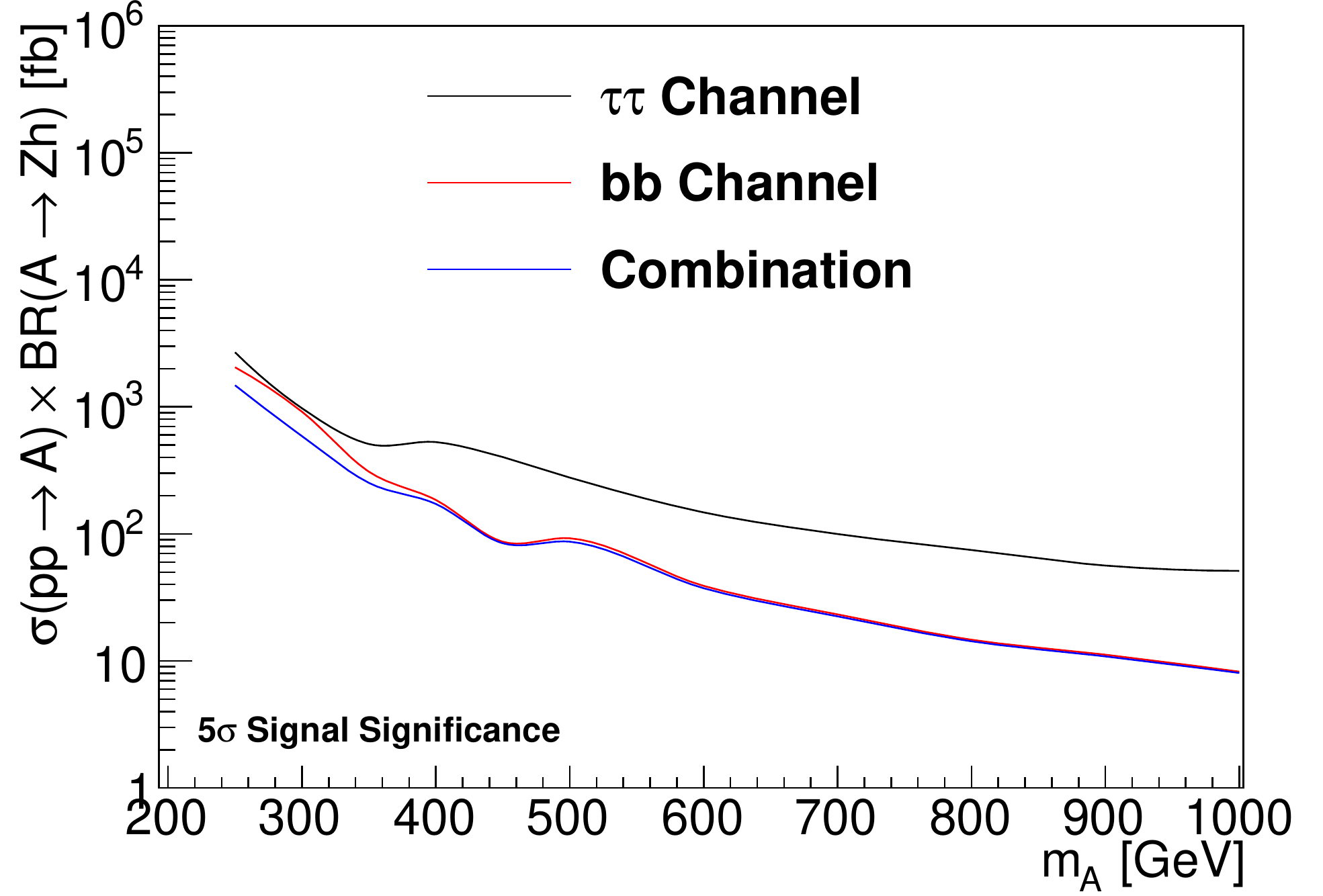}
\includegraphics[width=0.4\columnwidth,height=0.4\textheight,keepaspectratio=true]{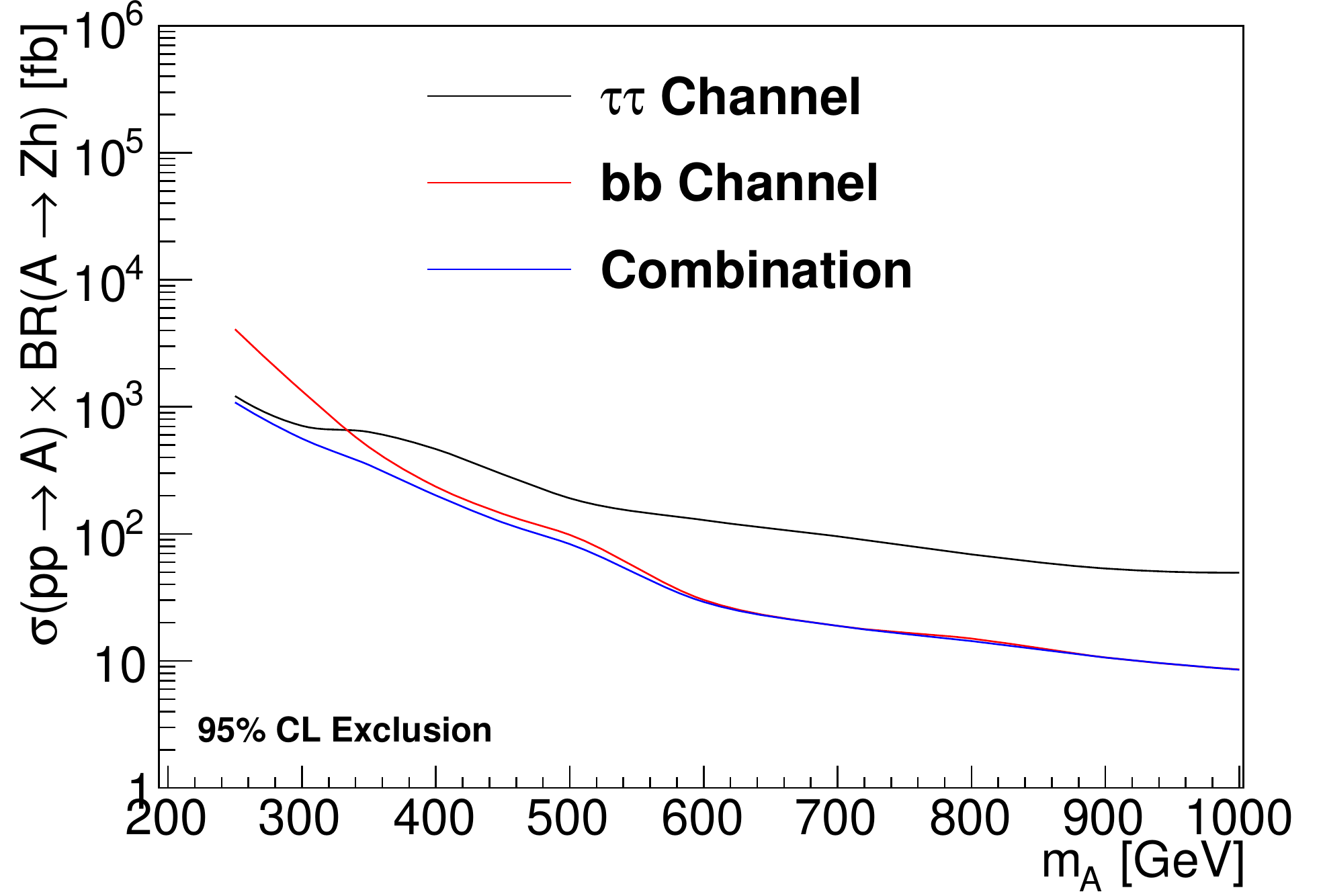}
\includegraphics[width=0.4\columnwidth,height=0.4\textheight,keepaspectratio=true]{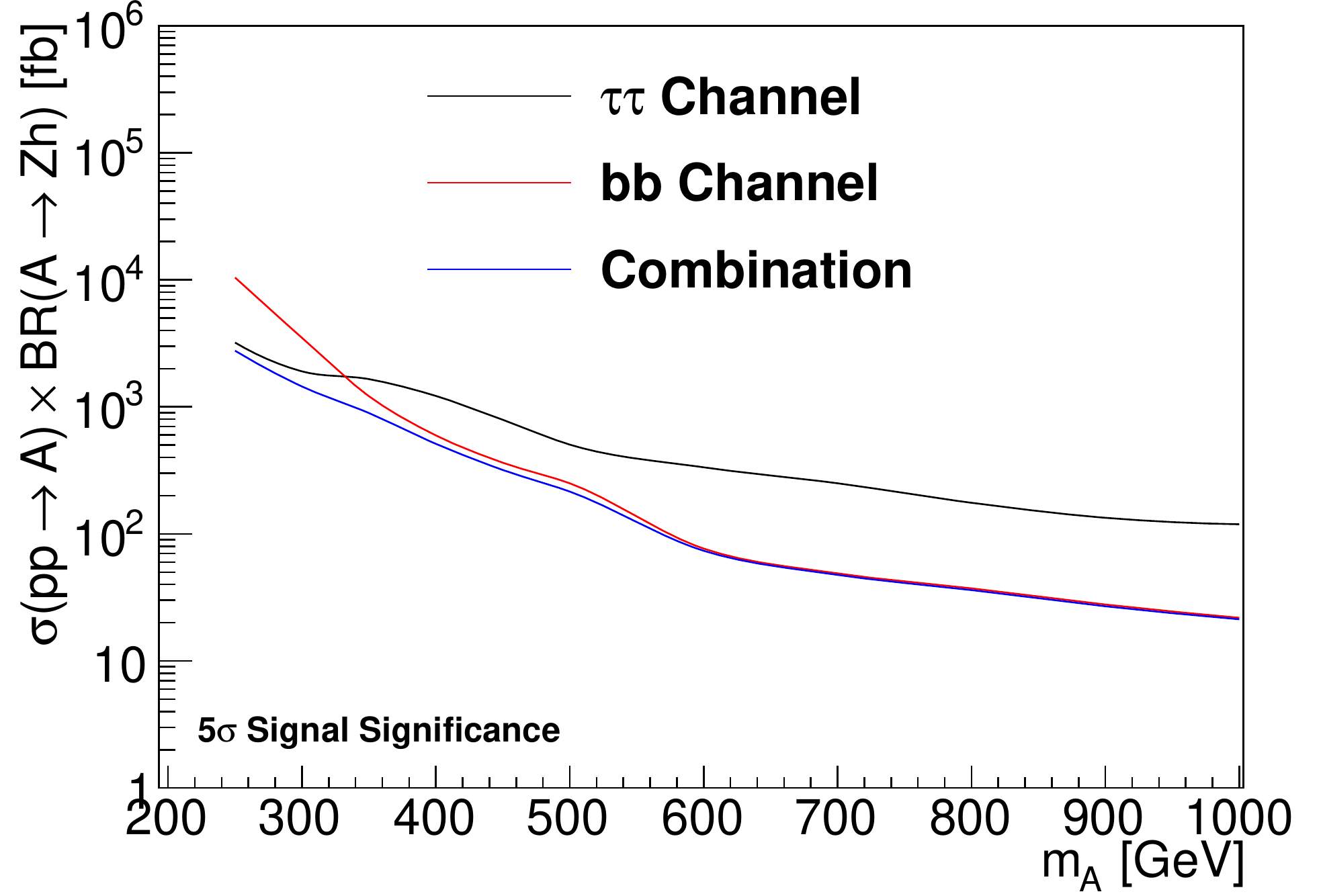}
\caption{The cross section which can be excluded at 95\% CL (left) and the cross section required for a 5$\sigma$ signal significance (right), 
for each $A$ mass hypothesis.  The $bb$ and \tautau~channels are shown separately, as well as the combination.  The top row shows results for $\int Ldt=$ 300~\ifb~at $\sqrt{s}=14$ TeV with $<N_{PU}>=$~50, the middle row for $\int Ldt=$ 3000~\ifb~at $\sqrt{s}=14$ TeV with $<N_{PU}>=$~140, and the bottom row for $\int Ldt=$ 3000~\ifb~at $\sqrt{s}=33$ TeV with $<N_{PU}>=$~140.}
\label{AZhCrossSectionLimitsByChannel}
\end{center}
\end{figure}

\end{document}

\end{document}